\pdfoutput=1

\documentclass{aa}   

\usepackage{graphicx,natbib,url,twoopt}
\usepackage[varg]{txfonts}           
\usepackage{hyperref}   
\usepackage{booktabs}
\usepackage{url}            
\usepackage{pdfcomment}              
\usepackage{acronym} 
\usepackage{epstopdf}                
\usepackage{bm}
\usepackage{longtable}
\usepackage{caption}
\usepackage{mathtools}
\usepackage{bm}
\usepackage{esvect}

\hypersetup{
  colorlinks=true,   
  urlcolor=blue,     
  linkcolor=red,     
}

\bibpunct{(}{)}{;}{a}{}{,}    

\makeatletter
\newcommand{\bibnote}[2]{\global\@namedef{#1note}{#2}}
\newcommand{\biblink}[2]{\global\@namedef{#1link}{#2}}
\newcommand{\LT}{$L_{\text{X}}-T$}
\newcommand{\YT}{$Y_{\text{SZ}}-T$}
\newcommand{\LbcgT}{$L_{\text{BCG}}-T$}
\newcommand{\LY}{$L_{\text{X}}-Y_{\text{SZ}}$}
\newcommand{\LxLbcg}{$L_{\text{X}}-L_{\text{BCG}}$}
\newcommand{\YLbcg}{$Y_{\text{SZ}}-L_{\text{BCG}}$}
\newcommand{\RL}{$R-L_{\text{X}}$}
\newcommand{\RY}{$R-Y_{\text{SZ}}$}
\newcommand{\RLbcg}{$R-L_{\text{BCG}}$}
\newcommand{\RT}{$R-T$}
\newcommand{\Lx}{$L_{\text{X}}$}
\newcommand{\Ysz}{$Y_{\text{SZ}}$}
\newcommand{\Lbcg}{$L_{\text{BCG}}$}
\newcommand{\ubf}{$u_{\text{BF}}$}
\newcommand{\nhtot}{$N_{\text{Htot}}$}
\newcommand{\Rapp}{$R_{\text{app}}$}
\newcommand{\Lbol}{$L_{\text{bol}}$}
\newcommand{\LbolT}{$L_{\text{bol}}-T$}
\newcommand{\LbolY}{$L_{\text{bol}}-Y_{\text{SZ}}$}
\makeatother

\makeatletter
 \newcommandtwoopt{\citeads}[3][][]{%
   \nonstopmode
   \href{http://adsabs.harvard.edu/abs/#3}%
        {\def\hyper@linkstart##1##2{}%
         \let\hyper@linkend\@empty\citealp[#1][#2]{#3}}
   \biblink{#3}{\href{http://adsabs.harvard.edu/abs/#3}{ADS}}%
   \errorstopmode}            
 \newcommandtwoopt{\citepads}[3][][]{%
   \nonstopmode
   \href{http://adsabs.harvard.edu/abs/#3}%
        {\def\hyper@linkstart##1##2{}%
         \let\hyper@linkend\@empty\citep[#1][#2]{#3}}
   \biblink{#3}{\href{http://adsabs.harvard.edu/abs/#3}{ADS}}%
   \errorstopmode}            
 \newcommandtwoopt{\citetads}[3][][]{%
   \nonstopmode
   \href{http://adsabs.harvard.edu/abs/#3}%
        {\def\hyper@linkstart##1##2{}%
         \let\hyper@linkend\@empty\citet[#1][#2]{#3}}
   \biblink{#3}{\href{http://adsabs.harvard.edu/abs/#3}{ADS}}%
   \errorstopmode}            
 \newcommandtwoopt{\citeyearads}[3][][]{%
   \nonstopmode
   \href{http://adsabs.harvard.edu/abs/#3}%
        {\def\hyper@linkstart##1##2{}%
         \let\hyper@linkend\@empty\citeyear[#1][#2]{#3}}
   \biblink{#3}{\href{http://adsabs.harvard.edu/abs/#3}{ADS}}%
   \errorstopmode}            
\makeatother

\newacro{ADS}{Astrophysics Data System}
\newacro{NLTE}{non-local thermodynamic equilibrium}
\newacro{NASA}{National Aeronautics and Space Administration}

\begin{document}


              

\title{Cosmological implications of the anisotropy of ten galaxy cluster scaling relations}

\author{K. Migkas$^1$, F. Pacaud$^1$, G. Schellenberger$^2$, J. Erler$^{1,3}$, N. T. Nguyen-Dang$^4$, T. H. Reiprich$^1$, M. E. Ramos-Ceja$^5$ and L. Lovisari$^{2,6}$}
\institute{$^1$ Argelander-Institut f{\"u}r Astronomie, Universit{\"a}t Bonn, Auf dem H{\"u}gel 71, 53121 Bonn, Germany \\ $^2$ Center for Astrophysics | Harvard \& Smithsonian, 60 Garden Street, Cambridge, MA 02138, USA \\ $^3$ Deutsches  Zentrum  f{\"u}r  Luft-  und  Raumfahrt  e.V.  (DLR)  Projekttr{\"a}ger, Joseph-Beuys-Allee 4, 53113 Bonn, Germany \\ $^4$ Institut f{\"u}r Astronomie und Astrophysik, Sand 1, D-72076 T{\"u}bingen, Germany  \\ $^5$ Max Planck Institute for Extraterrestrial Physics, Gie\ss enbachstra\ss e 1, 85748 Garching bei M{\"u}nchen, Germany \\ $^6$ INAF - Osservatorio Astronomico di Bologna, Via Piero Gobetti, 93/3, 40129 Bologna BO, Italy \\ \email{kmigkas@astro.uni-bonn.de}
}

\date{Received date} 

\abstract{The hypothesis that the late Universe is isotropic and homogeneous is adopted by most cosmological studies, including galaxy cluster ones. The cosmic expansion rate $H_0$ is thought to be spatially constant, while bulk flows are often presumed to be negligible compared to the Hubble expansion, even at local scales. Their effects on the redshift-distance conversion are hence usually ignored. Any deviation from this consensus can strongly bias the results of such studies, and thus the importance of testing these assumptions cannot be understated. Scaling relations of galaxy clusters can be effectively used for that. In previous works, we observed strong anisotropies in cluster scaling relations, whose origins remain ambiguous. By measuring many different cluster properties, several scaling relations with different sensitivities can be built. Nearly independent tests of cosmic isotropy and large bulk flows are then feasible. In this work, we make use of up to $570$ clusters with measured properties at X-ray, microwave, and infrared wavelengths, to construct 10 different cluster scaling relations (five of them presented for the first time to our knowledge), and test the isotropy of the local Universe. Through rigorous and robust tests, we ensure that our analysis is not prone to generally known systematic biases and X-ray absorption issues. By combining all available information, we detect an apparent $9\%$ spatial variation in the local $H_0$ between $(l,b)\sim ({280^{\circ}}^{+35^{\circ}}_{-35^{\circ}},{-15^{\circ}}^{+20^{\circ}}_{-20^{\circ}})$ and the rest of the sky. The observed anisotropy has a nearly dipole form. Using isotropic Monte Carlo simulations, we assess the statistical significance of the anisotropy to be $>5\sigma$. This result could also be attributed to a $\sim 900$ km/s bulk flow which seems to extend out to at least $\sim 500$ Mpc. These two effects are indistinguishable until more high$-z$ clusters are observed by future all-sky surveys, such as eROSITA.}



\titlerunning{Anisotropies of ten galaxy cluster scaling relations}
\authorrunning{K. Migkas et al. }

\maketitle

\section{Introduction} \label{intro}


The isotropy of the late Universe has been a question of great debate during the last decades, and a conclusive answer is yet to be given. As precision cosmology enters a new era with numerous experiments covering the full electromagnetic spectrum, the underlying assumption of isotropy for widely adopted cosmological models has to be scrutinized as well. The importance of new, independent tests of high precision cannot be understated. A possible departure of isotropy in the local Universe could have major implications for nearly all aspects of extragalactic astronomy. 

Galaxy clusters, the largest gravitationally bound objects in the Universe, can be of great service for this purpose. Due to the multiple physical processes taking place within them, different components of clusters can be observed almost throughout the full electromagnetic spectrum \citep[e.g.,][]{allen}. This provides us with several possible cosmological applications for these objects. 

Such a possible application for instance comes from the so-called scaling relations of galaxy clusters \citep[e.g.,][]{kaiser}. These are simply the correlations between the many cluster properties, and can be usually described by simple power-law forms. Some of their measured properties depend on the assumed values of the cosmological parameters (e.g. X-ray luminosity), while others do not (e.g. temperature). Utilizing scaling relations between properties of these two categories can provide us with valuable insights about different aspects of cosmology.  

More specifically, the cosmic isotropy can be investigated using such methods. In \citet{Migkas18} and in \citet{migkas20} (hereafter M18 and M20 respectively) we performed such a test with very intriguing results. We studied the isotropy of the X-ray luminosity-temperature ($L_{\text{X}}-T$) relation, which we used as a potential tracer for the isotropy of the expansion of the local Universe. In M20 we combined the extremely expanded HIghest X-ray FLUx Galaxy Cluster Sample \citep[eeHIFLUGCS,][Pacaud et al. in prep.]{eeHIF} with other independent samples, and detected a $\sim 4.5\sigma$ anisotropy toward the Galactic coordinates $(l,b)\sim (305^{\circ},-20^{\circ})$. The fact that several other studies using different cosmological probes and independent methods find anisotropies toward similar sky patches makes these findings even more interesting.

Multiple possible systematics were tested separately as potential explanations for the apparent anisotropies, but the tension could not be sufficiently explained by any such test. Therefore, the anisotropy of the $L_{\text{X}}-T$ relation seems to be attributed to an underlying, physical reason. There are three predominant phenomena that could create this: unaccounted X-ray absorption, bulk flows, and Hubble expansion anisotropies. Firstly, the existence of yet undiscovered excess X-ray absorption effects could bias our estimates. The performed tests in M20 however showed that this is quite unlikely to explain the apparent anisotropies. Further investigation is needed nonetheless to acquire a better understanding of these possibilities.

Secondly, coherent motions of galaxies and galaxy clusters over large scales, the so-called bulk flows (BFs), could also be the cause of the observed cluster anisotropies. The objects within a BF have a peculiar velocity component toward a similar direction, due to the gravitational attraction of a larger mass concentration such as a supercluster. These nonrandom peculiar velocities are imprinted in the observed redshifts. If not taken into account, they can result in a biased estimation of the clusters' redshift-based distances, and eventually their other properties (e.g., $L_{\text{X}}$). The necessary BF amplitude and scale to wash away the observed cluster anisotropies by far surpasses $\Lambda$CDM expectations, which predicts that such motions should not be present at comoving scales of $\gtrsim 200$ Mpc (see references below). If such a motion is confirmed, a major revision of the large scale structure formation models might be needed. BFs have been extensively studied in the past with various methods \citep[e.g.,][]{lauer,hudson04,kashl08,kashl10,colin2011,osborne,feindt,appleby-john,carrick,hoffman,scrimgeour,watkins2,peery,qin}. However, most of them used galaxy samples which suffer from the limited scale out to which they can be constructed. No past study has used cluster scaling relations to investigate possible BF signals to our knowledge. A more in-depth analysis to determine if such effects are the origin of the observed cluster anisotropies is thus necessary. 

The vast majority of cluster studies ignores the effects of BFs in the observed redshifts assuming the peculiar velocities to be randomly distributed. The frequent use of heliocentric redshifts in local cluster studies (instead of CMB-frame redshifts) might also amplify the introduced bias from BFs. Hence, such discovered motions could strongly distort the results for most cluster studies, and their cosmological applications. 

The third possible explanation for our results is an anisotropy in the Hubble expansion, and the redshift-distance conversion. To explain the observed anisotropies obtained in M20 solely by a spatial variation of $H_0$, one would need a $\sim 10\%$ variation between $(l,b)\sim (305^{\circ},-20^{\circ})$ and the rest of the sky. This scenario would contradict the assumption of the cosmological principle and the isotropy of the Universe, which has a prominent role in the standard cosmological model. This result was derived in M20 based on relatively low$-z$ data, with the median redshift of all the 842 used clusters being $z\sim 0.17$. Hence, for now it is not possible to distinguish between a possibly primordial anisotropy or a relatively local one purely from galaxy clusters. New physics that interfere with the directionality of the expansion rate would be needed in that case, assuming of course that no other underlying issues cause the cluster anisotropies. Many recent studies have tackled this question, with contradicting results \citep[e.g.,][see Sect. \ref{discussion} for an extended discussion]{bolejko,bengaly18,colin19,soltis,andrade19,hu,fosalba,salehi,secrest}.

The interplay of the different possible phenomena and the effects they may have on cluster measurements can make it hard to identify the exact origin of the anisotropies. Additionally, maybe a combination of more than one phenomena affects the cluster measurements, since, for instance, the existence of a dark gas cloud does not exclude the simultaneous existence of a large BF. In order to provide a conclusive answer to the problem, other tests with the same cluster samples must be utilized. These new tests should have somewhat different sensitivities than the \LT\ relation. This way, we can cross check if the anisotropies also appear in tests sensitive to X-ray effects only, BFs or cosmological anisotropies only, etc.

In this work, alongside \Lx\ and $T$, we also measure and use the total integrated Compton parameter \Ysz, the half-light radii of the clusters $R$ and the infrared luminosity $L_{\text{BCG}}$ of the brightest galaxy of each cluster (BCGs). This allows us to study 10 different scaling relations between these properties and test their directional dependance. Based on these results, we try to fully investigate the nature of the observed anisotropies and provide conclusive results. Throughout this paper we use a $\Lambda$CDM cosmology with $H_0=70\  \text{km}\  \text{s}^{-1} \text{Mpc}^{-1}$, $\Omega_{\text{m}}=0.3$ and $\Omega_{\Lambda}=0.7$ in order to constrain the scaling relation parameters, unless stated otherwise.

The paper is organized as follows: in Sect. \ref{sample} we describe the measurements of the used samples. In Sect. \ref{sc_rel_descr} we describe the modelling of the scaling relations and the statistical methods and procedures used to constrain their directional behavior. In Sect. \ref{mainresults}, the full sky best-fit parameters for the 10 scaling relations are presented. In Sect. \ref{xray_abs_anis}, we study the anisotropic behavior of scaling relations which are sensitive only to X-ray absorption effects, and not cosmological factors. In Sect. \ref{bf_cosmo_anisot}, the focal point of this work is presented, namely the anisotropy of scaling relations which are sensitive to cosmological anisotropies and BFs. In Sects. \ref{systematics} and \ref{mc_sim}, the possible systematic biases are discussed, and the comparison of our results to the ones from isotropic Monte Carlo simulations is presented. Finally, in Sects. \ref{discussion} and \ref{conclusions} the discussion and conclusions of this work are given.

\section{Sample and measurements}\label{sample}

As a general basis, we use the eeHIFLUGCS sample. The only exception is the use of the full Meta-Catalog of X-ray detected Clusters of galaxies \citep[MCXC,][]{mcxc} for the \LY\ relation. This is done in order to maximize the available clusters (since both \Lx\ and \Ysz\ are available beyond eeHIFLUGCS). From eeHIFLUGCS, a slightly different cluster subsample is used for different scaling relations (the vast majority of the subsamples naturally overlap), where all available measurements for both quantities that enter the scaling relation are considered. For the four scaling relations that include $T$, we use the same sample as in M20. In a nutshell, the sample includes 313 galaxy clusters, the vast majority of which are part of the eeHIFLUGCS sample. It is a relatively low redshift sample with median $z=0.075$ ($z\in [0.004,0.45]$), which includes objects with an X-ray flux of $f_{\text{X}, 0.1-2.4\ \text{keV}}\geq 5\times 10^{-12}$ erg/s/cm$^2$\footnote{Measured by ROSAT.}. The Galactic plane region ($b\leq |20^{\circ}|$) and the regions around the Virgo cluster and the Magellanic clouds were masked, and no clusters are considered from there. The spatial distribution of the sample in the rest of the sky is quite homogeneous. The X-ray luminosity \Lx, the temperature $T$, the redshift $z$, the $R_{500}$, the metallicities of the core $Z_{\text{core}}$ and of the $0.2-0.5\ R_{500}$ annulus $Z_{\text{out}}$ are obtained as described in M20. The only change we make in the sample is the spectral fit of NGC 5846. When $Z_{\text{out}}$ is left free to vary it results to unrealistically large values, affecting also the measured $T$ and making NGC 5846 a strong outlier in the \LT\ plane. Thus we repeat the spectral fitting with a fixed value of $Z_{\text{out}}=0.400\ Z_{\odot}$ (sample's median), which returns $T=0.927\pm 0.013$ keV. 

For the rest of the scaling relations, the number of clusters used depends on the availability of each measurement, which is described below and summarized in Table \ref{best-fit}. Finally, we also use the ASCA Cluster Catalog \citep[ACC,][]{horner}, after excluding all the common clusters with eeHIFLUGCS and after further cleaning its X-ray luminosities, and described in M18 and M20.

\subsection{Total integrated Compton parameter $Y_{5R500}$}\label{y500-extr}

Galaxy clusters can be observed in the submillimeter regime through the thermal Sunyaev-Zeldovich effect (tSZ). The tSZ is a spectral distortion of the CMB toward the sky positions of galaxy clusters caused by inverse Compton scattering of CMB photons by the hot electrons of the intracluster plasma. This causes a decrement (increment) in the observed temperature of the CMB at frequencies below (above) $\sim 217$ GHz. The amplitude of this spectral distortion is proportional to the Comptonization parameter $y$ defined as 
\begin{equation}
y(r)=\dfrac{\sigma _{\text{T}}}{m_{\text{e}} c^2}\int_{\text{l.o.s.}} k_{\text{B}}\ n_{\text{e}}(r)\ T_{\text{e}}(r)\ \mathrm{d} l,
\label{y-param}
\end{equation}
where $\sigma _{\text{T}}$ is the Thompson cross section, $m_{\text{e}}$ is the electron rest mass, $c$ is the speed of light, $k_{\text{B}}$ is the Boltzmann constant, $n_{\text{e}}$ the electron number density, $T_{\text{e}}$ the electron temperature and l.o.s. denotes the line of sight toward the cluster. The product of $k_{\text{B}}n_{\text{e}}\text{(r)}T_{\text{e}}\text{(r)}$ represents the electron pressure profile $P_{\text{e}}(r)$. We adopted the form of the latter from the widely used Generalized Navarro-Frenk-White (GNFW) pressure profile \citep{gnfw} with the "universal" parameter values obtained by \citet{arnaud2}. The required values of $R_{500}$ and $z$ were taken from M20 (when available) or MCXC.

A common technique for the extraction of the tSZ signal from multifrequency datasets are matched multifilters (MMFs). MMFs allow us to construct a series of optimal spatial filters that are build from the SED of the tSZ together with a spatial template computed from the expected pressure profile of clusters. MMFs have been widely used in blind cluster searches by the ACT, SPT, and \textit{Planck} collaborations \citep[][]{act,Hilton20,spt,Bleem20,Planck_PSZE,ade-c} and allow for the estimation of the integrated Comptonization parameter $Y_{5R500}$, where $Y_{5R500}=\int y\ \mathrm{d} \Omega$, with $\Omega$ being the chosen solid angle. 

We estimate $Y_{5R500}$ from the latest version of the 100, 143, 217, 353, 545 and $857 \, \mathrm{GHz}$ all-sky maps delivered by the \textit{Planck} High Frequency Instrument \citep[HFI; ][]{Planck_HFI}, which were published during the final data release in 2018 \citep[R3.00;][]{Planck_overview}. The $Y_{5R500}$ values are extracted from \textit{Planck} data with MMFs that are build by using the relativistic tSZ spectrum \citep[e.g.,][]{Wright79, Itoh98, Chluba12}, and have zero response to the kinematic SZ (kSZ) spectrum. Values of $Y_{5R500}$ obtained using standard MMFs derived from the nonrelativistic tSZ spectrum without additional kSZ removal were also tested, with completely negligible changes in the results. The uncertainties of $R_{500}$ (which are not provided by MCXC) are ignored, since their effect is negligible\footnote{The $R_{500}$ values are calculated through the X-ray luminosity-mass relation of \citet{arnaud2}. The only meaningful contribution to the $R_{500}$ uncertainties comes from the scatter of the scaling relation. However, due to the shallow dependence of $R_{500}$ to $L_{\text{X}}$ ($R_{500}\sim L_{\text{X}}^{0.2}$), a fluctuation of $L_{\text{X}}$ within the scatter would only cause a minor shift in $R_{500}$ (see M20). Additionally, as shown in Fig. \ref{Ysz-change}, even shifts of 6\% in $R_{500}$ would not cause a major change in $Y_{5R500}$, certaintly less than the statistical uncertainties. Thus, it is reasonable to ignore any uncertainties of $R_{500}$}. Further details of the processes used to extract the $Y_{5R500}$ values are presented in detail in \citet[][E19 hereafter]{erler} and in Appendix \ref{details_ysz}.

We measure $Y_{5R500}$ for all the 1743 entries in MCXC. We derive a signal-to-noise (S/N) of $>1$ for 1472 clusters , S/N$>2$ for 1094 clusters, S/N$>3$ for 746 clusters, and S/N$>4.5$ for 460 clusters. The latter matches the threshold set by the 2nd \textit{Planck} Sunyaev-Zeldovich Source Catalog \citep[PSZ2, ][hereafter A16]{ade-c}. We compare our derived $Y_{5R500}$ values with the ones from PSZ2 in Sect. \ref{psz2-comp}. For our work, we only kept clusters with S/N$>2$ to avoid using most measurements dominated by random noise, but without losing too many clusters\footnote{This S/N threshold is a safe choice since we already know galaxy clusters exist at these sky positions, through their X-ray detection.}. We always check if increasing the S/N threshold alters our results.

Finally, the cluster quantity that enters the cluster scaling relations is the total integrated Comptonization parameter \Ysz, which is given by
\begin{equation}
Y_{\text{SZ}}\left[\text{kpc}^2\right]=Y_{5R500}\left[\text{arcmin}^2\right]\times \left(\frac{\pi}{60\times 180}\right)^2 \times D_A^2,
\label{Ysz}
\end{equation}
where $D_A$ is the angular diameter distance of the cluster in kpc. The dependance of \Ysz\ on the cosmological parameters therefore enters through $D_A$.

\subsection{Half-light radius $R$}

The apparent angular size  \Rapp\ in the sky within which half of a cluster's total X-ray emission is encompassed, is a direct observable. By using a cosmological model, this observable (measured in arcmin) can be converted to a physical size of a cluster, the half-light radius $R$, where $R=R_{\text{app}}\times \dfrac{\pi }{60\times 180}\times D_A$. The size of a cluster correlates with many other cluster properties and can be used to construct scaling relations. We measured \Rapp\ for all eeHIFLUGCS clusters, and additionally for all MCXC clusters with $f_{\text{X}, 0.1-2.4\ \text{keV}}\geq 4.5\times 10^{-12}$ erg/s/cm$^2$. This led to 438 measurements. 

To measure \Rapp\ the ROSAT All-Sky Survey (RASS) maps were used. The count-rate growth curves were extracted for all eeHIFLUGCS clusters, and additionally for all MCXC clusters with $f_{\text{X}, 0.1-2.4\ \text{keV}}\geq 4.5\times 10^{-12}$ erg/s/cm$^2$. By a combination of applied iterative algorithms and visual validation by six different astronomers, the plateau of the count-rate growth curves (i.e., the boundaries of cluster X-ray emission) were determined. The radii \Rapp\ were then found, and converted to $R$ using the default cosmology. The exact details of this process will be presented in Pacaud et al. (in prep.). 

The point spread function (PSF) of the XRT/PSPC imager of ROSAT varies with the off-axis angle and the photon energies. In general, it is $\leq 1.5'$. To avoid strong biases due to PSF smearing, we excluded all the clusters with $R_{\text{app}}\leq 2'$. This left us with 418 cluster measurements with a median $R_{\text{app}}=3.76'$. Residual PSF smearing effects are still expected to affect the measurements. Since the scaling relations including $R$ are anyway currently inconclusive (see Sect. \ref{R-relations}), we neglect these effects for now. In future work, any PSF effects will be fully taken into account. eROSITA will also be able to provide $R_{\text{app}}$ values rather insensitive to PSF effects, due to its better spatial resolution.

\subsection{Near infrared BCG luminosity $L_{\text{BCG}}$}

The BCGs were found for the 387 clusters of the eeHIFLUGCS sample. To determine the BCG for all clusters, we used the optical/near infrared (NIR) data from the SDSS \citep{york2000sloan}, Pan-STARRS \citep{kaiser2002pan,kaiser2010pan}, VST ATLAS \citep{shanks2015vlt}, DES \citep{abbott2018dark}, 2MASS \citep{skrutskie2006two} and WISE \citep{wright2010wide} catalogs. The redshifts of the galaxies were either taken from the SDSS catalogue or the NASA/IPAC Extragalactic Database (NED)\footnote{\url{https://ned.ipac.caltech.edu/}}. All the galaxy magnitudes were corrected for Galactic extinction and the proper k-correction was applied. The exact details on the BCG selection are described in Appendix \ref{bcg_select}.

For this work, we use the magnitudes coming from the 2MASS catalog for two reasons. First and foremost, 2MASS returns the largest number of available BCGs for our sample. Out of the 387 clusters of eeHIFLUGCS, we detected the BCG in 2MASS for 331 of them. Secondly, the infrared BCG luminosities are not strongly sensitive to extinction effects, minimizing the potential risk of unaccounted absorption biases. We exclude all BCGs with $z<0.03$, as they appear to be systematically overluminous based on all \Lbcg\ scaling relations, indicating a flattening of the \Lbcg\ relations at low cluster masses. This flattening was also observed by \citet{bharad14}. 

Additionally, the redshift evolution of the BCG luminosity versus the other quantities is unknown. Due to the large scatter of the \Lbcg\ scaling relation as shown later in the paper, the applied evolution cannot be left free to vary simultaneously with the other scaling relation parameters, since no reliable constraints are obtained. Subsequently, we opt to also exclude all clusters with $z\geq 0.15$. This allows us to ignore any evolution during the model fitting. Even if this added a small bias in our estimates, it would not be expected to affect the anisotropy analysis, since the cluster redshift distributions of different sky regions are similar (see M20). As such, any potential bias would cancel out when comparing cluster subsamples from different sky regions. These criteria eventually leave us with 244 clusters with 2MASS \Lbcg\ measurements with a median redshift of 0.069.

\subsection{X-ray determined $N_{\text{H,Xray}}$}\label{xray_nh}

We determined the total hydrogen column density $N_{\text{H,Xray}}$ using the X-ray spectra of the 313 clusters from M20. The exact details of our methodology can be found in Appendix \ref{details_xray_nh}. Overall, we were able to obtain a safe estimation of $N_{\text{H,Xray}}$ for 156 clusters. Several systematics might creep in during the whole process, hence we approach the analysis done with these measurements conservatively.

\subsection{X-ray luminosity \Lx\ and redshift $z$ for clusters not included in M20}\label{Lx_from_mcxc}

For the 1430 MCXC clusters not included in the M20 sample, we started from their \Lx\ values given in MCXC which are corrected for the absorption traced by the neutral hydrogen column density only. We further corrected them in order to account for the total hydrogen absorption based on the \citet[][hereafter W13]{willingale} $N_{\text{Htot}}$ values. The procedure is exactly the same as the one followed in M20. The only difference here is the use of fixed $T=5$ keV and $Z=0.4\ Z_{\odot}$ values for all clusters in the \textsc{XSPEC} \citep{xspec} apec$\cdot$phabs model, since we do not have spectral measurements for these clusters. The redshift values were adopted from MCXC.

\subsection{ACC sample}

We measured \Ysz\ for the ACC sample \citep{horner} as well, using the same procedure as for our sample. We excluded all the clusters already included in any of our different subsamples. We also excluded the 55 clusters with \Ysz\ S/N$<2$. This results in 168 clusters with X-ray luminosity and temperature values, with 113 of them having a \Ysz\ measurement with S/N$>2$. Cross-checking the results of a completely independent cluster sample with our sample's results is crucial in order to understand the origin of the observed anisotropies (e.g., to exclude that sample selection effects may bias our results). The properties of the ACC sample as we use it are already given in detail by M18 and M20. In a nutshell, it mainly consists of massive galaxy clusters, spanning across $z\in [0.009, 0.839]$, with a median redshift of 0.226 (for the 113 clusters with \Ysz\ measurements). The temperatures are obtained by a single-thermal model for the whole cluster, while the X-ray luminosities $L_{\text{bol}}$ are given in the bolometric band, within the $R_{200}$ of the cluster (measured within 0.5-2 keV and within a "significance" radius, and then extrapolated). All measurements are performed by \citet{horner} and the ASCA telescope, while we corrected the $L_{\text{bol}}$ for the total absorption similarly to our sample's \Lx\ corrections. The necessary $R_{500}$ values to measure \Ysz\ were obtained using the mass-temperature scaling relation of \citet{reichert}. The latter mostly uses XMM-Newton-derived temperatures, which might differ from the ASCA temperatures in general. Thus, we compared the ACC temperatures with our temperatures for the common clusters between the two samples, and applied the necessary calibration factors before calculating $R_{500}$. 

\section{Scaling relations}\label{sc_rel_descr}

We study 10 cluster scaling relations in total, namely the \LY,  the \LT, the \YT, the \LxLbcg,  the \LbcgT, the \RY, the \RL, the \RLbcg, the \RT, and the \YLbcg\ relations. The first nine relations are sensitive to either additionally needed X-ray absorption corrections (1), BFs (2), or possible cosmological anisotropies (3). The \YLbcg\ cannot trace any of the above effects (4), as explained in Sect. \ref{ylbcg}. An observed anisotropy then could point to systematics in the measurements or methodology, which may affect the other nine scaling relations of interest.

In principle, it is quite challenging to distinguish cases (2) and (3) since their effects on the observed scaling relations are similar. One way to distinguish between the two is to perform a tomography analysis, analyzing redshift shells individually and see if the anisotropies persist at all scales. Evidently, the power with which each scaling relation traces a possible effect varies, since it depends on the relation's scatter and the exact way each effect might intervene with each measurement. The information of what effect each scaling relation can detect is given in Table \ref{effects}. The exact explanation on how a scaling relation detects (or not) an anisotropy origin is given in Sects. \ref{xray_abs_anis} and \ref{bf_cosmo_anisot}. 

\begin{table}[hbtp]
\caption{\small{Possible anisotropy causes that can be traced by each cluster scaling relation. 1: Unaccounted X-ray absorption effects. 2: Bulk flows. 3: Cosmological anisotropies. 4: Expected to look isotropic if no systematics exist in the measurements or methodology. The star $^*$ means that the detection of an underlying effect is rather weak, and cannot be achieved by current samples.}}
\label{effects}
\begin{center}
\renewcommand{\arraystretch}{1.1}
\small
\begin{tabular}{ c  c  c  c  c c }
\hline \hline

Measurement & \Lx & \Ysz & $R$ & \Lbcg & $T$ \\
\hline \hline

\Lx & & 1 & 2$^*$,3$^*$ & 1 & 1,2,3 \\ 
\Ysz &  &  & 2$^*$,3$^*$ & 4 & 1$^*$,2,3 \\ 
$R$ & & &  & 2$^*$,3$^*$ & 1$^*$,2,3 \\ 
\Lbcg &  &  &  &  & 1$^*$,2,3 \\ 

\hline

\end{tabular}
\end{center}
\end{table}

The form of the studied scaling relations between some measured cluster quantities $Y$ and $X$ is
\begin{equation}
\begin{aligned}
\dfrac{Y}{C_{Y}} E(z)^{\gamma _{YX}}=A_{YX}\times \left(\dfrac{X}{C_{X}}\right)^{B_{YX}},
\label{scal_rel}
\end{aligned}
\end{equation}
where $C_{Y}$ is the calibration term for the $Y$ quantity, the term $E(z)=[\Omega_\text{m}(1+z)^3+\Omega_{\Lambda}]^{1/2}$ accounts for the redshift evolution of the $Y-X$ relation, $\gamma _{YX}$ is the power index of this term, $A_{YX}$ is the normalization of the relation, $C_{X}$ is the calibration term for the $X$ quantity, and $B_{YX}$ is the slope of the relation. The calibration terms $C_{Y}$ and $C_{X}$ are taken to be close to the median values of $Y$ and $X$ respectively. They are shown in Table \ref{best-fit}, together with the assumed (self-similar) values of $\gamma_{YX}$. 

\subsection{Linear regression} \label{stat}

Similar to M20, in order to constrain the scaling relation parameters we perform a linear regression in the logarithmic space using a $\chi^2$ minimization procedure. We consider two separate cases. 

The first case is when we assume to know the universal, isotropic cosmological parameters (or when they do not matter due to canceling out between the two parts of the scaling relation) and wish to constrain the normalization, slope and scatter of a scaling relation. We then constrain the desired parameters by minimizing the expression
\begin{equation}
\chi^2_Y=\sum\limits_{i=1}^N\frac{\left(\log{Y'_i}-\log{A_{YX}}-B_{YX}\times \log{X'_i}\right)^2}{{\sigma _{\log{Y},i}}^2+{B_{YX}^2\times \sigma _{\log{X},i}}^2+{\sigma_{\text{int},YX}}^2},
\label{eq3}
\end{equation}
where $N$ is the number of clusters, $Y'=\dfrac{Y(z,H_0)}{C_{Y}}E(z)^{\gamma}$, $X'=\dfrac{X(z,H_0)}{C_{X}}$\footnote{$X$ does not depend on $z$ and $H_0$ when $X=$temperature $T$.}, $\sigma _{\log{Y}}$ and $\sigma _{\log{X}}$ are the Gaussian logarithmic uncertainties for $Y$ and $X$ respectively (derived as in M20), and $\sigma_{\text{int},YX}$ is the intrinsic scatter of the $Y-X$ relation with respect to $Y$, in orders of magnitude (dex). Following \citet{Maughan07}, $\sigma_{\text{int},YX}$ is iteratively increased and added in quadrature as an extra uncertainty term to every data point until the reduced $\chi^2_{\text{red}} \sim 1$\footnote{Our analysis and conclusions are rather insensitive to a (small) systematic over- or underestimation of $\sigma_{\text{int},YX}$. This was tested by repeating our anisotropy analysis using $20\%$ smaller or larger $\sigma_{\text{int},YX}$. The exact values of $\sigma_{\text{int},YX}$ are relevant only in Sect. \ref{MB-EB} (Malmquist bias), but a systematic small bias on $\sigma_{\text{int},YX}$ would again minimize the effects on this test as well.}. $\sigma_{\text{tot},YX}$ is the total scatter, equal to the average value of the denominator of Eq. \ref{eq3} for all considered clusters. Finally, we always choose $Y$ to be the quantity with the largest measurement uncertainties.

The second case applies only when there is a strong cosmological dependency on the best-fit scaling relations (i.e., all the $T$ scaling relations). Here we assume the normalization of the scaling relation to be known and direction-independent. This is a reasonable assumption since this quantity is associated with intrinsic properties of the clusters, and as such there is no obvious reason why it should spatially vary. In this case, the free parameters we wish to constrain are the Hubble constant $H_0$ or the BF amplitude and direction \ubf, while the slope is treated as a free, nuisance parameter. To do so, we minimize the following equation:
\begin{equation}
\chi^2_{D}=\sum\limits_{i=1}^N\frac{\left[D_{i,\text{obs}}(Y,X,A_{YX},B_{YX},\gamma)-D_{i,\text{th}}(H_0,z,{u_{\text{BF}}})\right]^2}{\sigma_{D_{i,\text{obs}}}^2+{\sigma_{\text{int},D}^2}},
\label{Dist_eq}
\end{equation}
where $D$ is either the luminosity distance $D_L$ (e.g., for \LT) or the angular diameter distance $D_A$ (e.g., for \YT), $D_{\text{obs}}$ is the observed distance given the measurements $Y$ and $X$ and their exact scaling relation $Y-X$, $D_{\text{th}}$ is the theoretically expected distance based on the cosmological parameters (e.g., $H_0$), the redshift $z$ and the existing BF \ubf, $\sigma_{D_{i,\text{obs}}}$ is the statistical uncertainty of the observed distance (which is a function of the measurement uncertainties of $Y$ and $X$), and $\sigma_{\text{int},D}$ is the intrinsic scatter of the relation in Mpc units. The observed distance $D_{\text{obs}}$ enters every scaling relation differently. Generally, it is given by
\begin{equation}
D_{\text{obs}}=\left[\dfrac{A_{YX}{X'}^{B_{YX}}}{Y'}\right]^{k}\times D_{H_0=70}.
\end{equation}
Here $D_{H_0=70}$ is the distance found for $H_0=70$ km/s/Mpc, which enters in the calculation of $Y$. Its use here cancels out this cosmological contribution to $Y$. For this, we need $k=\left(1/2,1/2,1,-\dfrac{1}{2B_{TL_{\text{BCG}}}}\right)$ for \LT, \YT, \RT, and \LbcgT\ respectively. This leaves us with X-ray flux (in the cluster's rest frame), apparent size in arcmin, $Y_{5R500}$, and BCG flux respectively. Moreover, $D_{\text{th}}$ is given by
\begin{equation}
D_{\text{th}}=\dfrac{(1+z)^{\pm 1}}{H_0} \int_0^{z}\frac{\mathrm{d}z'}{E(z')},
\label{dist_eq_new}
\end{equation}
 where the $(1+z)^{\pm 1}$ factor depends on if we are considering $D_L$ or $D_A$. Also, $z$ is the cosmological redshift which is given by $z=z_{\text{obs}}+\dfrac{u_{\text{BF}}\ (1+z)}{c}\cos{\phi}$ \citep[e.g.,][]{harrison,dai,feindt,springbob}, where $z_{\text{obs}}$ is the observed redshift (converted to the CMB rest frame), $u_{\text{BF}}$ is the amplitude of the BF, and $\phi$ is the angular distance between a cluster and the BF direction. Here we should note that the sign of $u_{\text{BF}}$ is changed from "$-$" to "$+$" to avoid confusion with negative velocities.

We should stress that when searching for cosmological anisotropies, the calculated $H_0$ variations express \emph{relative} differences between regions. The absolute $H_0$ values cannot be constrained by cluster scaling relations only since a fiducial $H_0$ value was assumed to calibrate the relation initially. Hence, the $H_0$ anisotropy range always extends around the initial $H_0$ choice. Apparent $H_0$ fluctuations could also mirror other underlying cosmological effects we are not yet aware of. 

\subsection{Bulk flow detection}\label{BF_detect}

We follow two different methods to estimate the best-fit BF. Firstly, we fit the scaling relations to all the clusters, adding a BF component in our fit, as in Eq. \ref{Dist_eq} and \ref{dist_eq_new}. The free parameters are the direction and amplitude of the BF, and the intrinsic scatter. $A_{YX}$ and $B_{YX}$ are left free to vary within their $1\sigma$ uncertainties of the considered (sub)sample, when no BF is applied. Every cluster's redshift (and thus $D_{\text{th}}$) is affected by the BF depending on the angle between the direction of the cluster and the BF direction. Since the spatial distribution of clusters is nearly uniform, any applied BF does not significantly affect the best-fit $A_{YX}$ and $B_{YX}$, but it does affect the scatter. This procedure essentially minimizes the average $Y$ residuals around each scaling relation (i.e., we get $\chi^2_{\text{red}}\sim 1$ for a smaller $\sigma_{\text{int},YX}$), and is thus labeled as "Minimum Residuals" (MR) method.  

The second method we follow to detect the best-fit BF is that of the "Minimum Anisotropies" (MA) method. When we add a BF component directly toward the most anisotropic region, other regions start altering their behavior and might appear anisotropic. Absolute apparent isotropy cannot then be achieved. Therefore, the amplitude and direction of the BFs are found so that the final, overall anisotropy signal of the studied relation is minimized. For this, we repeat the anisotropy sky scanning (Sects. \ref{anis_detect} and \ref{stat_sign}) every time for a different BF. This procedure is computationally expensive, and thus we reduce the number of bootstrap resamplings to 500 when estimating the uncertainties of the BF characteristics.

We consider both independent redshift shells (redshift tomography) and cumulative redshift bins to constrain the BF motions, with both the MR and MA methods. It is important to note that flux limited samples suffer from certain biases when determining BFs. Probably the most important one is the nonuniform radial distribution of objects. If this is not taken into account, the BFs found for spherical volumes with a large radius, will be strongly affected by low$-z$ objects and not clearly reflect the larger scale motions. This can be partially accounted for if the iterative redshift steps with which we increase the spherical volumes encompass similar numbers of clusters \citep[e.g.,][and references therein]{peery}. That way, the contribution of every scale to the overall BF signal is averaged out. Here, to minimize the biased contribution of low$-z$ systems to the larger volumes, we consider redshift bins with similar cluster numbers. Finally, the size and redshift range of our sample did not allow us to use the kSZ signal of the clusters to detect any BFs (see Sect. \ref{discussion}).

\subsection{Parameter uncertainties} \label{uncertainties}

To estimate the uncertainties of the best-fit parameters, we use a bootstrap resampling method with replacement. We randomly draw 5000 resamplings from the studied (sub)sample of clusters with the same size, simultaneously constrain their best-fit parameters, and obtain the final distribution of the latter. The quoted parameter uncertainties refer to the $68.3\%$ credible interval defined by the positive and negative $34.1_{\text{th}}$ percentile of the distributions with respect to to the best-fit value of the full (sub)sample. Each parameter value distribution is considered separately from the rest \footnote{Thus, this process is equivalent to marginalizing one parameter over the rest.}. This method provides more conservative and robust parameter uncertainties than the approach followed in M20, since it depends only on the true, random variation of the studied statistic and not on analytical expressions (e.g., $\Delta \chi ^2$ limits). 

\subsection{Detection of anisotropies, parameter sky maps, and direction uncertainties}\label{anis_detect}

The methodology followed here is described in detail in Sect. 4.3 of M20, and thus we direct the reader there. In a nutshell, we consider cones of various radii ($\theta = 60^{\circ}-90^{\circ}$) and point them toward every possible direction in the sky (with a resolution of $\leq 1^{\circ}$). Each time, we consider only the clusters within each cone, and we obtain the best-fit parameters with their uncertainties, as described in the two previous sections. We then create a color-coded full sky map based on the best-fit parameters of every direction.

There are only two differences with the method followed in M20. Firstly, in this work we leave the slope $B_{YX}$ free to vary, instead of fixing it to its best-fit value for the full sample. That way, the parameter of interest is marginalized over $B_{YX}$. In M20 we demonstrated that this choice does not significantly alter our results compared to the case where $B_{YX}$ is kept fixed, but it constitutes a more conservative approach and thus we make it the default. Secondly, we slightly change the applied statistical weight of the fitted clusters which depends on their angular distance from the center of the cone. In M20 we divided the statistical uncertainties $\sigma _{\log{Y}}$ and $\sigma _{\log{X}}$ with a normalized cosine factor that shifted from 1 (center of cone) to 0 (edge of cone), despite the actual radius of the cone. Under certain conditions however\footnote{Only when there are strongly up- or downscattered clusters close to the center of the cone, i.e. with high statistical significance, and the number of clusters is relatively small.}, this method can slightly overestimate the final statistical significance of the observed anisotropies. Even if this does not have an effect on the conclusions of M20, here we choose to follow a more conservative approach, dividing the uncertainties with the $\cos{\theta_1}$ term, where $\theta_1$ is the angular separation of a cluster from the center of the cone we consider. This is motivated by the fact that the effects of a BF or an $H_0$ anisotropy to the measured cluster distance scale with $\cos{\theta_1}$. Also, this results in the same weighting as in M20 for $\theta=90^{\circ}$, and in a weaker weighting for $\theta<90^{\circ}$.

To provide anisotropy direction uncertainties, we perform bootstrap resamplings similar to the ones described in Sect. \ref{uncertainties}. For every sample used to create a sky map, we create 1000 bootstrap resamplings and perform the sky scanning again for each one of them. The $68.3\%$ limits of the posterior distribution of the maximum anisotropy direction are reported as the $1\sigma$ limits.

\subsection{Statistical significance of anisotropies}\label{stat_sign}

We wish to assess the statistical significance of the observed differences of the scaling relations' behavior toward different directions. The procedure followed here is described in Sect. 4.4 of M20, with only minor differences, such as the way we calculate the parameter uncertainties (Sect. \ref{uncertainties}). Briefly, we obtain the best-fit values of the fitted parameters together with their uncertainties for every direction in the sky. We then identify the region that shares the most statistically significant anisotropy from the rest of the sky (similar to a dipole anisotropy). We assess the statistical significance of the deviation between them by
\begin{equation}
\centering
\text{No. of }\sigma=\dfrac{\mathbf{p_1}-\mathbf{p_2}}{\sqrt{\sigma^2_{\mathbf{p_1}}+\sigma^2_{\mathbf{p_2}}}},
\label{sigma_sig}
\end{equation}
where $\mathbf{p_{1,2}}$ are the best-fit values for the two independent subsamples and $\sigma_{\mathbf{p_{1,2}}}$ are their uncertainties derived by bootstrapping\footnote{In M20 we identified the two regions with the most extreme, opposite behaviors. Naturally, the currently adopted method leads to slightly reduced anisotropy signals, and is part of the most conservative approach we follow here.}. Finally, the statistical significance (sigma) maps are color-coded based on the observed anisotropy level between every region and the rest of the sky.

\subsection{Monte Carlo simulations}

To further validate the effectiveness of our methodology and the statistical significance of the observed anisotropies, we perform Monte Carlo (MC) simulations. We create isotropic cluster samples to which we apply the same procedure as in the real data. That way we estimate the frequency with which artificial anisotropies would be detected in an isotropic Universe, and compare this to the real data. More details are described in Sect. \ref{mc_sim}.

\subsection{Summary of statistical improvements compared to M20} \label{differences}

Here we summarize the improvements in the statistical analysis of this work compared to M20: 1) The parameter uncertainties at every stage of this work are found by bootstrap resampling instead of $\Delta \chi^2$ limits. 2) During the sky scanning for identifying anisotropies, all the "nuisance" parameters (e.g., the slope $B$) are left free to vary, instead of fixing them to their best fit values. 3) The statistical weighting of clusters during the anisotropy searching is relaxed, to avoid creating any artificial anisotropies. 4) Uncertainties of the anisotropy directions are provided. 5) Monte Carlo simulations are carried out to further assess the statistical significance of the anisotropies.

\section{General behavior of the 10 scaling relations}\label{mainresults} 

\begin{figure*}[hbtp]
               \includegraphics[width=0.33\textwidth, height=5cm]{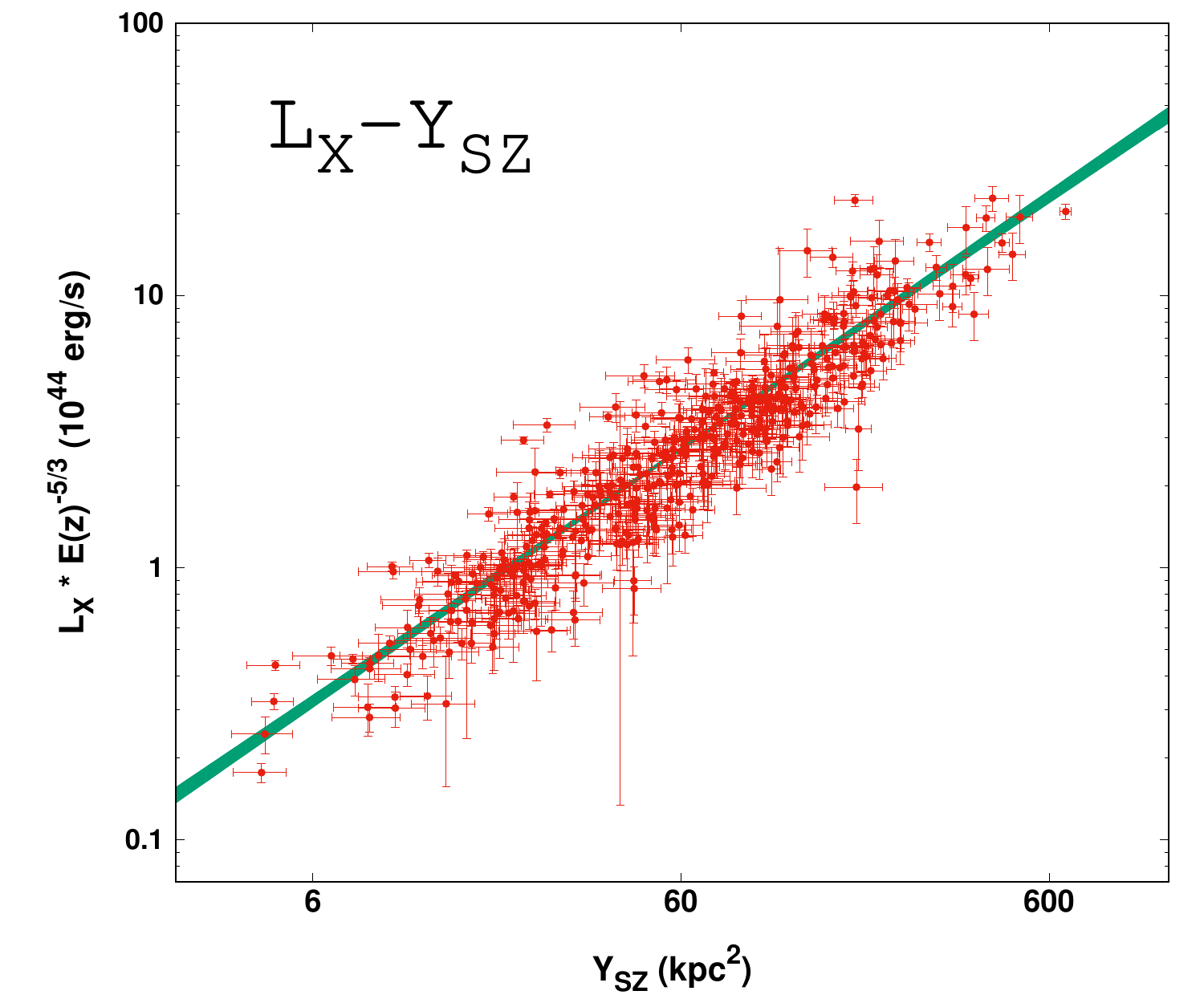}
                \includegraphics[width=0.33\textwidth, height=5cm]{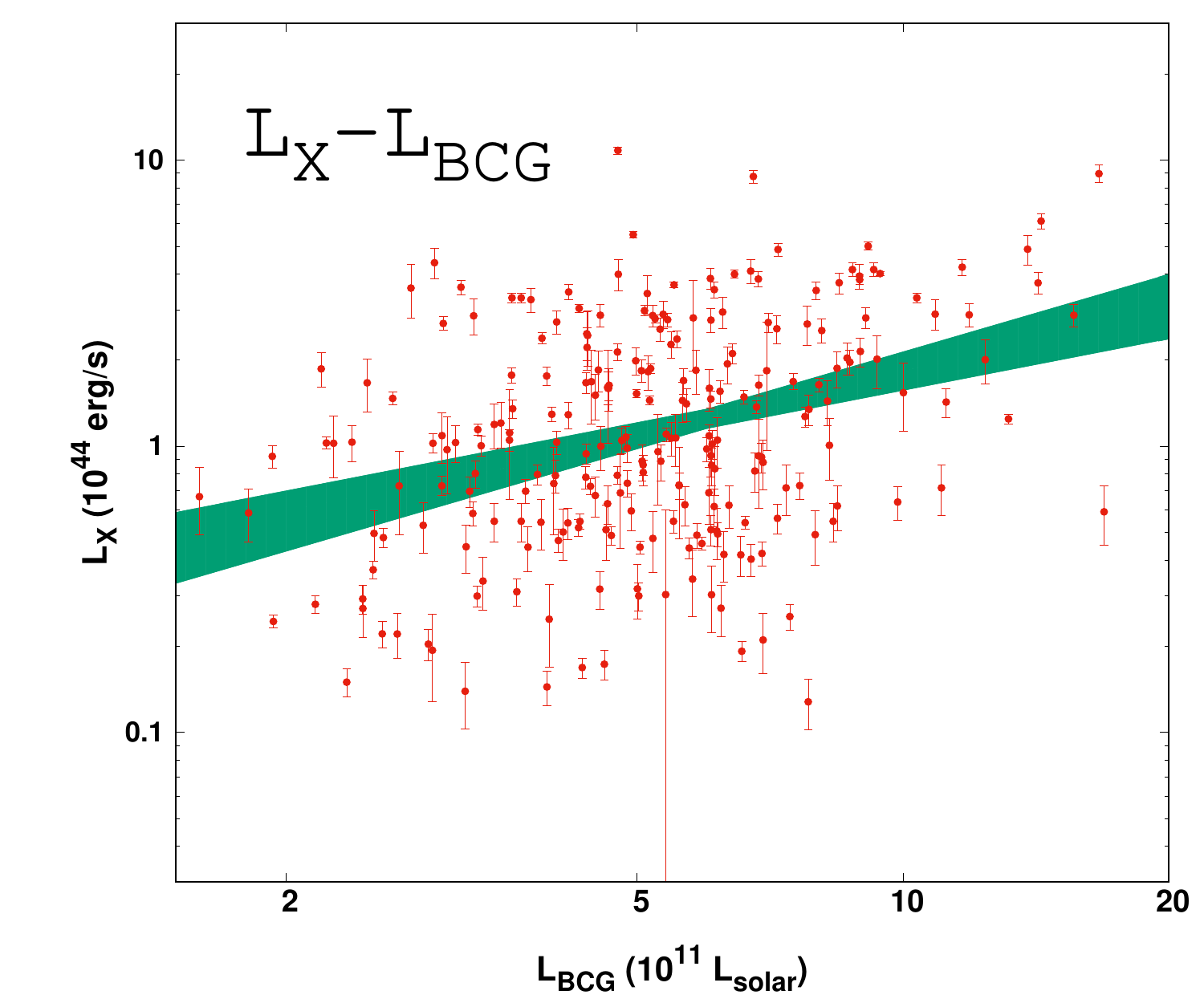}
               \includegraphics[width=0.33\textwidth, height=5cm]{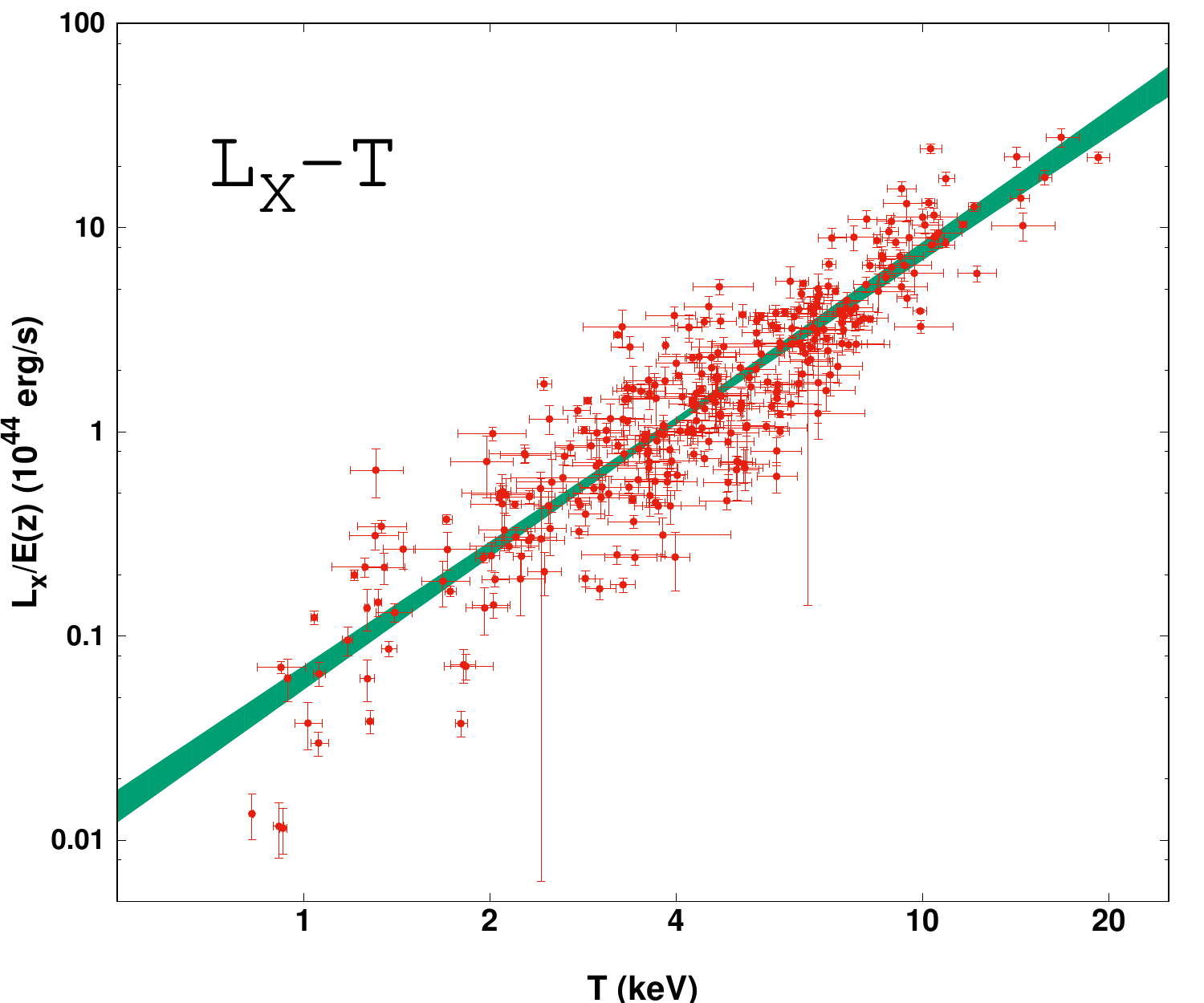}
                \includegraphics[width=0.33\textwidth, height=5cm]{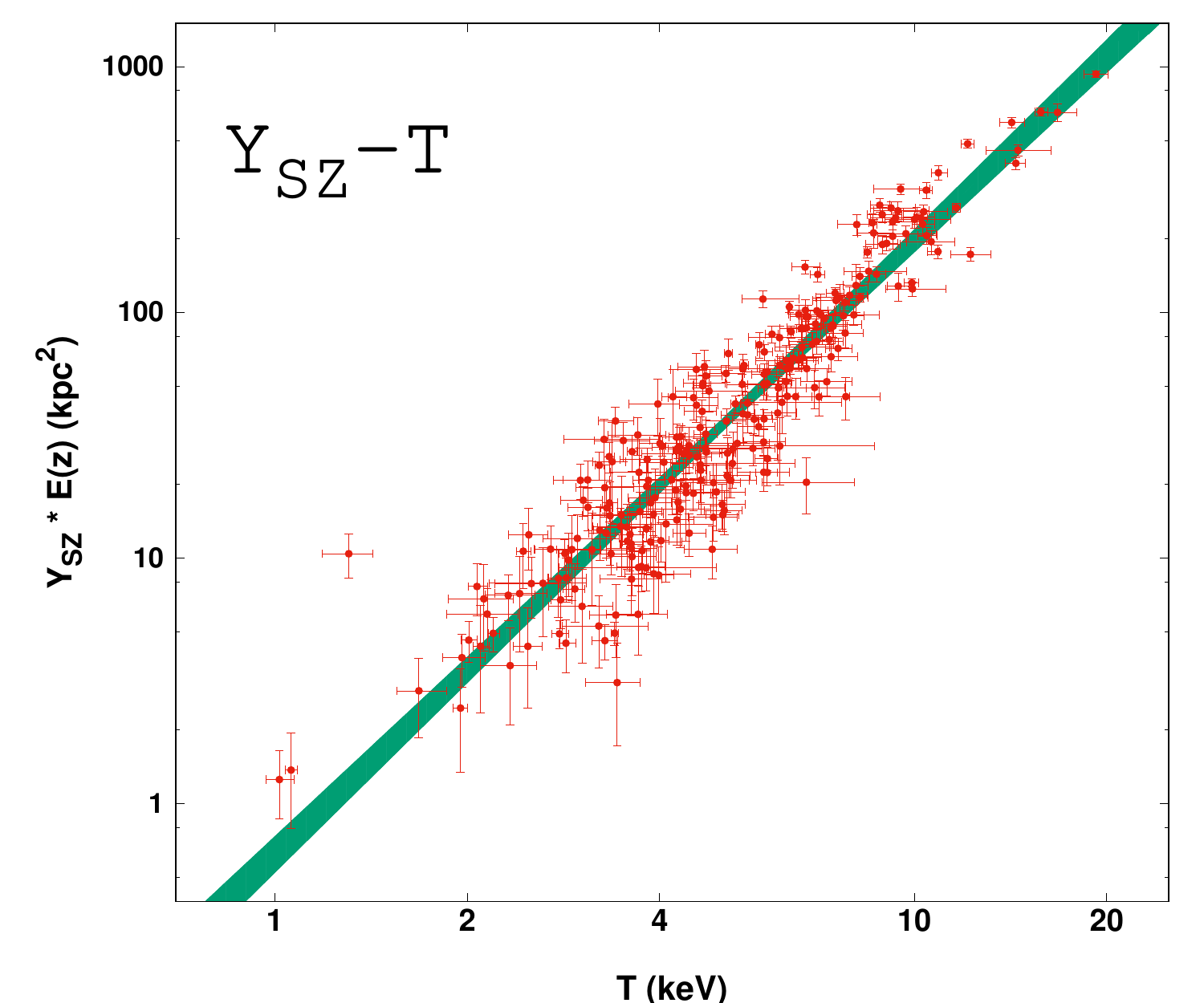}
                \includegraphics[width=0.33\textwidth, height=5cm]{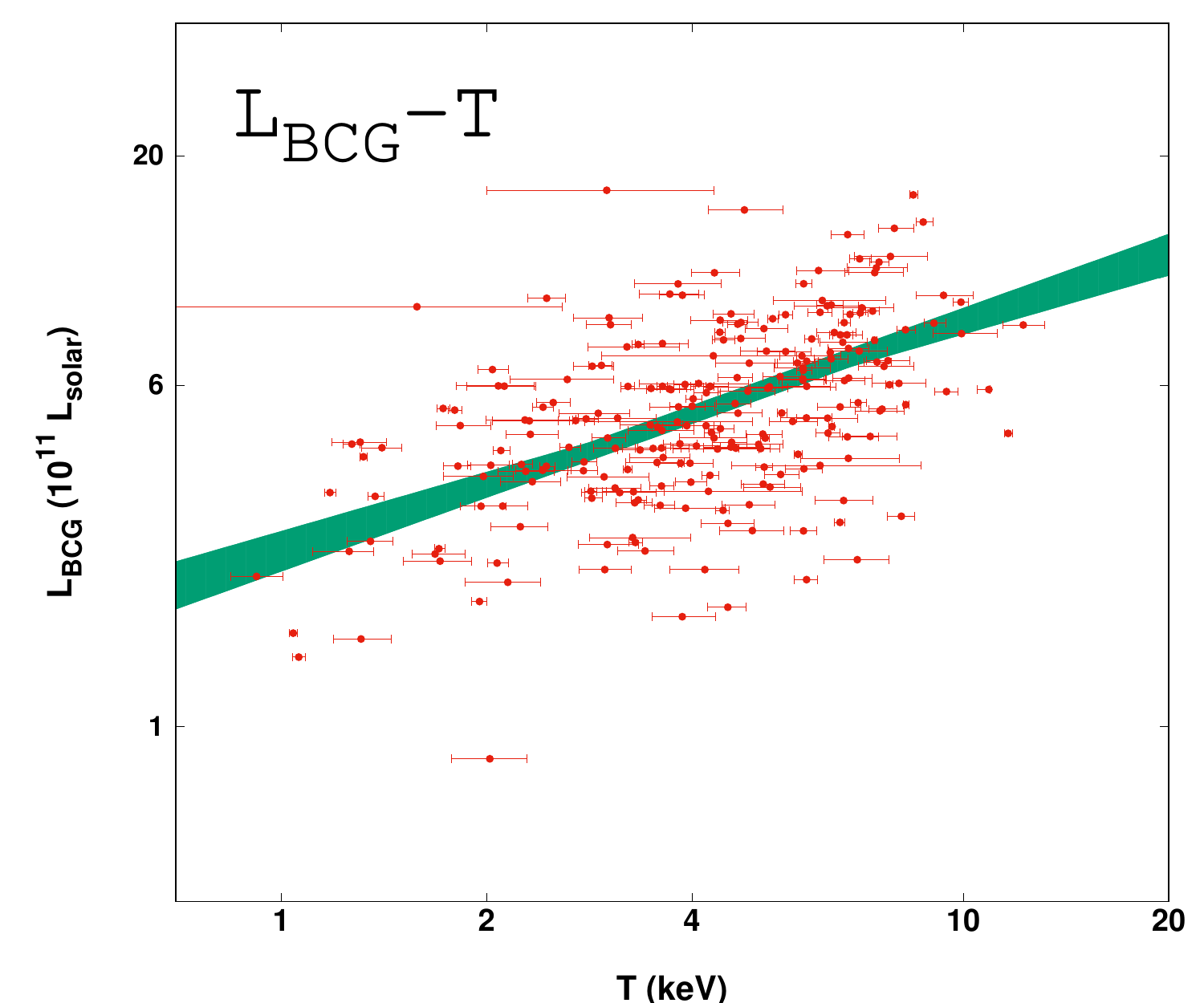}
               \includegraphics[width=0.33\textwidth, height=5cm]{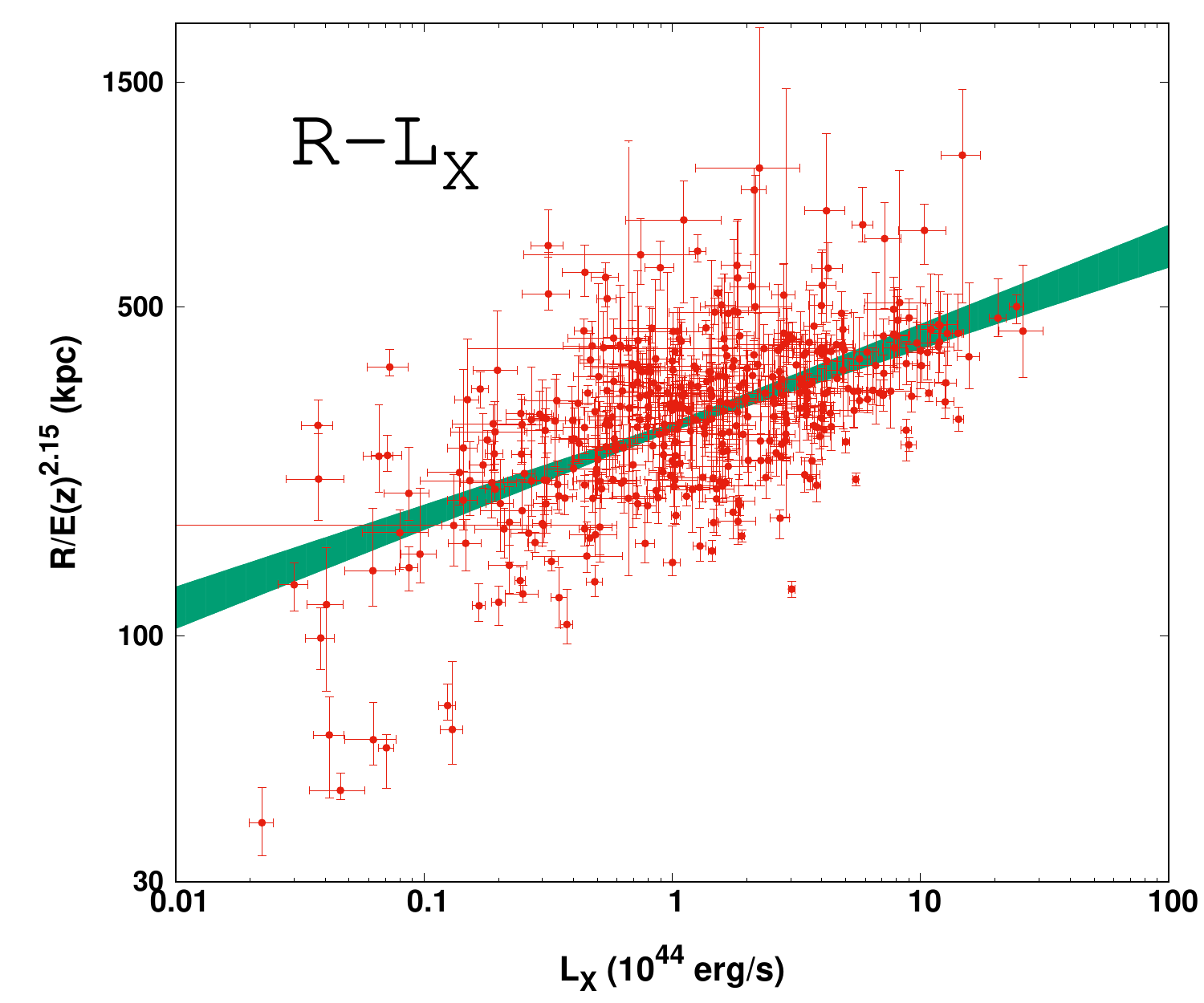}
               \includegraphics[width=0.33\textwidth, height=5cm]{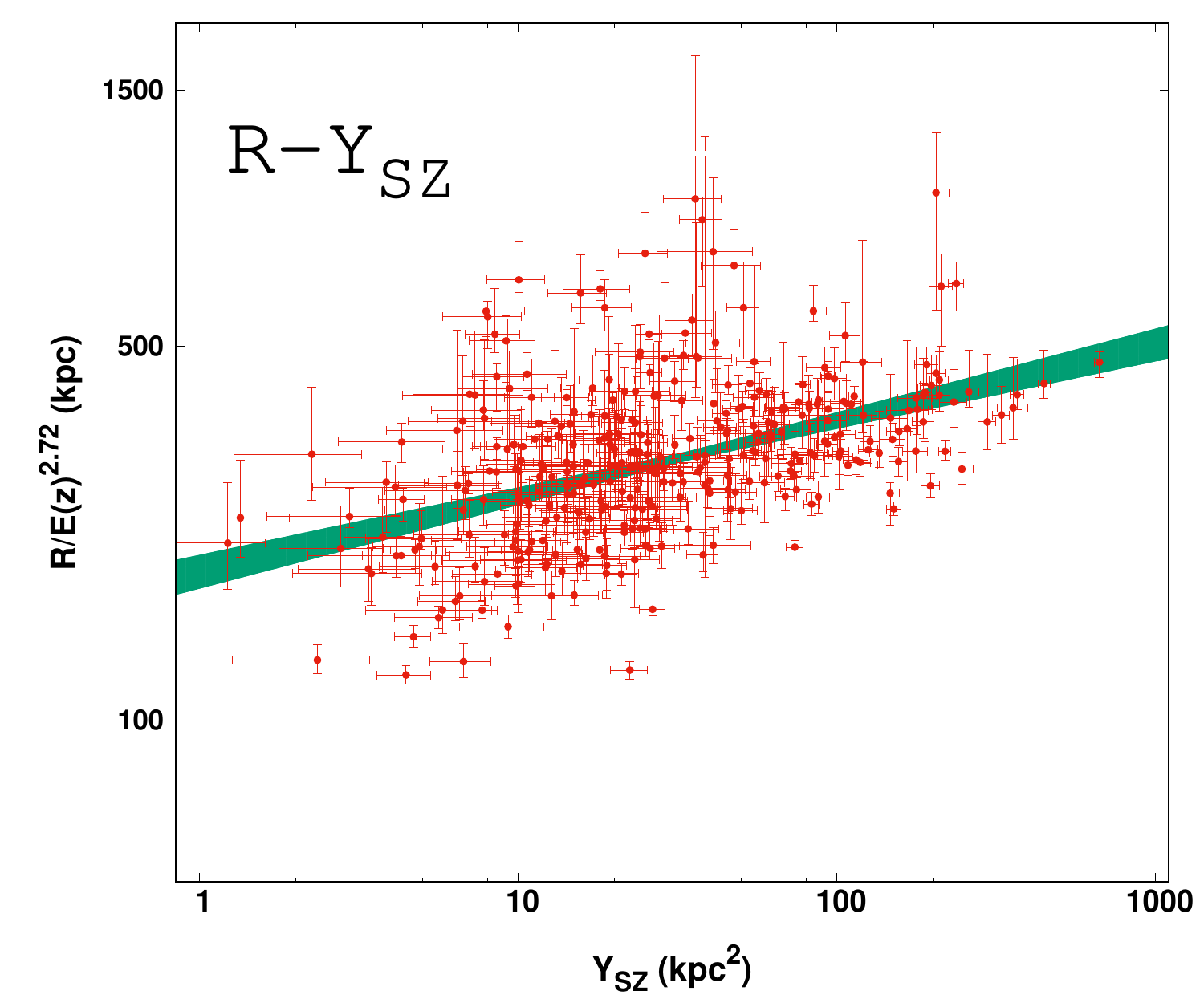}
               \includegraphics[width=0.33\textwidth, height=5cm]{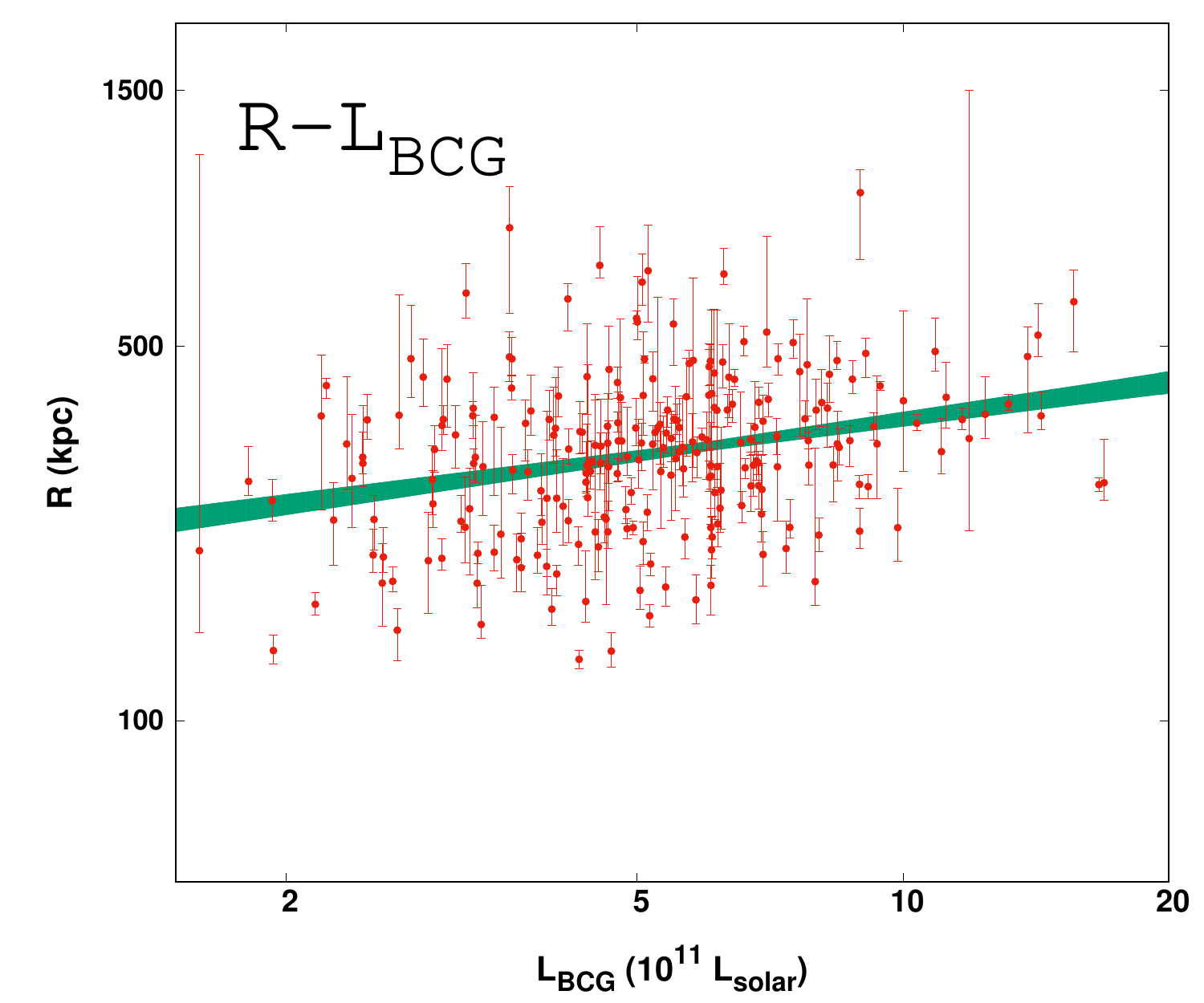}
               \includegraphics[width=0.33\textwidth, height=5cm]{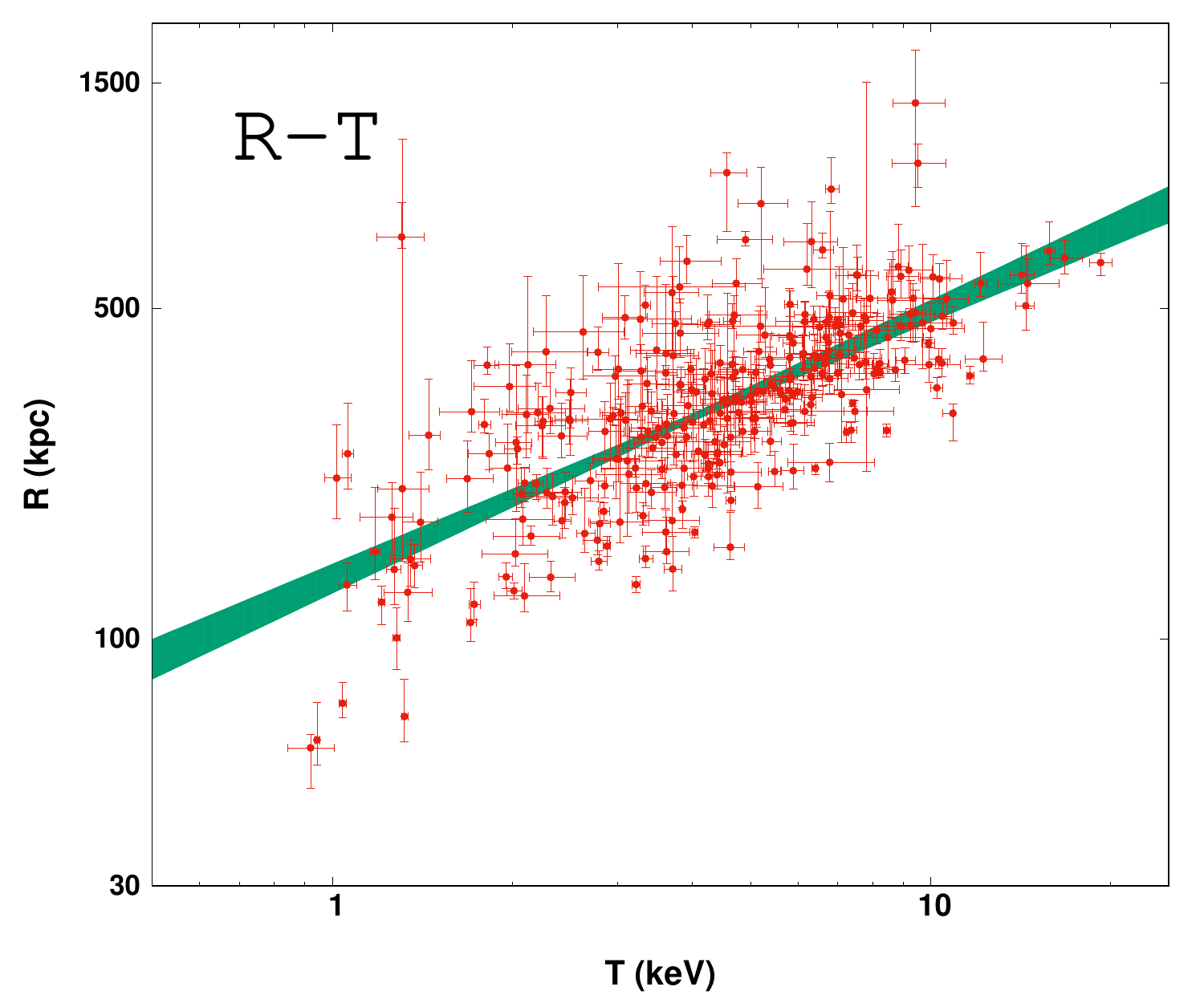}
               \includegraphics[width=0.33\textwidth, height=5cm]{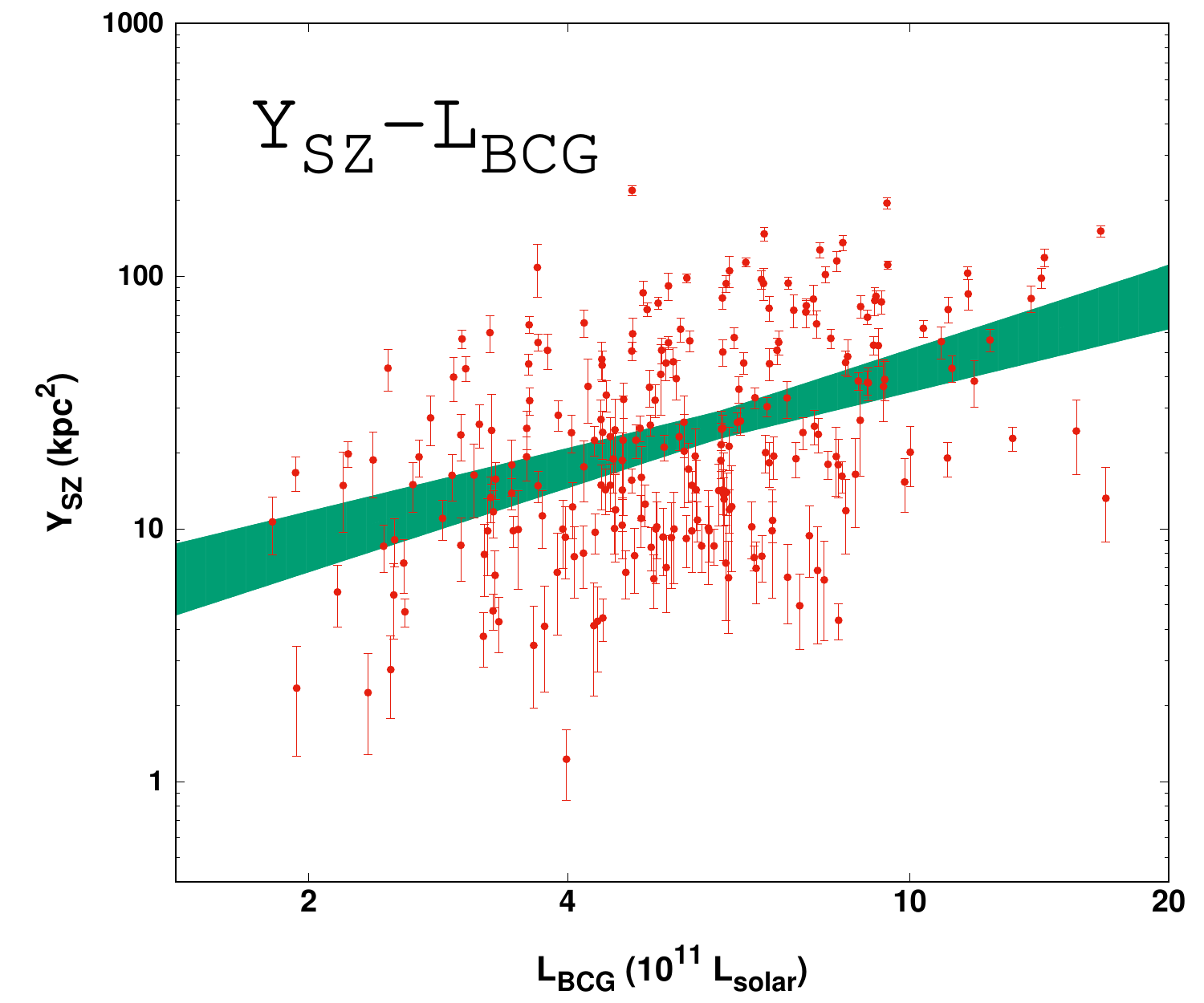}
                \includegraphics[width=0.33\textwidth, height=5cm]{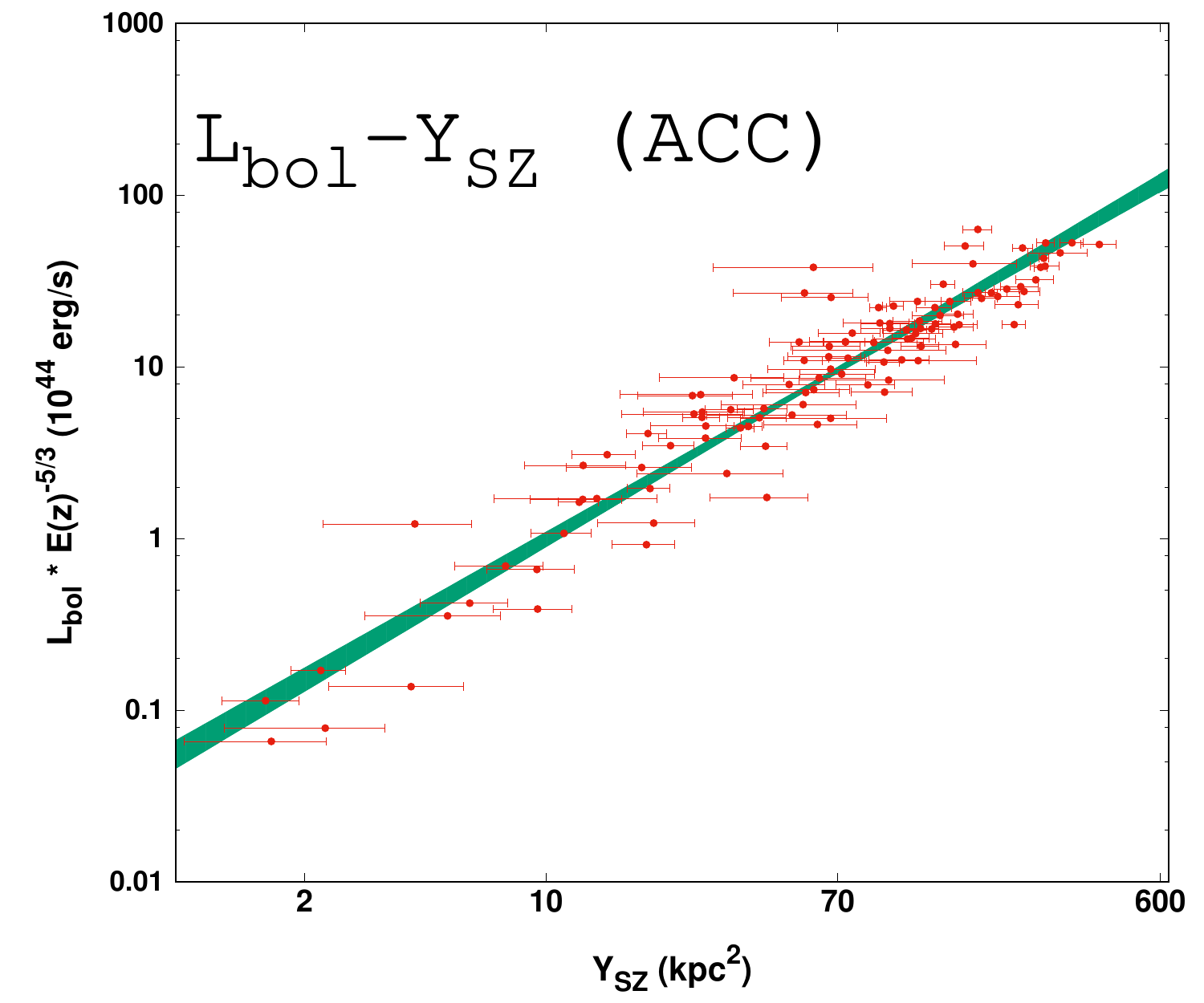}
               \includegraphics[width=0.33\textwidth, height=5cm]{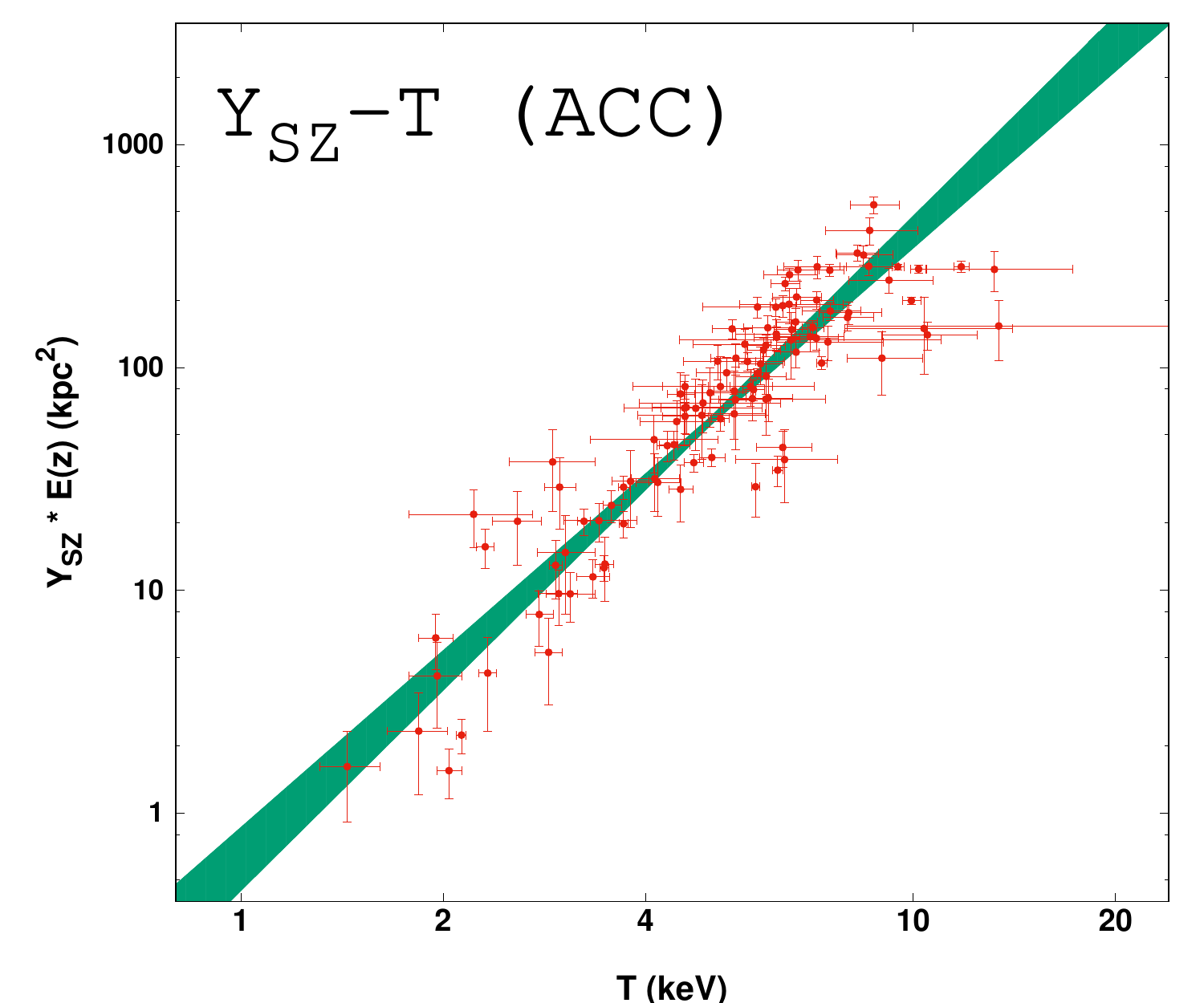}
               \caption{Best-fits of the 10 scaling relations studied in this work. The green area displays the $1\sigma$ limits of the best-fit area. From top left panel to the right: the \LY, \LxLbcg, \LT, \YT, \LbcgT, \RL, \RY, \RLbcg, \RT, \YLbcg, \RY, $L_{\text{bol}}-Y_{\text{SZ}}$ (ACC), and \YT\ (ACC) relations.}
        \label{bestfit-plots}
\end{figure*}

As a first step we constrain the overall behavior of the \emph{observed} scaling relations when one considers all the available data from across the sky. The effects of selection biases are discussed in Sect. \ref{select-bias}. The overview of the best-fit results of all the scaling relations is given in Table \ref{best-fit}, while the scaling relations themselves are plotted in Fig. \ref{bestfit-plots}.

\subsection{The \LT\ relation}

The full analysis of the \LT\ relation for our sample is presented in detail in M20. The only changes compared to the M20 results are the slightly changed $T$ for NGC 5846 (see Sect. \ref{sample}), and the use of bootstrap resampling for estimating the parameter uncertainties. Accounting for these changes, the best-fit values for the \LT\ relation remain fully consistent with M20 and with previous studies. For a detailed discussion see Sect. 5.1 of M20, while a more recent work confirming our results can be found in \citet{lorenzo20}. 

The $L_{\text{bol}}-T$ results for ACC are shown in M18 and M20. The only difference here is that we perform the $\chi^2$-minimization in the $L_{\text{bol}}$ axis, using bootstrap for estimating the parameter uncertainties. Since the bolometric X-ray luminosity is used for ACC (within $R_{200}$), both $A_{LT}$ and $B_{LT}$ are larger than the results of our sample. Also, $\sigma_{\text{int}}$ is $\sim 38\%$ larger for ACC.

\subsection{The \LY\ relation}

The \LY\ scaling relation has been studied in the past \citep[e.g.,][]{morandi,planck11,planck11b,demartino,ettori20,pratt20} mostly using \Ysz\ from \textit{Planck} and \Lx\ from ROSAT data and the MCXC catalog. Both of these quantities are efficient proxies of the total cluster mass, they also scale with each other. Their scatter with respect to to mass is mildly correlated \citep[e.g.,][]{aarti}. This results in \LY\ having the lowest scatter among all the scaling relations used in this study. 

As mentioned in Sect. \ref{y500-extr}, we measured \Ysz\ for 1095 MCXC clusters with S/N$>2$. Studying the \LY\ relation for these objects, we see that there are significant systematic differences between cluster subsamples based on their physical properties. For example, clusters with low $N_{\text{Htot}}$ or high $z$ tend to be significantly fainter on average than clusters with high \nhtot\ or low $z$ respectively. Surprisingly, the same behavior persists even when the original MCXC \Lx\ values and the \Ysz\ values from PSZ2 are used. As we increase the S/N threshold, these inconsistencies slowly fade out. More details on these effects can be found in Appendix \ref{differ_cuts}. Due to this, we choose to apply a low S/N$\geq 4.5$ threshold to the \Ysz\ values (same as in the PSZ2 sample), leaving us with 460 clusters with a median $z\sim0.14$. The clear benefits of this choice are that clusters with different properties now show fully consistent \LY\ solutions and no systematic behaviors are observed. Additionally, the intrinsic scatter of the \LY\ relation decreases drastically compared to cases with lower $S/N$ thresholds, allowing us to put precise constraints on the best-fit parameters and the possibly observed anisotropies of the relation. For these 460 clusters, the best-fit values are in full agreement with past studies within the uncertainties. The scatter we obtain however is lower than most past studies, most probably due to the use of the same $R_{500}$ between \Lx\ and \Ysz\ in our analysis. The observed slope $B_{LY}\sim0.93$ is slightly larger than the self-similar prediction of $B_{LY}=0.8$. 

For ACC, no trends for the \Ysz\ residuals are observed for S/N$>2$, with any of the cluster parameters. Therefore all 113 clusters can be used. The slope lies again close to unity, while the scatter is $\sim 32\%$ larger than when our sample is used. Nevertheless, the \LY\ scatter for ACC is sufficiently small to allow for precise constraints. We should note here that for S/N$>4.5$, the scatter of ACC is similar to our sample, but this cut would leave us with only 67 clusters, which are not enough for our purposes. 

\subsection{The \LxLbcg\ relation}

The \LxLbcg\ scaling relation has not been extensively studied in the past. \citet{mittal2} and \citet{bharad14} used 64 and 85 low-$z$ clusters respectively, to compare the \LxLbcg\ relation between cool- and noncool-core clusters, finding mild differences mostly for the slope. \citet{furnell} also studied the correlation of the stellar mass of BCGs (which is proportional to its luminosity) with the cluster's \Lx. Here we use significantly more clusters to study the \LxLbcg\ relation, namely 244. The best-fit slope of our analysis is less steep than the one \citet{bharad14} find, however they use the bolometric X-ray luminosity, contrary to us. The scatter we obtain is considerably smaller than theirs, although still significantly large. In fact, this is the scaling relation with the largest scatter out of the 10 we examine. The \Lx\ residuals do not show any systematic behavior as a function of cluster properties.

\subsection{The \RL\ relation}\label{R-L-relation}

The relation between $R$ and \Lx\ for galaxy clusters has not been investigated before to our knowledge. Here we use the 418 clusters with both measurements available to constrain this relation. The redshift evolution of \RL\ is unknown, thus we attempt to constrain it since the high number of clusters and the low scatter allow us to do so. We leave $\gamma_{RL_{\text{X}}}$ free to vary, simultaneously with the rest of the parameters. This results in $\gamma_{RL_{\text{X}}}=-2.15\pm 1.51$. The implied evolution is not statistically significant since the limited redshift range of the sample does not allow us to constrain it more efficiently. One should not expect the same evolution as in the $R_{500}-$\Lx\ relation (self-similar prediction of $\gamma\sim 1.3$), since a redshift evolution between the half-light radius $R$ and $R_{500}$ is also expected (e.g., due to the time-varying cool-core cluster fraction). 

Since $\gamma_{RL_{\text{X}}}=-2.15$ describes our data best, we fix $\gamma_{RL_{\text{X}}}$ to this value for the rest of our analysis. After performing the fit for all 418 clusters, we notice that clusters at $z<0.01$ are systematically downscattered compared to the rest. To avoid any biases during our anisotropy analysis, we exclude them. The final used subsample consists of the remaining 413 clusters. The \RL\ scatter residuals appear to be randomly distributed with respect to $z$, RASS exposure time, and \nhtot. However, a nonnegligible correlation is observed between the residuals and the apparent half-light radius \Rapp . The clusters with the lowest \Rapp\ appear to be downscattered in the \RL\ plane, and vice versa. A mild correlation of the $R$ residuals is also observed with the offset between the X-ray peak and the BCG position. The latter can be used as a tracer of the dynamical state of the cluster, which correlates with the existence (or not) of a cool core in the center of the cluster \citep[e.g.,][]{hudson,zitrin,rossetti,Lopes}. Cool-core clusters are expected to strongly bias the scaling relations involving $R$, since they emit most of their X-ray photons from near their centers. As a result, the half-light radius will be lower compared with noncool-core clusters at a fixed mass. More details on that can be found in Appendix \ref{R-relations}. 

Finally, the slope is lower than the self-similar prediction for the $R_{500}-$\Lx\ relation ($B\sim 0.33$), while there is a moderate scatter. Surprisingly, the latter is the largest one observed between all $R$ scaling relations. 

\subsection{The \YT\ relation}

The \YT\ relation has been previously studied by several authors, i.e., \citet{morandi,planck11b,Bender,ettori20} among others. It has never been studied before with such a large number of clusters as the one used in this work. Since the $T$ measurement is needed, we use the sample from M20. We retrieve 263 clusters with \Ysz\ measurements with S/N$>2$ from the M20 sample. No systematic differences in the \YT\ relation are observed for different cluster subsamples with different properties. The \Ysz\ residuals remain consistent with zero with increasing $T$, $z$, $N_{\text{Htot}}$, and other cluster parameters, while $A_{YT}$, $B_{YT}$ and $\sigma_{\text{int,YT}}$ also stay constant with an increasing S/N cut. These results clearly indicate that the applied S/N threshold does not introduce any strong biases to the \YT\ relation (see Appendix \ref{YT-cuts} for more details). The best-fit parameter values that we obtain for these clusters are in line with previous findings. The value of the slope agrees with the self-similar prediction ($B_{YT}=2.5$), while the scatter is lower than the one for the \LT\ relation. Finally, a single power law perfectly describes the relation since a change in $A_{YT,\text{all}}$ and $B_{YT,\text{all}}$ is not observed for a changing low $T$ cut.

When using the ACC sample, we obtain a similar scatter with our sample, but with a slightly steeper slope and larger normalization. This is due to the fact that temperatures are measured for ACC considering the entire cluster, leading to generally lower $T$ than our sample (where $T$ is measured within $0.2-0.5\ R_{500}$).

\subsection{The \YLbcg\ relation}

The \YLbcg\ scaling relation has not been studied in the past, and it is constrained for the first time in this work. The two quantities are expected to scale with each other since they both scale with the cluster mass. We have both measurements for 214 clusters with S/N$>2$ for the \Ysz\ measurement, and the applied redshift limits for the BCGs. When we performed the fit, we did not detect any strong systematic behavior of the residuals as a function of cluster parameters, with the exception of \Ysz\ S/N, with high \Ysz\ S/N clusters tending to be upscattered. This behavior persists even with an increasing S/N threshold. Although the best-fit $B_{YL_{\text{BCG}}}$ stays unchanged for different S/N cuts, the $A_{YL_{\text{BCG}}}$, varies by $\sim 3.4\sigma$ for S/N$>4.5$ (Fig. \ref{LY-ab-SN} in Appendix). Since the residuals are a function of S/N in any case, and since the \YLbcg\ relation offers only limited insights in the anisotropy analysis, we adopt S/N$>2$, but suggest caution because of the aforementioned dependance of $A_{YL_{\text{BCG}}}$. The slope lies close to linearity, while the scatter is $\sim 10\%$ smaller than the \LxLbcg\ scaling relation. 

\begin{table*}[hbtp]
\caption{\small{Best-fit parameters of the 10 scaling relations. Below we display the scaling relations, their number of used clusters $N$, their best-fit normalization $A$ and slope $B$, their intrinsic and total scatter $\sigma_{\text{int}}$ and $\sigma_{\text{tot}}$ respectively, their calibration terms $C_Y$ and $C_X$, and the power of their redshift evolution $\gamma$.}}
\label{best-fit}
\begin{center}
\renewcommand{\arraystretch}{1.3}
\small
\begin{tabular}{ c  c  c  c  c  c | c  c c}
\hline \hline

$Y-X$ & $N$ & $A_{YX}$ & $B_{YX}$ & $\sigma_{\text{int},YX}$ (dex) & $\sigma_{\text{tot},YX}$ (dex) & $C_Y$ & $C_X$ & $\gamma_{YX}$ \\
\hline \hline
 & & & & Our sample & & & & \\ \hline
\LT & 313 & $1.132^{+0.042}_{-0.046}$ & $2.086^{+0.073}_{-0.065}$ & $0.233\pm 0.016$ & $0.258\pm 0.018$ & $10^{44}$ erg/s & 4 keV & -1 \\ 
\LY & 460 & $2.737^{+0.043}_{-0.049}$ & $0.928^{+0.015}_{-0.018}$ & $0.108\pm 0.008$ & $0.143\pm 0.009$ & $10^{44}$ erg/s & 60 kpc$^2$ & -5/3 \\ 
\LxLbcg & 244 & $1.236^{+0.101}_{-0.091}$ & $0.775\pm 0.152$ & $0.372\pm 0.020$ & $0.379\pm 0.020$ & $10^{44}$ erg/s & $6\times 10^{11}\ L_{\odot}$ & - \\ 
\YT & 263 & $1.110^{+0.029}_{-0.033}$ & $2.546^{+0.071}_{-0.067}$ & $0.146\pm 0.013$ & $0.192\pm 0.015$ & $35$ kpc$^2$& 5 keV & 1\\ 
\YLbcg & 214 & $0.745^{+0.104}_{-0.094}$ & $0.868\pm 0.138$ & $0.328\pm 0.021$ & $0.341\pm 0.022$ & $35$ kpc$^2$ & $6\times 10^{11}\ L_{\odot}$ & - \\ 
\LbcgT & 196 & $0.847^{+0.020}_{-0.025}$ & $0.542^{+0.052}_{-0.110}$ & $0.170\pm0.008$ & $0.177\pm 0.009$  &4 keV & $6\times 10^{11}\ L_{\odot}$& - \\ 
\RL &  413 & $0.794\pm 0.016$ & $0.191^{+0.020}_{-0.017}$ & $0.142\pm 0.008$ & $0.156\pm 0.009$ & $10^{44}$ erg/s & 350 kpc & -2.15 \\ 
\RY & 347 & $0.895\pm 0.017$ & $0.141^{+0.015}_{-0.013}$ & $0.112\pm 0.008$ & $0.128\pm 0.008$ & 350 kpc & 35 kpc$^2$ & -2.72 \\ 
\RT &  308 & $0.829\pm 0.016$ & $0.568^{+0.036}_{-0.040}$ & $0.126\pm 0.008$ & $0.144\pm 0.010$ & 350 kpc & 4 keV & -1.98 \\ 
\RLbcg & 243 & $0.920\pm 0.019$ &  $0.232\pm 0.022$ & $0.128\pm 0.009$ & $0.140\pm 0.010$ & 350 kpc & $6\times 10^{11}\ L_{\odot}$ & -  \\ \hline
& & & & ACC & & & & \\ \hline
$L_{\text{bol}}-T$ & 168 & $1.074^{+0.067}_{-0.063}$ & $3.208\pm 0.122$ & $0.345\pm 0.028$ & $0.394\pm 0.033$ & $5\times 10^{44}$ erg/s & 5 keV & -1 \\ 
$L_{\text{bol}}-Y_{\text{SZ}}$ & 113 & $1.591^{+0.063}_{-0.060}$ & $1.168\pm 0.036$ & $0.149\pm 0.014$ & $0.195\pm 0.019$ & $5\times 10^{44}$ erg/s & 60 kpc$^2$ & -5/3 \\ 
\YT & 113 & $1.774\pm 0.084$ & $2.812\pm 0.150 $ & $0.159\pm 0.019$ & $0.227\pm 0.026$ & $35$ kpc$^2$& 5 keV & 1\\

 \hline

\end{tabular}
\end{center}
\end{table*}

\subsection{The \RY\ relation}

A relation between \Ysz\ and the X-ray isophotal radius $R$ is expected to exist since both quantities scale with cluster mass. Such a relation has not been observationally constrained however until now. For this work, both quantities were measured for 347 clusters. The redshift evolution of the relation is not known, however due to the small scatter it can be obtained observationally, similarly to the \RL\ relation. We find that the scatter is minimized for  $\gamma_{RY_{\text{SZ}}}=-2.72\pm 1.41$. Expectingly, the result is similar to $\gamma_{RL_{\text{X}}}$. It should be reminded that we look for the evolution describing our data best, and not necessarily for the true one. Fixing $\gamma_{RY_{\text{SZ}}}$ to its best-fit value, we repeat the fitting for the rest of the parameters. The residuals of $R$ behave in exactly the same way as for \RL. While no trend is seen with respect to $z$, the systematic behaviors as functions of \Rapp\ and of the X-ray-BCG offset persist. The observed scatter is the smallest one between all $R$ scaling relations, $\sim 20\%$ smaller than the \RL\ scatter. Finally, the slope of \RY\ is also the flattest compared to the other scaling relations.

\subsection{The \LbcgT\ relation}

Similarly to the \LxLbcg\ relation, the \LbcgT\ relation has not been the focus of many studies in the past. \citet{bharad14} studies the scaling of \Lbcg\ with the mass of clusters. The latter is however obtained through a cluster mass-temperature scaling relation. For the \LbcgT\ relation the exact redshift evolution is also not known, and constraining it from the existing data is not trivial. This is due to the large scatter of the \LbcgT\ relation and the other simultaneously fit parameters. Due to that, from the 259 clusters for which we have measured both $T$ and \Lbcg, here we again use only the 196 of them with $z<0.15$ in order to safely ignore any existing redshift evolution of the relation. The scatter of the \LbcgT\ relation is considerably large, namely $\sim 70\%$ larger than the \LT\ scatter when the latter is minimized with respect to $T$. The scaling between the two quantities however is clear, although with a less steep slope than the reversed \LT\ one ($\sim 2.8$). The slope has also a larger relative uncertainty ($\sim 10\%$) than the normalization ($\sim5\%$). 

\subsection{The \RLbcg\ relation}

The \RLbcg\ is another scaling relation that is presented for the first time in this work. Both measurements are available for 243 clusters, after the previously described \Lbcg\ redshift cuts. The intrinsic scatter of the relation is similar to the other $R$ relations, while the slope is slightly larger than the \RL\ and \RY\ relations. The same behavior for the $R$ residuals is observed as in the \RL\ and \RY\ scaling relations.

\subsection{The \RT\ relation}

The \RT\ relation has not been studied extensively in the past, although some studies with both observations and simulations were performed two decades ago \citep{mohr97,mohr,verde}. The relation has been used as a cosmological probe as well by \citet{mohr}, to constrain $\Omega_{\text{m}}$ and  $\Omega_{\Lambda}$. These authors however used only a few tens of clusters to constrain the relation. In this work, we use 308 clusters for which both $R$ and $T$ have been measured. In \citet{mohr} it is argued that there is no redshift evolution for the \RT\ relation. However, this conclusion specifically depends on their methodology for measuring $R$, and cannot be adopted for our method. Therefore, we choose to constrain the redshift evolution from our data, leaving it free to vary. We obtain $\gamma_{RT}=-1.98\pm 1.42$, similar to the other $R$ relations. Fixing $\gamma_{RT}=-1.98$, we obtain the final best-fit results. The slope is the largest ($\sim 0.57$) among all $R$ scaling relations. The same $R$ residual behavior as for the other $R$ scaling relations is observed. 

\section{Anisotropies due to unaccounted X-ray absorption effects}\label{xray_abs_anis}

In this section we study the scaling relations, whose observed anisotropies could be caused purely by previously unknown soft X-ray effects, such as extra absorption from "dark", metal-rich, gas and dust clouds. If the anisotropies observed in the \LT\ relation by M20 were due to such effects, one should detect the same anisotropies in the scaling relations studied in this section. We report the most anisotropic directions and the statistical significance of the observed tension and quantify the amount X-ray absorbing material that should exist to fully explain the discrepancy.

\subsection{\LY\ anisotropies}

\subsubsection{Our sample}

The anisotropies of the \LY\ relation are the focal point of the search for hidden X-ray effects that we were not aware of in the past. This relation exhibits the lowest scatter and the largest number of clusters among all scaling relations studied in this work. This allows for precise pinpointing of its anisotropies. Even more importantly, \LY\ is almost completely insensitive to any spatial $H_0$ variations, since both quantities depend on cosmological parameters in the same way. They also scale almost linearly with each other. Thus, if one changed $H_0$, no significant change in the best-fit $A_{LY}$ $B_{LY}$ would be observed. Analytically, as \Lx$\propto D_L^2= D_A^2 (1+z)^4$ and \Ysz$\propto D_A^2$, their ratio (considering their best-fit $B_{LY}=0.928$) would be 
\begin{equation}
\dfrac{L_{\text{X}}}{(Y_{\text{SZ}})^{0.928}}\propto A_{LY}E(z)^{5/3}\implies A_{LY}\propto \dfrac{(1+z)^{4} D_A^{0.144}}{E(z)^{5/3}}.
\end{equation}
Thus, a $\sim 15\%$ spatial variation of $H_0$ would cause a nondetectable $\sim 2\%$ variation in $A_{LY}$. Moreover, the \LY\ relation is quite insensitive to BFs as well. Based on the above calculation, if a BF of $\sim 1000$ km/s existed at $z\sim 0.05$ toward a sky region, this would only lead to a $< 2\%$ increase in $A_{LY}$ of this region. 

\begin{figure*}[hbtp]
               \includegraphics[width=0.49\textwidth, height=4.5cm]{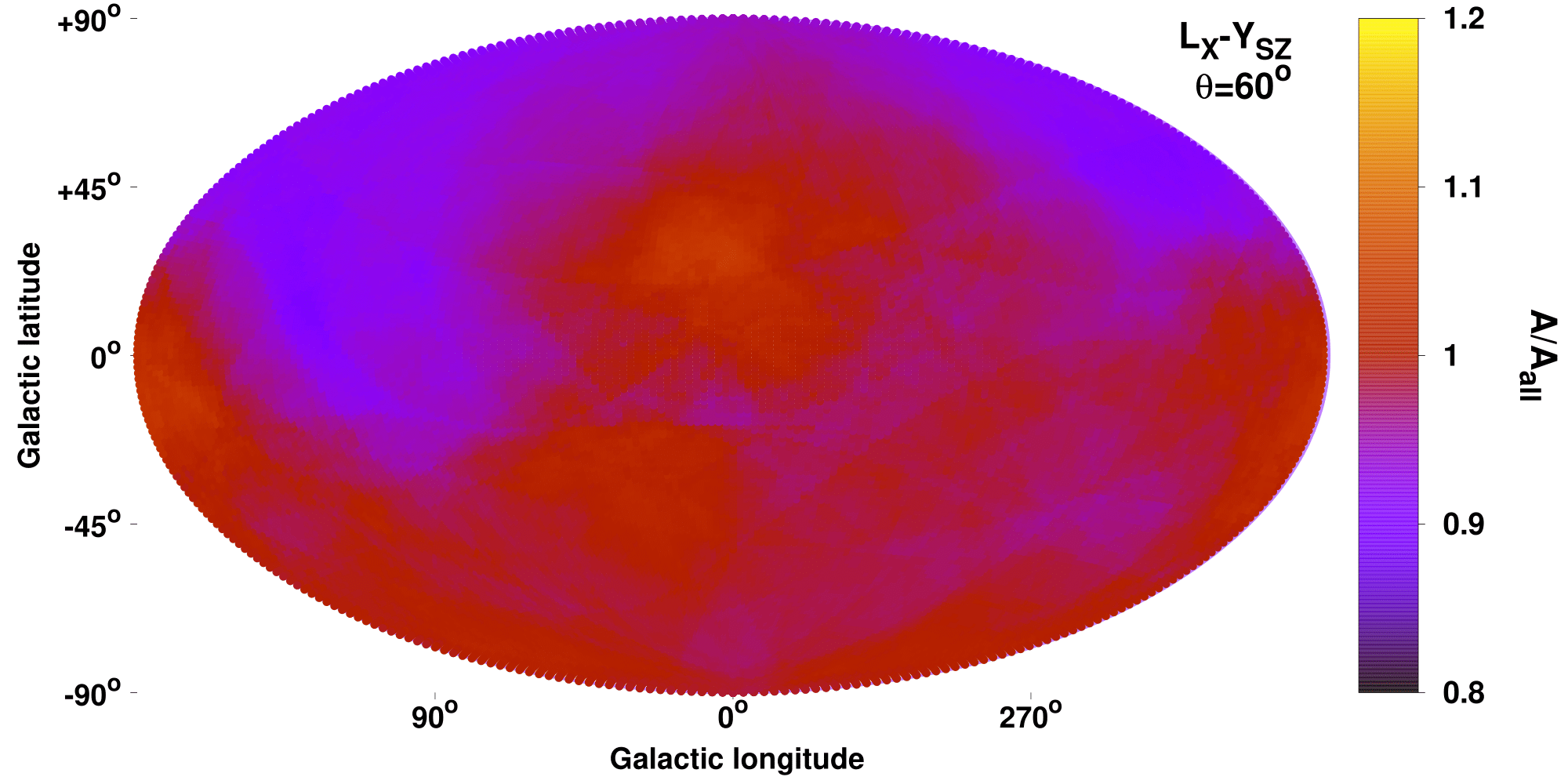}
               \includegraphics[width=0.49\textwidth, height=4.5cm]{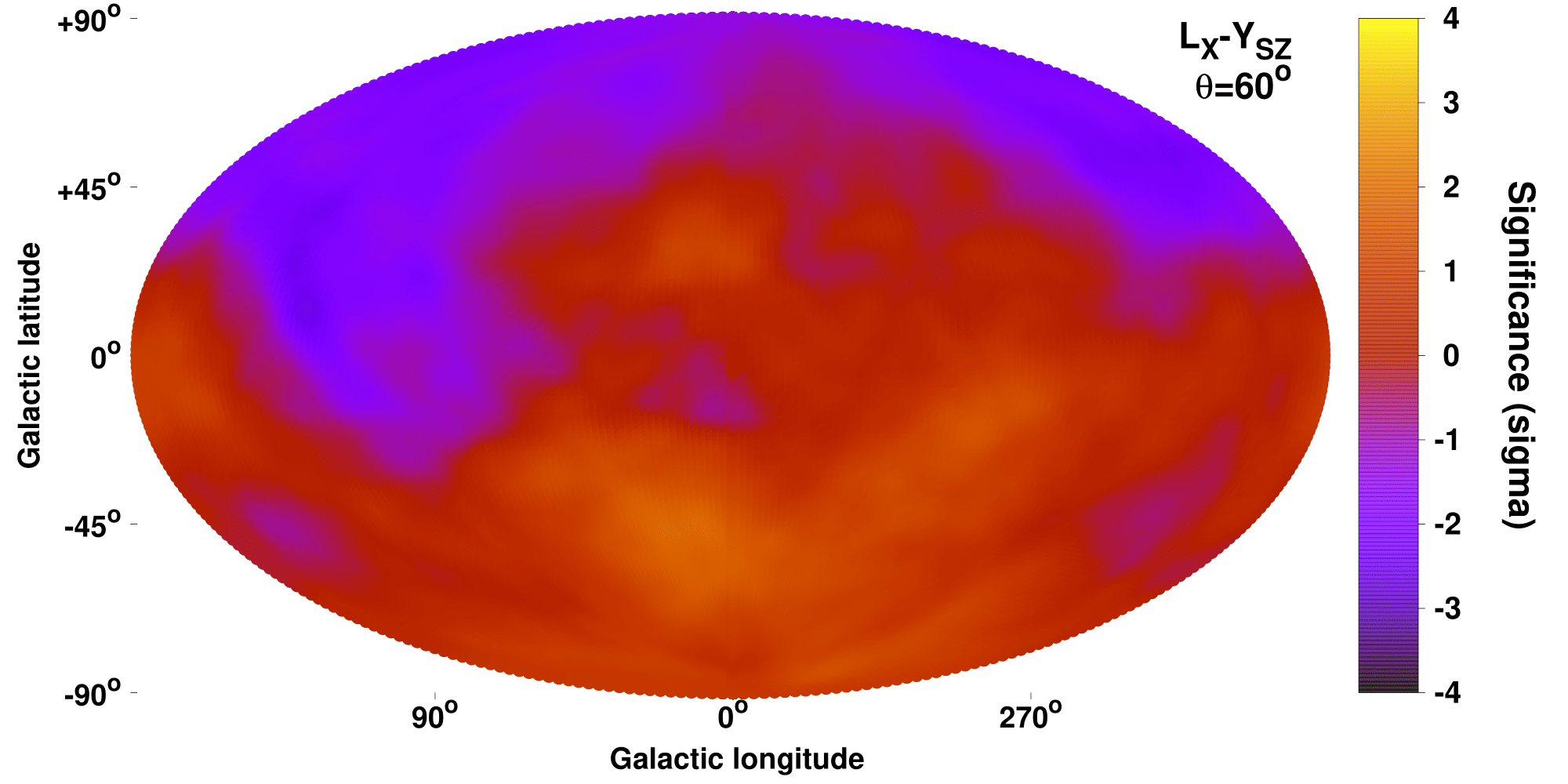}
                \includegraphics[width=0.49\textwidth, height=4.5cm]{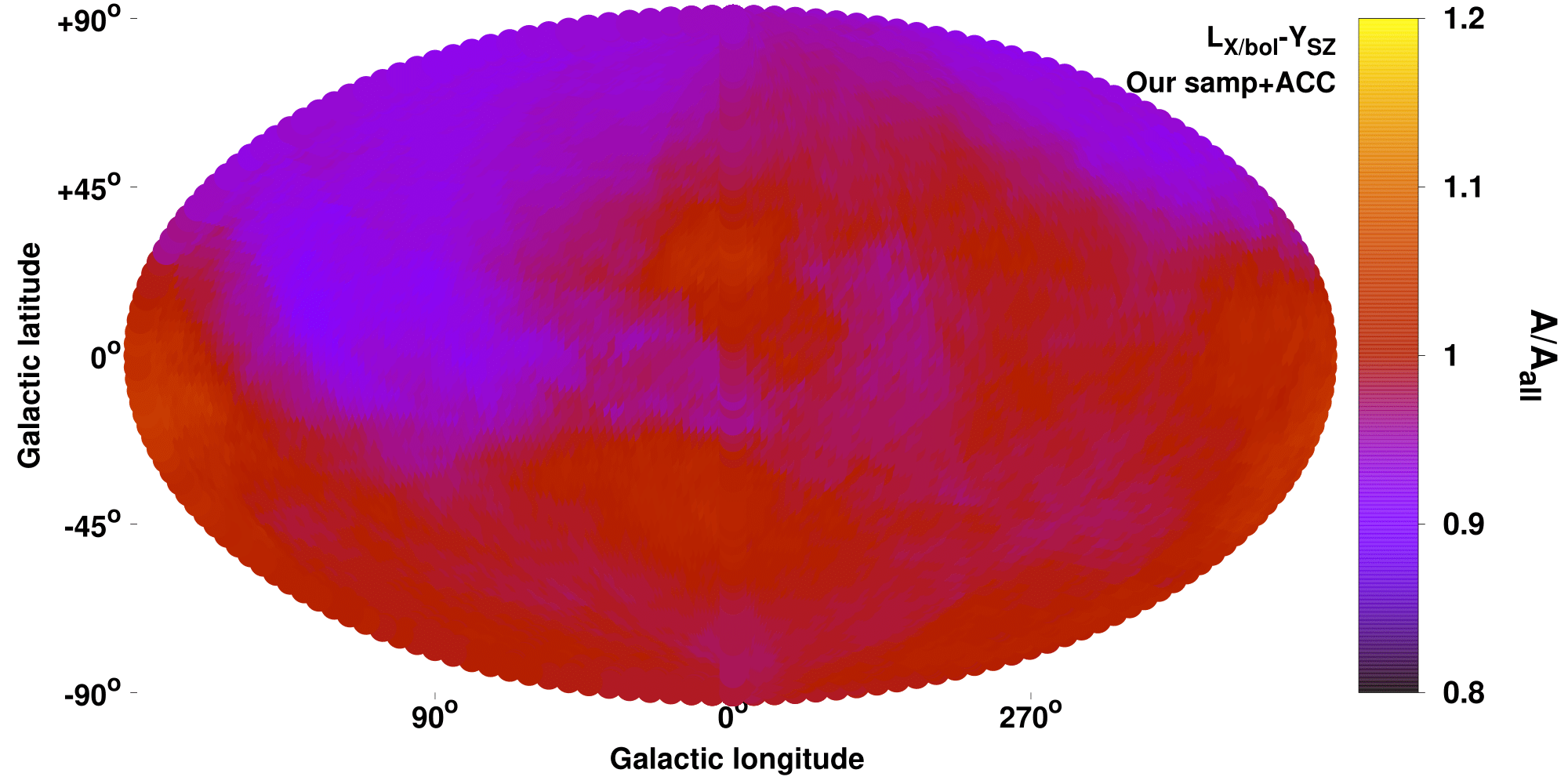}
               \includegraphics[width=0.49\textwidth, height=4.5cm]{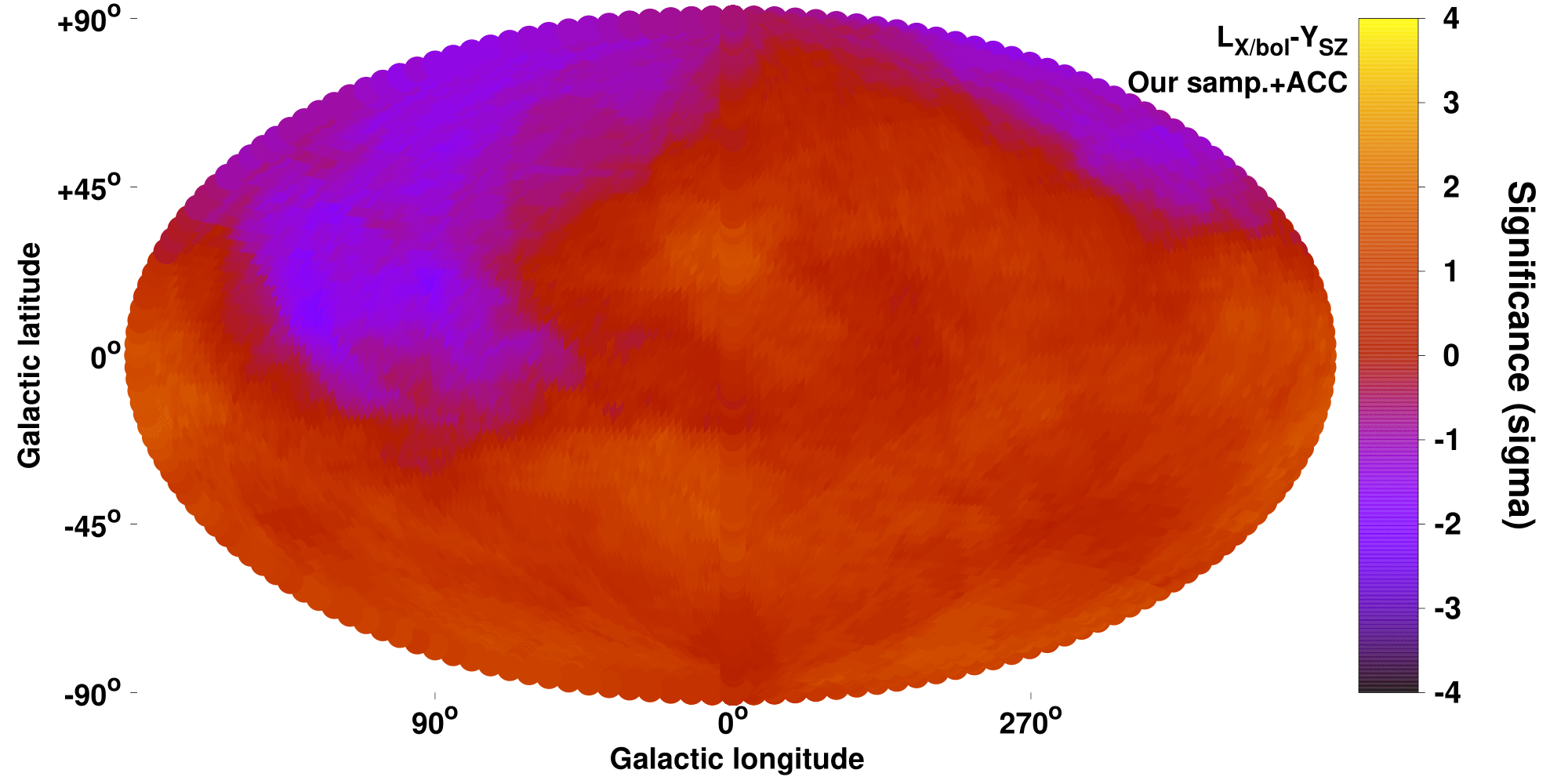}
               \includegraphics[width=0.49\textwidth, height=4.5cm]{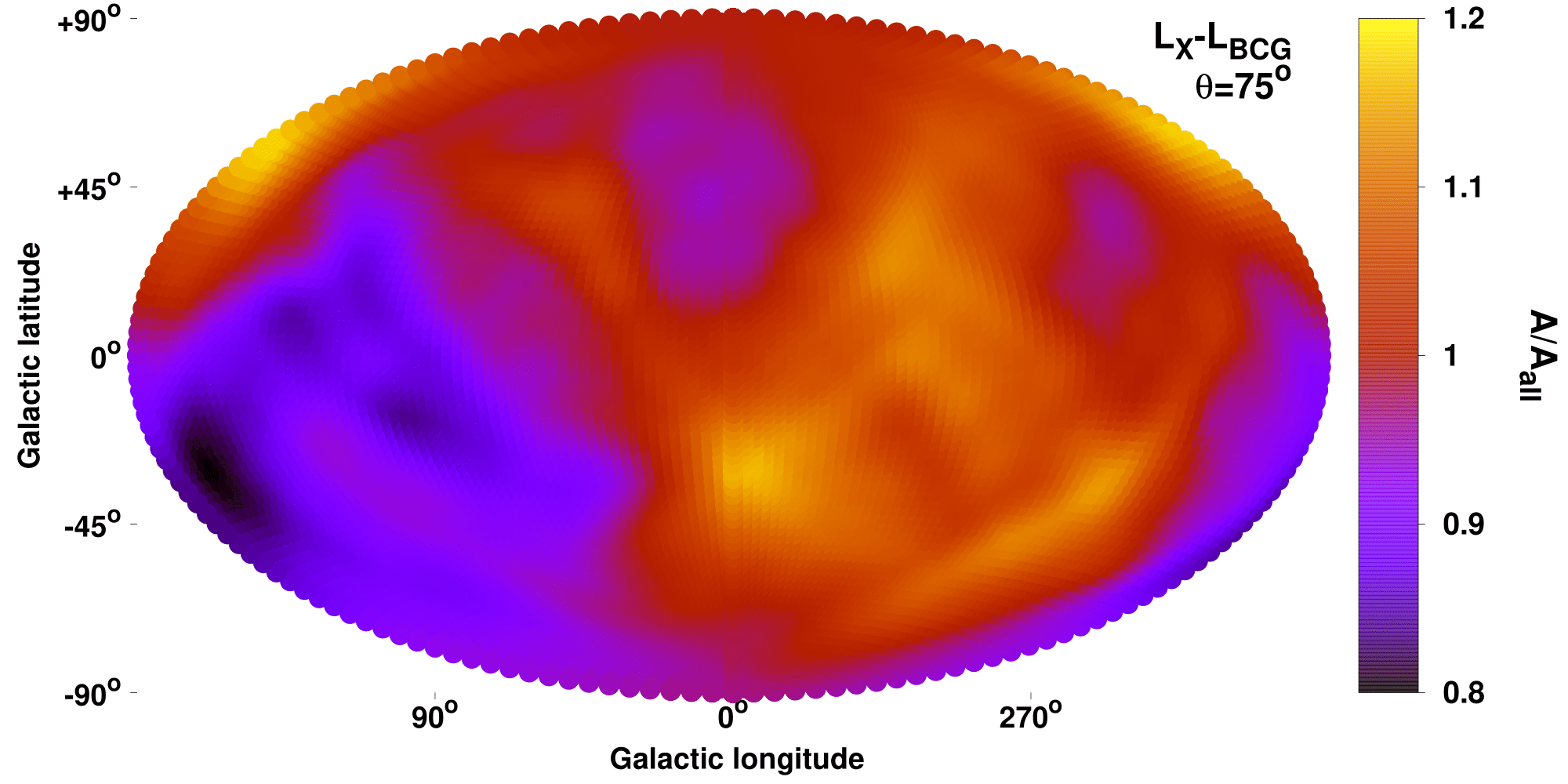}
               \includegraphics[width=0.49\textwidth, height=4.5cm]{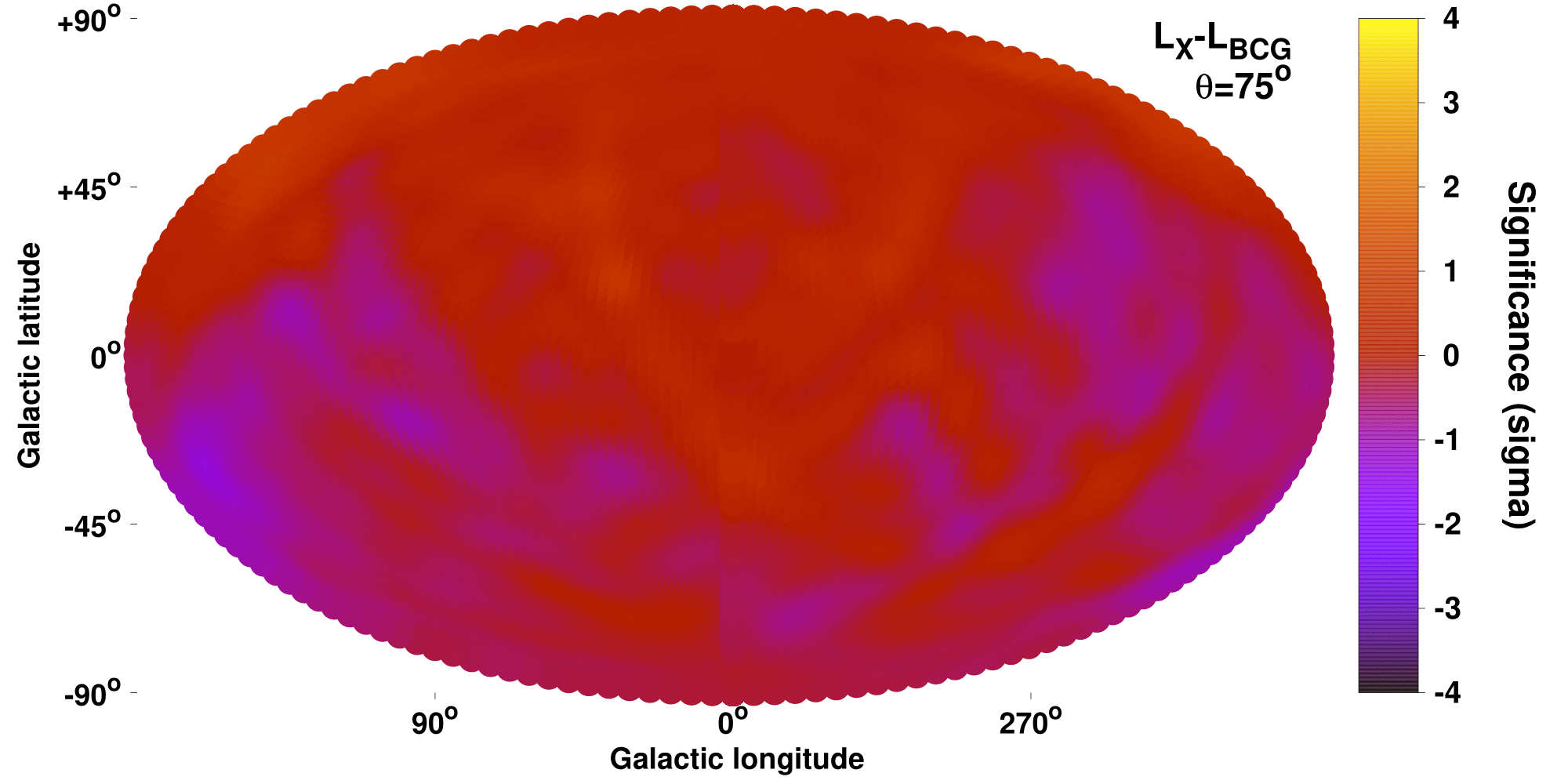}
               \caption{Normalization anisotropy maps (left) and the respective statistical significance maps of the anisotropies (right), for \LY\ (top), joint \LY\ (our sample $+$ ACC, middle), and \LbcgT\ (bottom). All the maps in this work are shown in a Hammer projection.}
        \label{anis-absor-plots}
\end{figure*}

Therefore, any statistically significant anisotropies in the \LY\ relation should mainly be caused by unaccounted X-ray absorption effects acting on \Lx. To scan the sky, we adopt a $\theta=60^{\circ}$ cone. This returns at least 72 clusters for each cone, which, considering the very low scatter of the relation, are sufficient to robustly constrain $A_{LY}$. The variation and significance maps are displayed in Fig. \ref{anis-absor-plots}. We detect the most anisotropic sky region toward $(l,b)=({118^{\circ}}^{+39^{\circ}}_{-31^{\circ}},{+7^{\circ}}^{+41^{\circ}}_{-12^{\circ}})$ which shares a $3.5\sigma$ anisotropy with the rest of the sky. The 100 clusters within this region appear to be $12\pm 3\%$ fainter in average than the rest of the sky. The extra $N_{\text{Htot}}$ needed to explain this discrepancy is $\sim 3.9\pm 1.1\times 10^{20}/$cm$^2$ (the uncertainties were symmetrized). If the assumed hydrogen quantity based on W13 was indeed the true one, then the metallicity of the absorbing material toward that direction would need to be $Z\sim 1.54\pm 0.16\ Z_{\odot}$ (currently assumed to be $Z=Z_{\odot}$). Considering that the specific region lies close to the Galactic plane, this metallicity value does not seem unlikely. The $A_{LY}$ and the anisotropy significance maps of \LY\ are displayed in Fig. \ref{anis-absor-plots}. One can see that there are no anomalously bright regions. This indicates that there are no regions with significantly lower-than-solar metallicities of the Galactic material. Assuming availability of a much larger number of clusters with \Lx\ and \Ysz\ measurements, ideally extending to low Galactic latitudes, this scaling relation could potentially be used as a new probe of the ISM metallicity.

It should be stressed that in M20, the most anisotropic direction for the \LT\ relation was found to be $(l,b)\sim (300^{\circ},-20^{\circ})$. Based on the above test, this region does not show any signs of extra, previously unaccounted absorption. Adding up to the numerous tests done in M20, this further supports the hypothesis that the observed \LT\ anisotropies are not caused by unmodeled Galactic effects.

\subsubsection{Joint analysis of the \LY\ relation for our sample and ACC}\label{LYjoint}

If the \LY\ anisotropies seen in our sample indeed originate by the unaccounted absorption effects of a yet undiscovered mass (or a higher interstellar gas and dust metallicity), then we should obtain similar results for the ACC sample. Before extrapolated to \Lbol, the flux of the ACC clusters was initially measured within the 0.5-2 keV energy range, and thus is sensitive to X-ray absorption effects. Therefore, jointly analyzing the two samples should provide us with better insights for any possible X-ray absorption issues.

To combine the results of the two independent cluster samples, we perform a joint likelihood analysis. The applied method is the one followed in M20 (Sect. 8.2) where we determined the overall apparent $H_0$ variation. Here, the joint parameter is the \LY\ normalization for every region over the normalization of the full sample ($A_{LY}/A_{LY,\text{all}}$), marginalized over the slope. We extract the posterior likelihood of $A_{LY}/A_{LY,\text{all}}$ for every sky region, for both samples. Then by multiplying the two posterior likelihoods, we obtain the combined, final one.

Performing the joint analysis, we find that the most anisotropic region lies toward $(l,b)=({122^{\circ}}^{+33^{\circ}}_{-34^{\circ}},{+8^{\circ}}^{+37^{\circ}}_{-12^{\circ}})$, where the clusters appear to be fainter in average by $11\pm 5\%$ than the rest of the sky. Our sample dominates the joint fitting due to the much higher number of clusters and lower scatter. However, since ACC does not show a strongly deviating behavior toward that region, the statistical significance of the anisotropy drops to just $2.1\sigma$. To fully explain the mild tension, an undetected \nhtot$\sim 3.5\pm 1.7\times 10^{20}$/cm$^2$ would be needed, or alternatively a $Z\sim 1.49\pm 0.23\ Z_{\odot}$ for the already-detected Galactic gas and dust.

As such, the tension could be attributed to chance and not necessarily to an unaccounted X-ray absorption on top of the already applied one. This is also indicated by the MC simulations later on. The normalization and sigma maps of the \LY\ anisotropies can be found in Fig. \ref{anis-absor-plots}.

\subsection{\LxLbcg\ anisotropies}\label{LxLbcg}

Following the same reasoning as for \LY, the \LxLbcg\ scaling relation cannot detect cosmological anisotropies or BFs, since both quantities depend on $D_L$ and the slope is close to unity. Also, \Lbcg\ is not expected to suffer from any excess extinction, since the near-infrared K$_{\text{S}}$ filter of 2MASS shows a nearly negligible sensitivity on extinction \citep{schlafly}, and the \Lbcg\ values were already corrected for the known extinction effects. Consequently, the only origin of any observed (statistically significant) anisotropies should be a stronger true X-ray absorption than the adopted one, affecting \Lx. We adopt a $\theta=75^{\circ}$ cone to scan the sky so each cone contains $\geq 70$ objects. Fewer clusters per cone might lead to strong cosmic \emph{and} sample variance effects which can result in overestimated anisotropic signals. Empirically, we conclude that 70 is a sufficient number of clusters to (mostly) avoid such effects. The anisotropy maps of \LxLbcg\ are displayed in the bottom panel of Fig. \ref{anis-absor-plots}. 

The most anisotropic region turns out to be again the one with the lowest $A_{L_{\text{X}}L_{\text{BCG}}}$, toward $(l,b)=({171^{\circ}}^{+29^{\circ}}_{-61^{\circ}},{-22^{\circ}}^{+23^{\circ}}_{-32^{\circ}})$. It appears to be $23\pm 14 \%$ dimmer than the rest of the sky. Its statistical significance however does not overcome $1.7\sigma$, and therefore the relation is consistent with being statistically isotropic. This might be due to the large scatter and parameter uncertainties of the relation and not necessarily due to the lack of anisotropy-inducing effects. The reported direction is also $57^{\circ}$ away from the faintest direction found by the joint \LY\ analysis, although within $\leq 1.5\sigma$. The necessary excess \nhtot\ to explain this mild discrepancy in the \LxLbcg\ relation is $\sim 8.7\pm 5.1\times 10^{20}/$cm$^2$. Alternatively, a metal abundance of $Z\sim 2.4\pm 0.8\ Z_{\odot}$ of the already-detected hydrogen cloud would also alleviate this small tension. Thus, the \LxLbcg\ relation does not show any indications of previously unknown X-ray absorption, although the large scatter of the relation limits the confidence of our conclusions. Finally, it is noteworthy that for once more, no apparent anisotropy exist toward $(l,b)\sim (300^{\circ},-20^{\circ})$.

\subsection{Comparison between $N_{\text{H,Xray}}$ and \nhtot}

As a final test for detecting potential excess absorption effects, we compare the X-ray determined $N_{\text{H,Xray}}$ with the \nhtot\ value given in W13, used in our default analysis. If a region shows a systematically larger $N_{\text{H,Xray}}$, it could indicate an extra, previously unaccounted X-ray absorption taking place toward there. This comparison is performed for the 156 clusters left after the cuts we apply, described in Sect. \ref{xray_nh}. The best-fit relation is $N_{\text{H,Xray}}=(7.14\pm 0.34)\times \left(\dfrac{N_{\text{Htot}}}{4\times 10^{20}\text{cm}^{-2}}\right)^{0.67\pm 0.06}\times 10^{20}/$cm$^2$. The X-ray based values are systematically higher than \nhtot, except for the high \nhtot\ range where the two values converge. This behavior marginally agrees with the findings of \citet{schellenberger}, when they used Chandra $T$ measurements and the same abundance table with us. However, this \nhtot\ discrepancy should not be taken at face value, since the overall comparison is biased by the exclusion of clusters with large $N_{\text{H,Xray}}$ uncertainties. As discussed in Appendix \ref{details_xray_nh}, these are mostly low $N_{\text{H,Xray}}$ clusters lying below the equality line since the $N_{\text{H,Xray}}$ measurement for these clusters is very challenging, due to the lower spectral cut at 0.7 keV. This selection effect would, therefore, tend to flatten the slope. Thus, this systematic upscatter of the remaining $N_{\text{H,Xray}}$ values, does not necessarily mean that the true X-ray absorption is higher than previously thought, but it is probably the result of unaccounted systematics. To avoid such issues as much as possible, we just compare the results of a region against the rest of the sky to evaluate if this region deviates more from the W13 values, compared to the rest of the sky. This assumes that any systematics should not be direction-dependent. It should be borne in mind that this is an approximate, complementary test due to its limitations, and not a stand-alone check for excess X-ray absorption.

\begin{figure}[hbtp]
               \includegraphics[width=0.47\textwidth, height=6cm]{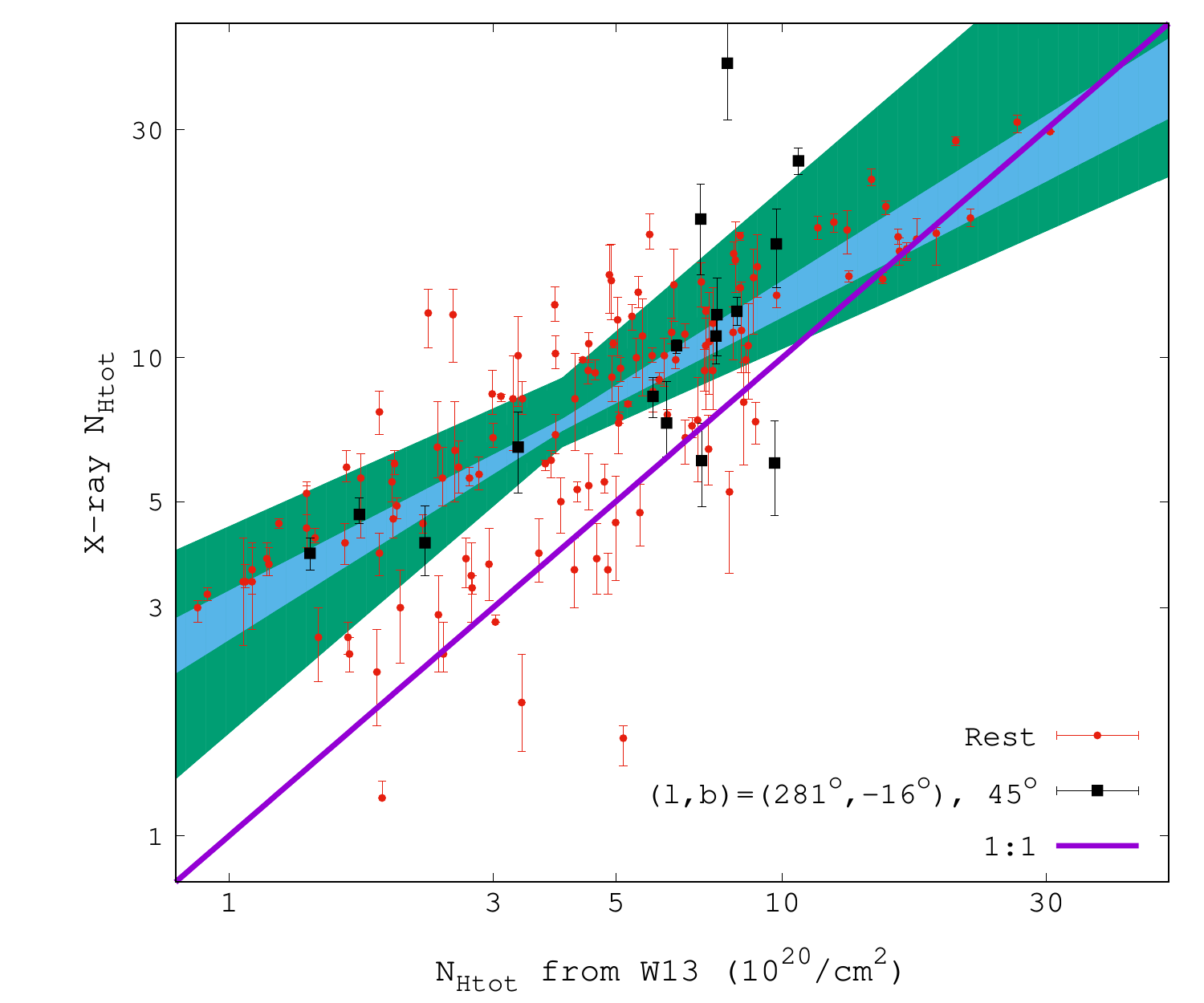}
               \includegraphics[width=0.47\textwidth, height=6cm]{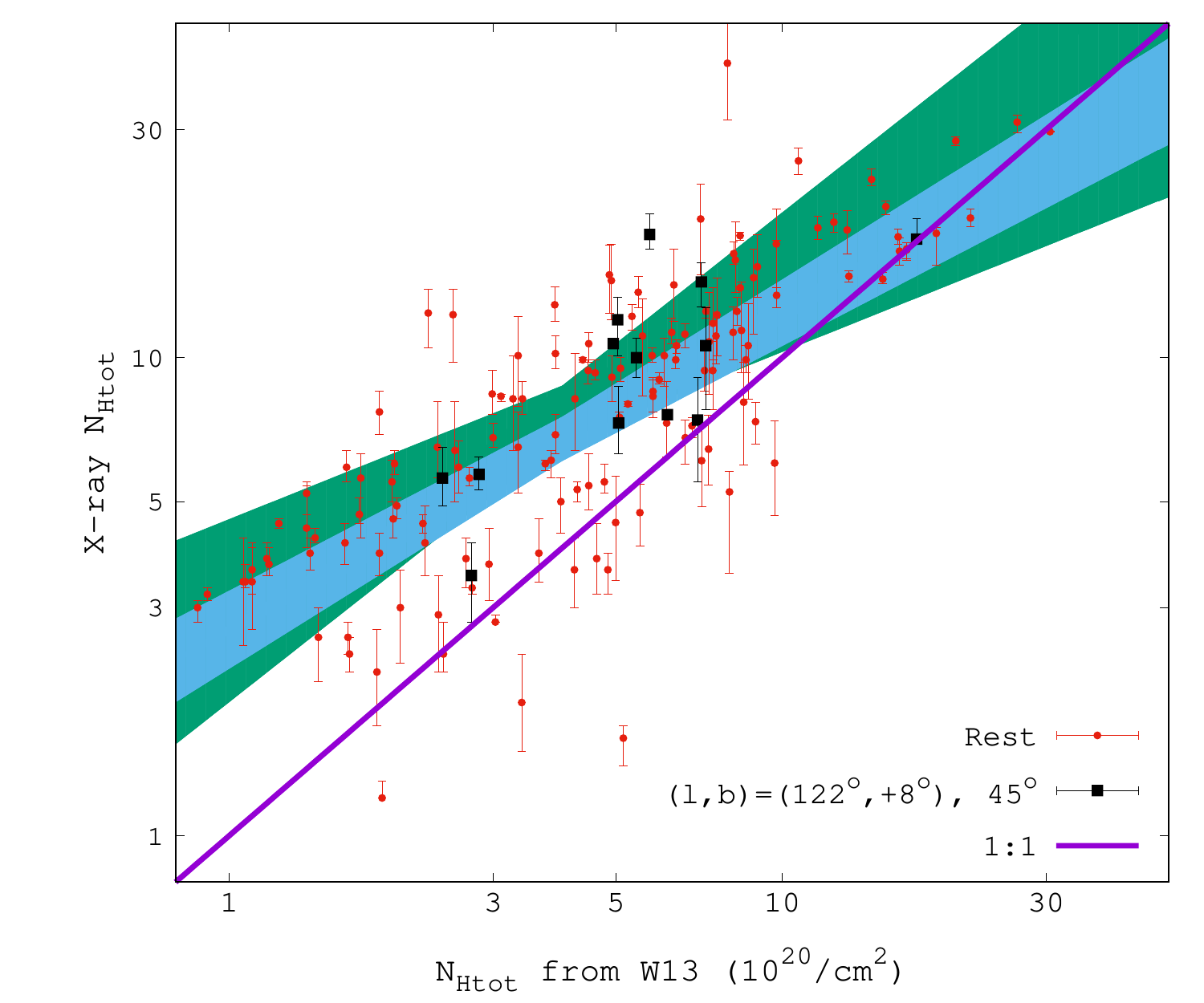}
               \caption{Comparison between the \nhtot\ from W13 and the \nhtot\ constrained by our X-ray spectral analysis. The equality line is displayed (purple), as well as the $1\sigma$ space of the best-fit function for the region of interest (green, black points) and for the rest of the sky (cyan, red points). The reason for the systematic difference is given in the main text. The region of interest is within $45^{\circ}$ from $(l,b)=(281^{\circ} ,-16^{\circ})$ (top) and from $(l,b)=(122^{\circ},+8^{\circ})$ (bottom).}
        \label{nh_comp}
\end{figure}

Firstly, we consider the region within $45^{\circ}$ around $(l,b)=(281^{\circ} ,-16^{\circ})$. This is the most anisotropic and faintest region of the \LT\ relation as found in M20, when only our sample is considered. For the 16 clusters lying within this region We find that the best-fit relation is $N_{\text{H,Xray}}=(7.64\pm 1.22)\times \left(\dfrac{N_{\text{Htot}}}{4\times 10^{20}\text{cm}^{-2}}\right)^{0.75\pm 0.25}\times 10^{20}/$cm$^2$. As also shown in the upper panel of Fig. \ref{nh_comp}, the behavior of this region is completely consistent with the rest of the sky. Assuming that the low \nhtot\ clusters indeed have a larger, true $N_{\text{H,Xray}}$, then the clusters of the tested region would be less biased, since they have a larger median \nhtot, compared to the rest of the sky. Consequently, there is no indication that an untraced, excess X-ray absorption is the cause behind the \LT\ anisotropies.

The second region we consider is within $45^{\circ}$ around $(l,b)=(122^{\circ} ,+8^{\circ})$. This region shows mild indications of uncalibrated true X-ray absorption that differs from the W13 values. Using its 11 clusters, we find $N_{\text{H,Xray}}=(7.56\pm 1.04)\times \left(\dfrac{N_{\text{Htot}}}{4\times 10^{20}\text{cm}^{-2}}\right)^{0.72\pm 0.24}\times 10^{20}/$cm$^2$. The results (Fig. \ref{nh_comp}) are again consistent with the rest of the sky, not revealing any signs of a biased applied X-ray absorption correction. 

\subsection{Overall conclusion on possible X-ray biasing effects on the anisotropy studies}\label{over_abs}

The main purpose of this section's analysis was to address if the strong observed \LT\ anisotropy found in M20, is caused by unaccounted soft X-ray absorption, not traced by the \nhtot\ values of W13. For instance, a yet undiscovered gas and dust cloud, or some galactic dust with oversolar metallicities, could cause such an effect. Additionally, we also wished to discover if there is such a region elsewhere in the sky.

All the different tests we performed, mostly independent to each other, did not show any signs of possible absorption biases toward $(l,b)\sim (300^{\circ},-20^{\circ})$. Future tests, particularly with the eROSITA All-Sky Survey (eRASS), will reveal more information on that topic. For now though, we can safely conclude that the anisotropies found in M20 are \emph{not} the result of any biases in the applied soft X-ray absorption correction. 

If a mysterious \emph{hard} X-ray absorbing material existed that did not affect soft X-rays, this would have an effect only on the measured $T$. However, this would lead to underestimated $T$ values, which would cause the opposite anisotropic behavior compared to the M20 results. So this hypothetical scenario is also rejected.

Finally, using the \LY\ relation for our sample and ACC, we identified a region, $(l,b)\sim (122^{\circ},+8^{\circ})$, that might imply an extra needed soft X-ray absorption correction for the clusters there. However, the statistical significance of this result is only $\sim 2\sigma$, while the \LxLbcg\ and the $N_{\text{H,Xray}}$ tests did not reveal any deviations toward that region. Future work will again give a clearer answer, but for now we conclude that no extra correction is needed for the \Lx\ values of the clusters lying within this region.

\section{Anisotropies due to bulk flows or $H_0$ variations}\label{bf_cosmo_anisot}

We have established that strong biases in the X-ray measurements related to previously uncalibrated absorption do not seem to exist, particularly toward $(l,b)\sim (300^{\circ},-20^{\circ})$. The anisotropies observed in M20 thus cannot arise from such issues. Two more possible origins of such anisotropies are BF motions or spatial variations of cosmological parameters. The first adds a systematic, nonnegligible, local velocity component on the measured redshift of the objects toward a particular direction. If not accounted for, it leads to a systematic over- or underestimation of the clusters' distances, and hence of their cosmology-dependent quantities. The same is true if the real underlying values of the cosmological parameters, e.g., $H_0$, vary from region to region. This spatial $H_0$ variation in that case only has to extend up to the redshift range of our samples, as the result of a local, unknown effect, while a convergence to isotropy at larger scales would be perfectly consistent with our results. We proceed to search for anisotropies in scaling relations sensitive to these two phenomena, and attempt to quantify the BF, or the needed $H_0$ variation to explain the observations.

\subsection{The \LT\ relation}

Our inference of \Lx\ would be strongly affected by either BFs or a cosmological anisotropy through the assumed luminosity distance (\Lx$\sim D_L^2$). At the same time, our $T$ measurement would remain relatively unchanged. This allows us to predict the \Lx\ values of the clusters across the sky based on their $T$ and the globally calibrated \LT, and attribute any directionally systematic deviations to BFs or $H_0$ anisotropies.

\subsubsection{Our sample}

The anisotropies of the \LT\ relation are extensively discussed in M20. Here we update our results based on the new statistical methods we follow, whose differences with M20 are described in Sect. \ref{differences}. Based on the observed scatter and the number of available clusters, we consider $\theta=75^{\circ}$ cones to scan the sky. The maximum anisotropy is detected toward $(l,b)=(274^{\circ}\pm 43^{\circ},-9^{\circ}\pm 32^{\circ})$ (120 clusters), deviating from the rest of the sky by $19\pm 7\%$ at a $2.8\sigma$ level. The decreased statistical significance of the anisotropy compared to the M20 results ($3.64\sigma$ there) is due to the more conservative parameter uncertainty evaluation which is now based on bootstrap resampling with marginalization over the slope, and the weaker statistical weighting of the clusters close to the center of the considered cones. Also, here we do not compare the two most extreme regions with opposite behavior, but only the most extreme one with the rest of the sky. The $A_{LT}/A_{LT,\text{all}}$ and the sigma maps are given in Fig. \ref{anis-cosmo-plots}. As expected, they are not significantly different than the ones obtained in M20. The most noticeable difference is the lack of the particularly bright region close to the Galactic center. Now the brightest region is located $\sim 150^{\circ}$ away from the faintest, indicating an almost dipolar anisotropy.

\paragraph{Cosmological anisotropies and bulk flows}

We assume that the cause of the observed tension is an anisotropic $H_0$ value, in a Universe without any BFs. One would need $H_0=66.0\pm 1.7 $ km/s/Mpc toward $(l,b)\sim (274^{\circ},-9^{\circ})$, and $H_0=72.1\pm 1.4$ km/s/Mpc for the rest of the sky. The respective Hubble diagrams are compared in the top panel of Fig. \ref{hub-diag}.

We now assume that a BF motion is the sole origin of the observed anisotropies, with $H_0$ being isotropic. Based on the MR method, we find a BF of \ubf$=980\pm 300$ km/s toward $(l,b)=(315^{\circ}\pm 34^{\circ},-10^{\circ}\pm 20^{\circ})$ for the full sample. This result is dominated however by lower $z$ clusters, since there are only 26 clusters with $z>0.2$. For the redshift range $z\in [0,0.06]$, we obtain \ubf$=1100\pm 410$ km/s toward $(l,b)=(318^{\circ}\pm 37^{\circ} ,-5^{\circ}\pm 23^{\circ})$. One retrieves a similar BF for clusters within 270 Mpc as for the full sample. Both the direction and amplitude of the BF, as well as its statistical significance, stay within the uncertainties as we consider iteratively larger volumes. The detailed results are given in Table \ref{BF_motions}. For the concentric redshift bins $z\in [0.06,0.12]$ and $z\in [0.12,0.3]$, one obtains \ubf$=1170\pm 400$ km/s toward $(l,b)=(262^{\circ}\pm 52^{\circ}, +2^{\circ}\pm 26^{\circ} )$, and \ubf$=1040\pm 570$ km/s toward $(l,b)=(253^{\circ}\pm 60^{\circ}, -18^{\circ}\pm 31^{\circ})$ respectively. The BF direction is consistent within $1\sigma$ between all redshift shells, not showing any convergence to zero even at $\gtrsim 500$ Mpc. The limited number of clusters beyond $\sim 500$ Mpc poses a challenge however for the precise pinpointing of the BF at larger scales. The results are in tension with $\Lambda$CDM which predicts much smaller BFs at scales of $\gtrsim 200$ Mpc \citep[e.g.,][]{li12,carrick,qin}. 

Using the MA method, we find a BF of \ubf$=600\pm 260$ km/s toward $(l,b)=(298^{\circ}\pm 25^{\circ},-21^{\circ}\pm 18^{\circ})$ for the full sample. When this BF is applied to our data, the \LT\ relation is consistent with isotropy within $1.4\sigma$ based on the usual sky scanning. Unfortunately, for the MA method we need to consider broader redshift bins than with the MR method, in order to have enough data available per sky patch. For the redshift bins $z\in [0,0.09]$\footnote{Here we go beyond the median $z$, since low $z$ clusters exhibit a larger scatter (due to galaxy groups), thus we need more than half our sample to obtain valuable constraints. For higher redshifts, the scatter reduces, so fewer clusters can also be used.} we find \ubf$=690\pm 300$ km/s toward $(l,b)=(268^{\circ}\pm 31^{\circ},-5^{\circ}\pm 23^{\circ})$, while iteratively increasing the cosmic volume does not significantly affect this result. For $z>0.09$, while the anisotropy toward $(l,b)\sim (270^{\circ} ,-25^{\circ} )$ persists at a $\sim 2\sigma$ level, the search for a BF is inconclusive due to the limited number of clusters, which leads to large uncertainties. To get an idea of the possible BF signal in the MA method purely at larger scales, one can exclude local clusters ($\leq 300$ Mpc, $z<0.067$)\footnote{This is the scale that many studies consider for studying BFs or local voids \citep[e.g.,][]{jla,carrick}, see later discussion.}, and only consider the 170 clusters at larger distances ($z>0.067$). For these, we obtain  \ubf$=620\pm 310$ km/s toward $(l,b)=(293^{\circ}\pm 39^{\circ},-12^{\circ}\pm 27^{\circ} )$. Although there is some overlap with the $z\in [0,0.09]$ results, these findings serve as a hint for BFs extending to larger scales. 

We see that both methods reveal large BFs, with no signs of fading when larger volumes or different redshift bins are considered. The direction appears to be relatively consistent between both methods, well within the $1\sigma$ uncertainties. The main difference is that the MA method returns a smaller amplitude for the BF, consistent within $1\sigma$ though with the MR method. Finally, one sees that the \LT\ anisotropies are not subject to a specific redshift bin but consistently extend throughout the $z$ range.

\subsubsection{Joint analysis of the \LT\ relation for our sample and ACC}

\begin{figure*}[hbtp]
               \includegraphics[width=0.49\textwidth, height=4.5cm]{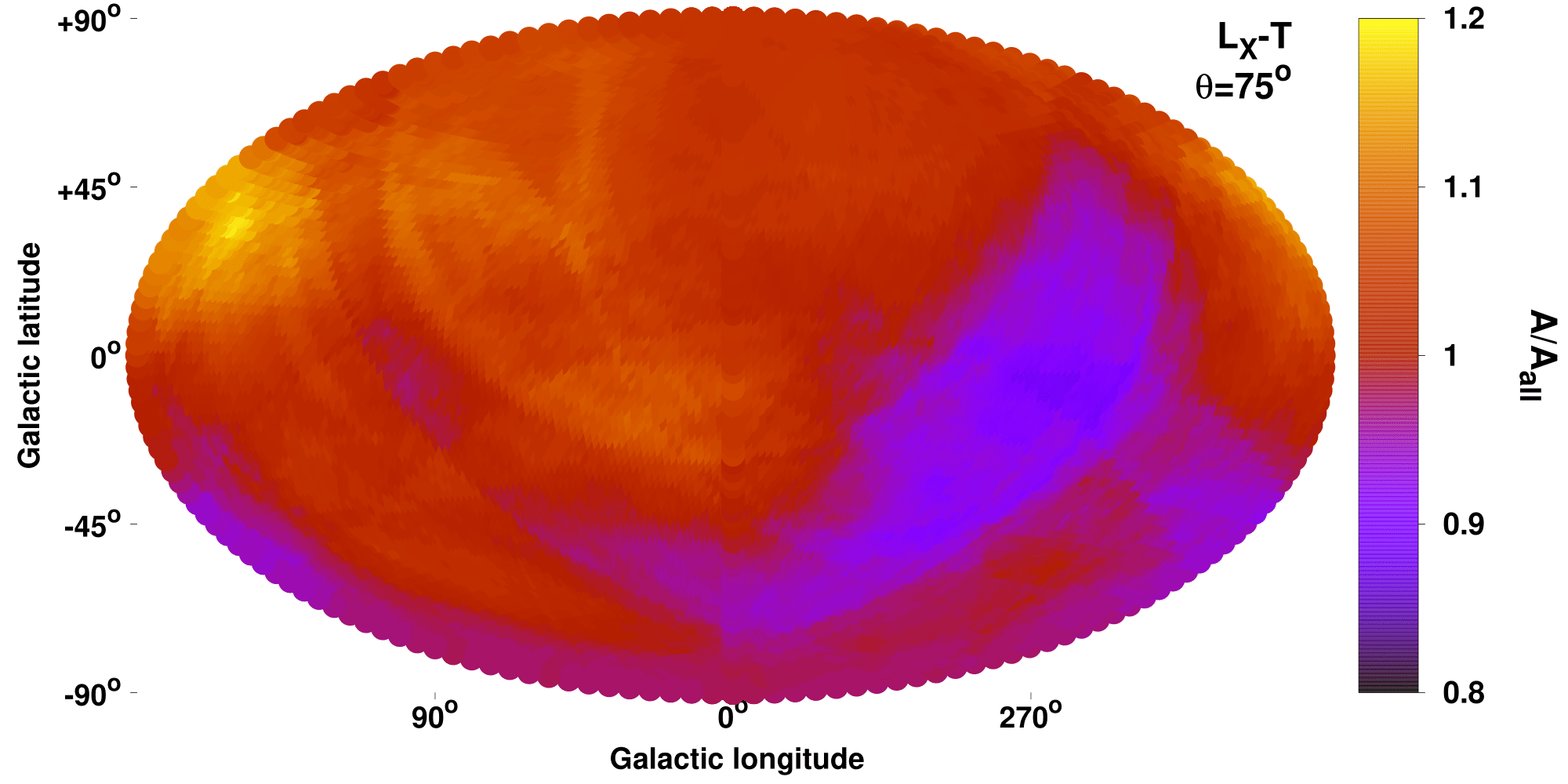}
               \includegraphics[width=0.49\textwidth, height=4.5cm]{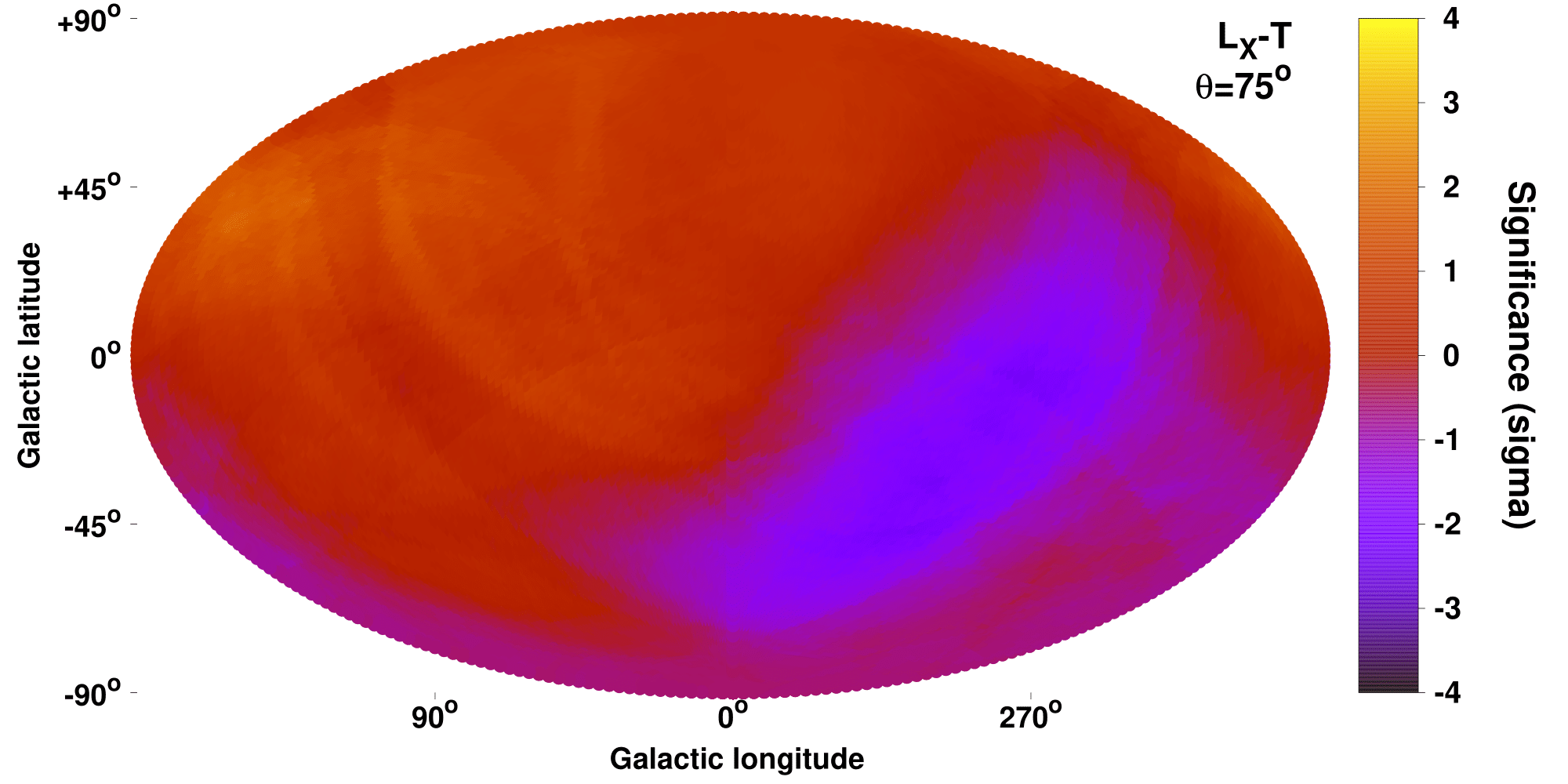}
                \includegraphics[width=0.49\textwidth, height=4.5cm]{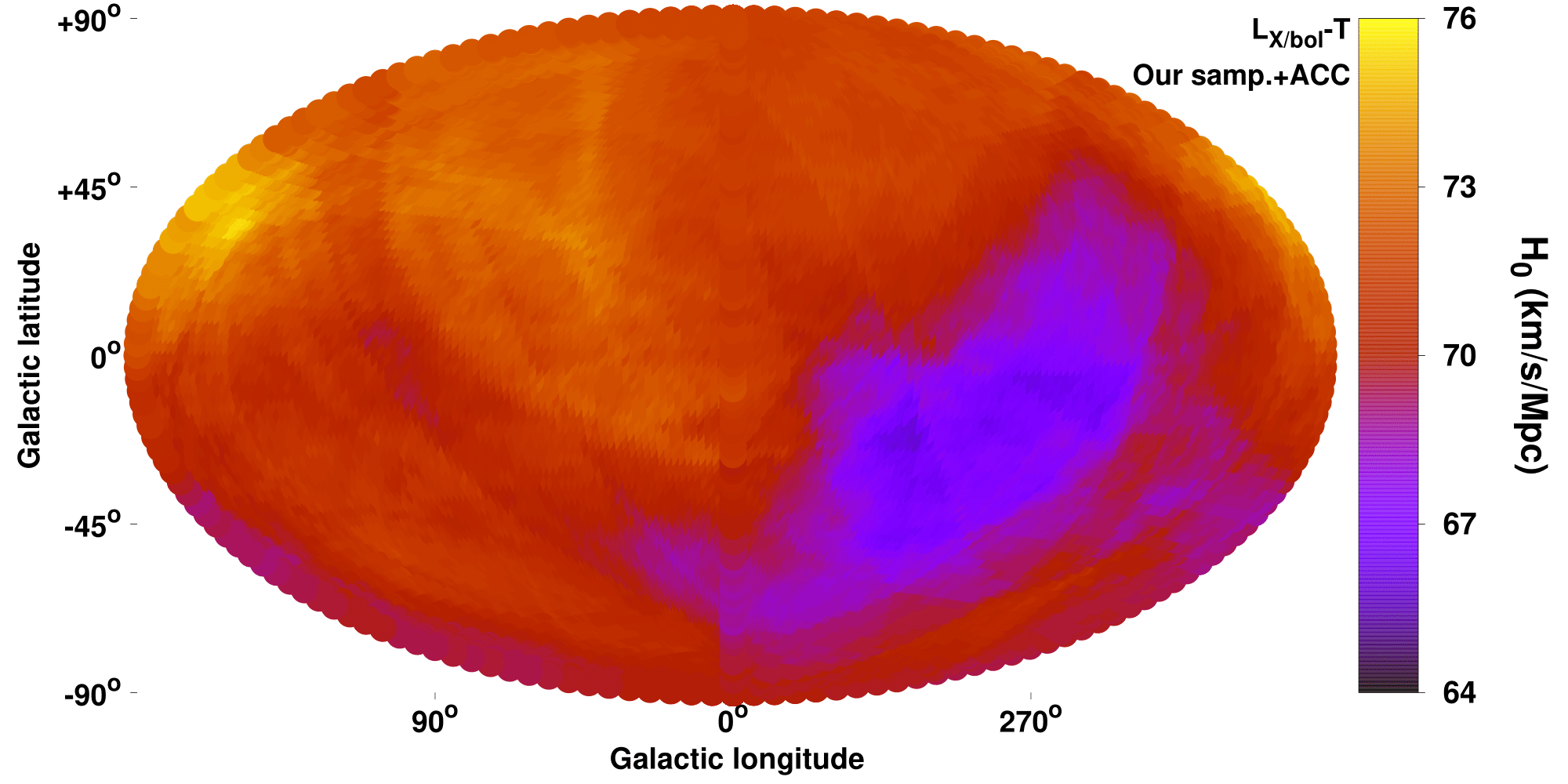}
               \includegraphics[width=0.49\textwidth, height=4.5cm]{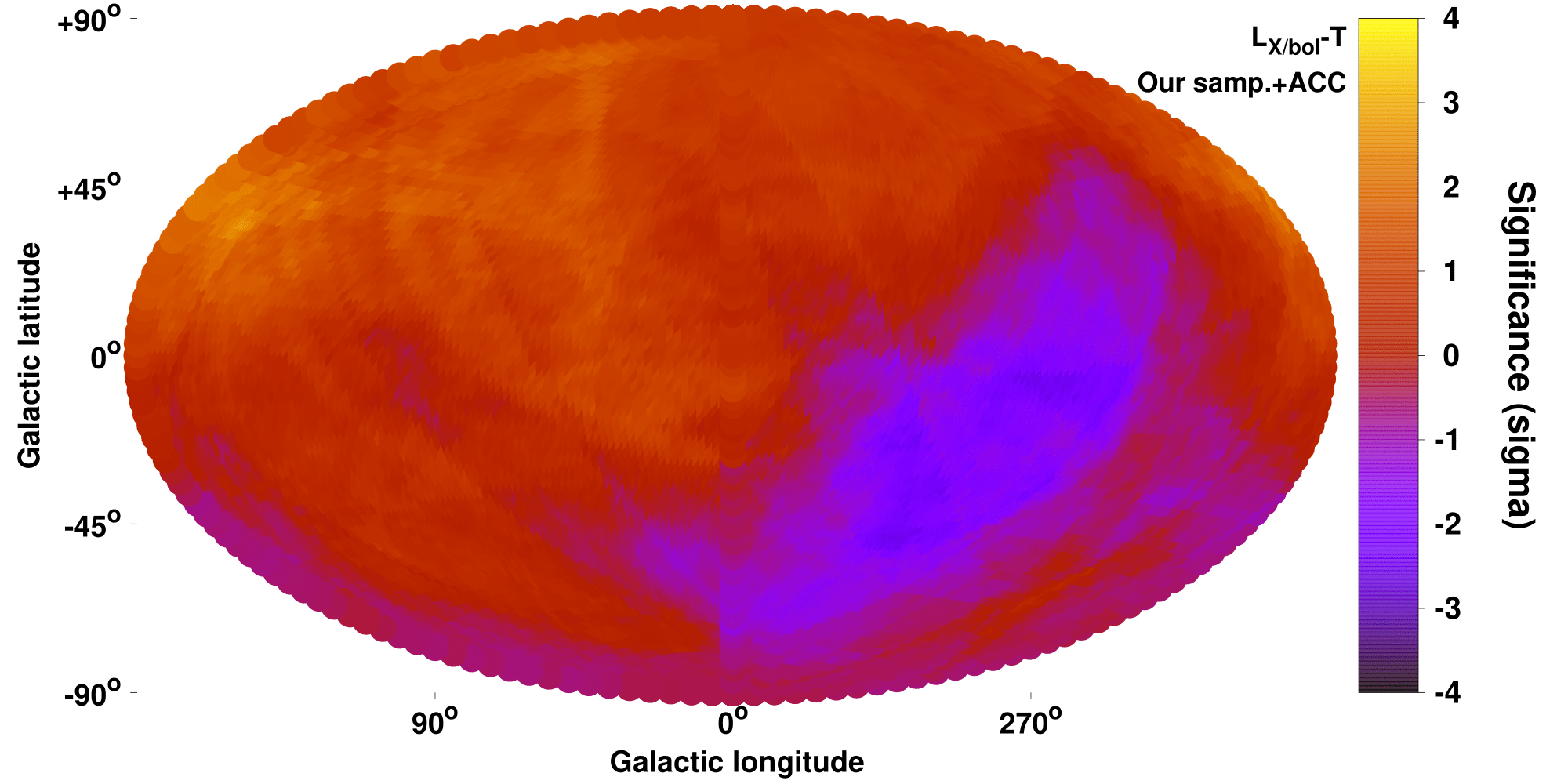}
               \caption{\textit{Top}: Same as in Fig. \ref{anis-absor-plots}, for \LT. \textit{Bottom}: The $H_0$ anisotropy map derived from the joint \LT\ (our sample+ACC).}
        \label{anis-cosmo-plots}
\end{figure*}

\begin{figure}[hbtp]
               \includegraphics[width=0.45\textwidth, height=7cm]{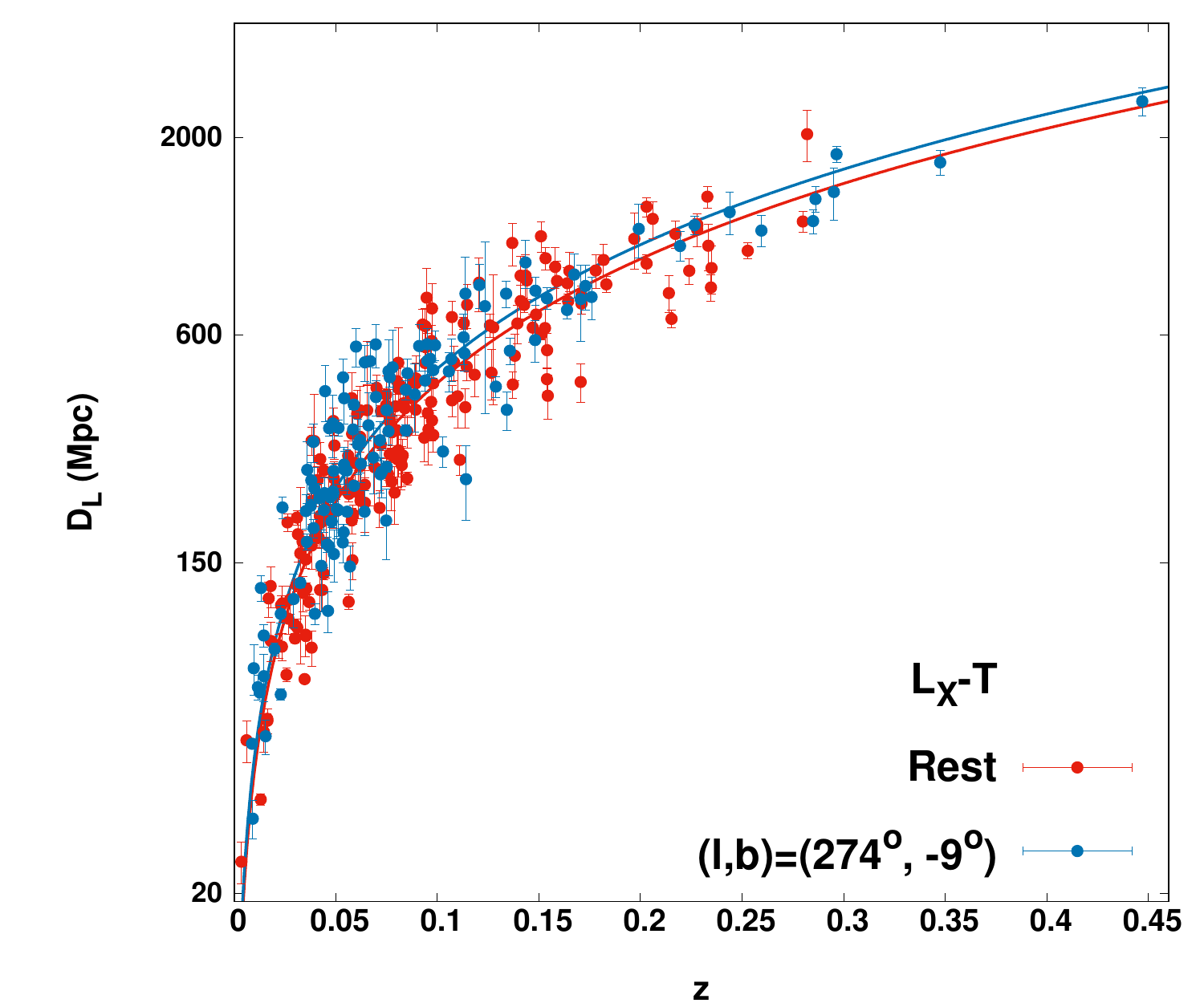}
               \includegraphics[width=0.45\textwidth, height=7cm]{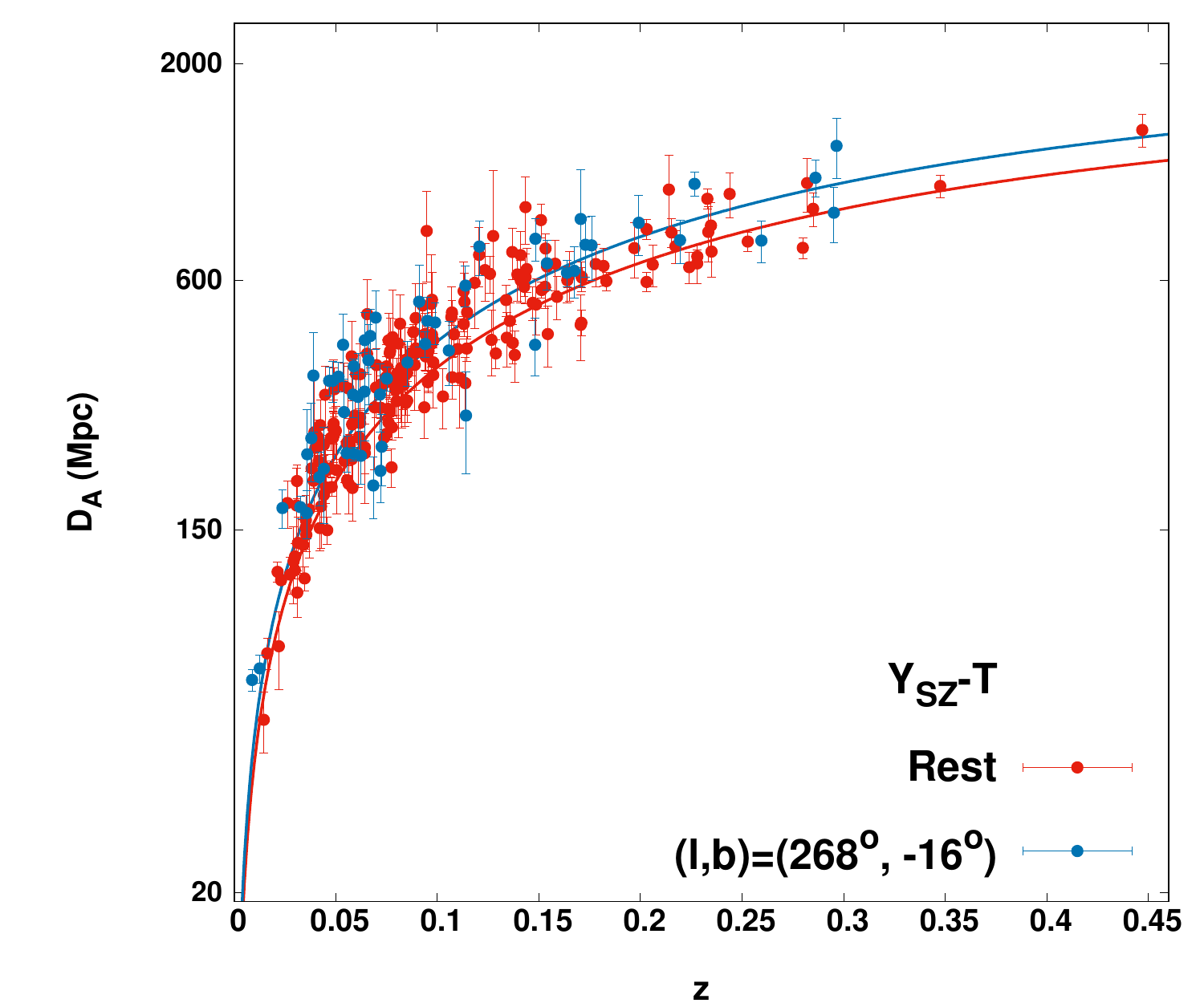}
               \caption{Hubble diagram of galaxy clusters as derived by the \LT\ (top) and the \YT\ (bottom) relations. The clusters from the most anisotropic region of each scaling relation are displayed (blue), together with the clusters from the rest of the sky (red). The best-fit lines are displayed with the same color.}
        \label{hub-diag}
\end{figure}

As shown in M20 (and in Appendix \ref{acc-LT}), ACC shows a similar anisotropic behavior with our sample, even though it is completely independent of the latter. We perform a joint likelihood analysis of the two independent samples, with the 481 individual clusters they include. We express the apparent anisotropies in terms of $H_0$, similar to Fig. 23 of M20. The results are plotted in Fig. \ref{anis-cosmo-plots}. $H_0$ seems to vary within $\sim 65-76$ km/s/Mpc. The most anisotropic region is found toward $(l,b)=({284^{\circ}}^{+31^{\circ}}_{-12^{\circ}} ,{-4^{\circ}}^{+7^{\circ}}_{-23^{\circ}})$. Its best-fit is $H_0=66.2\pm 1.6$ km/s/Mpc , while for the rest of the sky one gets $H_0=72.7\pm 1.5$ km/s/Mpc. While the posterior $H_0$ range is identical to the one found in M20, the significance of the anisotropies do not exceed $3\sigma$, due to the more conservative methodology followed here (and due to the nonuse of XCS-DR1 here). The peak anisotropy is close to our sample's result since it dominates the joint fit, something that also shows from the fact that the $\sigma$ level compared to our sample alone increases only slightly. There is a second, weaker peak on the maximum anisotropy position of ACC.

In terms of a BF motion, there is no meaningful way to combine the two independent datasets in an analytical way similar to the $H_0$ analysis, since any BF has meaning only within a certain $z$ range. The redshift distribution of the two samples differ however, and in every given redshift shell, one data set will dominate over the other.

\subsection{The \YT\ relation}

The anisotropies of the \YT\ relation are presented in this work for the first time. \Ysz\ strongly depends on the angular diameter distance (\Ysz$\sim D_A^2$), which is affected by BFs  and possible spatial changes of the cosmological parameters. At the same time $T$ is independent of these effects and thus the same reasoning as in the \LT\ relation applies. The advantages of the \YT\ relation are three. Firstly, the scatter is clearly smaller than the \LT\ relation allowing for more precise constraints. Secondly, \Ysz\ is unaffected by absorption issues, while $T$ (determined using spectra with photon energies of $>0.7$ keV) only has a weak dependence on uncalibrated absorption effects, with much less severe effects than \Lx. Thus, in practice no \YT\ anisotropies can occur from unaccounted Galactic absorption issues. Finally, due to the applied redshift evolution of \YT, the latter is more sensitive to BFs than \LT.

\subsubsection{Our sample}

Due to the low observed scatter and the 263 clusters with quality \Ysz\ measurements, we consider $\theta=60^{\circ}$ cones to scan the sky. The maximum anisotropy is detected toward $(l,b)=(268^{\circ}\pm 34^{\circ},-16^{\circ}\pm 29^{\circ})$ (57 clusters), deviating from the rest of the sky by $27\pm 7\%$ at a $4.1\sigma$ level. The direction agrees remarkably with the results from the \LT\ relation. The amplitude of the anisotropy is larger, due to the narrower cones, which limit the "contamination" of unaffected clusters in each cone. The statistical significance of the anisotropies is also considerably larger, as a result of the smaller \YT\ scatter and the narrower scanning cones.

\begin{figure*}[hbtp]
               \includegraphics[width=0.49\textwidth, height=4.5cm]{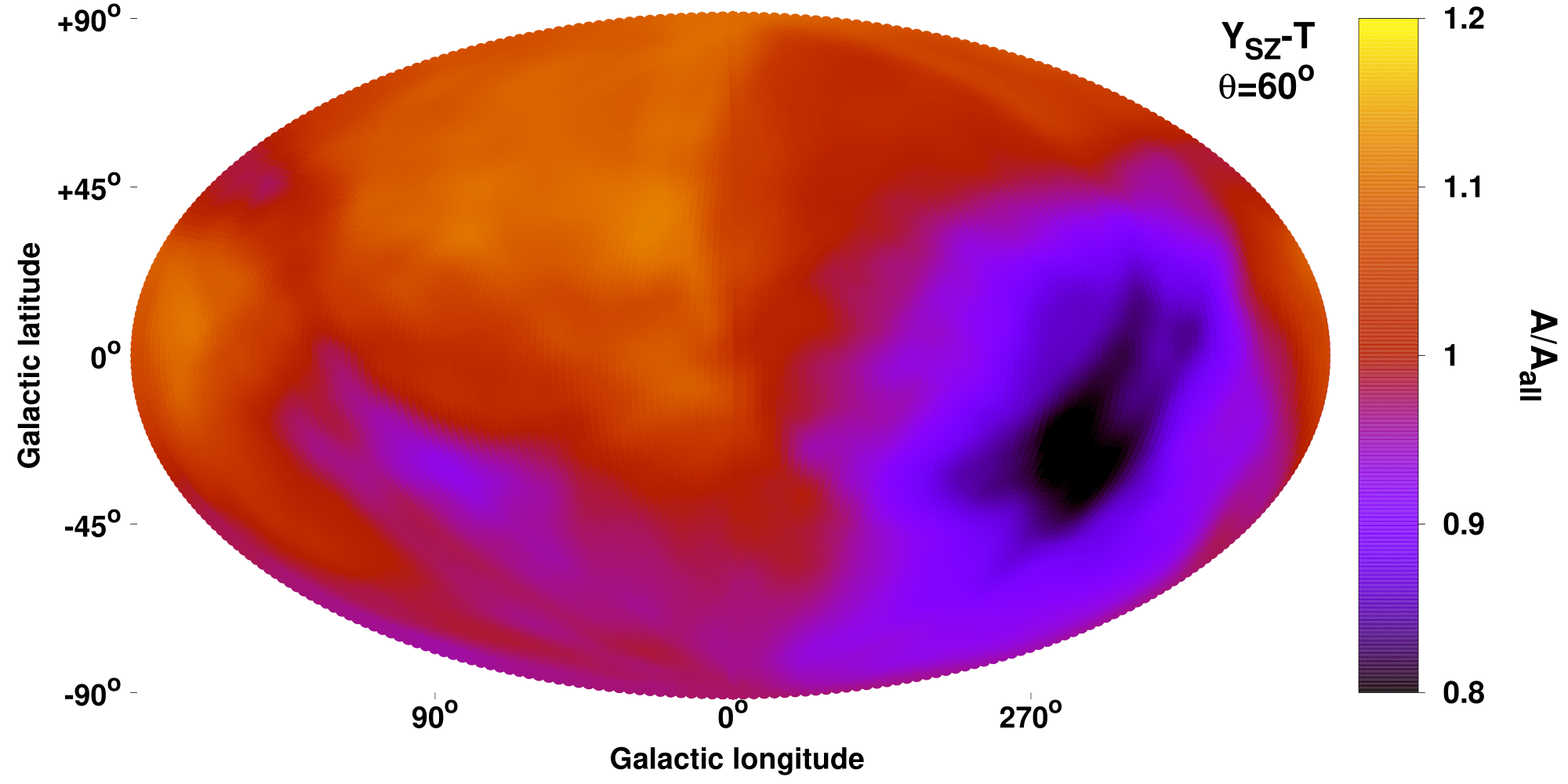}
               \includegraphics[width=0.49\textwidth, height=4.5cm]{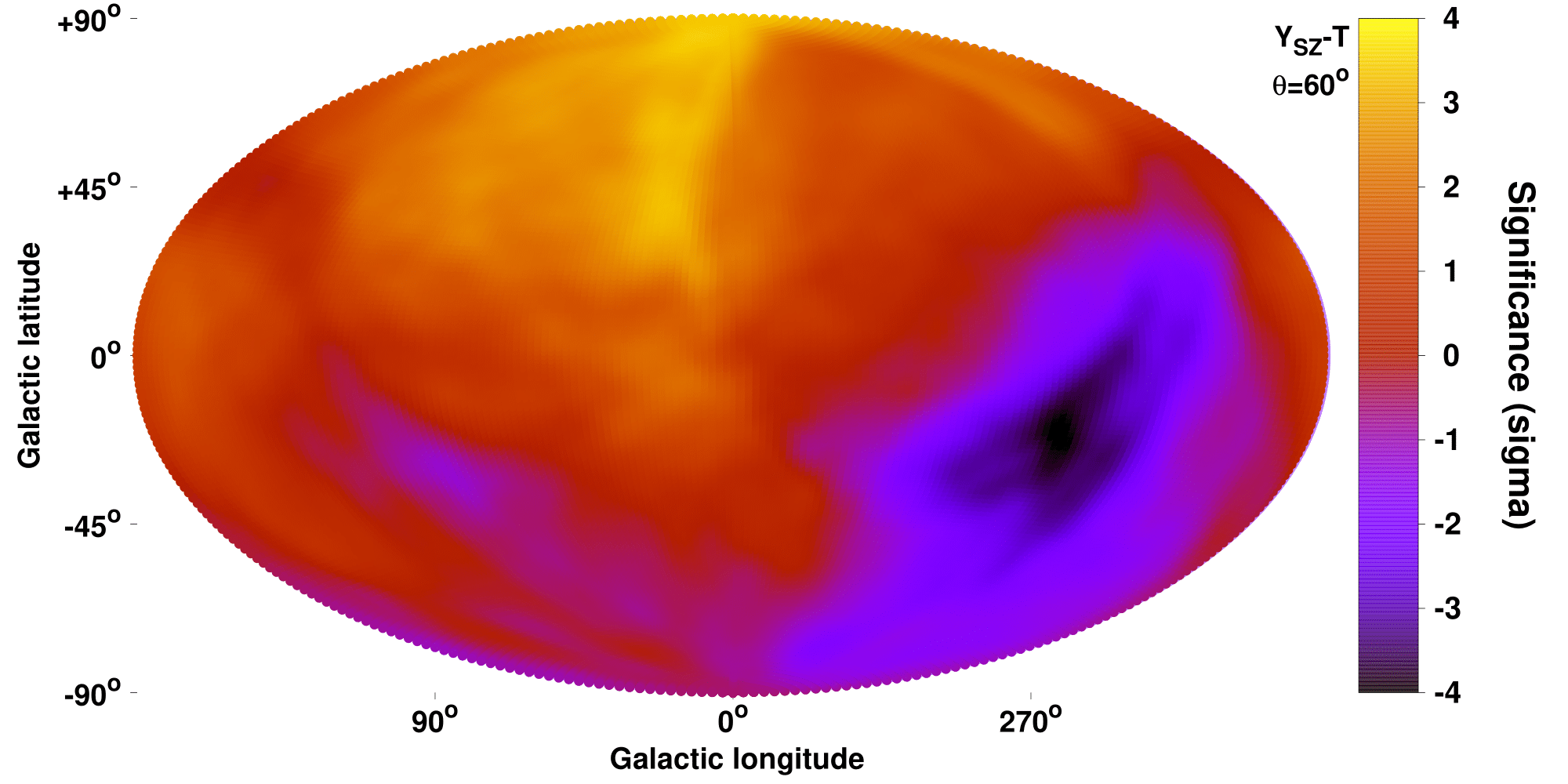}
                \includegraphics[width=0.49\textwidth, height=4.5cm]{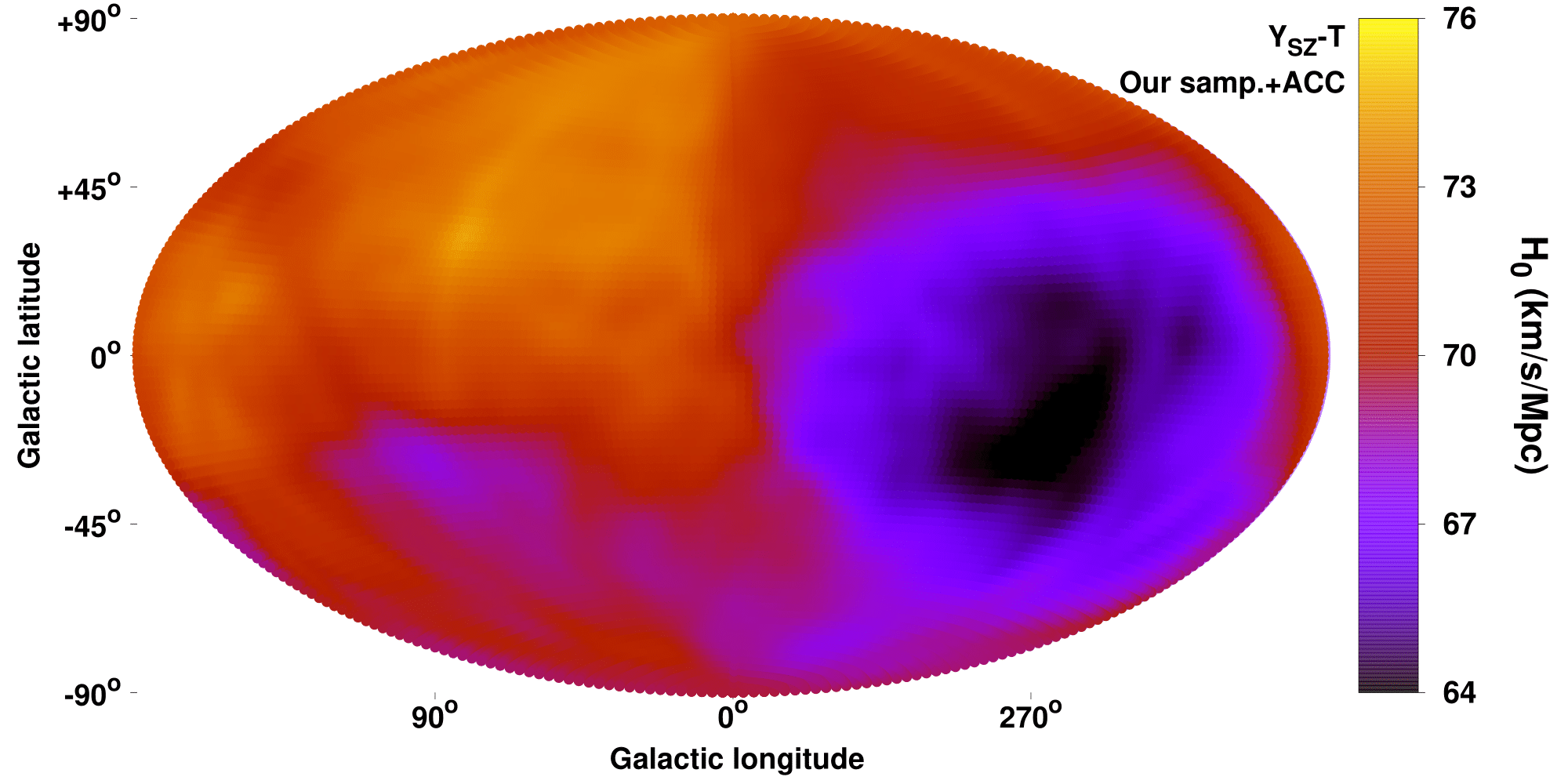}
               \includegraphics[width=0.49\textwidth, height=4.5cm]{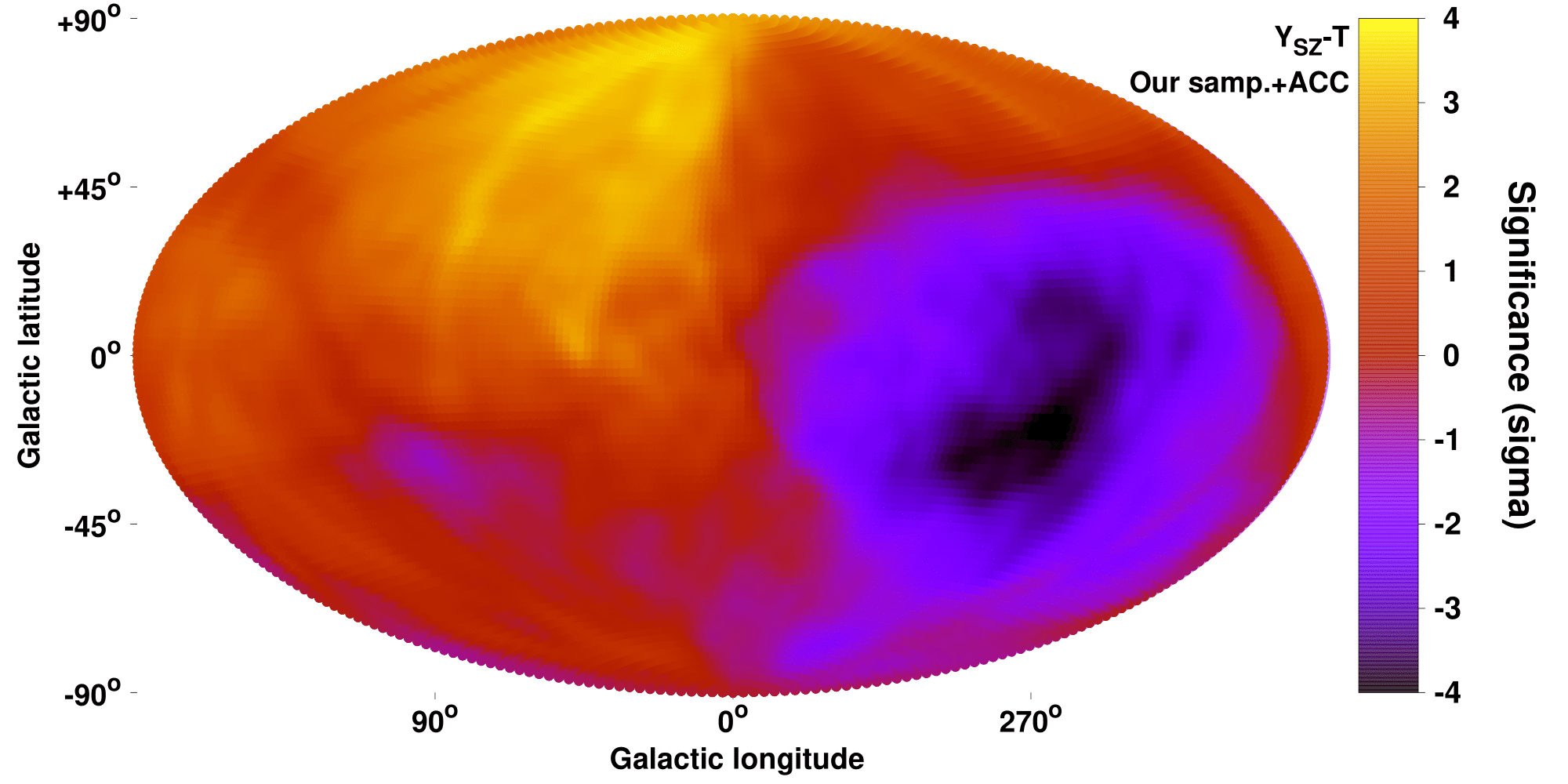}
               \caption{\textit{Top}: Same as in Fig. \ref{anis-absor-plots} for \YT. \textit{Bottom}: The $H_0$ anisotropy map (left) and the respective significance map (right), derived from the joint \YT\ (our sample+ACC, bottom). }
        \label{anis-cosmo-YT}
\end{figure*}

This result strongly demonstrates that the observed \LT\ anisotropies are not due to X-ray absorption issues. The $A_{YT}/A_{YT,\text{all}}$ and the sigma maps are shown in Fig. \ref{anis-cosmo-YT}. Here we should note that there is a mild correlation between the \Lx\ and \Ysz\ scatter compared to $T$ due to the physical state of galaxy clusters. If the origin of the anisotropies was sample-related (e.g., a surprisingly strong archival bias), one would expect to indeed see similar anisotropies in \LT\ and \YT. However, this would still not explain the large statistical significance of the \YT\ anisotropies, or the fact that we see a similar effect in ACC. Nonetheless, we further explore this possibility later in the paper (Sects. \ref{mc_sim} and \ref{scatter_corr}), and confirm that this is not the reason for the agreement of the two relations. Finally, if one uses the \Ysz\ values from PSZ2 instead, one obtains similar anisotropic results (Sect. \ref{psz2-test}).

\paragraph{Cosmological anisotropies and bulk flows}

We now investigate the necessary $H_0$ variations to fully explain the observed anisotropies. One would need $H_0=62.3\pm 2.2 $ km/s/Mpc toward $(l,b)\sim (268^{\circ},-16^{\circ})$, and $H_0=72.1\pm 0.9$ km/s/Mpc for the rest of the sky. The result is consistent within $1.4\sigma$ with the $H_0$ value obtained from the joint \LT\ analysis. The corresponding Hubble diagram is shown in the bottom panel of Fig. \ref{hub-diag}. 

The observed anisotropies cannot be caused by a local spatial variation of $\Omega_{\text{m}}$. The \YT\ relation is rather insensitive to such changes due to the low $z$ of the clusters and the opposite effect that an $\Omega_{\text{m}}$ change would have on the \Ysz\ value, and in the redshift evolution of the relation. However, the anisotropies appear to be even stronger now compared to the \LT\ relation. 

Assuming the apparent anisotropies are due to a BF affecting the entire sample, using the MR method one finds \ubf$=950\pm 340$ km/s toward $(l,b)=(263^{\circ}\pm 39^{\circ},-22^{\circ}\pm 20^{\circ})$. For the redshift bin $z\in [0,0.07]$ we obtain \ubf$=1060\pm 390$ km/s toward $(l,b)=(254^{\circ}\pm 42^{\circ},-17^{\circ}\pm 19^{\circ})$, similar to the full sample's results. By gradually expanding the redshift range, the best-fit results remain within the uncertainties, without an amplitude decay. For the $z\in [0.07,0.12]$ bin, we obtain \ubf$=840\pm 490$ toward $(l,b)=(312^{\circ}\pm 61^{\circ},-34^{\circ}\pm 31^{\circ})$. The BF drifts by $\sim 50^{\circ}$, but remains within the (large) uncertainties and within the marginal anisotropic region. The amplitude is also slightly decreased, but still well above the $\Lambda$CDM prediction for such scales. For the $z>0.12$ clusters we find \ubf$=1110\pm 670$ km/s toward $(l,b)=(321^{\circ}\pm 103^{\circ},-42^{\circ}\pm 45^{\circ})$. There is some indication that the BF persists toward a similar direction up to scales larger than $\sim 500$ Mpc, however the poor constraining power of the $z>0.12$ subsample does not allow for robust conclusions.

From the MA method, we obtain \ubf$=960\pm 290$ km/s toward $(l,b)=(267^{\circ}\pm 22^{\circ},-28^{\circ}\pm 16^{\circ})$, completely consistent with the MR method. Within $z\leq 0.09$, the constrained BF is \ubf$=1200\pm 350$ km/s toward $(l,b)=(254^{\circ}\pm 22^{\circ},-18^{\circ}\pm 16^{\circ})$. For $z>0.09$, the amplitude is reduced (\ubf$=720\pm 380$ km/s), while its direction $(l,b)=(242^{\circ}\pm 87^{\circ},-13^{\circ}\pm 22^{\circ})$ is rather uncertain (pointing however toward a similar sky patch).

One sees that the \YT\ anisotropies reveal similar BFs than the \LT\ case. In both cases, the MR and MA methods agree on the full sample and the low $z$ regime, while the results are uncertain for higher redshifts. Both the scale out to which the apparent BF extends, and its amplitude, by far exceed the $\Lambda$CDM expectations. 

\subsubsection{Joint analysis of the \YT\ relation for our sample and ACC}

The ACC sample shows a very similar anisotropic \YT\ behavior to our sample (Appendix \ref{acc-YT}). We combine the two independent samples and their 376 different clusters, to jointly constrain the apparent $H_0$ anisotropies with the \YT\ relation. The $H_0$ spatial variation with the sigma maps are given in the bottom panel of Fig. \ref{anis-cosmo-YT}. The combined maximum anisotropy direction is found toward $(l,b)=(276^{\circ}\pm 26^{\circ},-14^{\circ}\pm20^{\circ} )$, with $H_0=63.4\pm 2.5$ km/s/Mpc, at a $4.3\sigma$ tension with the rest of the sky ($13\pm 3\%$). The statistical significance is slightly increased compared to our sample alone, while the direction is mostly determined by our sample. The obtained anisotropies remarkably agree with the joint \LT\ results. Finally, a dipole form of the anisotropy is apparent in the sigma maps. 

\subsection{The \LbcgT\ relation}

The final scaling relation that can potentially trace cosmological anisotropies and BFs is the \LbcgT\ relation. Absorption effects are rather irrelevant for this relation, as explained in Sect. \ref{LxLbcg}. Furthermore, the latter strongly depends on the luminosity distance (\Lbcg$\sim D_L^2$). The large scatter of the relation and the fewer number of clusters compared to \LT\ and \YT\ constitute the main disadvantages of \LbcgT . Due to that, $\theta=90^{\circ}$ scanning cones are considered. 

Despite the aforementioned disadvantages, the \LbcgT\ relation can offer additional insights on the observed anisotropies. Indeed, a $1.9\sigma$ anisotropy is detected toward $(l,b)=(257^{\circ}\pm 55^{\circ},-12^{\circ}\pm 38^{\circ})$, where the BCGs appear to be $19\pm 10\%$ fainter than the rest of the sky. The normalization anisotropy map is displayed in the bottom panel of Fig. \ref{LbcgT-tests}. This mild tension does not provide sufficient statistical evidence for a deviation of isotropy. However, the agreement of the direction with the \LT\ and \YT\ results offers additional confirmation of the existence of the physical phenomenon causing this. 

\paragraph{Cosmological anisotropies and bulk flows}

In terms of $H_0$, one obtains $H_0=66.2\pm 2.5$ km/s/Mpc for that direction, and $H_0=72.6\pm 2.3$ km/s/Mpc for the opposite hemisphere. The $H_0$ values are consistent with the other scaling relations, for similar directions.

To search for any BFs, we can only consider the full sample, due to the restricted $0.03<z<0.15$ range, and the few available clusters. For the MR method, we obtain \ubf$=580\pm 370$ km/s toward $(l,b)=(293^{\circ}\pm 50^{\circ},+2^{\circ}\pm 29^{\circ})$. With the MA method, we find similar results as well, displayed in detail in Table \ref{BF_motions}. Even though the direction is rather uncertain, the BF results are consistent with the ones found in \LT\ and \YT. 

\subsection{Combined anisotropies of \LT, \YT, and \LbcgT}\label{joint_all}

We now combine the information from the \LT, \YT, and \LbcgT\ anisotropies, for both our sample and ACC into one single $H_0$ anisotropy map, ignoring the peculiar velocities of clusters within the CMB reference frame. The joint analysis procedure is the same as before, where the different $H_0$ posterior likelihoods for every region, from every scaling relation are combined. 

This method shows certain limitations that need to be kept in mind before we present the joint results. Firstly, it assumes that the underlying effect causing the anisotropies does not affect $T$, but only \Lx, \Ysz, and \Lbcg. This is of course true for cosmological anisotropies and BFs. This also requires that there are no unaccounted X-ray absorption effects that bias the $T$ measurement, which should be the case according to the analysis presented in Sect. \ref{xray_abs_anis}. If a strong temperature outlier exists due to chance, it would not significantly affect the results by propagating to all scaling relations. The effect of the outlier would be "buried" under the average behavior of the rest of the clusters in its region, if no reason for an anisotropy exists. The posterior $H_0$ likelihood from the entire region will be then combined with the same results from the other scaling relations.

 Secondly, a moderately correlated scatter exists between \Lx\ and \Ysz\ which relates to the physical state of galaxy clusters. We discuss this in detail in Sect. \ref{scatter_corr}. When the average properties of clusters are similar from sky region to sky region (and any anisotropies arise purely from cosmological effects), this correlated scatter is not expected to bias our combined results. If, however, a strongly inhomogeneous spatial distribution of cluster populations (e.g., cool-core and noncool-core clusters) exist, then the correlation between \Lx\ and \Ysz\ will artificially boost the statistical significance of the observed anisotropy. Nevertheless, we account for this effect in the MC simulations later on (Sect. \ref{mc_sim}) and show that the obtained statistical significance of the anisotropies remains rather unchanged.

Thirdly, if our methodology suffers from a systematic bias in the estimation of the statistical significance, this would affect all three scaling relations. Since their results are combined, this bias would be amplified in the final, joint anisotropy estimation. However, in Sects. \ref{mc_sim} and \ref{zoa-sect} we quantify this bias and take it into account, showing that it has no significant effect on the final conclusions.

Bearing in mind the above, we proceed to combine all the available information together to construct the final map for the apparent $H_0$ spatial variation.

\subsubsection{Apparent $H_0$ anisotropy from joint analysis}

\begin{figure*}[hbtp]
                \includegraphics[width=0.89\textwidth, height=7.5cm]{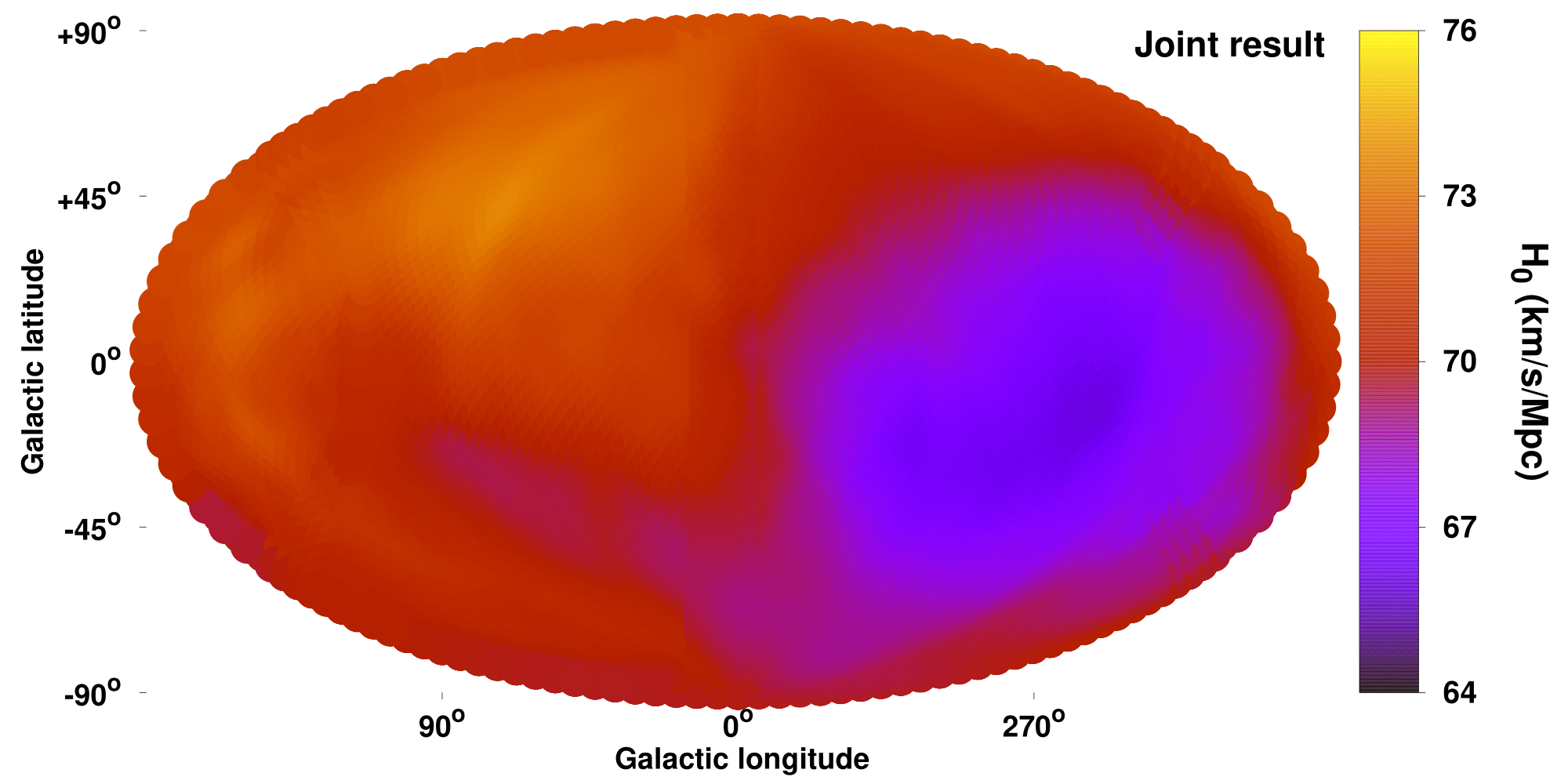}
               \includegraphics[width=0.89\textwidth, height=7.5cm]{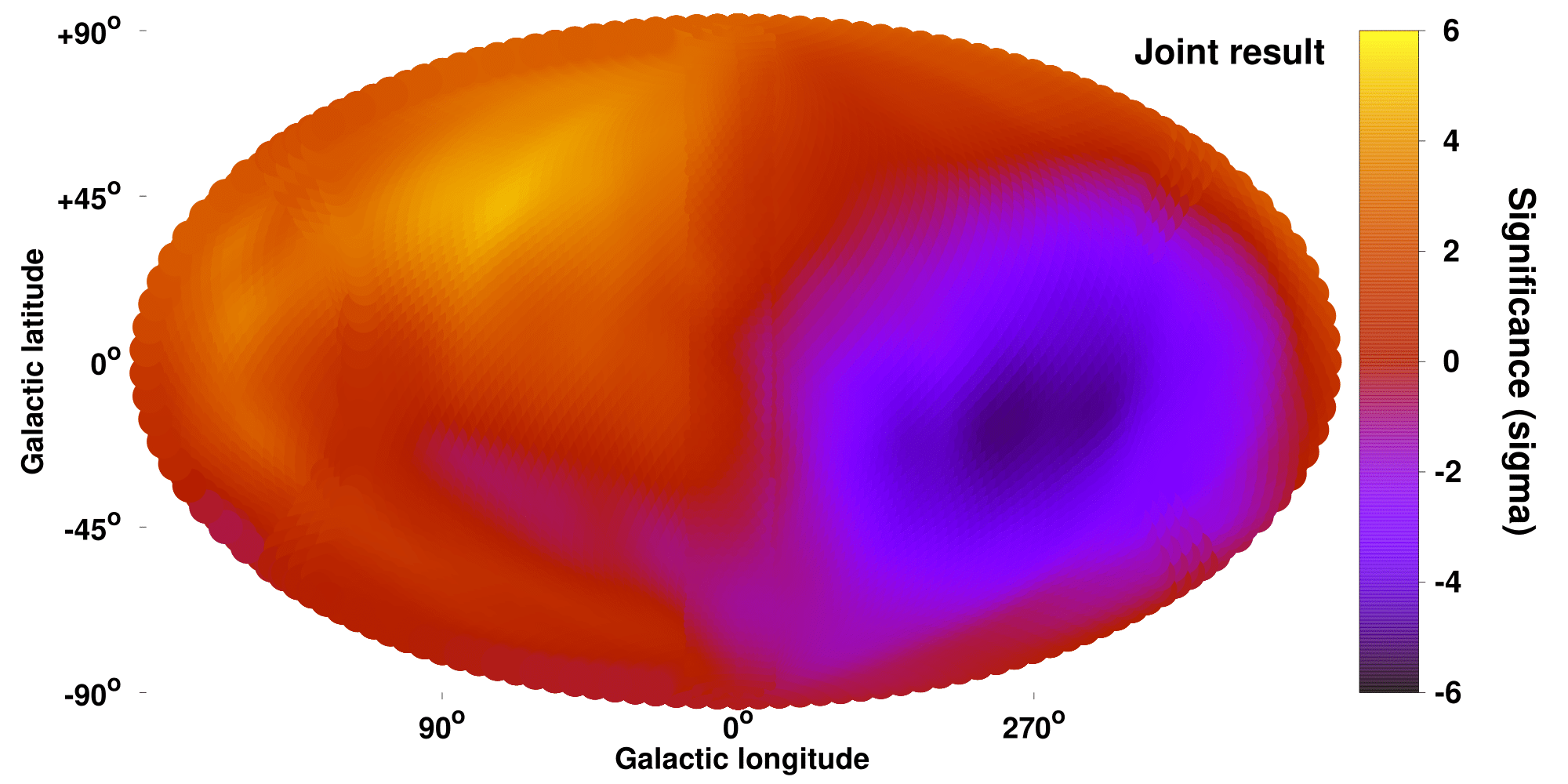}
               \caption{$H_0$ anisotropy map as derived from the joint analysis of \LT, \YT, and \LbcgT\ relations for both samples.}
        \label{anis-cosmo-joint-all}
\end{figure*}

When \LT, \YT, and \LbcgT\ results for both our sample and ACC are combined, we obtain $H_0=66.5\pm 1.0$ km/s/Mpc toward $(l,b)=(273^{\circ}\pm 40^{\circ},-11^{\circ}\pm 27^{\circ})$, and  $H_0=72.8\pm 0.6$ km/s/Mpc for the rest of the sky. There is a $5.4\sigma$ tension between the two values, demonstrating the high statistical significance of the detected anisotropies. 

The small $H_0$ uncertainties are not surprising if one considers that five nearly independent results (that generally agree) have been combined into one map and as such, there is plethora of information for every sky patch. Moreover, the normalization of the scaling relations (which relates to the calibration of true cluster distances) is assumed to be perfectly known, as discussed in Sect. \ref{stat}. This removes a dominant source of $H_0$ uncertainty from which probes that attempt to put \emph{absolute} constraints on $H_0$ suffer. However, it is not needed for the relative spatial differences to which we are interested in. The joint $H_0$ map together with the statistical significance map are displayed in Fig. \ref{anis-cosmo-joint-all}. The observed anisotropy forms a dipole.

\section{Comparison with isotropic Monte Carlo simulations}\label{mc_sim}

The well-established statistical methods used up to now provide reliable parameter uncertainties and a robust estimation of the rareness of the observed anisotropies. Nonetheless, biases we are not aware of, can still be present in our analysis and lead to overestimated anisotropy signals. The same is true for cosmic variance. To further investigate this, we need to apply our analysis to simulated isotropic MC simulations. For the \LT, \YT, and \LbcgT\ scaling relations, we create 10000 simulated isotropic samples similar to the real ones. Analyzing these samples with the same procedures as in the main analysis, we test the frequency with which anisotropies equal or larger than the ones observed in the real data are retrieved. We also calculate the frequency with which the directions of the anisotropies for different scaling relations are randomly found to be as close as in the real data.

\subsection{Constructing the isotropic simulated samples}

To build every isotropic MC realization, we create the same number of clusters as in the real sample (for both our sample and ACC). We start from the \LT\ relation. We keep the coordinates, redshifts, and $T$ (together with $\sigma_{\log{T}}$) of the simulated clusters fixed to the real values. This is done to incorporate all the possible effects that could create anisotropies, including the spatial distribution of the real clusters. Based on the best-fit scaling relation for the full, real sample, we calculate the predicted \Lx\ value. We then add a random offset to the latter, drawn from a log-normal distribution with a standard deviation of $\sqrt{\sigma_{\text{int},LT}^2+\sigma _{\log{L_{\text{X}}}}^2+B_{LT}^2\times \sigma _{\log{T}}^2}$. Since low mass systems might exhibit a larger intrinsic scatter, we consider three different values of $\sqrt{\sigma_{\text{int},LT}}$. We divide the sample in three equally sized subsamples according to their $T$ value, and for every subsample we constrain $\sigma_{\text{int},LT}$\footnote{One finds  $\sigma_{\text{int},LT}$=(0.29,0.23,0.16) for the $T<3.5$ keV, $3.5$ keV$<T<5.8$ keV, and $T>5.8$ keV clusters respectively.} given the best-fit $A_{LT}$ and $B_{LT}$ for the entire sample. Based on the $T$ value of each simulated cluster, we then use the respective $\sigma_{\text{int},LT}$ to draw the random offset of the \Lx\ value. We ensure that the posterior distribution of the best-fit values of the simulated samples follow the input values for every parameter. 


After that, we need to simulate \Ysz\ for every cluster with S/N$>2$ by taking into account the correlated scatter of \Ysz\ and \Lx\ with $T$. We first predict \Ysz\ for every cluster, based on the best-fit \YT\ relation and $T$. Then, we further add the expected scatter of \Ysz\ based on the observed best-fit correlation in Sect. \ref{scatter_corr} and the random, simulated \Lx\ scatter from before. On top of that, further noise is added on the \Ysz\ value, based on the scatter of this correlation (drawn from a log-normal distribution combining the statistical uncertainties of the observed \Lx\ and \Ysz\ residuals and the intrinsic scatter). We constrain the best-fit $A_{YT}$, $B_{YT}$, and $\sigma_{\text{intr},YT}$ for all 10000 samples and confirm their distribution follow the values from the real samples.

A simulated, isotropic \Lbcg\ value is also drawn for the 196 clusters with $0.03<z<0.15$, in the same way as for \LT. The infrared BCG luminosity does not show any correlation in its scatter with \Lx\ and \Ysz, and therefore no extra procedures are needed. We finally repeat the \LbolT\ and \YT\ simulations for ACC as well.

\subsection{Results}

\begin{figure*}[hbtp]
               \includegraphics[width=0.33\textwidth, height=5cm]{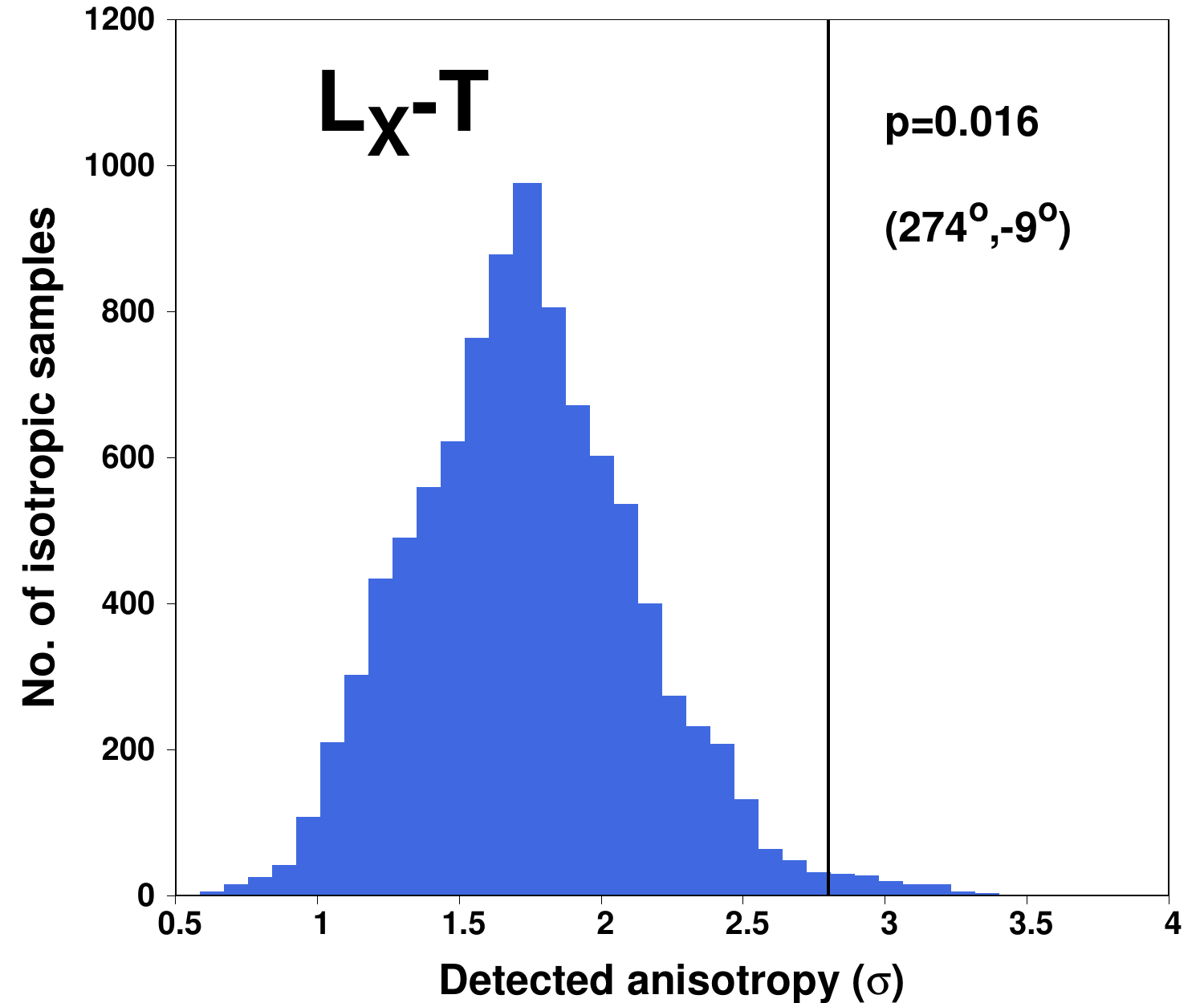}
                \includegraphics[width=0.33\textwidth, height=5cm]{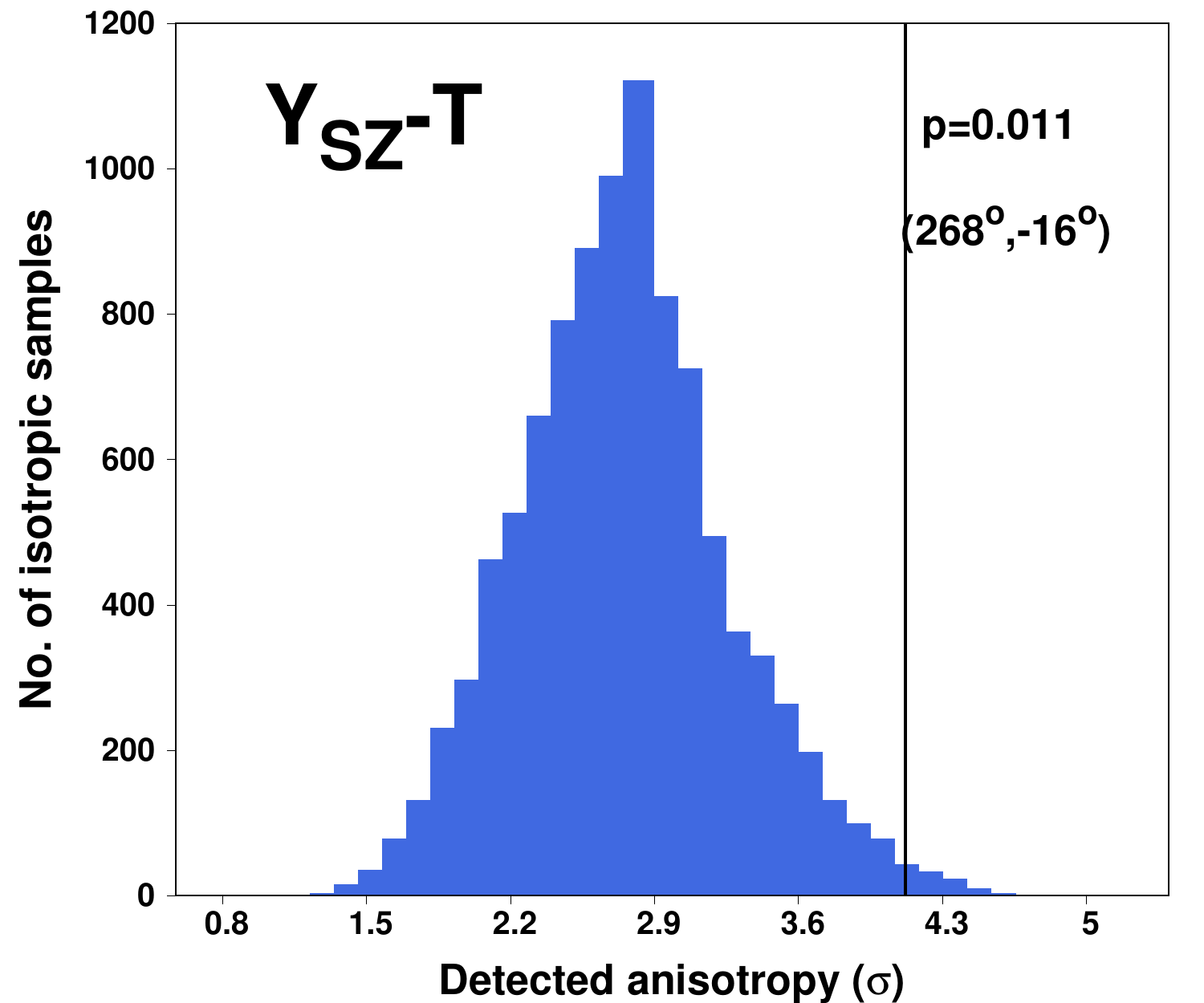}
               \includegraphics[width=0.33\textwidth, height=5cm]{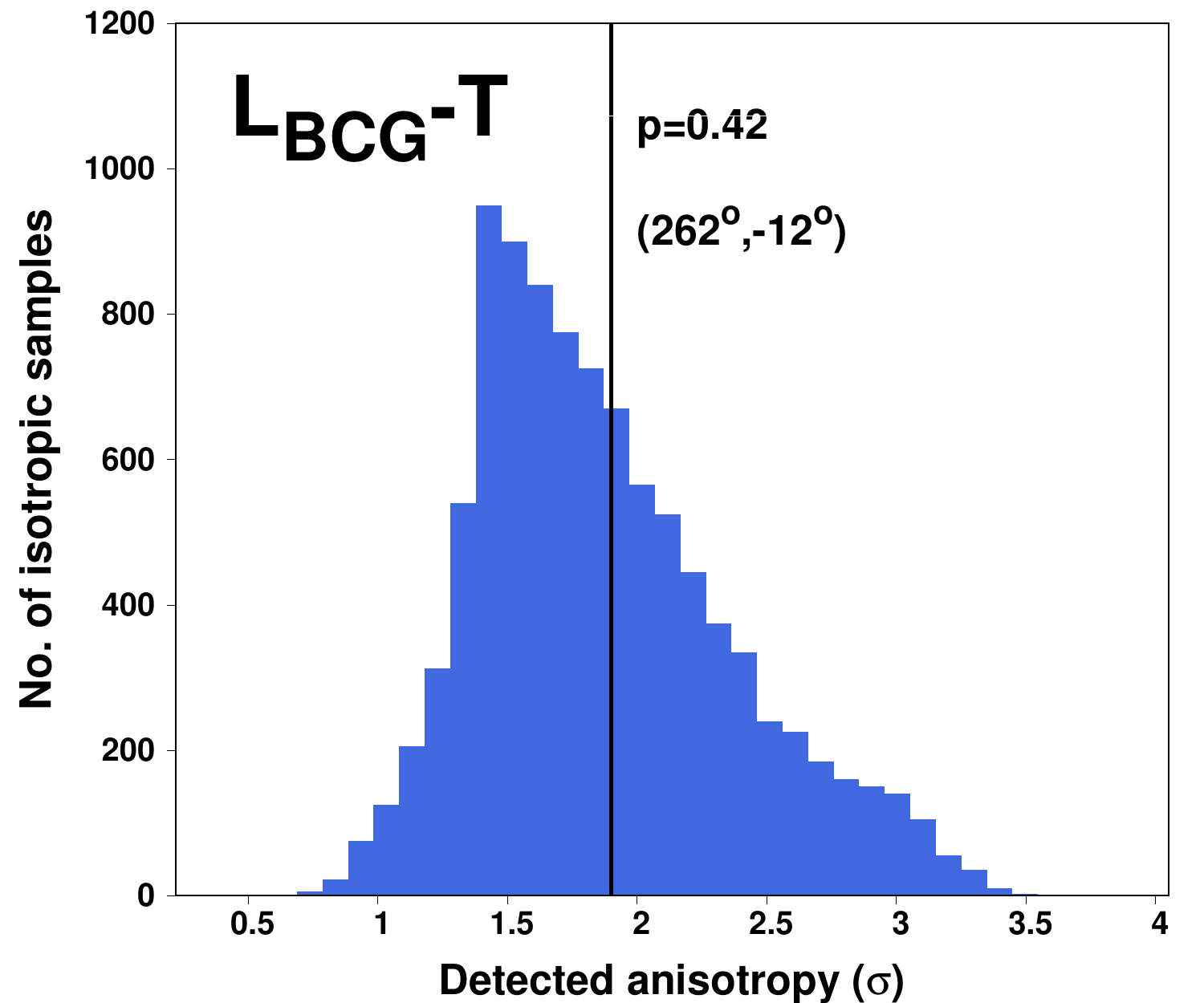}
               \includegraphics[width=0.33\textwidth, height=5cm]{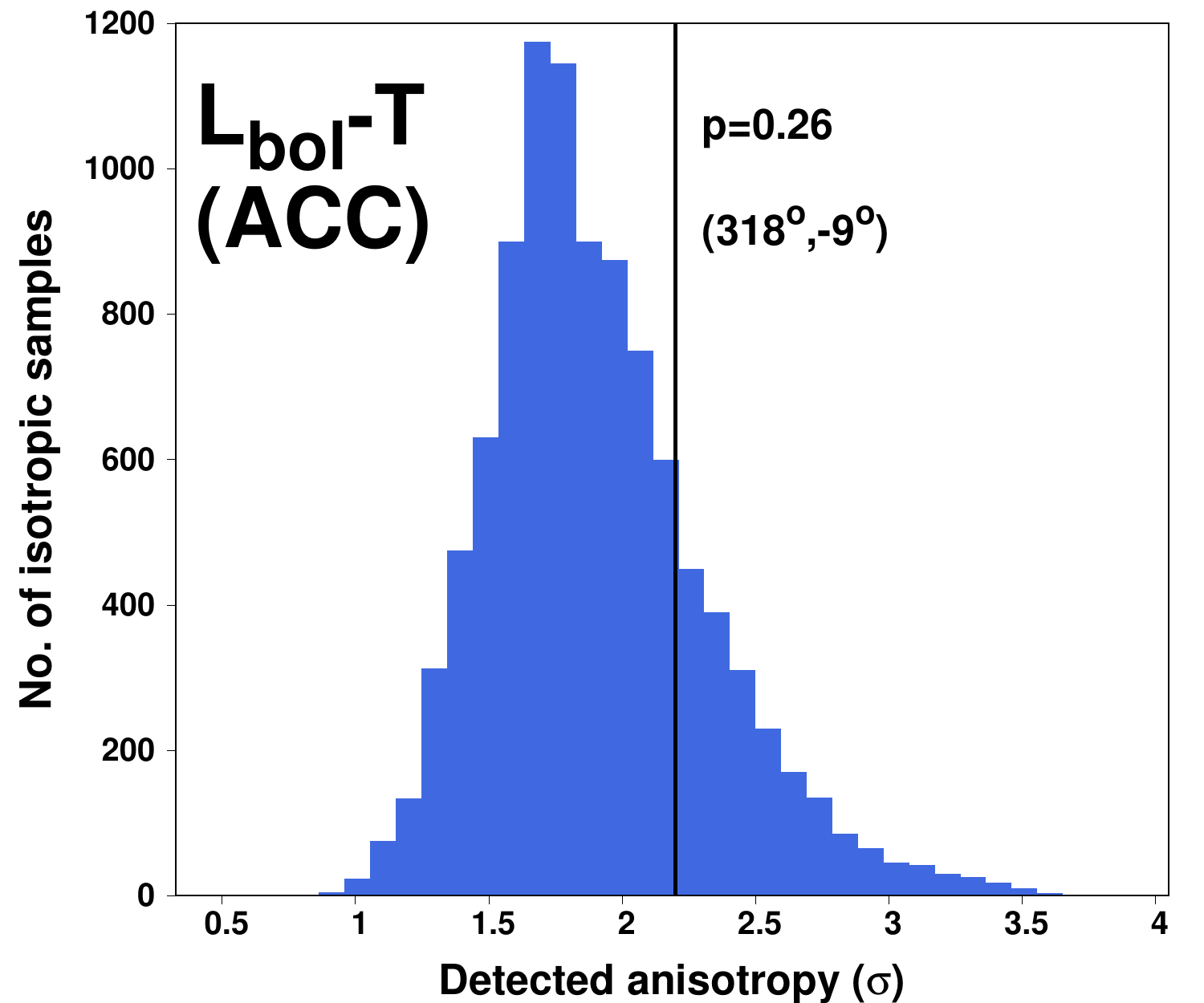}
               \includegraphics[width=0.33\textwidth, height=5cm]{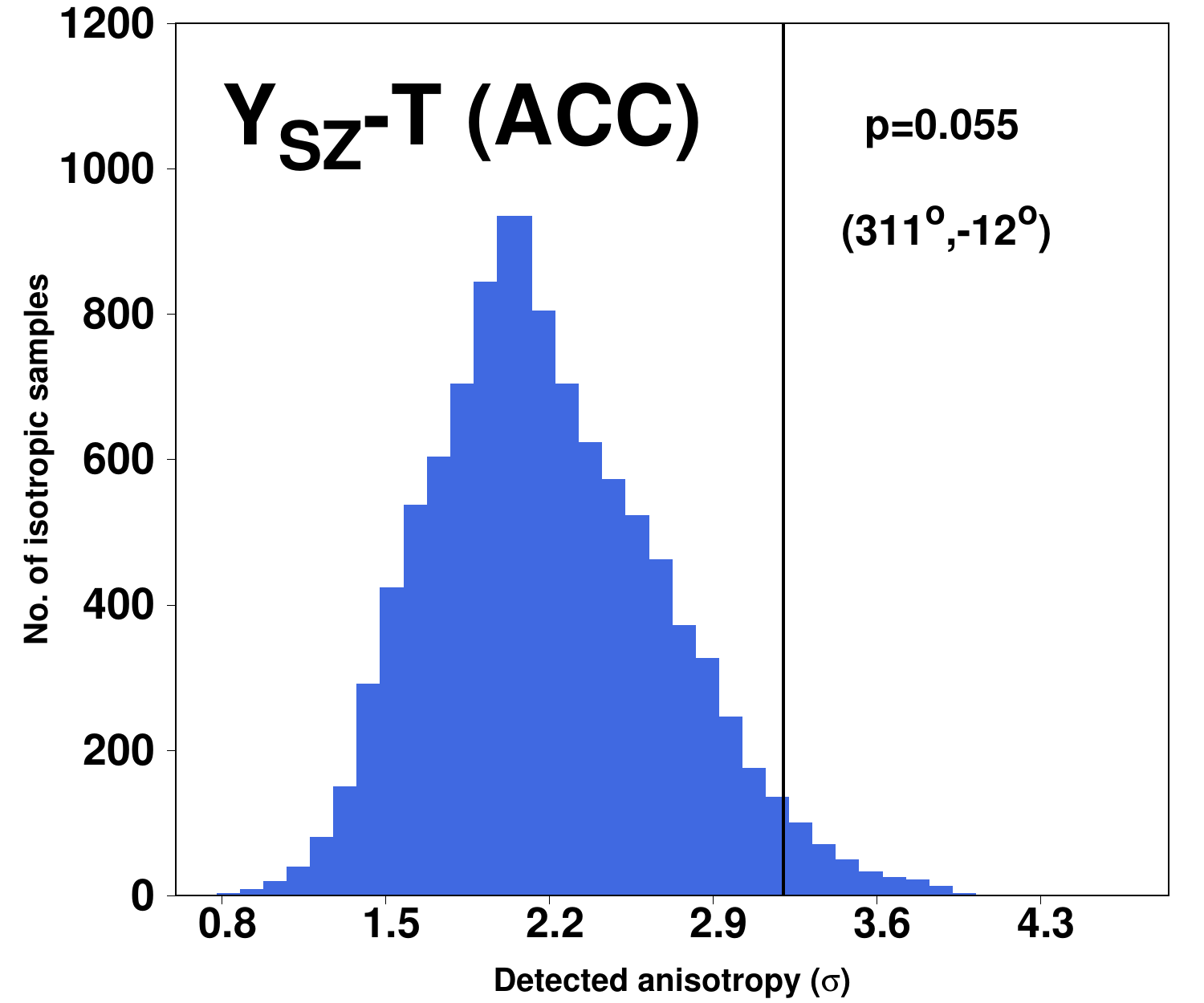}
               \caption{Histograms of the statistical significance of the maximum anisotropy as detected in $10,000$ isotropic MC simulations for the \LT, \YT, \LbcgT, \LbolT\ (ACC), and \YT\ (ACC) scaling relations. The vertical black line represents the results from the real data, with the $p-$value and the direction also shown. }
        \label{mc-hist}
\end{figure*}

We firstly explore our sample and ACC separately, for the available scaling relations. At the end, we combine the results to estimate the overall probability that our entire findings are due to chance. The histograms of the maximum anisotropy detected in the MC samples for every scaling relation, together with the results from the real samples, are displayed in Fig. \ref{mc-hist}.

\subsubsection{Our sample}

For the \LT\ relation, we find that there is a 1.6\% probability ($p=0.016$) to observe a $\geq 2.8\sigma$ anisotropy within an isotropic Universe, using our methodology. One sees that the statistical significance of the two methods does not vary significantly. 

For the \YT\ relation, the chance to randomly observe a $\geq 4.1\sigma$ anisotropy in an isotropic Universe is $p=0.011$. This is significantly more probable than implied by standard analysis and demonstrates how apparent anisotropies might be present due to noise and cosmic variance. The effect of the latter are probably enhanced by the narrower adopted scanning cone for \YT\ than for \LT, which results in fewer clusters per cone. Nevertheless, the probability of such results to occur randomly is still quite low and does not meaningfully change with increasing cone size.\footnote{If we adopt a $\theta=75^{\circ}$ scanning cone for the \YT\ relation, the default analysis returns a $3.1\sigma$ anisotropy, while the MC simulations return $p=0.013$, which is in closer agreement than before. Therefore, the effects of cosmic and sample variance can indeed overestimate the default statistical significance, but this is sufficiently taken into account by the MC isotropic simulations. Finally, the statistical significance is anyway expected to drop slightly by increasing the cone width, since more clusters that are less affected by the anisotropy are included per cone.}

If we consider the probability that the observed \LT\ and \YT\ anisotropies emerge simultaneously for a simulated sample, then the probability dramatically drops to $p=0.001$. On top of that, one needs to consider the direction agreement of the anisotropies of the real samples, which is within $10^{\circ}$. Out of the 10000 isotropic MC samples, only one exhibits this level of anisotropy for both scaling relations and within such a narrow cone ($p=10^{-4}$). It is important to stress again that the correlation between the \Lx\ and \Ysz\ values has been taken into account as explained before. 

For the \LbcgT\ relation, a $\geq 1.9\sigma$ anisotropy is detected for $42\%$ of the samples, mostly due to the large scatter. Although this result alone is consistent with isotropy, when combined with the \LT\ and \YT\ results, it returns $p=4.2\times 10^{-4}$, without accounting for the three similar anisotropy directions. When this is also included, one obtains $p=3.5\times 10^{-6}$.\footnote{Here we find the maximum angular distance $\theta$ among the most anisotropic directions of the three scaling relations. We then see how often all three directions lie within $\theta$. This is done for all simulated samples. We then multiply this probability with the probability which comes purely from the amplitude of the anisotropies (i.e., $p=4.2\times 10^{-4}$).}.

To sum up, from our sample alone, there is a 1 in $\sim 286000$ probability that an observer would obtain our results due to statistical noise, cosmic variance, or scatter correlation between \Lx, \Ysz, and \Lbcg.

\subsection{Joint probability}

Before we assess the overall probability of our results considering ACC as well, we wish to test ACC alone. For the \LbolT\ relation, the observed anisotropy amplitude is detected in the isotropic samples with $p=0.26$, much less rare than implied by the default methods and consistent with isotropy. The observed behavior of the \YT\ relation yields $p=0.055$ alone, more statistical significant than \LbolT. The joint probability to observe such anisotropies simultaneously in a sample is $p=0.028$. If they were also to be separated by only $<8^{\circ}$ as in the real data, then the probability significantly drops to $p=0.003$. One sees that there are strong indications of an existing anisotropy from the ACC alone, although not enough to completely exclude the scenario of a random event.  

Since the two samples are completely independent, the overall probability of our results is given by multiplying the distinct probabilities from the two samples. This yields $p=1.2\times 10^{-8}$. Furthermore, one needs to account for the agreement in the anisotropy direction between our sample and ACC. The maximum angular separation between the results of the two samples, for any two scaling relations, is $50^{\circ}$. Since the two samples are independent, one would expect a uniform distribution in the angular separation of their simulated maximum anisotropy directions. Thus, a $\leq 50^{\circ}$ separation would have a $\sim 28\%$ chance to occur. 

Finally, combining all the available information, we find that our results are practically impossible to occur randomly within an isotropic Universe without an underlying physical cause, since the probability for this to happen is $p=3.4\times 10^{-9}$.\footnote{Even if one completely ignores the directions of the anisotropies and only consider the amplitudes, the probability still is $p=2.5\times 10^{-5}$.}

\section{Possible systematics}\label{systematics}

Several possible biases could jeopardize the interpretation of the observed cluster anisotropies. A large number of them, including numerous X-ray and sample-related issues were tested and discussed for the \LT\ relation in M20. Here we explore some additional effects that might undermine the significance of our results. 

\subsection{Cool-core and morphologically relaxed clusters}\label{cc-bias}

Cool-core (CC) clusters have a strong central peak in their surface brightness profile, and they are known to be intrinsically brighter in X-rays than noncool-core (NCC) clusters \citep[e.g.][]{mittal}. This bias propagates to scaling relations when the two physical quantities are not similarly affected by the (N)CC nature of clusters. Previous studies have indeed found differences for such scaling relations between CC and NCC, or morphologically relaxed (i.e., regular) and disturbed clusters \citep[e.g.,][]{Maughan07,pratt,zhang,maughan,bharad,lorenzo20}. Environmental effects have been also found to mildly correlate with some cluster properties \citep[M18,][]{maria}.

For classifying clusters as CC or NCC, one needs a robust proxy of the dynamical state of the clusters, such as the central cooling time, the shape of the surface brightness profile, the concentration parameter, etc. Future work will soon provide such information for eeHIFLUGCS, which will allow for a more precise calibration of our scaling relations. For now, we use the offset between the X-ray peak and the BCG position (XBO) to categorize clusters as morphologically relaxed or disturbed. This categorization is not the same as CC and NCC, however there is a rather strong correlation between them. Specifically, the XBO has been shown to approximately correlate with the existence of a CC in the center of the cluster \citep[e.g.,][]{hudson,zitrin,rossetti,Lopes}. We consider as morphologically relaxed, and possibly CC, the clusters with XBO$<0.01\ R_{500}$. We also consider as disturbed, possibly NCC clusters, the ones with XBO$>0.08\ R_{500}$. Each of these subsamples constitutes $\sim 30\%$ of our sample.

\begin{figure}[hbtp]
                \includegraphics[width=0.49\textwidth, height=4.5cm]{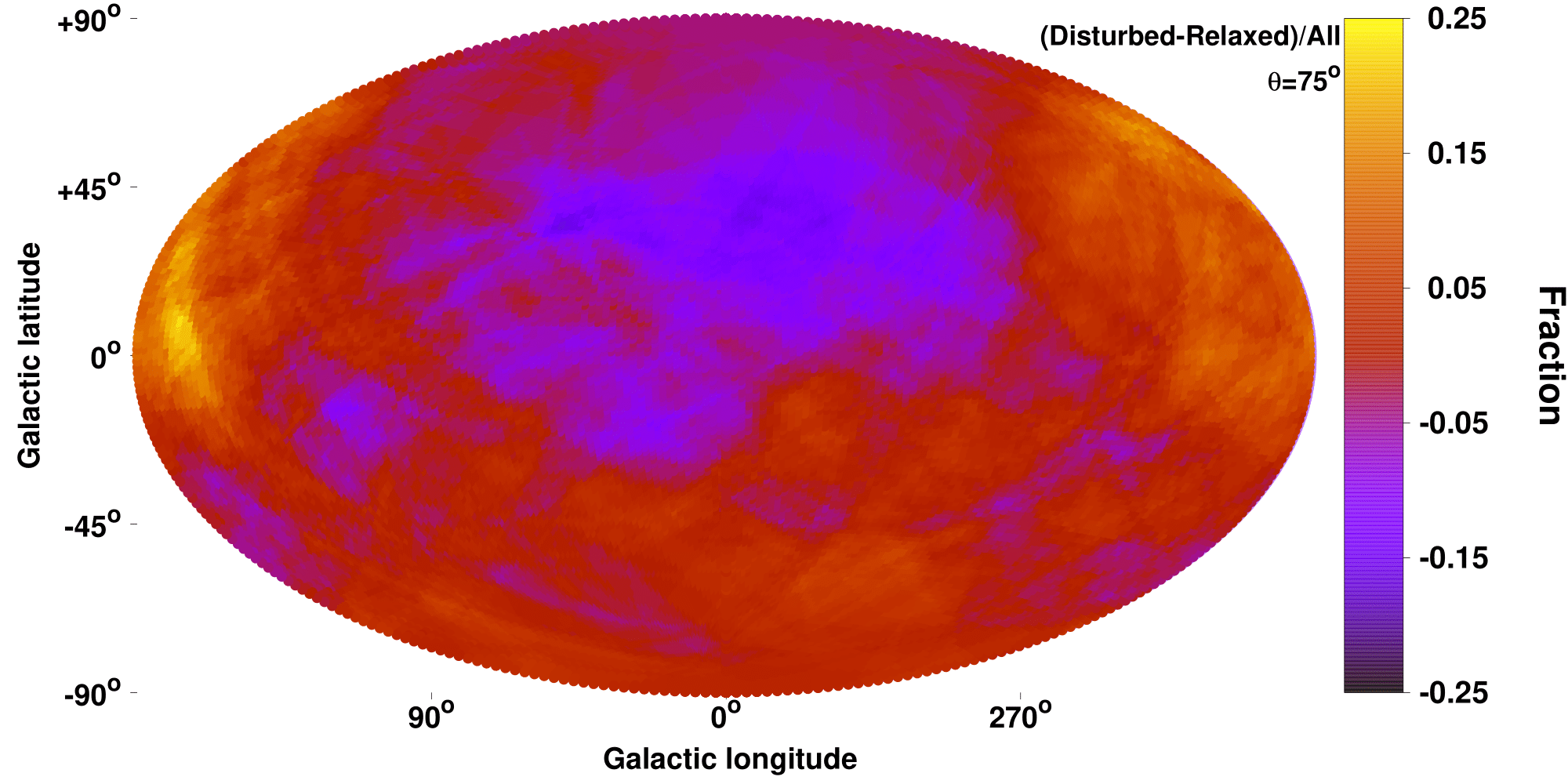}
               \caption{Fractional difference between disturbed and relaxed clusters over all the clusters for every sky patch of the extragalactic sky.}
        \label{cc-sky-plots}
\end{figure}

For any anisotropy study, this bias is only relevant if the spatial distribution of morphologically relaxed and disturbed clusters is not relatively uniform. Due to the homogeneous selection of our sample however, one  sees only mild, random variations in the fraction of such clusters across the sky (Fig. \ref{cc-sky-plots}). There is a small excess of relaxed clusters toward $(l,b)\sim (60^{\circ},+30^{\circ})$. This mild distribution imbalance strongly correlates with the directional behavior of the $R$ scaling relations (see Appendix \ref{R-relations} for more details).

Finally, it is important to stress that the $(l,b)\sim (280^{\circ},-15^{\circ})$ region shows an average behavior in its cluster population, and hence no relevant bias is expected there (as is further shown later).

\subsubsection{Lack of bias in \LT, \YT, and \LbcgT}

\begin{figure*}[hbtp]
                \includegraphics[width=0.33\textwidth, height=5cm]{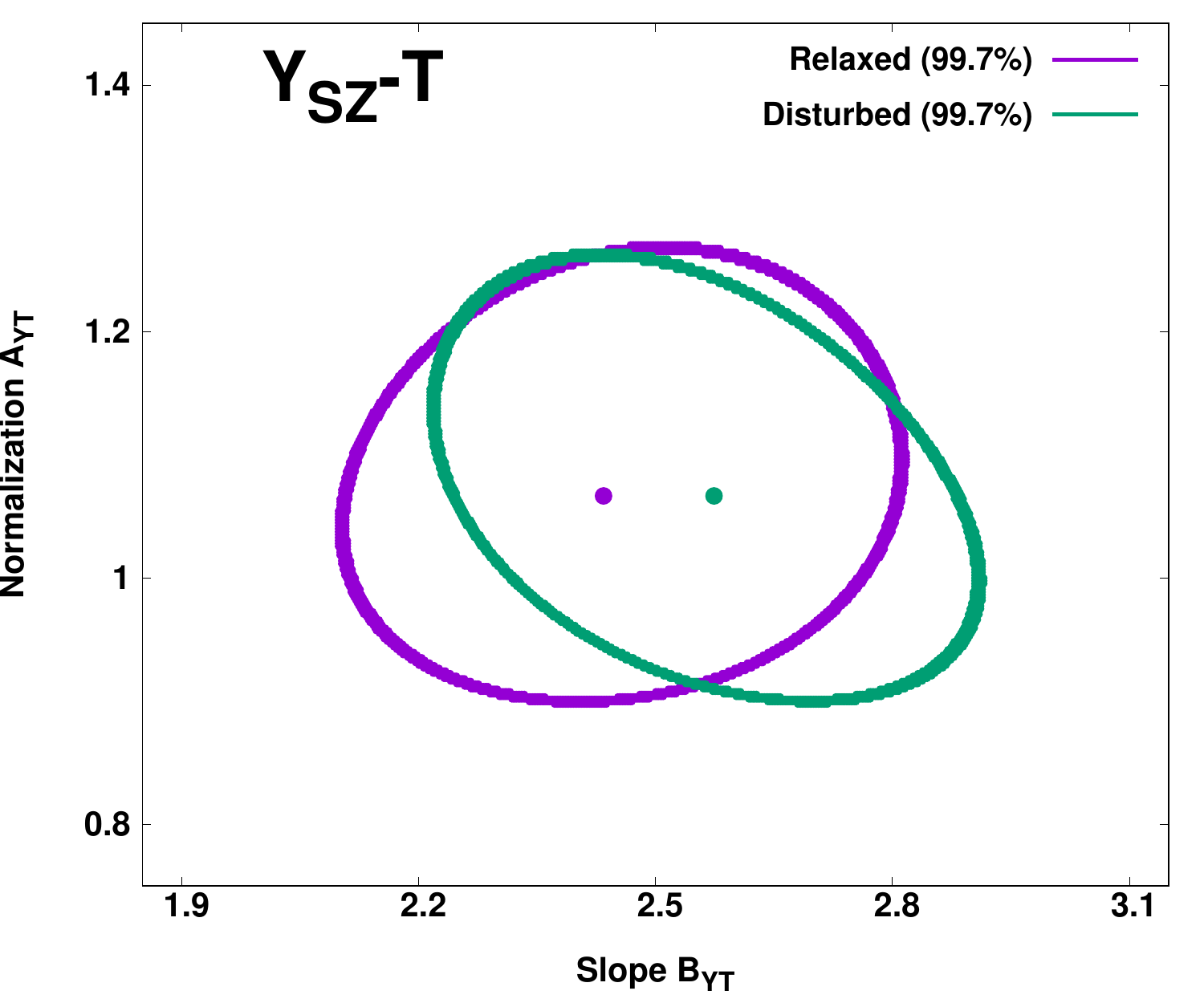}
                \includegraphics[width=0.33\textwidth, height=5cm]{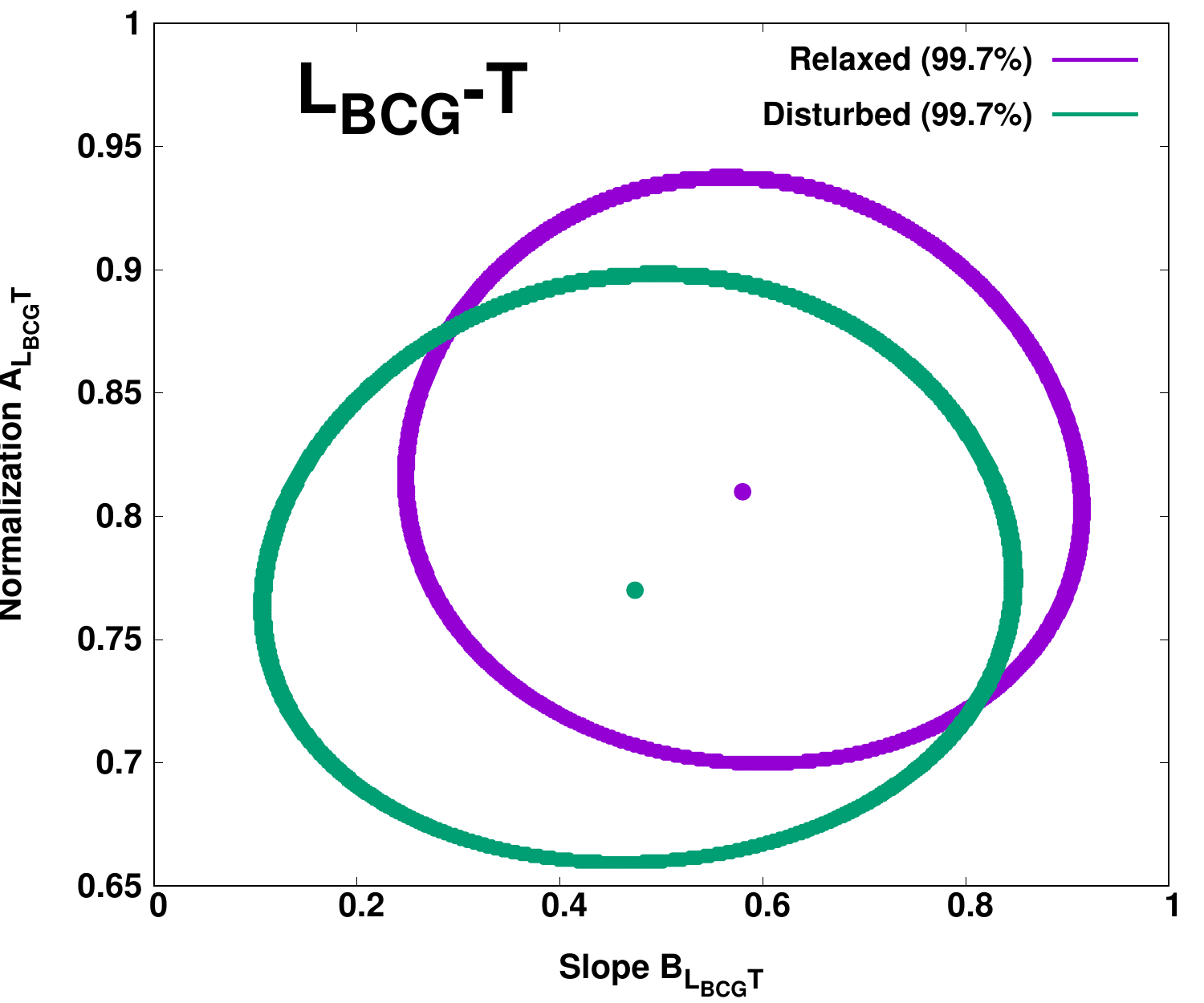}
                \includegraphics[width=0.33\textwidth, height=5cm]{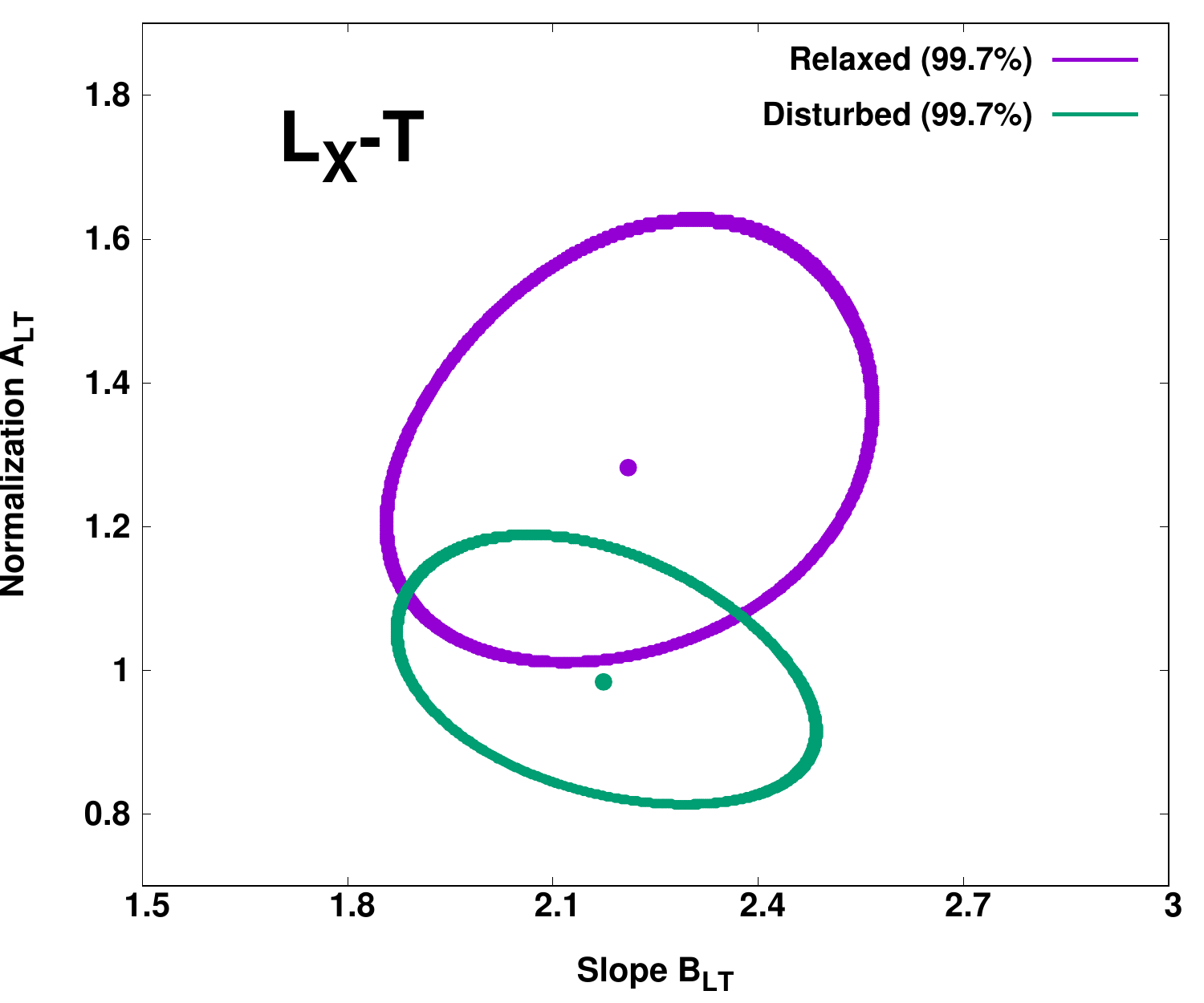}
               \caption{$3\sigma$ (99.7\%) parameter space of the normalization and slope of the \YT\ (left), \LbcgT\ (middle), and \LT\ (right) relations, for relaxed (purple) and disturbed (green) clusters.}
        \label{a-b-relax-dist}
\end{figure*}

Relaxed and disturbed clusters do not show any meaningful difference in their \YT\ and \LbcgT\ normalization, as displayed in Fig. \ref{a-b-relax-dist}. Therefore, the possible CC bias is irrelevant for these two scaling relations. For \LT, relaxed clusters appear $26\pm 10\%$ brighter than the disturbed ones (right panel of Fig. \ref{a-b-relax-dist}), which is generally expected. Of course, all these clusters are not found in only one region, but they are distributed sparsely across the sky. Thus, their effects will be hardly detectable over the possible cosmological effects. Even more importantly, there is only an average number of relaxed and disturbed clusters toward the \LT\ anisotropy region, hence no considerable bias is expected. 

This was indeed shown in M18, where using core-excised \Lx\ did not have any effect on the \LT\ anisotropies of the HIFLUGCS sample \citep{reiprich}. Excluding the possibly relaxed clusters and considering only clusters within superclusters (which are more likely to be disturbed), did not affect the anisotropies in M18 either. 

\begin{figure}[hbtp]
                \includegraphics[width=0.45\textwidth, height=5.3cm]{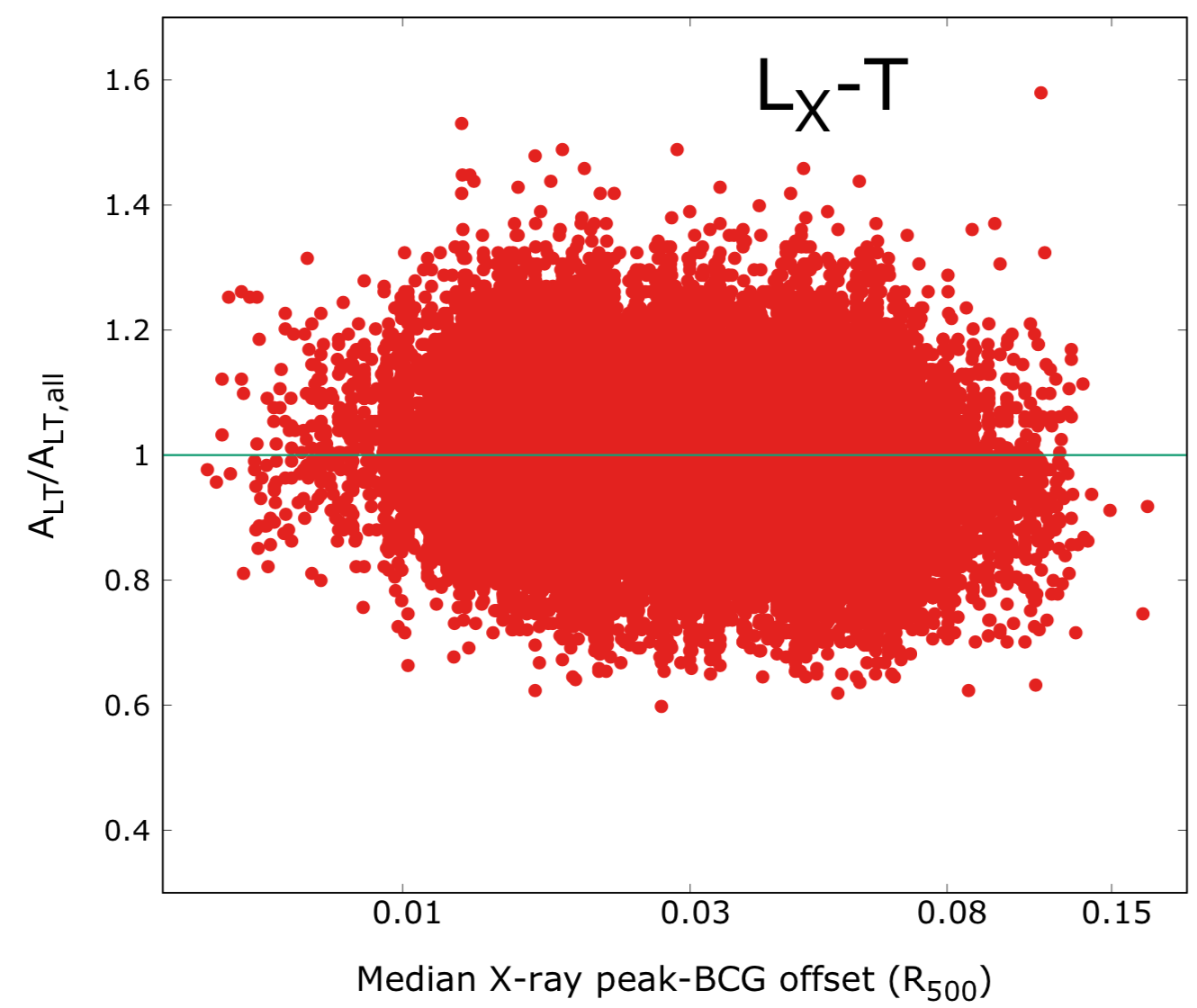}
               \caption{Correlation between the best-fit $A_{LT}$ (over the full sample's best-fit value) and the median XBO for every of the $10^5$ bootstrap subsamples.}
        \label{bcg-L-corr}
\end{figure}

To provide further evidence that the dynamical state of clusters does not have an effect on the detected \LT\ anisotropies, we create $10^5$ randomly drawn bootstrap subsamples (same process as in M20), independent of direction. We investigate the correlation between the $A_{LT}$ and the median XBO for every subsample. In Fig. \ref{bcg-L-corr}, one sees that there is only a quite weak anticorrelation between the sample's median XBO and $A_{LT}$ (Pearson's correlation coefficient $r=-0.11$). This simply shows that it is extremely improbable for enough relaxed clusters to be included in a subsample to make a noticeable difference in $A_{LT}$. 

From all the above, it is evident that an inhomogeneous distribution of morphologically relaxed and disturbed clusters is not the reason behind the observed anisotropies.

\subsection{Malmquist and Eddington biases}\label{MB-EB}

\subsubsection{Malmquist bias}

Flux-limited samples like eeHIFLUGCS, suffer from the so-called Malmquist bias (MB). Clusters that are intrinsically brighter than the others, are overrepresented in the sample, especially close to the flux limit. If not taken into account, this results in a biased estimation of the scaling parameters of the true, underlying cluster population \citep[e.g.,][]{hudson,mittal,eckert}. This bias is expected to be stronger for scaling relations including \Lx, since the selection of the sample was conducted based on the clusters' X-ray flux. At the same time, the effects of the MB should be relatively weaker for the other scaling relations. We wish to assess if the MB could affect the statistical significance of the detected anisotropies. We focus on the \LT\ relation since this is probably the most affected one, and the one we used in M20. 

If the MB influences all regions equally, then there is no effect on the detected anisotropies. This is indeed the naive expectation, since the MB is directly linked to the scatter of the \LT\ relation (for decreasing scatter, the effects of the MB are also weakened). The scatter of a relation is an intrinsic cluster property, since it mainly depends on the physical state of these objects. In an isotropic Universe, there is no obvious reason why such an intrinsic cluster property would spatially vary. A risk factor that would make the MB relevant is if the cluster population differs from region to region (e.g., due to archival bias). For large enough, homogeneously selected samples like our own, one does not expect significant differences, but nevertheless we try to quantify the possible bias in our anisotropy estimates. 

For \LT, the faintest, maximum anisotropy region lies toward $(l,b)\sim (274^{\circ}, -9^{\circ})$, containing 120 clusters with $\sigma_{\text{int}}=58.9\pm 4.6\%$. The rest of the 193 clusters are $19\%$ brighter with a $2.8\sigma$ tension, and with $\sigma_{\text{int}}=50.9\pm 3.5\%$\footnote{Here we display $\sigma_{\text{int}}$ in terms of percentile and not dex, in order to make the following calculations clearer. $\sigma_{\text{int}}$ mirrors the standard deviation of the distribution of clusters around the best-fit line. The relative difference of the two subsamples though corresponds to the standard error difference, which is the uncertainty of the mean of this distribution (i.e., the best-fit normalization). The statistical significance of this relative difference (i.e., anisotropy) accounts of course for $\sigma_{\text{int}}$ through the normalization uncertainties.}. The scatter of the faint region is actually larger than the rest of the sky, at a $1.4\sigma$ level. According to \citet{vikh09}, the observed scatter estimate is also the true, underlying scatter for X-ray flux-limited samples. 

The MB correction to be applied in the normalization of the \LT\ for low$-z$, flux-limited samples like our own is MB$_{\text{corr}}$$\approx \exp(-1.5\sigma_{\text{int}}^2)$ \citep{vikh09}. Consequently, for the $(l,b)\sim (274^{\circ}, -9^{\circ})$ region we have MB$_{\text{corr}}$$\approx 0.594\pm 0.048$, and for the rest of the sky MB$_{\text{corr}}$$\approx 0.678\pm 0.036$. Therefore, there is a $84\%$ chance that after the proper MB correction, the faint region will become even more anisotropic compared to the rest of the sky. Based on the MB$_{\text{corr}}$ uncertainties, there is only a $0.007\%$ (4$\sigma$) probability that the full amplitude of the \LT\ anisotropy is a result of the MB.  

Similarly to the effects of the scatter, one might expect that larger \Lx\ uncertainties could lead to larger normalizations, and vice versa. However, the faintest region has again a slightly larger average \Lx\ uncertainty than the rest of the sky, while the median value is similar between the two subsamples.

This result strongly suggests that the MB is not the reason behind the \LT\ anisotropies, even though there is a small chance $(<16\%)$ that it might lead to a slight overestimation of the statistical significance of the findings. However, considering the even larger relative anisotropy of the \YT\ relation and the weaker effects that MB has there, one concludes that the latter is probably irrelevant for our analysis.

\subsubsection{Eddington bias}

The Eddington bias refers to the fact that at large distances, only luminous, massive clusters will exceed the flux limit of a survey. Thus, they are overrepresented at large $z$. This leads to a different \Lx, \Ysz, and $T$  (i.e., mass) distributions between low and high $z$. Low and high mass systems might share slightly different scaling laws. If the $z$ distributions of the compared cluster subsamples were significantly different, this could lead to a biased comparison and artificial anisotropies. However, in M20 we showed that the $z$ and $T$ distributions of the $(l,b)\sim (280^{\circ}, -15^{\circ})$ are very similar to the rest of the sky, as expected for a homogeneously selected sample. This is further confirmed by our physical properties test in Sect \ref{prop-comb}. Subsequently, the Eddington bias is rather irrelevant to our analysis. 

\subsection{Zone of Avoidance bias}\label{zoa-sect}

The fact that the direction of the observed anisotropies mostly lies close to the Galactic plane gap, or the Zone of Avoidance (ZoA, $|b\leq 20^{\circ}|$) is another issue that might raise concerns about their underlying origin. In Sect. \ref{xray_abs_anis}, we demonstrated that uncalibrated X-ray absorption issues are not the reason behind \LT\ anisotropies, and definitely cannot explain the \YT\ anisotropies. Subsequently, the only possible bias that might lead to the artificial detection of anisotropies close to the ZoA is the applied methodology, or some unaccounted archival bias toward that region. 

We use two independent tests to ensure that our applied methodology is not biased toward this region. Firstly, we repeat the sky scanning for the \LT, \YT\, and \LbcgT\ relations of our sample, this time \emph{without} applying any weighting on the cluster uncertainties (these weights were based on each cluster's distance from the center of each cone). As a result, the scanning algorithm does not take into account the spatial distribution of clusters, and does not "see" the ZoA gap. The only information it reads is the number of clusters in each cone. The cones close to the most anisotropic region happen to have average numbers of objects (Fig. \ref{no-clusters} in Appendix). Even if fewer clusters were included in the cones, in M20 we showed that the best-fit normalization is completely independent of the number of clusters. 

When repeating the analysis for \LT\, we obtain an anisotropy of $2.7\sigma$ toward $(l,b)=(284^{\circ}\pm 44^{\circ},-4^{\circ}\pm 33^{\circ})$. The result is similar to the default analysis both in terms of the anisotropy direction and amplitude. In a similar manner, for the \YT\ relation we obtain a $3.9\sigma$ anisotropy toward $(l,b)=(262^{\circ}\pm 33^{\circ} ,-22^{\circ}\pm 30^{\circ})$, again similar to the default case. The anisotropies of the \LbcgT\ anisotropies are practically not affected. These results establish that the observed anisotropies in our data are not an artefact of the ZoA effect on our analysis. It also shows that the applied weighting does not significantly alter the anisotropy results. 

The second test we utilize is twofold. Firstly, we consider the 10000 isotropic MC simulated samples used in Sect. \ref{mc_sim}, which are drawn from the same distribution as the real sample. There the positions of the clusters are kept fixed, and thus the ZoA remains empty. For the same scaling relations as before, we measure how many times the maximum anisotropy is found within the ZoA. We then compare this to the random expectation based on the fraction of the full sky area the ZoA covers, namely $\sim 33\%$. We find that $\sim 43\%$ of the maximum anisotropy directions lie within the ZoA. This indicates that the applied statistical (distance) weighting of the clusters during the sky scanning can introduce a small bias in the direction of the detected anisotropies. However, this bias is quite small, and completely disappears if one repeats the procedure with uniform weighting. The real data anisotropies however do not drift in this case. Finally, in Sect. \ref{mc_sim} we found that 1.6\% and 1.1\% of the isotropic simulated samples show larger anisotropies than the real data, for the \LT\ and \YT\ relations respectively. The probability that such anisotropies are found \emph{within} the ZoA for an isotropic simulated sample is 1\% and 0.7\% respectively.

For the next part of this test, we randomly fill in the empty ZoA area of these 10000 MC samples with $\sim 130-150$ simulated clusters. These clusters have the same number density as the rest of the sky, and the same $T$ and scatter distributions. We measure again the level of the anisotropies, and compare it with the case where the ZoA is excluded. That way we wish to see if excluding the clusters within the ZoA in the real data introduces a bias in our anisotropy estimates. We find that the amplitude of the anisotropies decreases by $14\pm 8\%$, which is expected due to the larger number of isotropic data. Additionally, $\sim 36\%$ of the most deviate directions are located within ZoA, slightly decreased compared to the case when the ZoA clusters are excluded, and consistent with the random expectation. Hence, the gap of the ZoA can weakly affect the direction of the detected anisotropies, but not enough to compromise our results.

These tests combined strongly suggest that there is no significant bias in our methodology that favors and amplifies anisotropic signals close to the ZoA. The minor changes that can be indeed caused are much smaller than the measured direction uncertainties. Relative plots for these tests can be found in Appendix. \ref{zoa_bias}. Future surveys will offer a more robust cluster detection toward that region, which will help us pinpoint the clusters anisotropies slightly more accurately.

\subsection{Anomalous combination of cluster properties}\label{prop-comb}

In M20 we have investigated if a single average physical property of cluster subsamples can be associated with an anomalous behavior of $A_{LT}$. Here we wish to find out if such a behavior of $A_{LT}$ (or $A_{YT}$) could also result from a certain combination of average cluster properties. To do so, we construct $10^6$ random bootstrap cluster subsamples of random size ($10-60\%$ of the total sample's size), as done in M20. Except for the best-fit $A_{LT}$, $B_{LT}$ and $\sigma_{\text{int},LT}$, 12 more average properties of each subsample are also derived\footnote{These are the average redshift, temperature, $<0.2\ R_{500}$ (core) temperature, flux, core metallicity, $0.2-0.5\ R_{500}$ metallicity, \nhtot, RASS time exposure, X-ray peak-BCG offset, number of clusters, original source catalog (REFLEX/BCS=1, NORAS=2), and instrument used for $T$ measurement (XMM-Newton=1, Chandra=2).}. We express $A_{LT}$ as a function of all the parameters $p_i$ (normalized by their sample mean $p_{\text{mean}}$), and their power-law index $v_i$, as follows:
\begin{equation}
\begin{aligned}
&A_{LT}=1.132+ \sum_{i=1}^{N=14} v_i \log{\left(p_i/p_{\text{mean}}\right)}.
\label{ALT-pred}
\end{aligned}
\end{equation}
The 1.132 term corresponds to the best-fit value for the full sample. 

To understand which combination of properties could lead to a significantly altered $A_{LT}$, one needs to constrain $v_i$. Knowing these, one can predict the best-fit $A_{LT}$ based just on the physical properties of the clusters. To do so, we perform a Markov-Chain Monte Carlo (MCMC) fitting to the $10^6$ bootstrap subsamples. The exact details of the fitting process together with the relevant plots can be found in Appendix \ref{mcmc_details}. $A_{LT}$ shows a negligible dependence on 11 out of the 14 parameters. There is a weak anti-correlation with the best-fit $B_{LT}$ ($v_B= -0.40\pm 0.22$), a moderate anti-correlation with mean $T$ ($v_T=-0.87\pm 0.21$), and a moderate correlation with mean $z$ ($v_z= +0.82\pm 0.19$). In flux-limited samples and due to the Eddington bias, high$-z$ clusters are also high$-T$ clusters. As a result, the contribution of these two terms in the predicted $A_{LT}$ balances out. In general, subsamples with local, hot clusters tend to appear fainter than average, and vice versa. 

To determine if the observed \LT\ anisotropies are caused by the physical cluster properties of the $(l,b)\sim (274^{\circ}, -9^{\circ})$, we predict its $A_{LT}$ using Eq. \ref{ALT-pred} and its mean cluster properties. We obtain $A_{LT}\sim 1.12$, and $A_{LT}\sim 1.14$ for the rest of the sky. The two results are similar and so we confirm that the strong apparent anisotropies are not caused by a possible bias due to an inhomogeneous distribution of different cluster populations. This result was expected, since in M20 we showed that the average physical cluster properties are similar across the sky. Repeating the analysis for \YT\ returns similar results.

\subsection{Temperature calibration issues}\label{temp-calib}

If the measured cluster temperatures suffered from calibration issues and were biased toward higher values within the $(l,b)\sim (280^{\circ}, -10^{\circ})$ region, this would create apparent anisotropies in all scaling relations that include $T$. This would eventually lead to an overestimation of the statistical significance of the anisotropies, when we combine all the $T$ scaling relations. 


However, this scenario is highly unlikely for several reasons. Firstly, if $T$ was overestimated toward $(l,b)\sim (280^{\circ}, -10^{\circ})$, the same anisotropies should appear in the $R-T$ scaling relation as well, but they do not (Sect. \ref{R-relations}). Secondly, there is no obvious reason why the ACC sample would show similar anisotropies to our sample, when its temperatures were measured two decades ago with a different X-ray telescope and data analysis process. Thirdly, clusters in that region were not observed within a certain, narrow time interval, but throughout many years. As such, any calibration issues could not have affected only these clusters, but it should affect clusters from the entire sky. One would not observe strong anisotropies then, since one needs several clusters from one region to show a systematic behavior in order for a statistically significant signal to appear.

One possible effect that could bias the directionality of $T$ scaling relations, is the fraction between clusters with measured $T$ from Chandra (237 in total) or XMM-Newton (76). As shown in \citet{schellenberger} and further confirmed in M20, $T$ values measured with the two telescopes slightly (and systematically) differ. They follow a clear, low-scatter relation though, which we used to convert XMM-Newton temperatures to the equivalent Chandra ones. Therefore, one should not expect any dependance of the results on the fraction of Chandra and XMM-Newton clusters. This is confirmed by the analysis in Sect. \ref{prop-comb}, where this fraction does not strongly correlate with the best-fit $A_{LT}$. Nevertheless, we test if the spatial variation of this fraction correlates with the observed anisotropies in \LT, \YT, and \LbcgT. We find that the $(l,b)\sim (280^{\circ}, -10^{\circ})$ does not show any strong imbalance (Fig. \ref{xmm-chandra}). There is an excess of Chandra clusters toward $(l,b)\sim (150^{\circ}, +40^{\circ})$, while an excess of XMM-Newton clusters is found toward $(l,b)\sim (334^{\circ}, -50^{\circ})$. Moreover, when only Chandra clusters were used in M20, the same \LT\ anisotropies were observed.

Based on all the above, one can safely conclude that the observed anisotropies are not due to $T$ measurement calibration issues. Finally, eRASS will provide cluster observations from a single instrument, analyzed within a strict time interval, and thus avoid such possible systematics.

\subsection{Correlation of \Lx\ and \Ysz\ scatter with $T$}\label{scatter_corr}

\begin{figure}[hbtp]
               \includegraphics[width=0.4\textwidth, height=5cm]{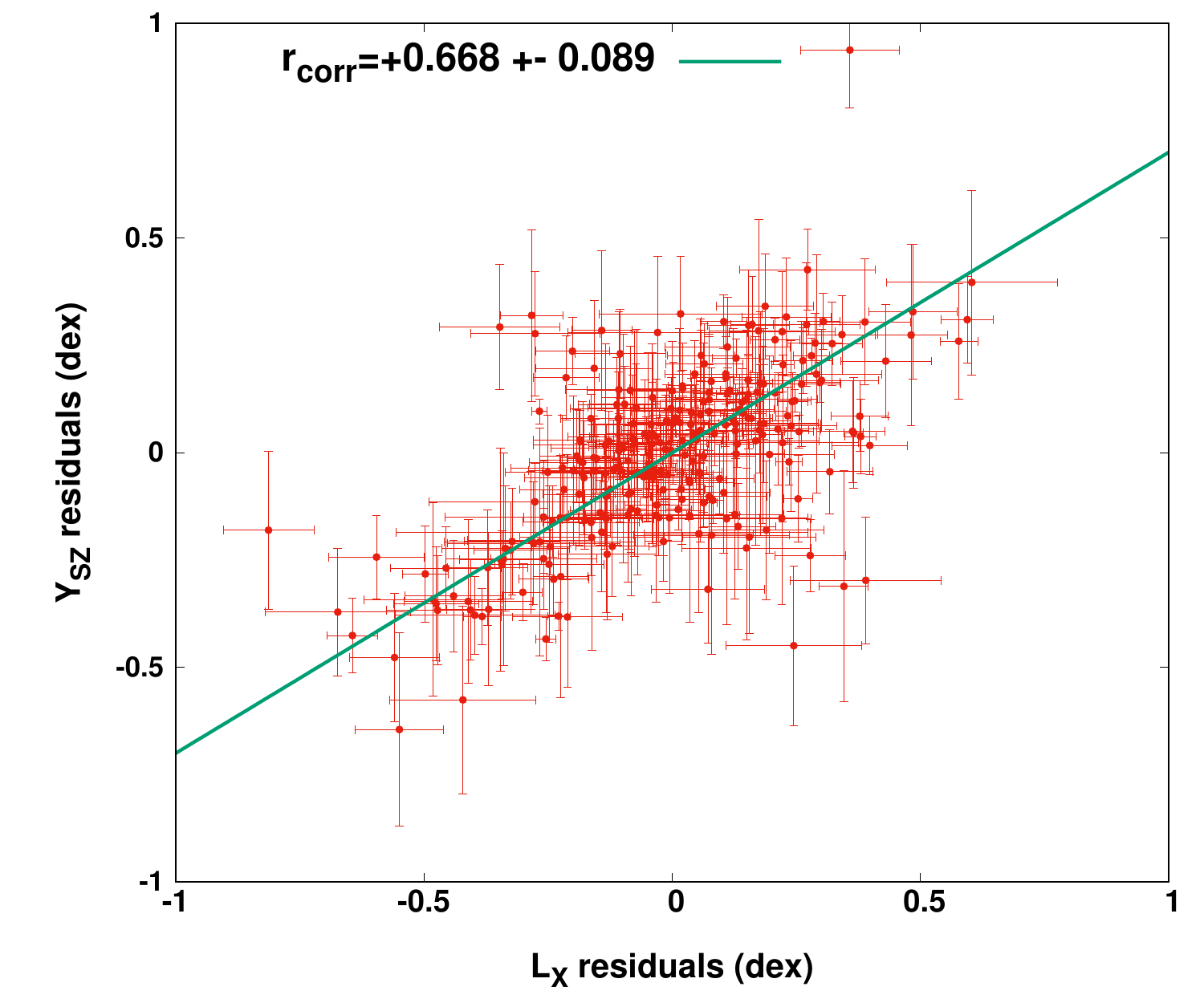}
               \caption{Correlation between the \Ysz\ scatter from the \YT\ relation, and the \Lx\ scatter from the \LT\ relation.}
        \label{L-Y-sc-corr}
\end{figure}

Past studies \citep[i.e.][and references therein]{aarti} have reported a positive correlation between the scatter of \Lx\ and \Ysz\ at fixed mass (or $T$ equivalently). Here we test this correlation with much larger samples, as well as the correlation of the scatter between these two quantities and \Lbcg. Considering the 263 clusters with both \Ysz\ and \Lx\ measurements, we confirm that their residuals with respect to $T$ are correlated, with a Pearson's correlation coefficient of $r_{\text{corr}}=0.668\pm 0.089$. This is shown in Fig. \ref{L-Y-sc-corr}. The best-fit line has a slope of $\sim 0.63$, while the total scatter of the relation is $\sim 0.16$ dex. 

We should note that this correlation has no effect on the statistical significance values extracted for every scaling relation individually, although it can affect the combined statistical significance of the anisotropies as obtained in Sect. \ref{joint_all}. In practice, this means that both \YT\ and \LT\ would exhibit similarly biased anisotropies in the case of an inhomogeneous distribution of CC clusters or mergers for instance. Their results then could not be considered independent and combined as in Sect. \ref{joint_all}. Our previous analysis showed that such a CC (or merger) bias is quite unlikely to cause the observed anisotropies, since no correlation was found between the $A_{LT}$ and $A_{YT}$ and the average dynamical state of a subsample's clusters. We fully explore the effect that this correlation has in our results however in the next section where MC simulations are carried out.

Similar results are obtained for the ACC sample, for both the correlation coefficient and the slope of the relation. Finally, the scatter of \Lbcg\ with respect to $T$ does not show any meaningful correlation with the scatter of \Lx\ or \Ysz.

\subsection{Selection biases}\label{select-bias}

As extensively argued in M20, selection biases are not expected to significantly affect our anisotropy results. We attempt to constrain the \emph{relative} deviations between sky regions and thus we use the overall best-fit scaling relations as a frame of reference. The bias-corrected scaling relations are not a necessity for that, as long as the cluster properties across the sky do not vary significantly. This was shown to be true in both M20 and in this work. Our results however are still directly comparable with similar literature results since many studies focus on the observed scaling relations. Nonetheless, we performed several tests to evaluate how some selection biases can affect the statistical significance of the anisotropies. The results from Sects. \ref{cc-bias} and \ref{MB-EB} support the notion that the observed anisotropies are not overestimated by selection biases. More tests are performed in Appendix \ref{differ_cuts}, yielding the same conclusion. The bias-corrected scaling relations will be presented in future work.

Overall, the results of Sect. \ref{systematics} further suggest that the observed anisotropies do not emerge due to unaccounted biases.

\section{Discussion and conclusions}\label{discussion}

The study of the (an)isotropic behavior of galaxy cluster scaling relations opens a very promising window for testing the cosmological principle. Galaxy cluster samples are much more spatially uniform than SNIa ones, and they extent to much larger scales than the galaxy samples which are used to study BFs. They can also be observed in several wavelengths and allow us to measure many different properties of them. This provides us with plenty of nearly independent tests that trace cosmological phenomena, and that can be carried out with the same sample of objects. This has been accomplished in this work.

\subsection{Known systematics do not explain the observed anisotropies}

The \LT\ anisotropy toward $(l,b)\sim (280^{\circ},-15^{\circ})$ that was originally observed in M20, could in principle be due to uncalibrated X-ray effects, such as excess X-ray absorption toward this sky direction. With our new data and analysis strategy, especially from the \LY\ relation which shows no signs of extra X-ray absorption toward the region in question, we can now decisively exclude this possibility. The lack of excess X-ray absorption effects toward that region is further confirmed by the analysis of ACC, and by the \LxLbcg\ relation of our sample. The comparison between the X-ray-determined \nhtot\ and the W13-based \nhtot\ also does not indicate such an effect. Furthermore, the same anisotropies persist in scaling relations that are insensitive to unaccounted X-ray absorption, such as \YT\ and \LbcgT. Thus, one sees that this is not the reason for the observed anisotropy.



Another possible solution to the observed anisotropies was the existence of biases related to the sample or the followed methodology, such as selection and archival biases, the effect of the Zone of Avoidance gap, biases introduced by cluster morphology, temperature calibration issues, and correlations between cluster properties. Our current results clearly exclude these possibilities as well. We showed that the Malmquist bias is more likely to underestimate the observed anisotropies rather than explain them. Also, it would not have such a strong effect for \YT, \LbcgT, and the ACC sample (which is not flux-limited). It was also shown that the fraction of morphologically relaxed and disturbed clusters does not spatially vary strongly enough to have any effect in the apparent anisotropy of the \LT\ relation. Even if it did, this would have no strong effect on the \YT\ anisotropies, or the results from the other samples. By using isotropic simulated data we further showed that the ZoA gap does not affect the amplitude of the anisotropies, while it has a very small effect in the direction of the latter. Even when we adjust our methodology to eliminate the effects of this gap, the results are the same. The isotropic MC simulations we performed, verify the high ($>5\sigma$) statistical significance of the observed anisotropies, when all the cluster scaling relations that are sensitive to cosmological phenomena are combined. In future work, anisotropic simulations, due to both $H_0$ anisotropies and BFs, will be employed to better understand the precision and accuracy with which our methodology can detect such phenomena.

Despite testing and rejecting the generally known biases in cluster analysis as the source of the observed anisotropies, the existence of currently unknown systematics cannot be excluded. Although it would be surprising for such a significant, direction-dependent, and yet undiscovered bias to exist, all possibilities should be thoroughly scrutinized. In case such a systematic is discovered in the future, its effects on past studies with similar cluster samples as our own should be investigated.

\begin{table*}[hbtp]
\caption{\small{Maximum anisotropy direction for every scaling relation, together with the needed $H_0$ relative variation in order to fully explain the anisotropy. The statistical significances derived by the default Bootstrap method and by the MC isotropic simulations are also displayed. The MC $p-$values account for both the amplitude of the anisotropies, and the agreement in their directions. The Gaussian $\sigma$ values that correspond to the $p-$values are also displayed, for easier comparison.}}
\label{summary_anisot}
\begin{center}
\renewcommand{\arraystretch}{1.3}
\small
\begin{tabular}{ c  c  c  c c}
\hline \hline

Scaling & Max. anisot. &  $H_0$  & Bootsrap & MC\\
relation & direction $(l,b)$  & variation ($\%$)  & significance ($\sigma$)& $p-$value ($\sigma$)  \\
\hline \hline
 & &  Our sample & & \\ \hline
\LT & $({274^{\circ}}^{+43^{\circ}}_{-41^{\circ}},{-9^{\circ}}^{+33^{\circ}}_{-31^{\circ}})$ & $8.7\pm 3.1\%$ & $2.8\sigma$ & $p=0.016$ ($2.4\sigma$)   \\ 
\YT & $({268^{\circ}}^{+34^{\circ}}_{-31^{\circ}},{-16^{\circ}}^{+29^{\circ}}_{-30^{\circ}})$ & $14.0\pm 3.4\%$ & $4.1\sigma$ & $p=0.011$ ($2.6\sigma$) \\ 
\LbcgT & $({257^{\circ}}^{+58^{\circ}}_{-49^{\circ}},{-12^{\circ}}^{+38^{\circ}}_{-40^{\circ}})$ & $9.1\pm 4.8\% $ & $1.9\sigma$ & $p=0.420$ ($0.8\sigma$) \\ 

\hline
& &  Our sample$+$ACC & &  \\ \hline
\LT & $({284^{\circ}}^{+31^{\circ}}_{-12^{\circ}},{-4^{\circ}}^{+7^{\circ}}_{-23^{\circ}})$ & $9.3\pm 3.1\% $ & $3.0\sigma$  & $p=8.8\times 10^{-4}$ ($3.3\sigma$)   \\ 
\YT & $({276^{\circ}}^{+29^{\circ}}_{-23^{\circ}},{-14^{\circ}}^{+19^{\circ}}_{-21^{\circ}})$ & $13.3\pm 3.1 \%$ & $4.3\sigma$ & $p=1.9\times 10^{-4}$ ($3.8\sigma$)\\ 
\hline
\textbf{All combined} & $\mathbf{({273^{\circ}}^{+42^{\circ}}_{-38^{\circ}},{-11^{\circ}}^{+27^{\circ}}_{-27^{\circ}})}$ & $\mathbf{9.0\pm 1.7\%}$ & $\mathbf{5.4\sigma}$ & $\mathbf{p=3.4\times 10^{-9}}$ ($5.9\sigma$) \\

 \hline

\end{tabular}
\end{center}
\end{table*}

\begin{table*}[hbtp]
\caption{\small{Best-fit bulk flows for every scaling relation, method, and redshift bin that was considered in this work. MR and MA methods are explained in Sect. \ref{BF_detect}.}}
\label{BF_motions}
\begin{center}
\renewcommand{\arraystretch}{1.3}
\small
\begin{tabular}{ c  c  c  c | c  c c }
\hline \hline

Scaling relation  & Redshift & Amplitude& Direction & Redshift   & Amplitude  & Direction \\
(method)  & bin  & (km/s) & $(l,b)$ & bin  & (km/s)  & $(l,b)$\\
\hline \hline
&  &  Our sample & & & \\ \hline
\LT\ (MR)&  $z<0.06$ & $1100\pm 410$ & $({318^{\circ}}^{+37^{\circ}}_{-37^{\circ}},{-5^{\circ}}^{+24^{\circ}}_{-23^{\circ}})$ & $z<0.1$ & $1080\pm 380$  & $({322^{\circ}}^{+34^{\circ}}_{-33^{\circ}},{-13^{\circ}}^{+23^{\circ}}_{-22^{\circ}})$ \\ 
 &$0.06<z<0.12$ & $1170\pm 400$ & $({262^{\circ}}^{+54^{\circ}}_{-50^{\circ}},{+2^{\circ}}^{+24^{\circ}}_{-28^{\circ}})$ & $z<0.16$ & $1030\pm 350$  & $({309^{\circ}}^{+34^{\circ}}_{-31^{\circ}},{-20^{\circ}}^{+21^{\circ}}_{-20^{\circ}})$\\ 
&  $0.12<z<0.3$ & $1040\pm 570$ & $({253^{\circ}}^{+61^{\circ}}_{-60^{\circ}},{-18^{\circ}}^{+33^{\circ}}_{-29^{\circ}})$ & $z<0.25$ & $980\pm 310$ &  $({311^{\circ}}^{+32^{\circ}}_{-31^{\circ}},{-18^{\circ}}^{+19^{\circ}}_{-19^{\circ}})$\\ 
&  All & $980\pm 300$ & $({315^{\circ}}^{+33^{\circ}}_{-35^{\circ}},{-10^{\circ}}^{+20^{\circ}}_{-21^{\circ}})$ & & &\\ 
\LT\ (MA)&  $z<0.09$ & $690\pm 300$ & $({268^{\circ}}^{+32^{\circ}}_{-30^{\circ}},{-5^{\circ}}^{+24^{\circ}}_{-22^{\circ}})$ & $z<0.16$ & $670\pm 270$ & $({278^{\circ}}^{+29^{\circ}}_{-28^{\circ}},{-12^{\circ}}^{+23^{\circ}}_{-22^{\circ}})$\\ 
 &$z>0.067$ & $620\pm 310$ & $({293^{\circ}}^{+38^{\circ}}_{-40^{\circ}},{-12^{\circ}}^{+27^{\circ}}_{-27^{\circ}})$ & $z<0.25$  & $620\pm 150$ & $({291^{\circ}}^{+26^{\circ}}_{-27^{\circ}},{-21^{\circ}}^{+21^{\circ}}_{-21^{\circ}})$\\ 
&  All & $600\pm 260$ & $({298^{\circ}}^{+25^{\circ}}_{-25^{\circ}},{-21^{\circ}}^{+19^{\circ}}_{-18^{\circ}})$ &  &\\ 
\YT\ (MR)&  $z<0.07$ & $1060\pm 390$ & $({254^{\circ}}^{+45^{\circ}}_{-40^{\circ}},{-17^{\circ}}^{+19^{\circ}}_{-19^{\circ}})$ & $z<0.1$ & $1020\pm 380$  & $({257^{\circ}}^{+25^{\circ}}_{-24^{\circ}},{-28^{\circ}}^{+25^{\circ}}_{-21^{\circ}})$ \\ 
 &$0.07<z<0.12$ & $840\pm 490$ & $({312^{\circ}}^{+59^{\circ}}_{-62^{\circ}},{-34^{\circ}}^{+32^{\circ}}_{-29^{\circ}})$ & $z<0.16$ & $980\pm 370$  & $({265^{\circ}}^{+34^{\circ}}_{-31^{\circ}},{-27^{\circ}}^{+24^{\circ}}_{-21^{\circ}})$\\ 
&  $0.12<z<0.3$ & $1110\pm 670$ & $({321^{\circ}}^{+75^{\circ}}_{-103^{\circ}},{-42^{\circ}}^{+45^{\circ}}_{-29^{\circ}})$ & $z<0.25$ & $940\pm 360$ &  $({267^{\circ}}^{+32^{\circ}}_{-31^{\circ}},{-27^{\circ}}^{+22^{\circ}}_{-22^{\circ}})$\\ 
&  All & $950\pm 340$ & $({263^{\circ}}^{+33^{\circ}}_{-35^{\circ}},{-22^{\circ}}^{+20^{\circ}}_{-21^{\circ}})$ & & &\\ 
\YT\ (MA)&  $z<0.09$ & $1200\pm 350$ & $({254^{\circ}}^{+22^{\circ}}_{-22^{\circ}},{-28^{\circ}}^{+14^{\circ}}_{-18^{\circ}})$ & $z<0.16$ & $1090\pm 340$ & $({270^{\circ}}^{+25^{\circ}}_{-23^{\circ}},{-30^{\circ}}^{+24^{\circ}}_{-19^{\circ}})$\\ 
 &$z>0.09$ & $720\pm 380$ & $({242^{\circ}}^{+94^{\circ}}_{-79^{\circ}},{-13^{\circ}}^{+20^{\circ}}_{-24^{\circ}})$ & $z<0.25$  & $1000\pm 310$ & $({264^{\circ}}^{+24^{\circ}}_{-23^{\circ}},{-23^{\circ}}^{+22^{\circ}}_{-23^{\circ}})$\\ 
&  All & $960\pm 290$ & $({267^{\circ}}^{+23^{\circ}}_{-21^{\circ}},{-28^{\circ}}^{+17^{\circ}}_{-15^{\circ}})$ &  &\\ 
\LbcgT\ (MR)&  $0.03<z<0.15$ & $580\pm 370$ & $({293^{\circ}}^{+50^{\circ}}_{-51^{\circ}},{+2^{\circ}}^{+29^{\circ}}_{-30^{\circ}})$ & &   &  \\ 

\LbcgT\ (MA)&  $0.03<z<0.15$ & $600\pm 340$ & $({261^{\circ}}^{+47^{\circ}}_{-52^{\circ}},{-25^{\circ}}^{+30^{\circ}}_{-22^{\circ}})$ &  &  & \\

\hline
&  & & ACC & & \\ \hline
\LbolT\ (MR)&  $z<0.2$ & $960\pm 590$ & $({277^{\circ}}^{+66^{\circ}}_{-59^{\circ}},{+2^{\circ}}^{+29^{\circ}}_{-31^{\circ}})$ & $z<0.35$ & $870\pm 510$  & $({244^{\circ}}^{+69^{\circ}}_{-52^{\circ}},{+20^{\circ}}^{+24^{\circ}}_{-33^{\circ}})$ \\ 
&  All & $850\pm 410$ & $({254^{\circ}}^{+60^{\circ}}_{-44^{\circ}},{+18^{\circ}}^{+24^{\circ}}_{-30^{\circ}})$ & & &\\ 
\LbolT\ (MA)&  All & $810\pm 400$ & $({324^{\circ}}^{+39^{\circ}}_{-51^{\circ}},{-3^{\circ}}^{+32^{\circ}}_{-33^{\circ}})$ & &  &\\ 

\YT\ (MR)&  $z<0.2$ & $930\pm 540$ & $({281^{\circ}}^{+61^{\circ}}_{-52^{\circ}},{-7^{\circ}}^{+39^{\circ}}_{-37^{\circ}})$ & $z<0.35$ & $800\pm 410$  & $({264^{\circ}}^{+59^{\circ}}_{-52^{\circ}},{+6^{\circ}}^{+27^{\circ}}_{-29^{\circ}})$ \\ 
&  All & $800\pm 400$ & $({268^{\circ}}^{+45^{\circ}}_{-39^{\circ}},{+6^{\circ}}^{+29^{\circ}}_{-29^{\circ}})$ & & &\\ 
\YT\ (MA)&  All & $810\pm 370$ & $({273^{\circ}}^{+43^{\circ}}_{-35^{\circ}},{-11^{\circ}}^{+21^{\circ}}_{-20^{\circ}})$ & &  & \\ 

\hline
 
\end{tabular}
\end{center}
\end{table*}
 
 \subsection{Cosmological phenomena behind the tension}
 
\begin{figure*}[hbtp]
               \includegraphics[width=0.49\textwidth, height=6cm]{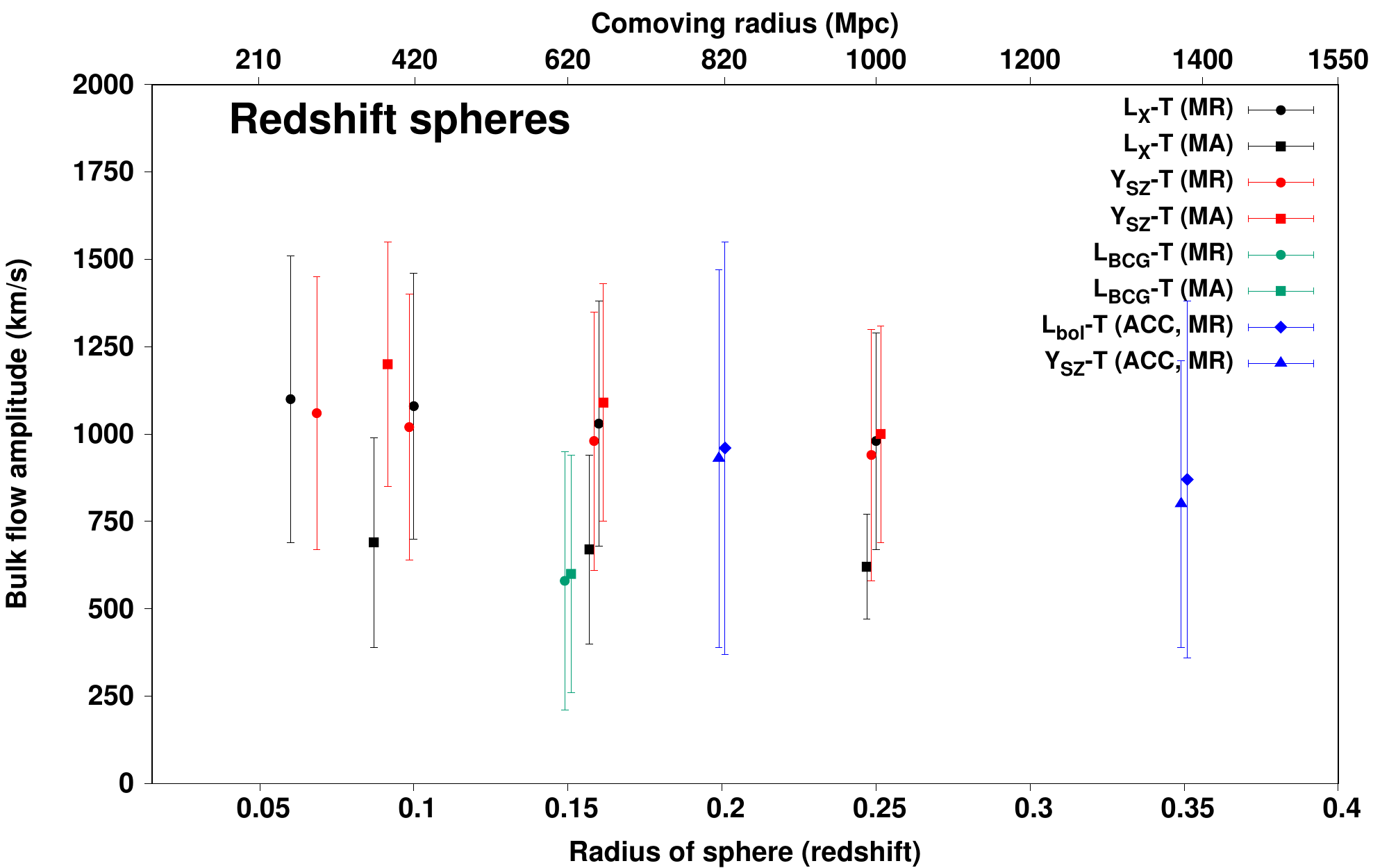}
                \includegraphics[width=0.49\textwidth, height=6cm]{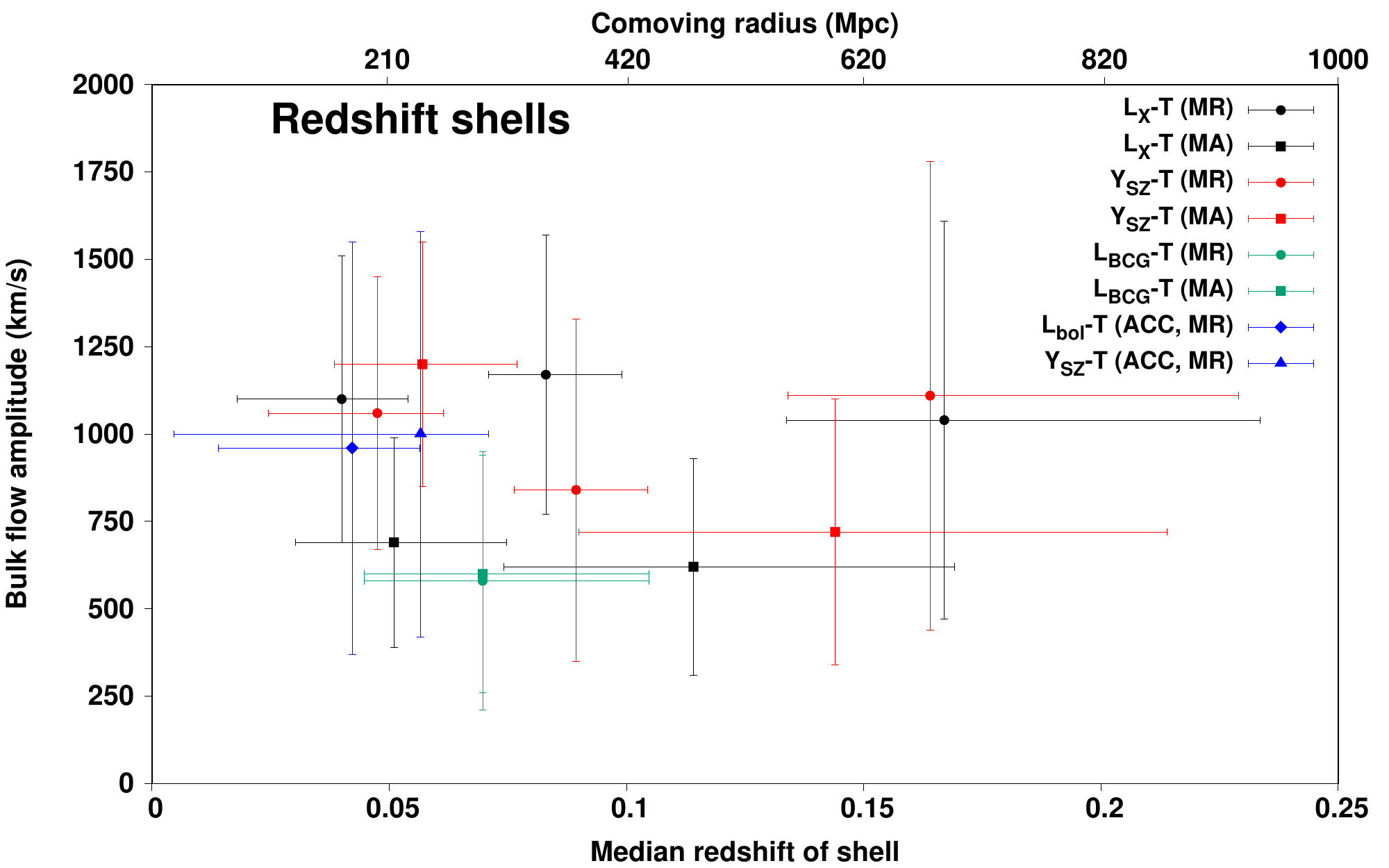}
               \includegraphics[width=0.95\textwidth, height=8cm]{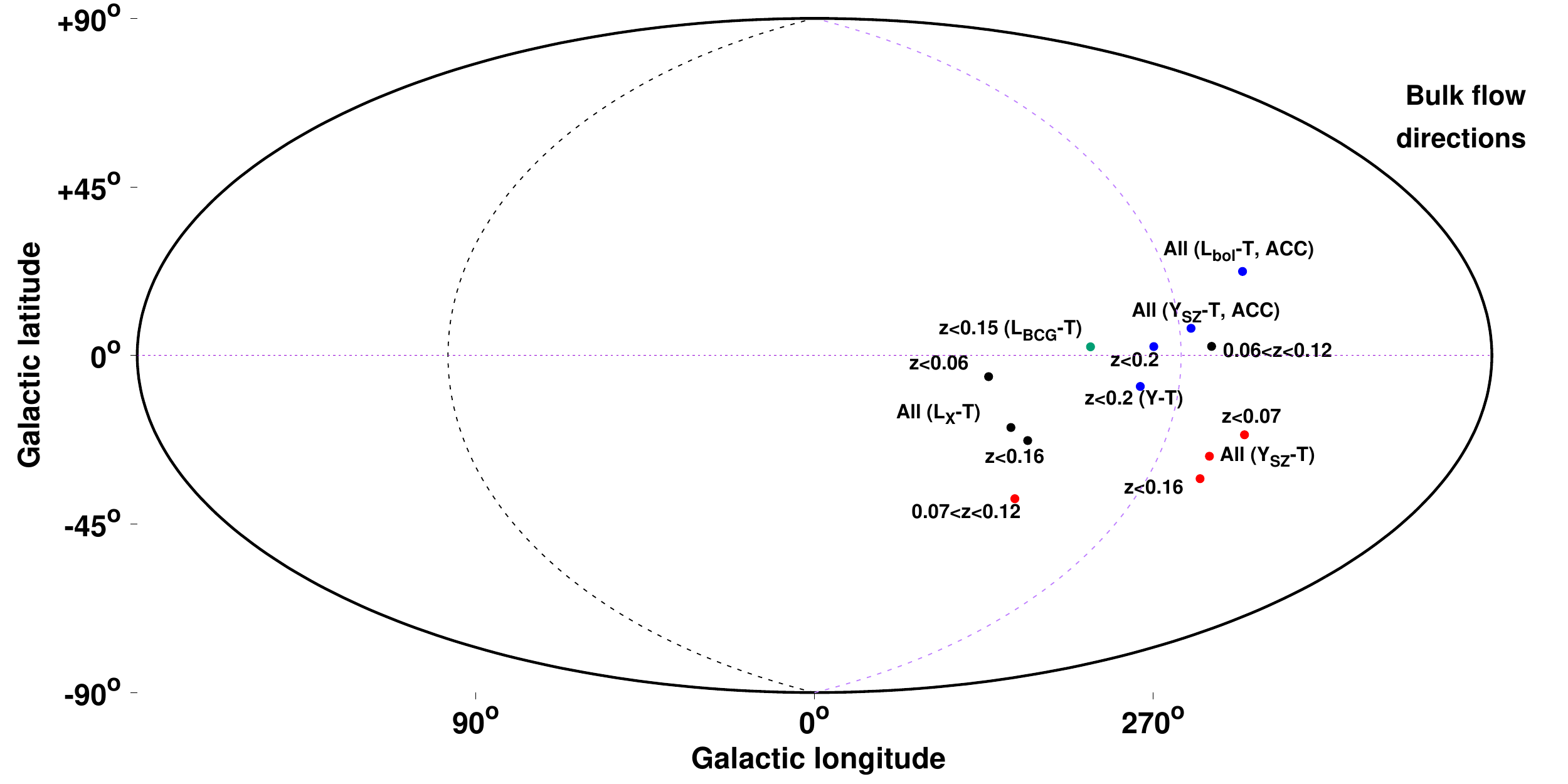}
               \caption{\textit{Top left}: Bulk flow amplitude and its 68.3\% uncertainty as a function of the redshift radius of the used spherical volumes (or the comoving distance radius). Circles and squares correspond to the MR and MA methods respectively. Black, red, and green points correspond to the \LT, \YT, and \LbcgT\ relations respectively. Blue diamonds and triangles correspond to the ACC \LbolT\ and \YT\ relations respectively. Results above $z\gtrsim 0.16$ are expected to be dominated by more local clusters and thus being overestimated. \textit{Top right}: Bulk flow amplitude and its 68.3\% uncertainty as a function of the median redshift of each shell (or the comoving distance), together with the standard deviation of the redshift distribution. The color coding is the same as before. Data points with both the same color and shape, are independent to each other. \textit{Bottom}: Examples of bulk flow directions as found for different redshift bins and methods. The color coding is the same as before. All directions agree with each other within $1\sigma$.}
        \label{BF-plots}
\end{figure*}

For now, the only obvious remaining explanations for the persisting cluster scaling relation anisotropies are two. The first one is a spatial variation of $H_0$ at a $9\%$ level (assuming no BF motions), due to a primordial anisotropy or new physics at $z\lesssim 0.2$ scales. The second one is a BF motion of $\sim 900$ km/s extending to $\gtrsim 500$ Mpc scales (assuming $H_0$ is isotropic). Another way to interpret this is that the matter rest frame is not identical to the CMB rest frame within the investigated redshift range. All of these scenarios are in tension with $\Lambda$CDM and the generally adopted isotropic assumption after transitioning to the CMB rest frame. The results for every scaling relation that traces such phenomena are summarized in Tables \ref{summary_anisot} and \ref{BF_motions}. In Fig. \ref{BF-plots}, the amplitude and directions of the detected BFs are plotted, for different redshift bins. 

Unfortunately, the above explanations are not distinguishable with the current data sets and more high$-z$ clusters are needed. The upcoming eRASS catalogs will provide us with thousands of such clusters. These distant objects will be of crucial importance when applying the techniques presented in this work, since we will be able to tell if, and at which scale, the cluster behavior converges to isotropy. The kSZ effect, which is caused by the peculiar motion of clusters in the CMB frame, can also be utilized to differentiate cosmological anisotropies from BFs. The limited number of clusters in our sample, together with the low redshifts and the large angular cluster sizes, did not allow us to extract useful information for this effect by stacking filtered kSZ maps of objects that presumably move toward a common direction. The reason for this lies in the large power of the primary CMB anisotropies at the scales that are relevant to our sample, as well as the limited angular resolution and sensitivity provided by \textit{Planck}. Next-generation SZ instruments will significantly improve over \textit{Planck's} sensitivity and angular resolution and allow for direct measurements of BFs using the kSZ effect. 

\subsubsection{Anisotropic Hubble expansion}

In the case of an anisotropic Hubble expansion, a primordial anisotropy that extends to very large scales might be present. Such an anisotropy might correlate with the CMB dipole, if the latter is not purely of kinematic origin. Many claims for the detection of such cosmological anisotropies have been made recently\footnote{The existing literature is too large to be fully included here, therefore we focus only on the most recent results, or on results that were not already mentioned in M20. We direct the reader there for many additional studies that find consistent results with this work.}. \citet{fosalba} found highly statistically significant anisotropies in the cosmological parameter constraints from the CMB using \textit{Planck}, toward a region consistent with our results. \citet{secrest} also found a large-scale anisotropy on the matter distribution, rejecting cosmic isotropy and the solely kinematic interpretation of the CMB dipole at a $4.9\sigma$ level. If their result was caused only due to our local motion within the matter rest frame, this would correspond to a $\sim 800$ km/s velocity, similar with our BF results.

Similar large-scale anisotropies in the distribution of high$-z$ radio and infrared sources have also been found \citep[e.g.,][]{tiwari,colin, bengaly18, rameez, siewert}, usually implying a $\sim 600-1500$ km/s motion compared to the matter rest frame. In other words, there are very strong indications that the matter rest frame differs from the CMB one, which is assumed to be the ultimately isotropic one. If true, this would have crucial implications on the standard model of cosmology. Recently, several theoretical frameworks that accumulate such anisotropies were developed \citep[e.g.,][]{mariusz,paliathanasis,das,spallici}. On the other hand, results that do not show any strong evidence for departure for statistical isotropy have also been presented \citep[e.g.,][]{andrade19,bengaly19}. Future eRASS catalogs will also shed more light on this question since they will provide previously unmatched catalogs of millions of X-ray point sources, tracing the matter distribution out to very large scales. 

In addition, much work has been done on testing the isotropy using SNIa. \citet{colin19} detected a $3.9\sigma$ anisotropy in the deceleration parameter using a maximum-likelihood analysis. In contrast, \citet{soltis} found no evidence of departure from $H_0$ isotropy using a novel nonparametric methodology. Except for different SNIa samples, the two papers also follow different approaches in the treatment of the data. Many other studies \citep[e.g.,][]{chang15,deng,sun2,zhao19,salehi,hu} also used SNIa combined with other probes, identifying an anisotropic direction in impressive agreement with our results. The statistical significance of their findings however is marginally consistent with an isotropic Universe. They also argue that the main problem of such SNIa studies is still the highly inhomogeneous distribution of the data across the sky, while the effect of peculiar velocity corrections has also been studied \citep[e.g.,][]{huterer20,mohayaee}. 

Finally, other kinds of nonlocal tensions with the cosmological principle have been recently observed \citep{horvath,shamir}.

\subsubsection{Large scale bulk flows}

The consistent observation of cluster anisotropies across several scaling relations could hint to the existence of large BFs that by far exceed the BF scales predicted by $\Lambda$CDM. In Fig. \ref{BF-plots}, one sees that a $\gtrsim 600$ km/s motion is consistently detected with respect to the CMB rest frame, with no signs of fading at $\gtrsim 500$ Mpc. The observed BF amplitude of spherical volumes of much larger radius might be overestimated since clusters at lower distances are expected to dominate the BF signal. However, this plays no role for the iterative redshift shells results at $z>0.12$, which seem to hint that the large BF might indeed persist farther (with a small statistical significance of $\sim 2\sigma$). More high$-z$ clusters are necessary to derive safe conclusions at these scales.

$\Lambda$CDM predict negligible BF amplitudes at scales of $\gtrsim 250$ Mpc. Flow motions of similar, large amplitudes have been reported in the past for galaxy clusters \citep[e.g.,][]{lauer,hudson99,kashl08,kashl10,atrio-bar}, while \citet{osborne,mody,planck14} fail to see any large scale BFs in the CMB kSZ data. Studies using thousands of galaxies, consistently detect a BF toward the same direction with us, but with a $\sim 3-5$ times smaller amplitude. However, it also seems to extend to scales larger than $\sim 300$ Mpc, up to where the galaxy surveys are usable. For instance, \citet{watkins} and \citet{magoulas} found a $\sim 400$ km/s BF at $\sim 100-150 h^{-1}$ Mpc scales toward $(l,b)\sim (295^{\circ},+10^{\circ})$, which is moderately discrepant with the standard cosmological model. \citet{carrick} and \citet{boruah} found a $\sim 170$ km/s BF toward $(l,b)\sim (303^{\circ},+3^{\circ})$ that extends to $z>0.067$, but not particularly inconsistent with the standard expectations. \citet{lavaux} used the kSZ effect on galaxy halos and found a BF of $\sim 285$ km/s for $200h^{-1}$ Mpc scales. \citet{watkins2} also found a good agreement with $\Lambda$CDM at small scales, but a moderate disagreement at larger ones. Furthermore, \citet{peery} find only a $2\%$ chance for their detected BF at scales $150h^{-1}$ Mpc to occur within $\Lambda$CDM. At $z<0.07$ scales, the reason behind the mild disagreement in the BF amplitude between our analysis and studies of peculiar motions of galaxies remains unknown. At larger scales, a direct comparison cannot be made yet, due to the lack of BF measurements based on galaxies.

Studies have tried to constrain the BF with SNIa data as well, which extend to much larger scales, but are highly inhomogeneous \citep[][and references therein]{colin2011,feindt,appleby-john,mathews,salehi2}. Such studies usually find similar results to the galaxy surveys, where mild BFs extend to larger-than-expected scales. However, they again stress the limitations of current SNIa samples for such studies and that conclusive answers cannot be given. Other studies have stressed that the usual Newtonian treatment of peculiar velocity fields can significantly underestimate the inferred bulk flows \citep[e.g.,][]{tsagas,tsaprazi}. From all that it is evident that it is still not quite clear if the detected BF motions are entirely in agreement with the standard model, and out to what scale the BF persists. 

Substantial, local matter inhomogeneities such as local voids or overdensities, could contribute to giving rise to unexpected large BFs. If we are located away from the void center, this could cause the apparent expansion rate to be lower toward the more matter-dominated side. There have been claims for the existence of such large voids (or overdensities) that could create outflows out to $\sim 400$ Mpc scales \citep[][and references therein]{keenan,rubart14,whitbourn,shanks19b,shanks19,tully19,bohringer20,kazantzidis,moritz}, and their effects on the measured $H_0$ have been studied. These scales are close to the median distance distribution of our cluster samples, and thus could indeed affect our results. The predicted $H_0$ variation due to these voids is slightly lower than the one we observe, but generally agrees. Of course, the fact that they offer a possible (at least partial) explanation for the local cluster anisotropies, does not alleviate the problem, since the existence of such large voids is in direct disagreement with the $\Lambda$CDM.

\section{Summary} \label{conclusions}

In this work, we applied a scrutinized test for the isotropy of the local Universe. We studied the anisotropy of 10 galaxy cluster scaling relations, utilizing observations in X-rays, infrared, and submillimeter. Using the \LT, \YT, and \LbcgT\ scaling relations of eeHIFLUGCS, and combining with the completely independent ACC sample, we detected a $\sim 5.5\sigma$ anisotropy toward $(l,b)\sim ({280^{\circ}}^{+35^{\circ}}_{-35^{\circ}},{-15^{\circ}}^{+20^{\circ}}_{-20^{\circ}})$. This high statistical significance is further confirmed by applying our methods to isotropic Monte Carlo simulated samples. Considering the low median redshift of the cluster data ($z\sim 0.1$), this direction agrees with a plethora of past studies of several probes. We robustly showed that our data do not suffer from unknown X-ray absorption issues, and our results do not originate from any kind of known biases. Moreover, it would be difficult for any "typical" bias to simultaneously explain the similar anisotropic results across nearly independent scaling relations, in different wavebands. Nevertheless, more work is needed to further confirm our work is not prone to any currently unknown systematics.

If the observed anisotropies were the result of an anisotropic expansion rate, one would need a $\sim 9\%$ spatial variation of $H_0$ to reconcile with the observations. Alternatively, one would need a bulk flow motion of $\sim 900$ km/s, possibly extending beyond $500$ Mpc, to explain the obtained cluster anisotropies. Due to the low redshift range of our samples, these two phenomena are currently inseparable for a typical observer. However, both of these scenarios are in tension with the standard assumptions in $\Lambda$CDM. Since these are currently the only available explanations for our observations, we are faced with a severe problem that needs to be solved.  The future eRASS catalogs will help us understand these anisotropies better, and determine if there is a scale of convergence with isotropy, or if this anomaly extends to much larger scales. 

\begin{acknowledgements}  
We thank the anonymous referee for their constructive comments that helped us improve our manuscript. KM is a member of the Max-Planck International School for Astronomy and Astrophysics (IMPRS) and of the Bonn-Cologne Graduate School for Physics and Astronomy (BCGS), and thanks for their support. GS acknowledges support through NASA Chandra grant GO5-16137X. LL acknowledges financial contribution from the contracts ASI-INAF Athena 2019-27-HH.0, ``Attivit\`a di Studio per la comunit\`a scientifica di Astrofisica delle Alte Energie e Fisica Astroparticellare'' (Accordo Attuativo ASI-INAF n. 2017-14-H.0), and from INAF ``Call per interventi aggiuntivi a sostegno della ricerca di mainstream di INAF''.
This research has made use of the NASA/IPAC Extragalactic Database (NED)
which is operated by the Jet Propulsion Laboratory, California Institute of Technology, under contract with the National Aeronautics and Space Administration.
\end{acknowledgements}

\newpage

\bibliographystyle{aa} 
\bibliography{XXX}          

\newpage

\appendix

\section{Measurements of \Ysz, \Lbcg, $R$, and $N_{\text{H, Xray}}$}

\subsection{Details on the \Ysz\ measurement}\label{details_ysz}

The MMFs algorithms are applied to the \textit{Planck} data in steps that are analogous to the ones presented by A16 (Sect. 2.1). Fields with a size of $10^{\circ} \times 10^{\circ}$ are extracted around the X-ray coordinates of each cluster at each of the \textit{Planck} HFI bands. The Low Frequency Instrument (LFI) channels are excluded because of their low spatial resolution and lower sensitivity. The six \textit{Planck} HFI maps for each cluster are then processed with a MMF, which filters and combines them into a single map. The HFI beams are approximated as two-dimensional Gaussians with a solid angle that is equivalent to that of the effective beams. The corresponding FWHMs are taken from Table 12 of \citet[][]{Planck_HFI}. The required spectral energy distribution (SED) of the tSZ effect has been computed for both the non-relativistic case as well as with relativistic corrections \citep[computed with SZpack;][]{Chluba12} as explained later. In all cases, the instrumental impact on the shape of the SED has been taken into account by computing the bandpass-corrected spectra (see Eq. 13, Eq. A1, and Table A1 in \citealt{erler18}).  

Applying the filters and coadding the filtered maps yields a map of the deconvolved central Comptonization parameter $y_0^{\text{MMF}}$ in each pixel of the map. We determine the tSZ-measured centre of each cluster by the position of the brightest pixel within a $15\arcmin \, \times \, 15\arcmin$ box around the X-ray coordinates. Using the normalized cluster template, the value of $y_0^{\text{MMF}}$ is converted to $Y_{5R500}$ by
\begin{equation}
	Y_{5R500} \ \left[\mathrm{arcmin}^2\right] = y_0^\mathrm{MMF} 2 \pi \int_0^{5\theta_{500}} \mathrm{d}\theta \, y(\theta) \, \theta,
\end{equation}
where $y(\theta)$ is the $y$-parameter profile of the cluster that has been normalised to unity, and $\theta$ is the radial angular coordinate in the plane of the sky, and $\theta _{500}$ being the apparent $R_{500}$ in arcmin.

A total of four different $Y_{5R500}$ values are extracted for every cluster. The first value is obtained via a MMF approach using the nonrelativistic spectrum of the tSZ effect. This method is analogous to the MMF1 and MMF3 pipelines used by A16. The second value is obtained via a MMF approach using the relativistic spectrum of the tSZ effect computed using the M20 estimates for $T$. When the latter is not available, $T$ is calculated through the $L_{\text{X}}-T$ relation found in M20, where $L_{\text{X}}$ is given by MCXC, after it has been corrected for the X-ray absorption based on the \citet{willingale} values (following the same procedure as in M20). The third value is obtained via a constrained MMF approach (CMMF, E19) using the nonrelativistic spectrum of the tSZ effect and the spectrum of the kSZ effect. The latter is used to remove any bias introduced by the kSZ effect. The fourth and last value is obtained identically to the third value, but using a relativistic spectrum of the tSZ effect. 

The four $Y_{5R500}$ values do not differ significantly to each other $(<7.2\%)$, and any selection for the default $Y_{5R500}$ used in this work results in the same conclusions with only minimal numerical fluctuations. We choose as the default $Y_{5R500}$ the result of the fourth approach, where a CMMF approach is followed as described in E19, correcting for relativistic effects and attempting to remove any kSZ bias. The change in the $Y_{5R500}$ values due to the removal of the kSZ effect is usually much smaller than the 1$\sigma$ uncertainties of the central values. This has nearly no effect in any BF detection from our analysis, since the major contribution of BFs to \Ysz\ comes from $D_A$ and the (biased) redshift-distance conversion.

\begin{figure}[hbtp]
\centering
               \includegraphics[width=0.5\textwidth, height=6cm]{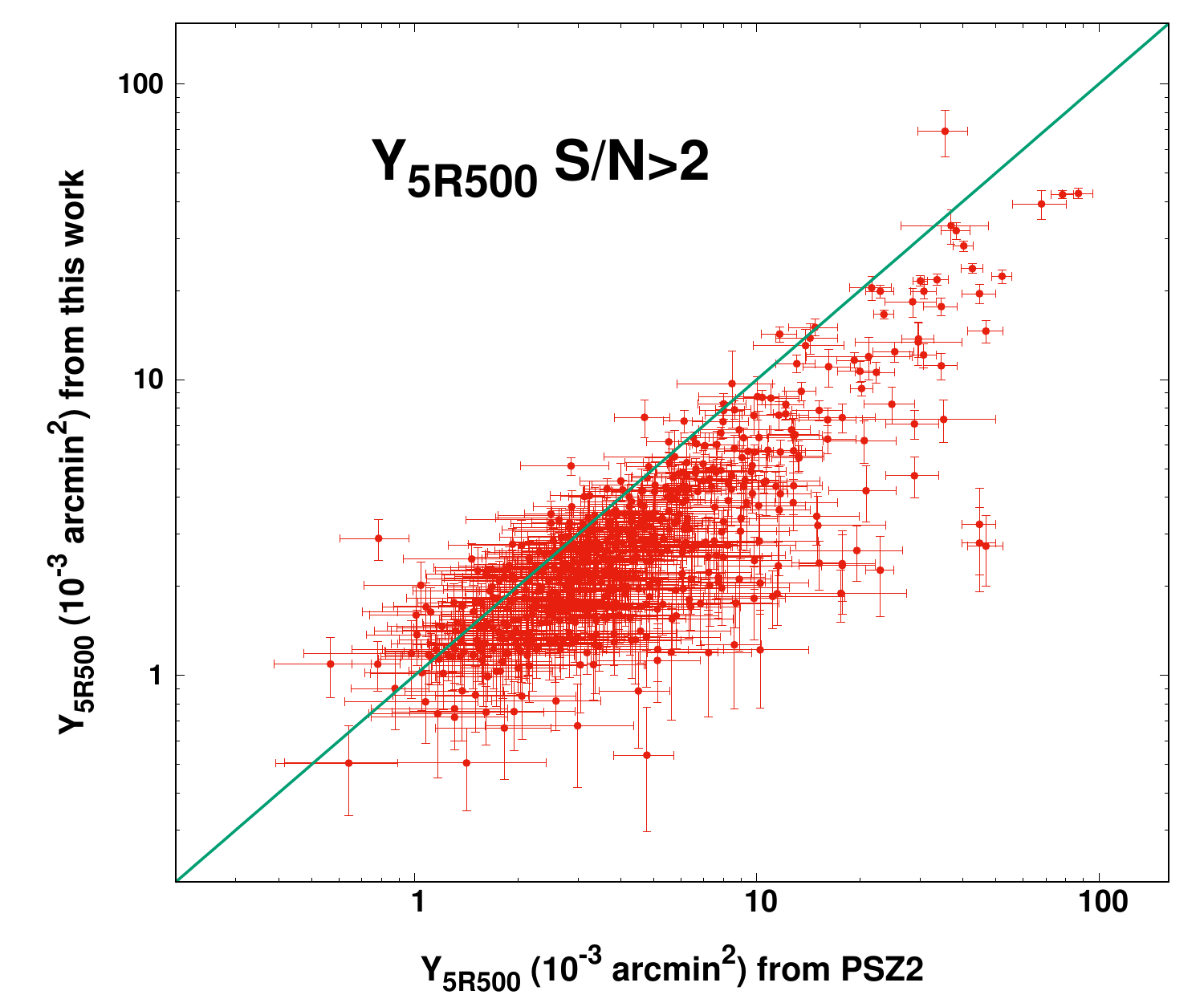}
               \caption{Comparison between the default $Y_{5R500}$ values we use in this work based on the method of E19 and the $Y_{5R500}$ values as given in the PSZ2 catalog for the 566 clusters in common. The equality line is displayed in green.}
        \label{psz2-jens}
\end{figure}

\subsubsection{Comparison with the PSZ2 values}\label{psz2-comp}

In Fig. \ref{psz2-jens} we compare the values of our default $Y_{5R500}$ with the ones from PSZ2 for 566 common clusters\footnote{The matching criteria were that the angular separation of the two given cluster centers should be $\leq 1.5^{\circ}$ and the redshift difference $\Delta z\leq 0.01$ between the two catalogs.}. One can see there is a linear relation between the two, which demonstrates the robustness of our method. Nevertheless, a systematic departure from the values reported by the \textit{Planck} Collaboration is observed, especially in higher $Y_{5R500}$ values. This partially arises from the fact that A16 estimate the size of clusters directly from the SZ data by mapping the significance of the obtained values for $Y_{5R500}$ as a function of MMF filter templates of different cluster size, parameterized through the scale radius ${\theta}_s={\theta}_{500}/c_{500}$, where $c_{500}$ is the cluster's concentration parameter. However, for the vast majority of clusters, the $\theta_s-Y_{5R500}$ plane is left largely unconstrained, for which reason the \textit{Planck} Collaboration employs an XMM-Newton derived $\theta-Y_{5R500}$ prior to break the degeneracy between the two parameters (e.g. Fig. 16 of A16). Additionally, small differences in the adopted centers might contribute to the observed scatter as well\footnote{The used SZ center is the position of the brightest pixel within a $15\times 15$ arcmin$^2$ box centered around the X-ray peak.}. Although the data employed here are identical and our filtering is based on the same MMF technique as used by A16, differences in the reported values for $Y_{5R500}$ can therefore still appear. The comparison does not significantly change with the use of $Y_{5R500}$ values based on the other three E19 approaches we described in Appendix \ref{details_ysz}. 

\subsection{Details on obtaining the near infrared BCG luminosity $L_{\text{BCG}}$}\label{bcg_select}

The main criteria for a galaxy to be selected as the BCG were the following. Firstly, it must be the brightest in the \textit{r}-band of the optical regime or \textit{Ks} and \textit{W1} in 2MASS and WISE of the NIR domain. Secondly, it should be located within a distance of $R_{500}$ from the X-ray center of the cluster. Both of these values are taken by the MCXC catalog. Thirdly, the BCG's redshift must be within a certain range $\Delta z$ of the cluster redshift. If the BCG redshift was estimated via photometric measurements, then $\Delta z\leq 0.02$, whilst the allowed redshift difference for spectroscopic redshifts was $\Delta z\leq 0.01$. If multiple BCG candidates were suggested for one cluster, the following criteria were considered. The BCG should possess a rather extended envelope and it should be selected by optical surveys over NIRs. 

Two different corrections took place before selecting the BCG of every cluster. The galaxy magnitudes were corrected for galactic extinction, using the Schlegel maps \citep{schlegel1998maps} and assuming an extinction law \citep{fitzpatrick1999correcting} with $R_V$= 3.1. The photometric magnitudes were also corrected to the rest frame of individual galaxies by applying an appropriate k-correction. The k-correction code provided by \citet{chilingarian2011universal} was used for all data except for WISE, as the coefficients for WISE filters are not included in this program. For WISE galaxies, we instead used the k-correction code from Blanton \citep{blanton2007k}. 

\subsection{Details on the X-ray determined $N_{\text{H,Xray}}$}\label{details_xray_nh}

To determine $N_{\text{H,Xray}}$, we used the exact same spectral fitting procedure as described in M20, when the X-ray redshift was fitted. This time the $N_{\text{H,Xray}}$ is left free to vary instead. Its value was linked between the spectra from the $0-0.2\ R_{500}$ region and the $0.2-0.5\ R_{500}$ region. The redshift is kept fixed at its M20 value, while the temperature, metallicity and normalization values are left free to vary for each region separately. We used the 0.7-7.0 keV range to fit the spectra (to be consistent with the results from M20). The \citet{aspl} abundance table was used for the fits.

We were able to constrain $N_{\text{H,Xray}}$ for 213 out of 237 Chandra clusters and 27 out of 76 XMM-Newton clusters. When we performed the $N_{\text{H,Xray}}-$\nhtot\ comparison, we saw that the data residuals compared to the best-fit model behaved strongly as a function of the $N_{\text{H,Xray}}$ measurement uncertainty. Clusters with large statistical uncertainties were significantly downscattered, increasing the scatter of the relation. These were mostly low $N_{\text{H,Xray}}$ clusters. The 0.7 keV cut removes most of the power for detecting \nhtot\ (which shows up more strongly at lower energies). So when the \nhtot\ effects are already small, a constraint is nearly impossible. To sufficiently restrict the noise, we excluded the $35\%$ of the sample with the largest (average) uncertainties, leaving us with 156 clusters with an $N_{\text{H,Xray}}$ measurement uncertainty of $<25\%$. Although these cuts were chosen arbitrarily, they took place strictly before any anisotropy test, to avoid any confirmation bias.

\section{ACC results}\label{acc_results}

In this section, we only use the ACC sample to study the anisotropic behavior of the \LbolY, \LbolT, and \YT\ relations. Due to the much smaller number of clusters, we consider a $\theta=90^{\circ}$ scanning cones for all scaling relations, so at least 35 clusters lie within each cone. 

\subsection{The \LY\ relation}

We study the \LbolY\ relation for ACC alone. The results are shown in Fig. \ref{ACC-plots}. 

The maximum anisotropy is found toward $(l,b)=(24^{\circ}\pm 42^{\circ} ,+22^{\circ}\pm 21^{\circ} )$ at a $2.3\sigma$ level. These 59 clusters are $26\pm 11\%$ dimmer than the rest. This level of anisotropy can be considered marginally consistent with an isotropic \LbolY\ relation. If however there was some extra X-ray absorption taking place toward that direction, it would correspond to an extra \nhtot$\sim 7.9\pm 3.4\times 10^{20}/$cm$^2$. This value is consistent with the \LY\ result of our sample within $1.1\sigma$. The maximum anisotropy region however is located $90^{\circ}$ away from the \LY\ direction. This could indicate that the normalization variation is due to statistical noise and does not reveal any unknown absorption. Alternatively, it could be attributed to the incapability of the $\theta=90^{\circ}$ cones to accurately pinpoint the direction where the hidden X-ray absorption takes place, especially since the direction uncertainties are considerably large. Based on all the above, the case that the mild anisotropies seen in the ACC sample are due to chance cannot be excluded. 

Finally, ACC further confirms our previous conclusion about the \LT\ anisotropies found in M20. There is no indication that they appear due to previously unaccounted X-ray effects, since the region toward $(l,b)\sim (300^{\circ},-20^{\circ})$ shows a completely consistent behavior with the rest of the sky.

\subsection{The \LbolT\ relation}\label{acc-LT}

\begin{figure*}[hbtp]
               \includegraphics[width=0.49\textwidth, height=4.5cm]{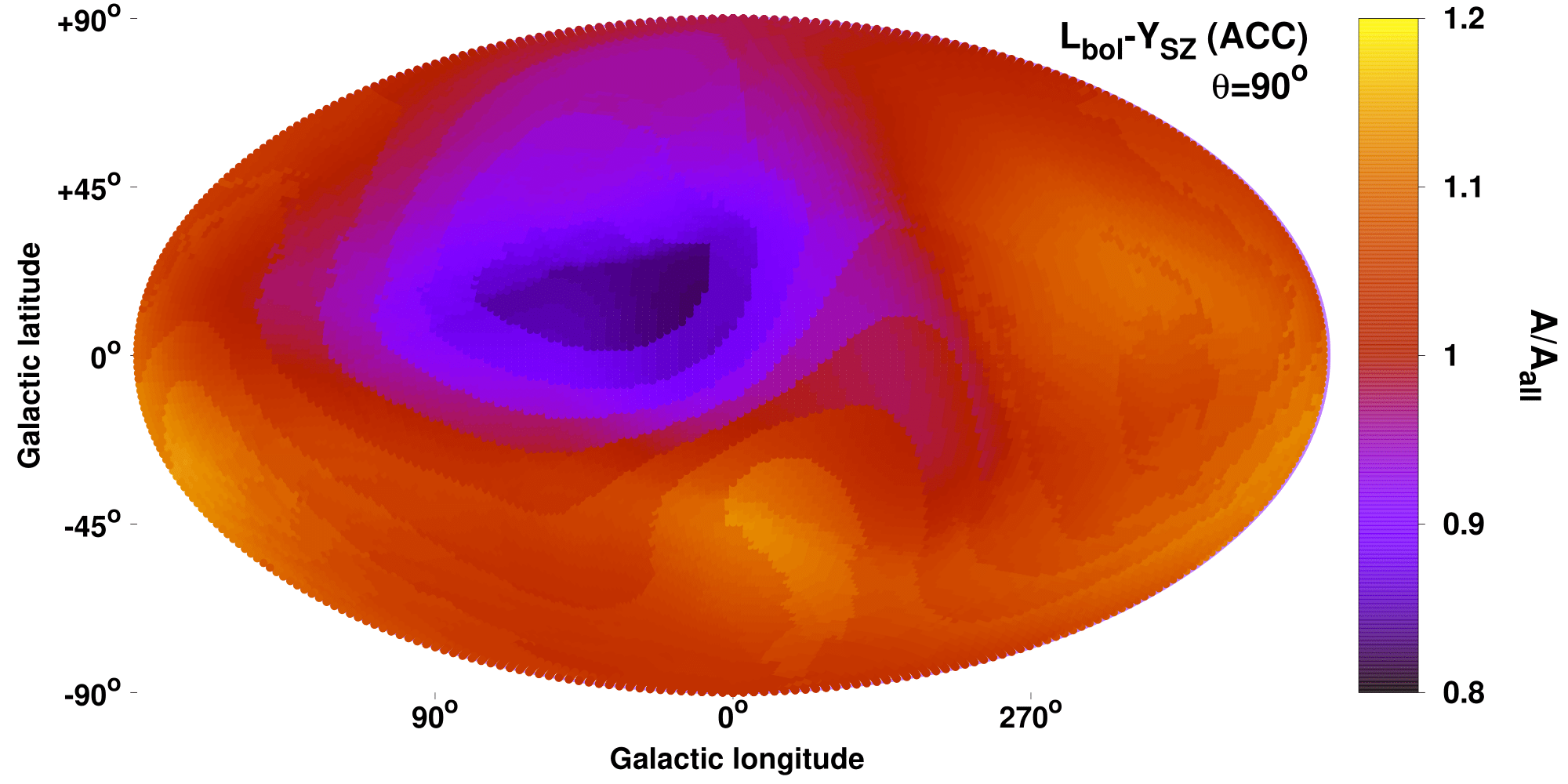}
               \includegraphics[width=0.49\textwidth, height=4.5cm]{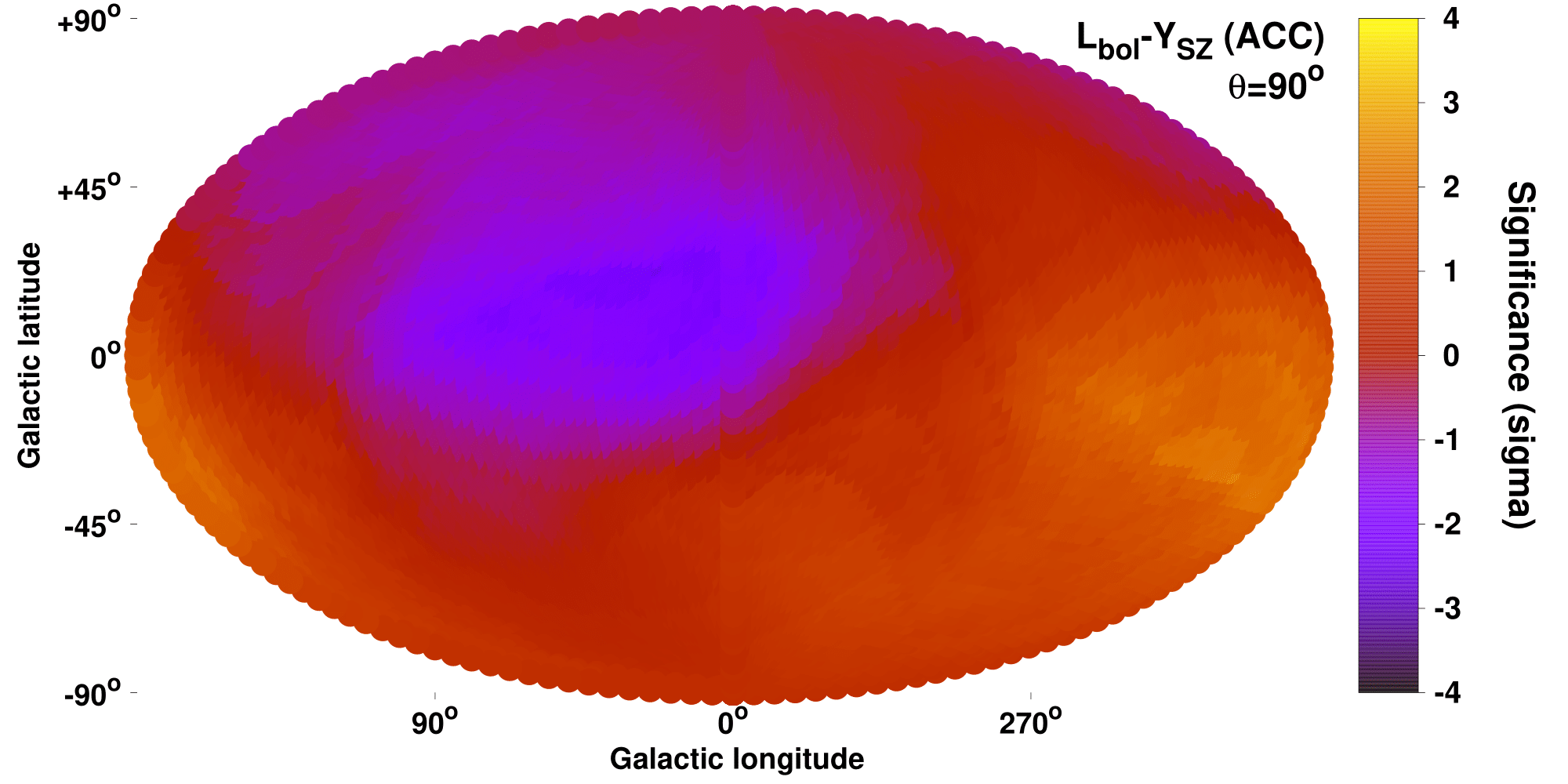}
               \includegraphics[width=0.49\textwidth, height=4.5cm]{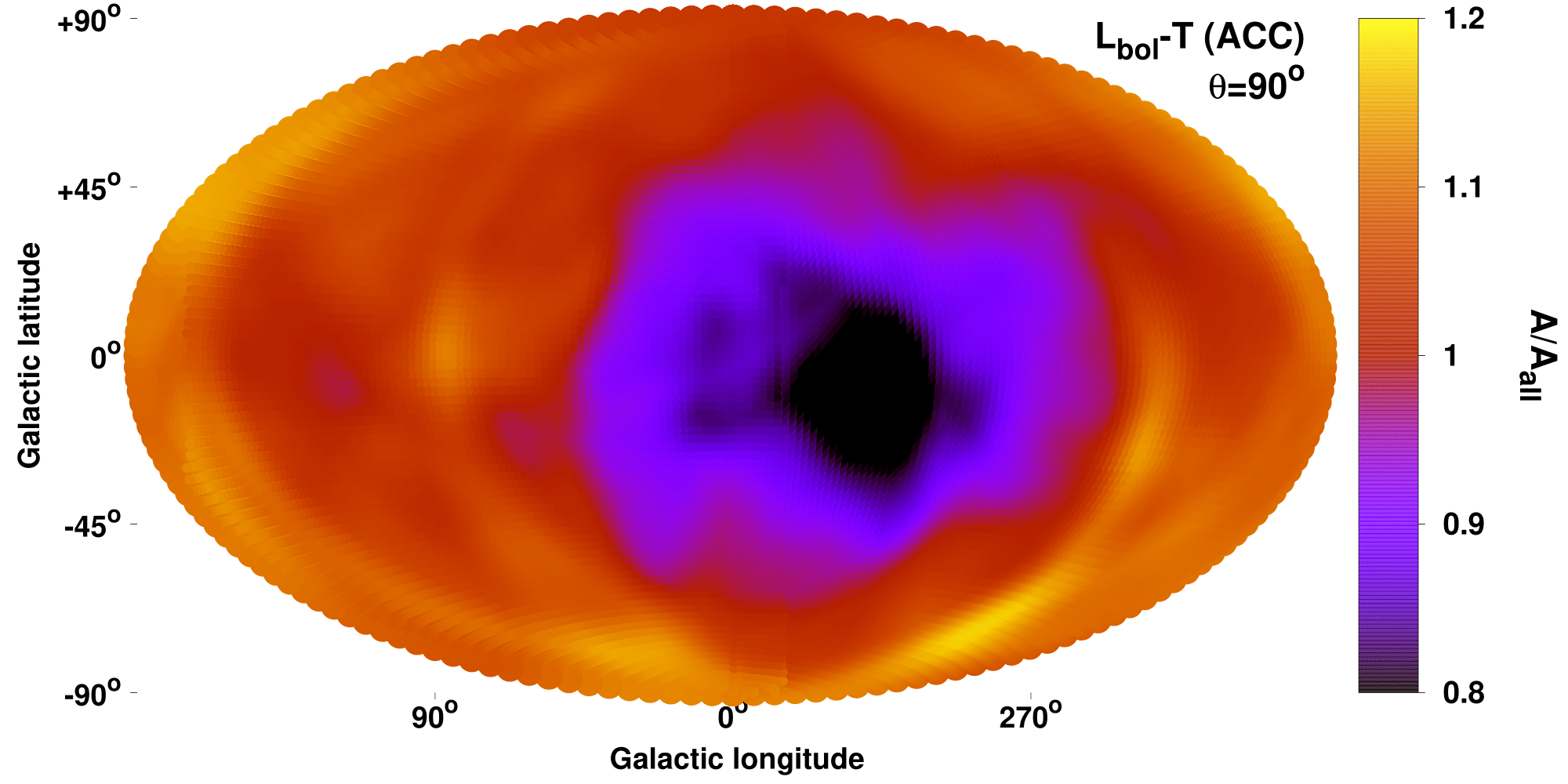}
               \includegraphics[width=0.49\textwidth, height=4.5cm]{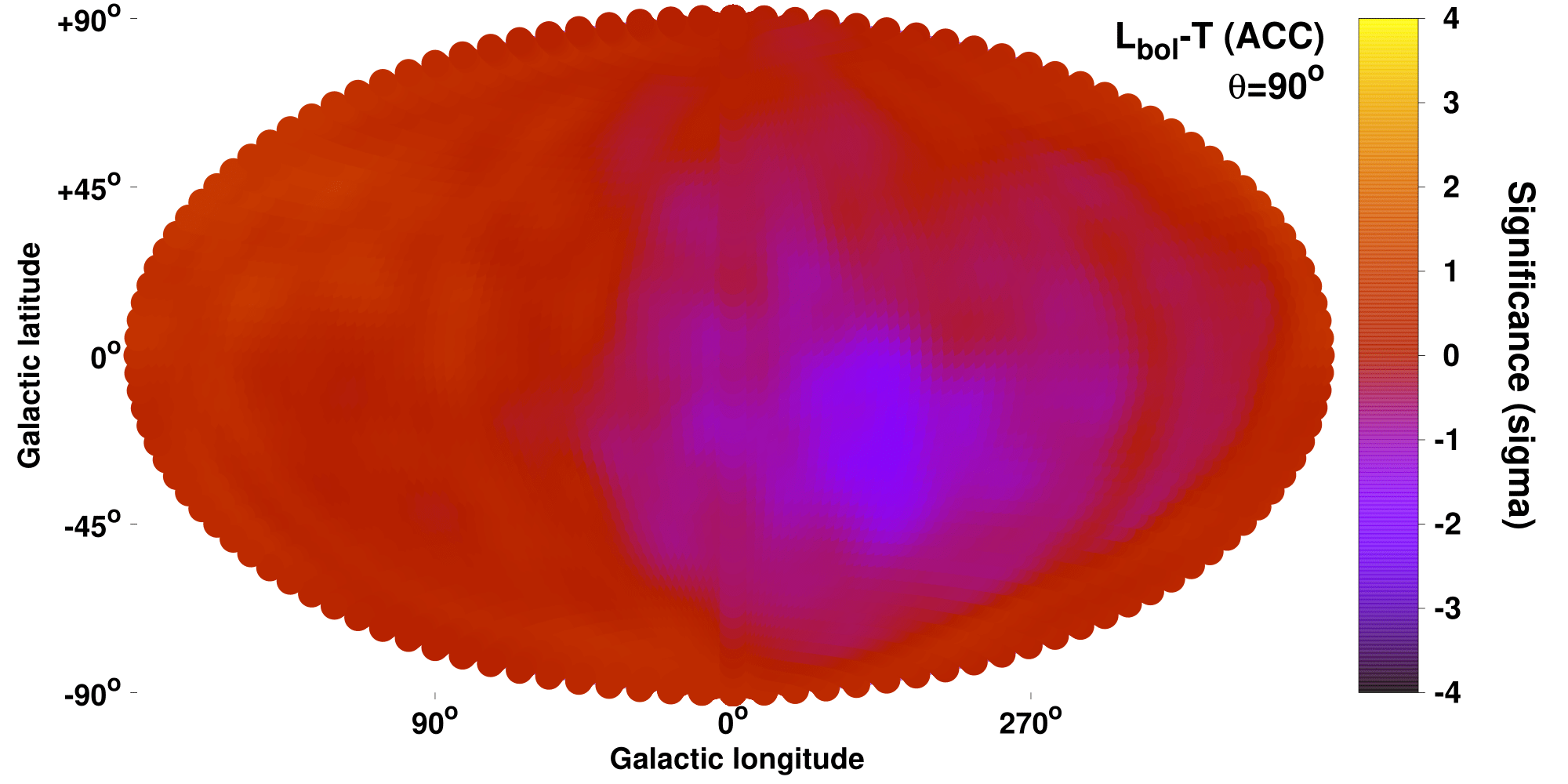}
               \includegraphics[width=0.49\textwidth, height=4.5cm]{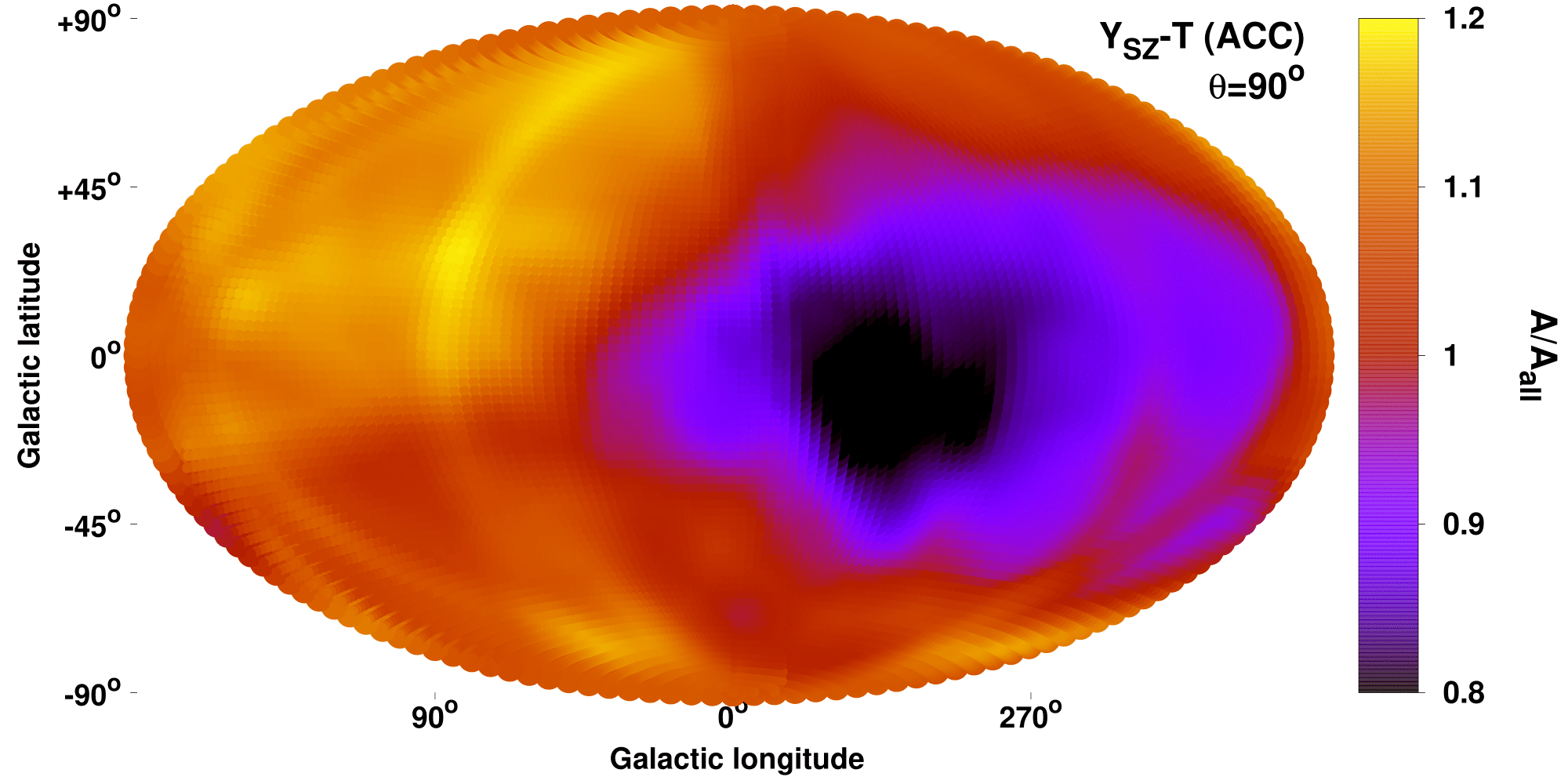}
               \includegraphics[width=0.49\textwidth, height=4.5cm]{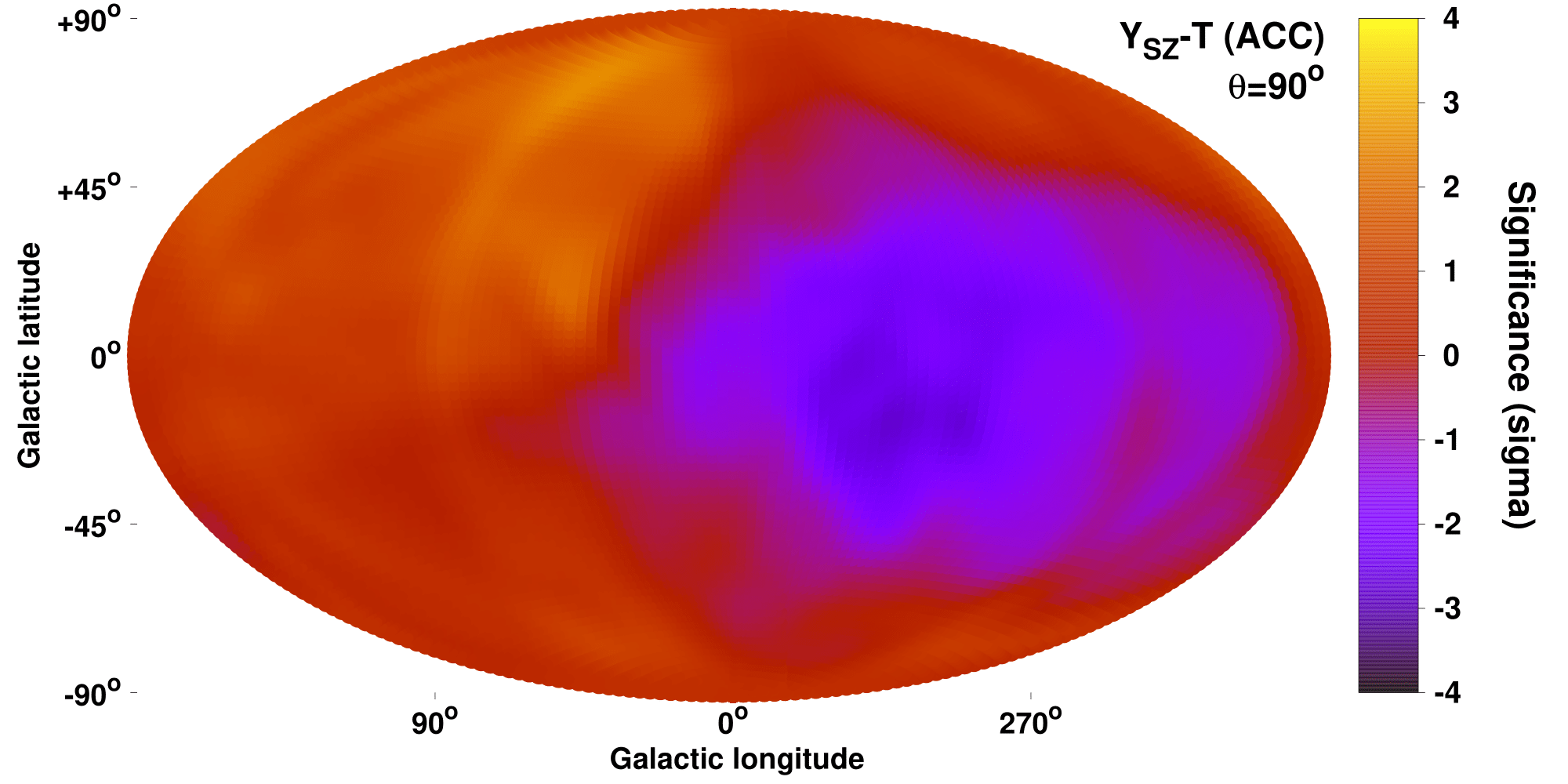}
               \caption{Normalization anisotropy maps and the respective statistical significance maps of the anisotropies for the \LbolY\ (top), the $L_{\text{bol}}-T$ (middle), and the \YT\ (bottom) scaling relations for the ACC sample. The \LbolY\ traces only unaccounted X-ray absorption effects, while the \LT\ and \YT\ behavior mirrors cosmological anisotropies and BFs.}
        \label{ACC-plots}
\end{figure*}

For the \LbolT\ relation of ACC, the maximum anisotropy is found toward $(l,b)=(318^{\circ}\pm 45^{\circ},-9^{\circ}\pm 37^{\circ})$, in a $2.2\sigma$ tension with the rest of the sky\footnote{In M20 we considered $\theta=75^{\circ}$, because the slope was kept fixed, and fewer cluster per regions were needed to sufficiently determine the normalization. Here the slope is left free to vary.}. This region appears fainter than the rest by $\sim 32\pm 13\% $ on average. Its behavior is almost identical to the results of M20, as seen in Fig. \ref{ACC-plots}. However, the decreased statistical significance suggests that the \LbolT\ relation for ACC alone is marginally consistent with isotropy. The angular separation between the most anisotropic regions of ACC and our sample is $40^{\circ}$, and well within the uncertainties.

\subsubsection{Cosmological anisotropies and bulk flows}

For ACC, one would need $H_0=61.3\pm 4.2 $ km/s/Mpc toward $(l,b)\sim (318^{\circ},-9^{\circ})$ and $H_0=72.2\pm 2.4 $ km/s/Mpc for the rest of the sky. The obtained $H_0$ value agrees within $1\sigma$ with the independent result from our sample.


For the BF explanation, applying the MR method to ACC, we find a BF of \ubf$=850\pm 410$ km/s toward $(l,b)=(254^{\circ}\pm 52^{\circ},+18^{\circ}\pm 27^{\circ})$. This direction is separated by $31^{\circ}$ from the CMB dipole, and by $67^{\circ}$ from the maximum anisotropy region, although this difference is within $1\sigma$. Due to the limited number of data and the large scatter, we can only divide the sample into two independent redshift bins, $z<0.2$ and $z>0.2$. For the former we practically find the same BF as for the full sample, while for the latter we find no statistical significant evidence of a BF. However, the results are inconclusive due to the limited number of available clusters in this sample, and the subdominant effect BFs have at these redshifts compared to the intrinsic scatter.



For the MA method, we obtain \ubf$=810\pm 400$ km/s toward $(l,b)=(324^{\circ}\pm 45^{\circ},-3^{\circ}\pm 32^{\circ})$. This direction is more in agreement with the maximum anisotropy direction of ACC, while the BF amplitude is similar. The sample size is not sufficient to consider individual redshift bins that would provide different insights than the full sample. 

\subsection{The \YT\ relation}\label{acc-YT}

We repeat the \YT\ anisotropy analysis with only the 113 ACC clusters with \Ysz\ S/N$>2$, with a median $z\sim 0.22$. The scatter is similar to our sample, but the limited number of clusters forces us to consider $\theta=90^{\circ}$ cones. This way we ensure we have at least 35 clusters in each cone (which inevitably leads to large uncertainties). The normalization and sigma maps are displayed in Fig. \ref{ACC-plots}. We find an anisotropy of $3.2\sigma$ toward $(l,b)=(311^{\circ}\pm 57^{\circ} ,-12^{\circ}\pm 39^{\circ} )$, $42^{\circ}$ away from the direction of our sample and well within the $1\sigma$ uncertainties. The relative difference of $A_{YT}$ of this region compared to the rest of the sky is $35\pm 11\%$. The direction is also identical to the one obtained with the \LbolT\ relation, with a larger statistical significance.

\subsubsection{Cosmological anisotropies and bulk flows}

In terms of $H_0$, one obtains $H_0=60.6\pm 3.6$ km/s/Mpc toward the most anisotropic region, and $H_0=73.4\pm 1.9$ km/s/Mpc for the opposite hemisphere. These values are consistent within $<1.4\sigma$ with our sample's results, and with the joint \LT\ analysis results. 

For the BF scenario and the MR method, we obtain \ubf$=800\pm 400$ km/s toward $(l,b)=(268^{\circ}\pm 42^{\circ},+6^{\circ}\pm 29^{\circ})$ for the full sample. Once again for ACC, the direction is close to the CMB dipole, but slightly shifted compared to the maximum anisotropy direction. The BF has its usual amplitude, even though the used sample has a large median $z\sim 0.22$. For $z<0.2$ we obtain a slightly larger BF than for the full sample. For $z>0.2$, the BF points toward $(l,b)\sim (320^{\circ},+15^{\circ})$, however it is poorly constrained and not statistical significant.

For the MA method, we find \ubf$=810\pm 370$ km/s toward $(l,b)=(273^{\circ}\pm 38^{\circ},-11^{\circ}\pm 21^{\circ})$ for the full sample, in agreement with the MR method.

\section{Anisotropies in the other scaling relations} \label{other_scal}

The \RL, \RT, \RY, \RLbcg, and \YLbcg\ scaling relations cannot currently provide meaningful insights on the possible origin of the detected anisotropies. However, they may prove to be very useful for future tests with larger samples, and with a better characterization of the dynamical state of the clusters. Here we discuss their results, limitations, and future potential.

\subsection{Anisotropies of the \RL, \RT, \RY, and \RLbcg\ scaling relations}\label{R-relations}

The scaling relations of the cluster effective radius are potentially very interesting, once the systematic biases are properly handled and the sample sizes are further increased. For the \RL, \RY, and \RLbcg\ relations, both quantities depend on the cosmological parameters, but due to the rather flat slopes, the normalization weakly depends on the angular diameter distance ($A_{RX}\propto D_A^{(0.5-0.7)}$). Thus, a $15\%$ spatial variation of $H_0$ would result in a $7-10\%$ variation of $A_{RX}$. Much smaller effects are expected due to BFs. This level of anisotropy would not be detectable over random noise combined with the mild, existing systematic issues. This is the most crucial limitation of these relations. Future samples, such as the eRASS catalogs, will dramatically increase the sample sizes, which will reduce the random noise and allow the detection of $\sim 10\%$ anisotropies.

For the \RT\ relation, there is an $A_{RT}\propto D_A$ dependency. For the same underlying effect, \RT\ will show half the normalization variation than \LT, \YT\, and \LbcgT. Despite the weaker signal, it can still be a valuable test to detect cosmological anisotropies and BFs when applied to larger cluster samples, and when a more precise modelling of the systematics is feasible.

$R$ is also rather insensitive to unaccounted X-ray absorption effects. The latter can only affect the determination of $R_{500}$, which is based on the absorption-affected flux. However, \Lx$\propto R_{500}^{4.8}$, so a $20\%$ bias in \Lx\ would only cause a $<4\%$ bias in $R_{500}$, which is negligible. To confirm this, we remeasure $R$ for all clusters, after changing the input \nhtot\ by $\pm 50\%$. Almost all clusters ($99.5\%$ of the sample) show an $R$ change of $<2.5\%$.

\begin{figure}[hbtp]
               \includegraphics[width=0.49\textwidth, height=6cm]{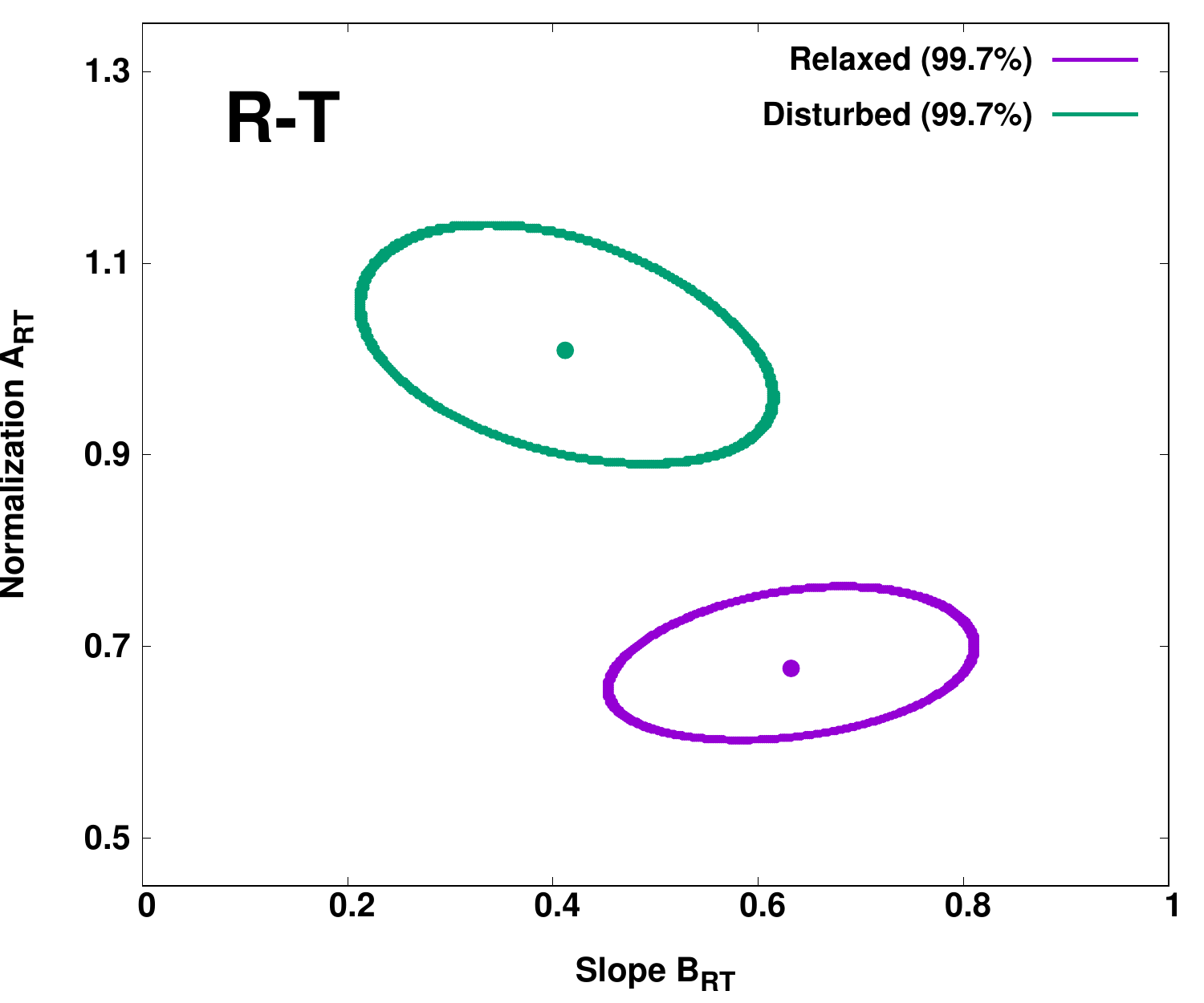}
               \caption{$3\sigma$ (99.7\%) parameter space of the normalization and slope of the \RT\ relation, for relaxed (purple) and disturbed (green) clusters.}
        \label{R-a-b}
\end{figure}

The main systematic effect compromising the $R$ scaling relations is the effect of the CC clusters on the measurements. This was already discussed in Sect. \ref{R-L-relation}, and is more extensively discussed here. The directional behavior of these relations clearly correlates with the dynamical state of the clusters. The latter affects their surface brightness profiles, and consequently their half-light radii. For the \RT\ relation, the relaxed systems appear to be $45\pm 7\%$ "smaller" (lower normalization) than the disturbed ones. This constitutes a $6.5\sigma$ deviation (Fig. \ref{R-a-b}). It is clear then that the anisotropies possibly detected in these scaling relations will be entirely driven by the slightly inhomogeneous spatial distribution of such clusters, and not by cosmological phenomena. As discussed before, the $R$ scaling relations (contrary to \LT\ and \YT) are not sensitive enough to cosmological effects, to overcome the bias coming from CC clusters. 

\begin{figure*}[hbtp]
               \includegraphics[width=0.49\textwidth, height=4.5cm]{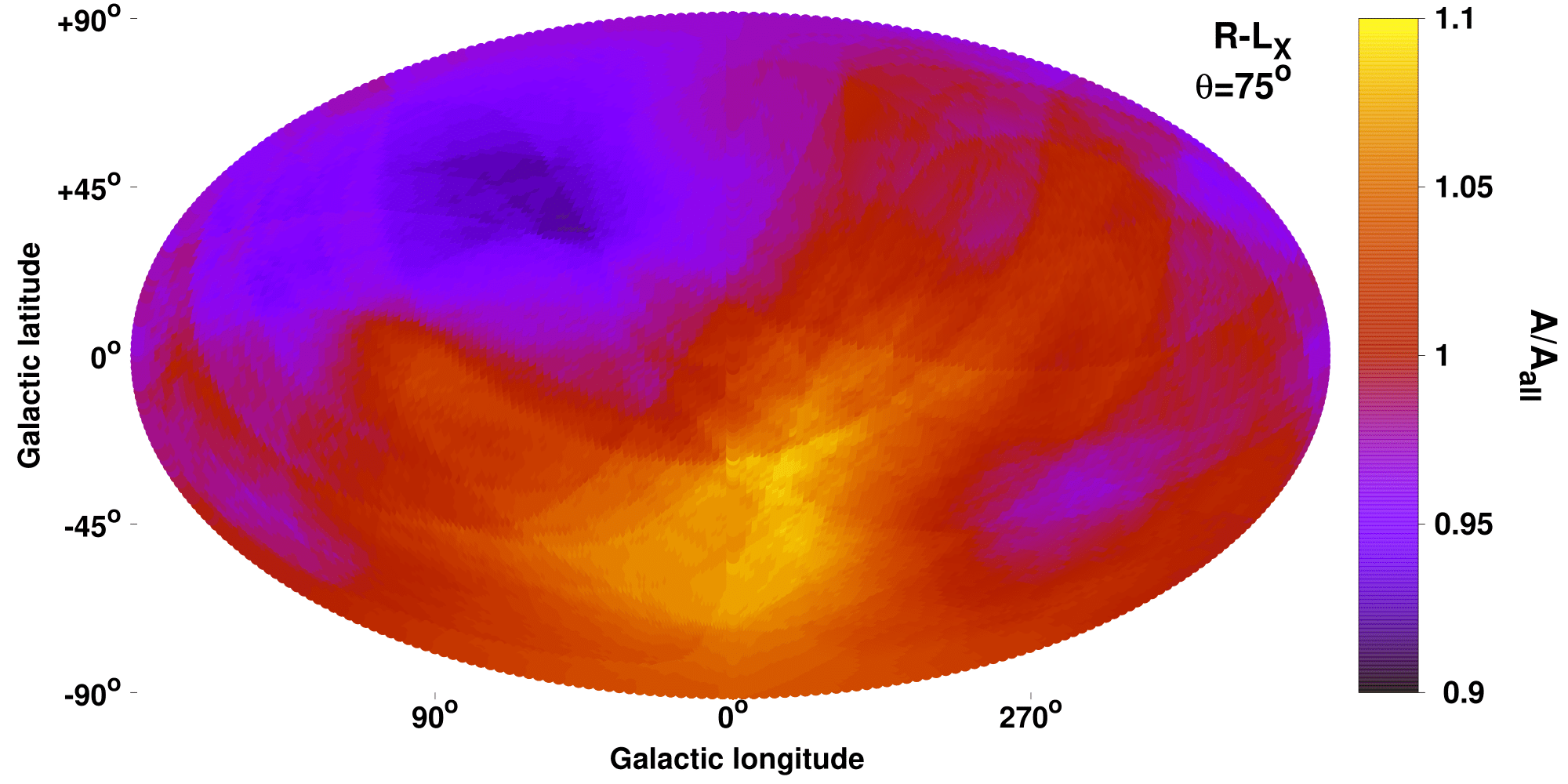}
               \includegraphics[width=0.49\textwidth, height=4.5cm]{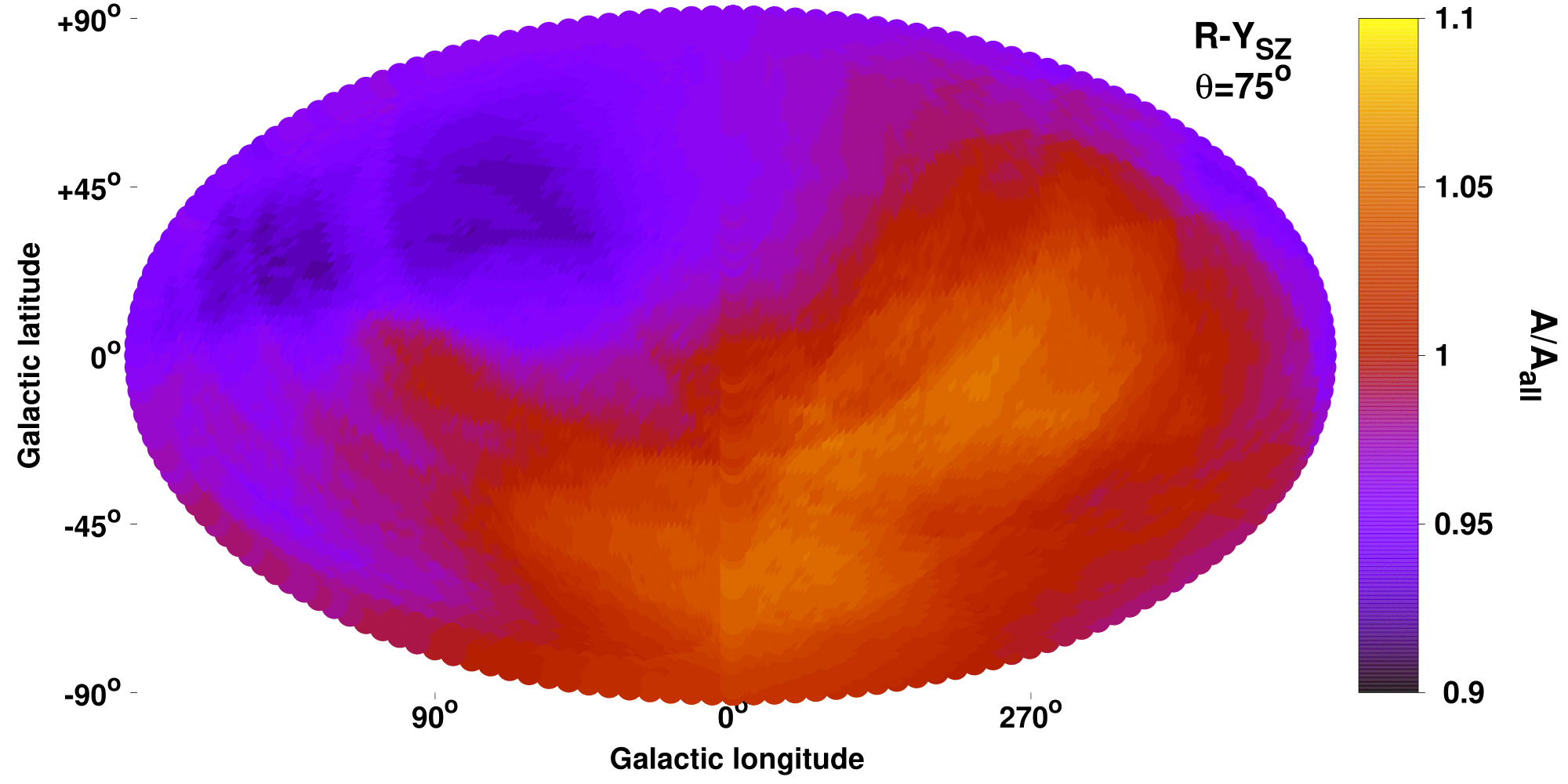}
               \caption{Normalization anisotropy map for \RL\ (left) and \RY\ (right).}
        \label{R-maps}
\end{figure*}

Indeed, when one attempts to scan the sky with a $\theta=75^{\circ}$ cone, one obtains the same sky pattern for all four scaling relations. The normalization anisotropy maps for \RL\ and \RY\ are displayed in Fig. \ref{R-maps}. The map for \RLbcg\ is not displayed since they are very similar to the other $R$ maps. The smallest clusters are consistently found within $12^{\circ}$ from $(l,b)\sim (64^{\circ}, +32^{\circ})$. This direction strongly correlates with the region where the highest fraction of relaxed clusters in our sample is found, where there is also a lack of disturbed clusters, as can be seen by Fig. \ref{cc-sky-plots}. The maximum anisotropy does not exceed $2.3\sigma$ for any relation, with a $\lesssim 14\%$ variation of $A_{RX}$. As such, the effect of this bias is not extremely strong, but sufficient to dominate over the effect of the possible cosmological anisotropies and BFs.

\subsubsection{Calibration of \RT\ anisotropies for dynamical state of clusters}

To study the underlying cosmological effects, one would need larger samples to further smooth out the sky distribution of different dynamical types of clusters. Also, one could attempt to calibrate the scaling relations for this systematic bias. This requires an observable proxy that effectively traces the existence of CC clusters. Unfortunately, such an observable is not currently available for our sample. Future work will soon provide a remarkable characterization of the core state of the eeHIFLUGCS clusters using numerous independent measurements. Proper calibration of the $R$ relations will be then possible, followed by their application for the search of cosmological anisotropies.

\begin{figure}[hbtp]
                \includegraphics[width=0.45\textwidth, height=5.3cm]{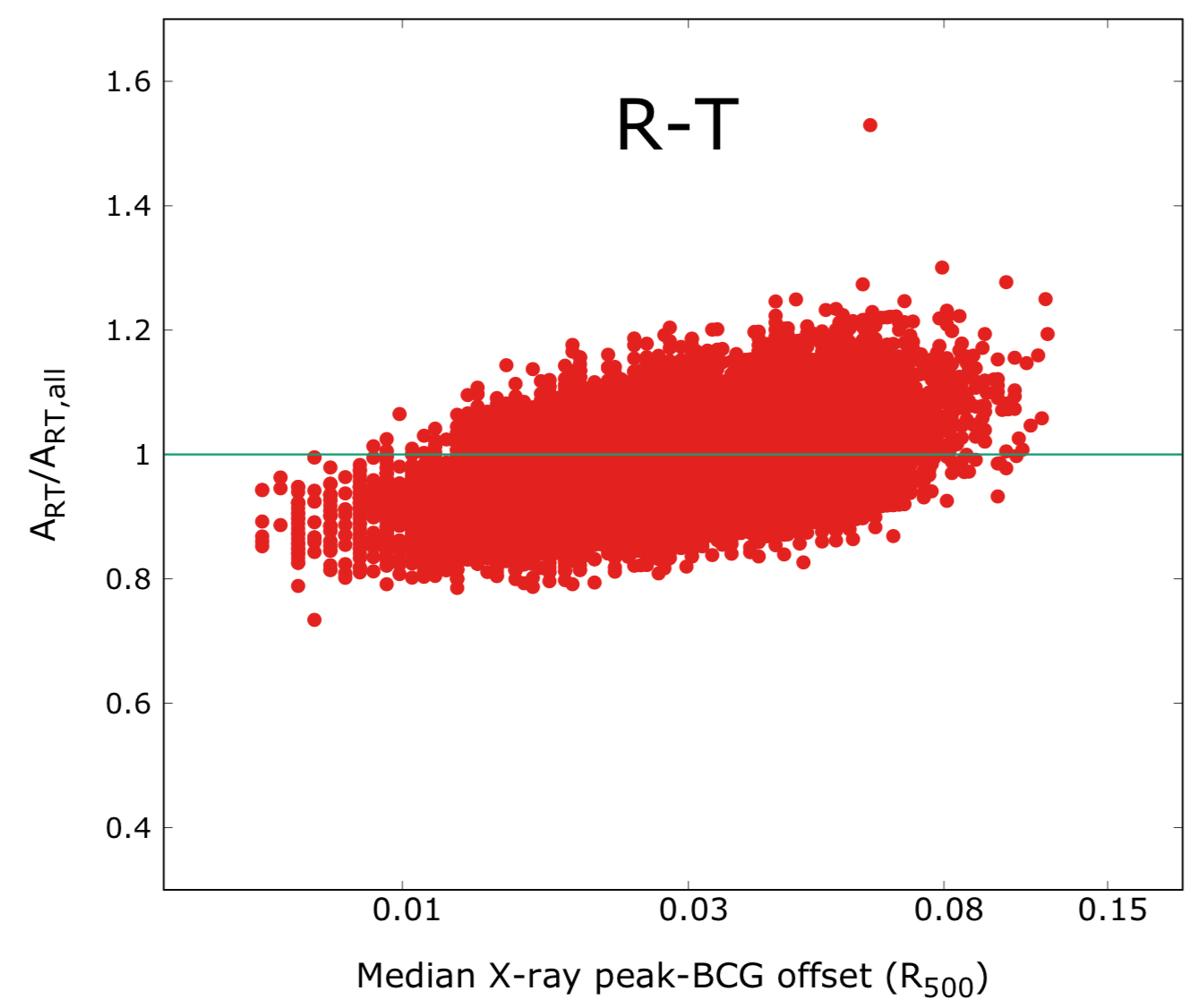}
               \includegraphics[width=0.49\textwidth, height=4.5cm]{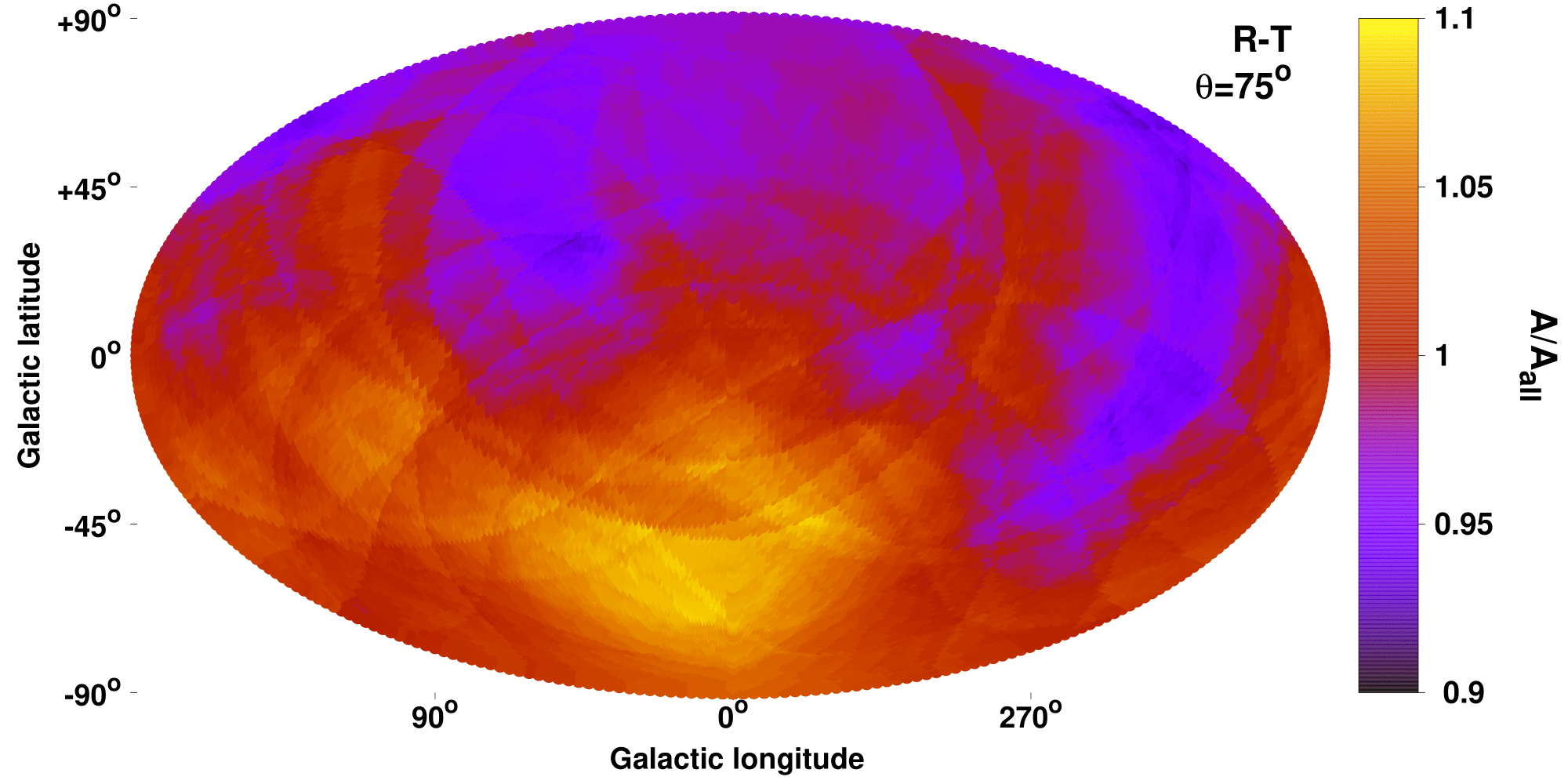}
               \caption{\textit{Top}: Correlation between the best-fit normalization $A_{RT}$ (over the full sample's best-fit value) and the median XBO for every of the $10^5$ bootstrap subsamples. \textit{Bottom}: $A_{RT}$ anisotropy map after calibrating for the existing correlation with the XBO.}
        \label{bcg-R-calib}
\end{figure}

For now, we use XBO as a proxy for relaxed clusters, as already discussed. We attempt to calibrate the $R-T$ relation and its anisotropy map. We create $10^5$ randomly drawn bootstrap subsamples (same process as in M20), independent of direction. We investigate the correlation between the $A_{RT}$ and the median XBO for every subsample. The results are shown in Fig. \ref{bcg-R-calib}. We then calibrate the $A_{RT}$ anisotropy map based on the median XBO of every sky region and the observed correlation between the two. Although the induced uncertainties due to this calibration are too large to allow for any meaningful conclusions, this offers a useful example for potential future applications of the $R$ relations. Finally, the behavior of the apparent \RT\ anisotropies after this correction is applied is very interesting. The previously anisotropic region now becomes milder. The $(l,b)\sim (260^{\circ},-10^{\circ})$ direction starts showing a lower $A_{RT}$ behavior, consistent with the previous results and the scenario of a cosmological anisotropy or a large BF.

\subsection{The \YLbcg\ relation}\label{ylbcg}

Both quantities of this relation are unaffected by absorption effects due to the infrared and submillimeter wavelengths, while they depend on the cosmological parameters in the same manner. Accounting for the overall best-fit slope, a very weak dependance of the normalization on the angular diameter distance remains ($A_{YL_{\text{LBCG}}}\sim D_A^{0.26}$). In detail, a $15\%$ variation in $H_0$ would only lead to a $<3\%$ variation in $A_{YL_{\text{LBCG}}}$. Similarly, a $1000$ km/s BF at $z=0.05$ would only cause a $0.5\%$ change in $A_{YL_{\text{LBCG}}}$. It is clear then that only sample-related biases can create observable anisotropies in this scaling relation.

\begin{figure}[hbtp]
                \includegraphics[width=0.49\textwidth, height=4.5cm]{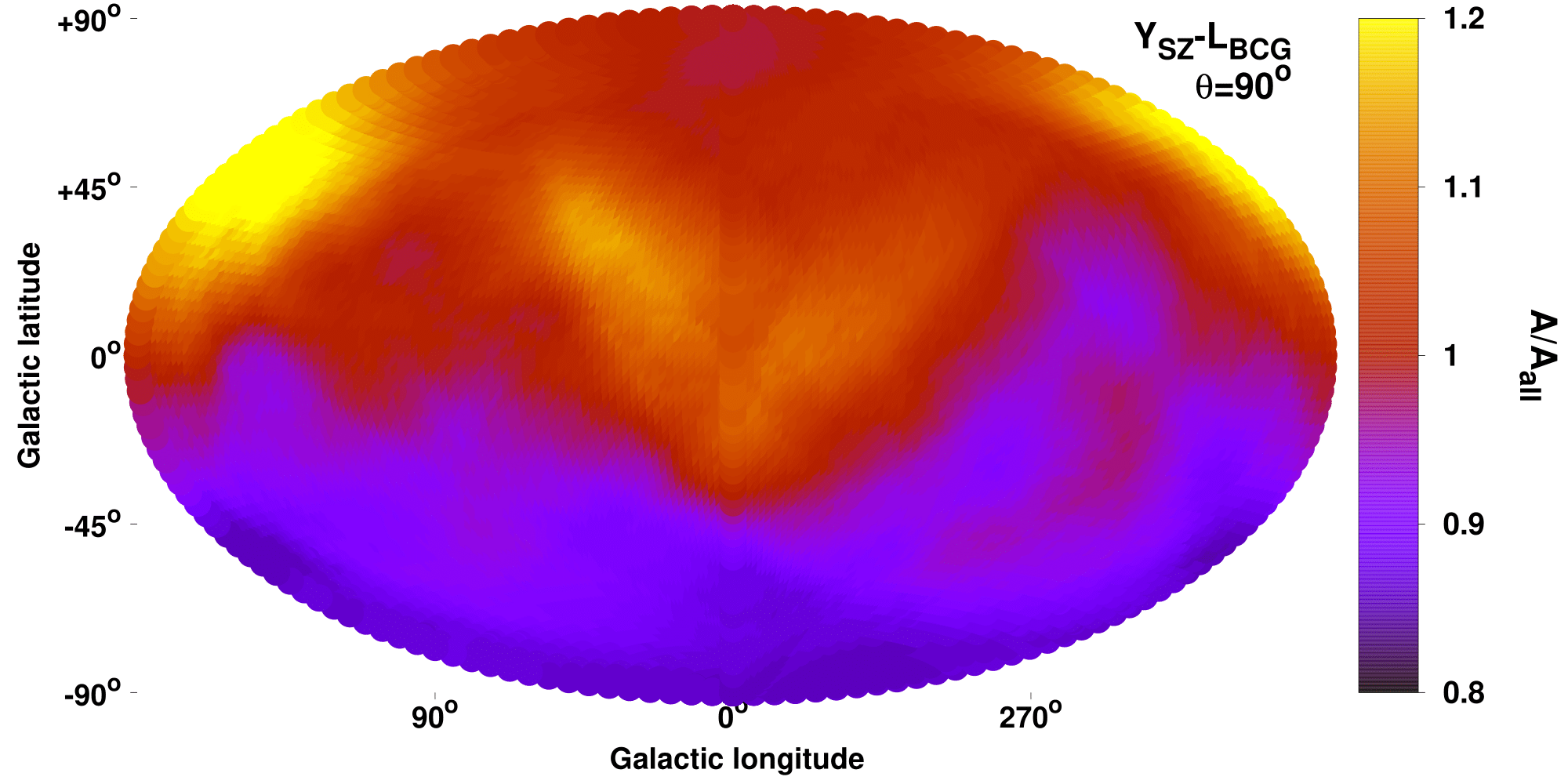}
               \caption{Normalization anisotropy map for the \YLbcg\ relation. The statistical significance of the observed anisotropies is $\leq 1.4\sigma$, and thus the relation is statistically isotropic.}
        \label{YLbcg-anis}
\end{figure}

Due to the large scatter and the limited number of clusters, we consider $\theta=90^{\circ}$ cones to scan the sky. The anisotropy of the \YLbcg\ normalization is displayed in Fig. \ref{YLbcg-anis}. The relation shows a $\sim 30\%$ variation mostly between the Northern and the Southern Galactic hemispheres, but with a negligible statistical significance of $1.4\sigma$. The relation can be considered statistically isotropic, as expected, with no strong biases toward a direction. The normalization map is displayed in Fig. \ref{YLbcg-anis}.

\section{Scaling relations and anisotropies as functions of different selection cuts}\label{differ_cuts}

In this section, we repeat the analysis for the \LY, \YT, and \LbcgT\ relations, for different \Ysz\ S/N and redshift cuts. The summarized result is that our main conclusions remain unchanged.

\subsection{The \LY\ relation}

\begin{figure}[hbtp]
                \includegraphics[width=0.49\textwidth, height=6cm]{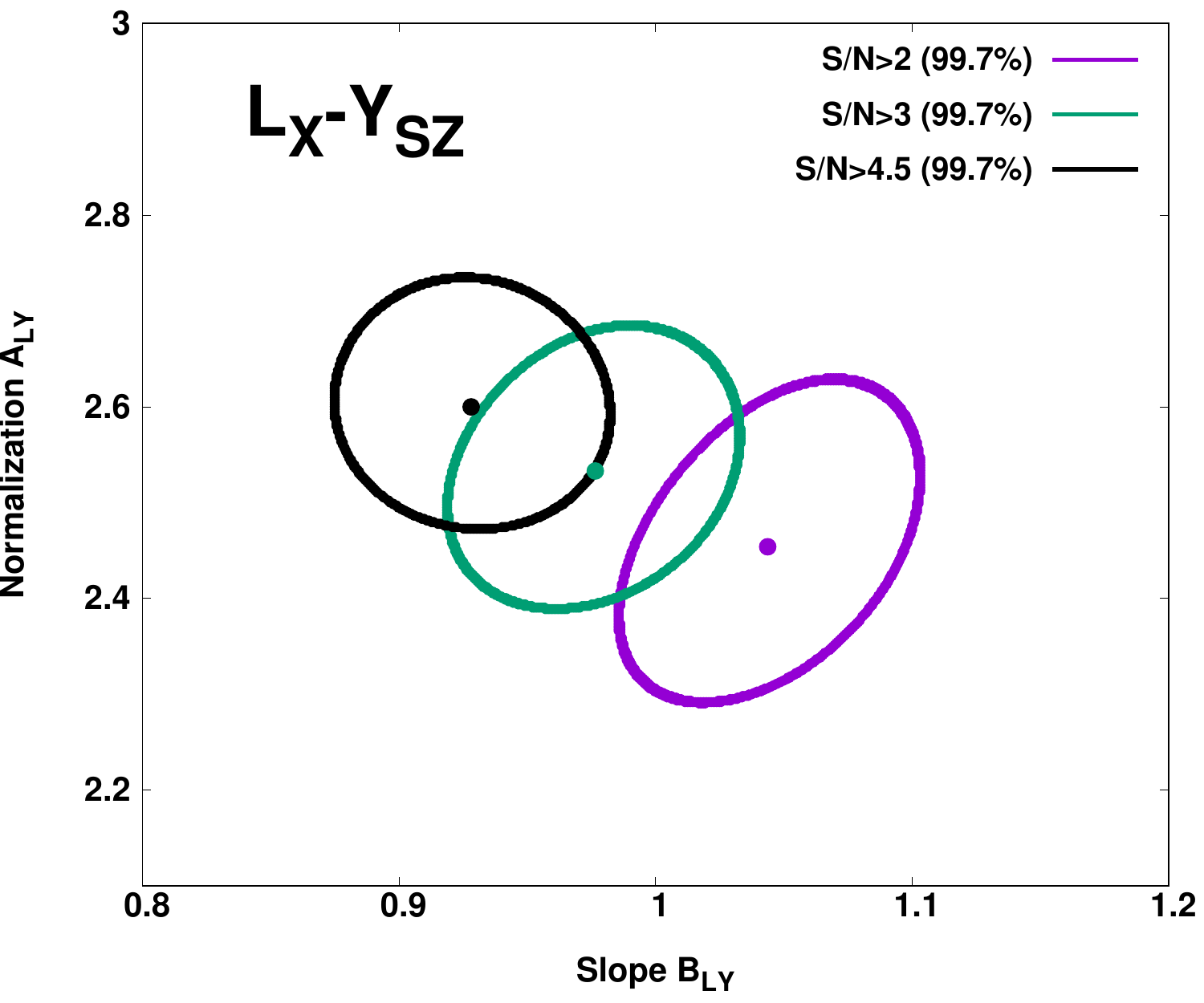}
                \includegraphics[width=0.49\textwidth, height=6cm]{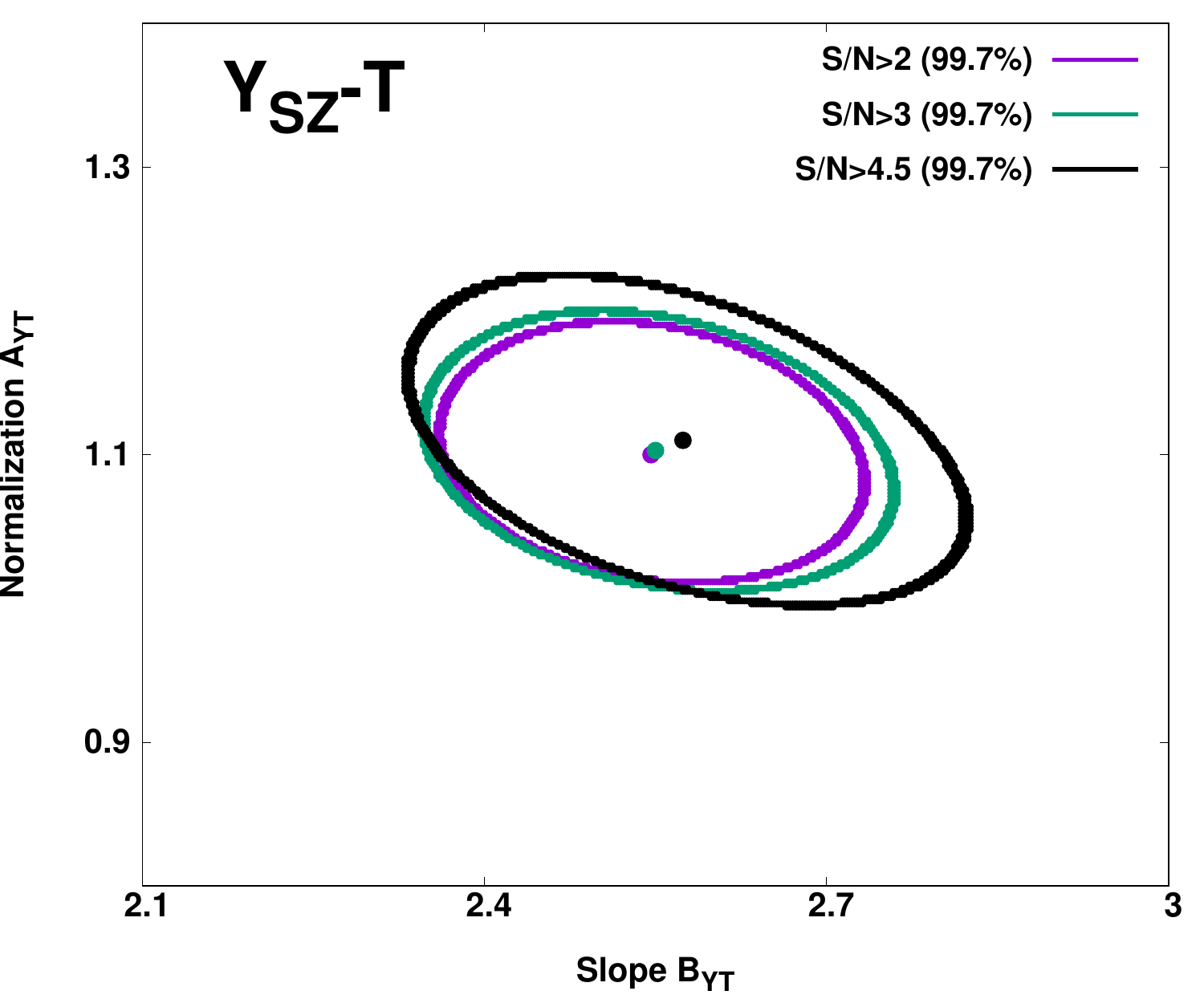}
                \includegraphics[width=0.49\textwidth, height=6cm]{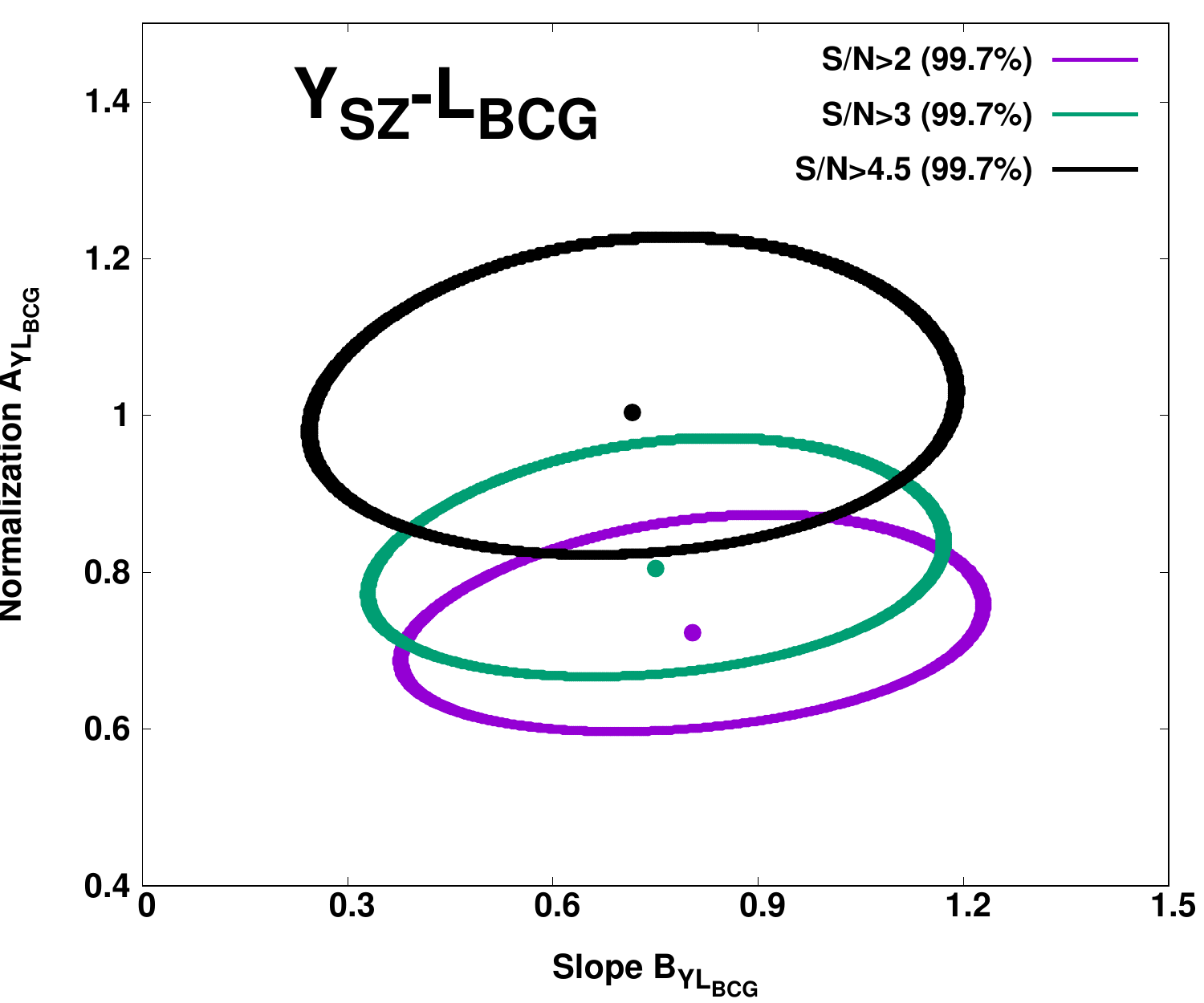}
               \caption{$3\sigma$ (99.7\%) parameter space of the normalization and slope of the \LY\ (top), \YT\ (middle), and \YLbcg\ (bottom) relations, for S/N$>2$ (purple), S/N$>3$ (green), and S/N$>4.5$ (black).}
        \label{LY-ab-SN}
\end{figure}

\begin{figure*}[hbtp]
                \includegraphics[width=0.33\textwidth, height=5cm]{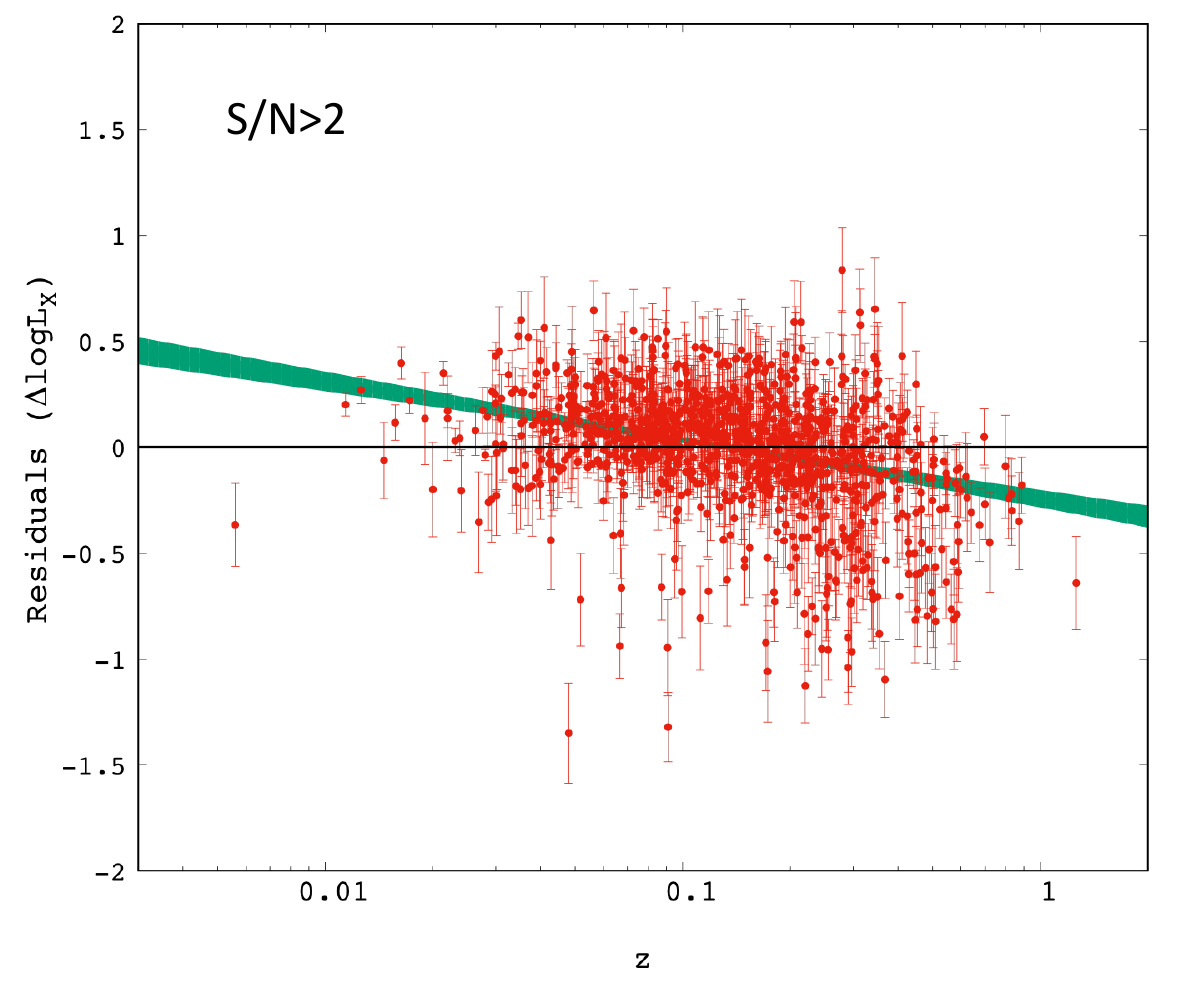}
                \includegraphics[width=0.33\textwidth, height=5cm]{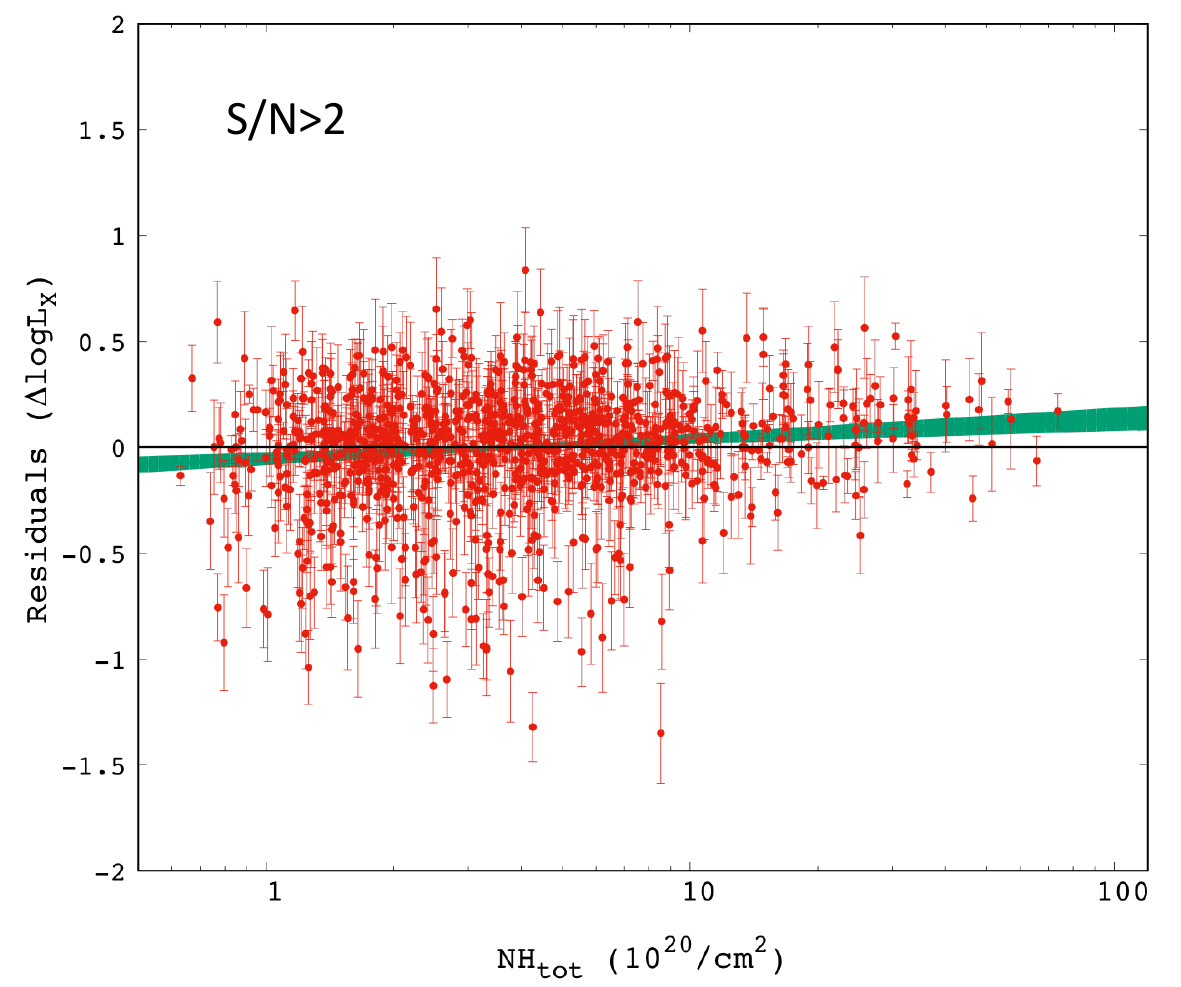}
                \includegraphics[width=0.33\textwidth, height=5cm]{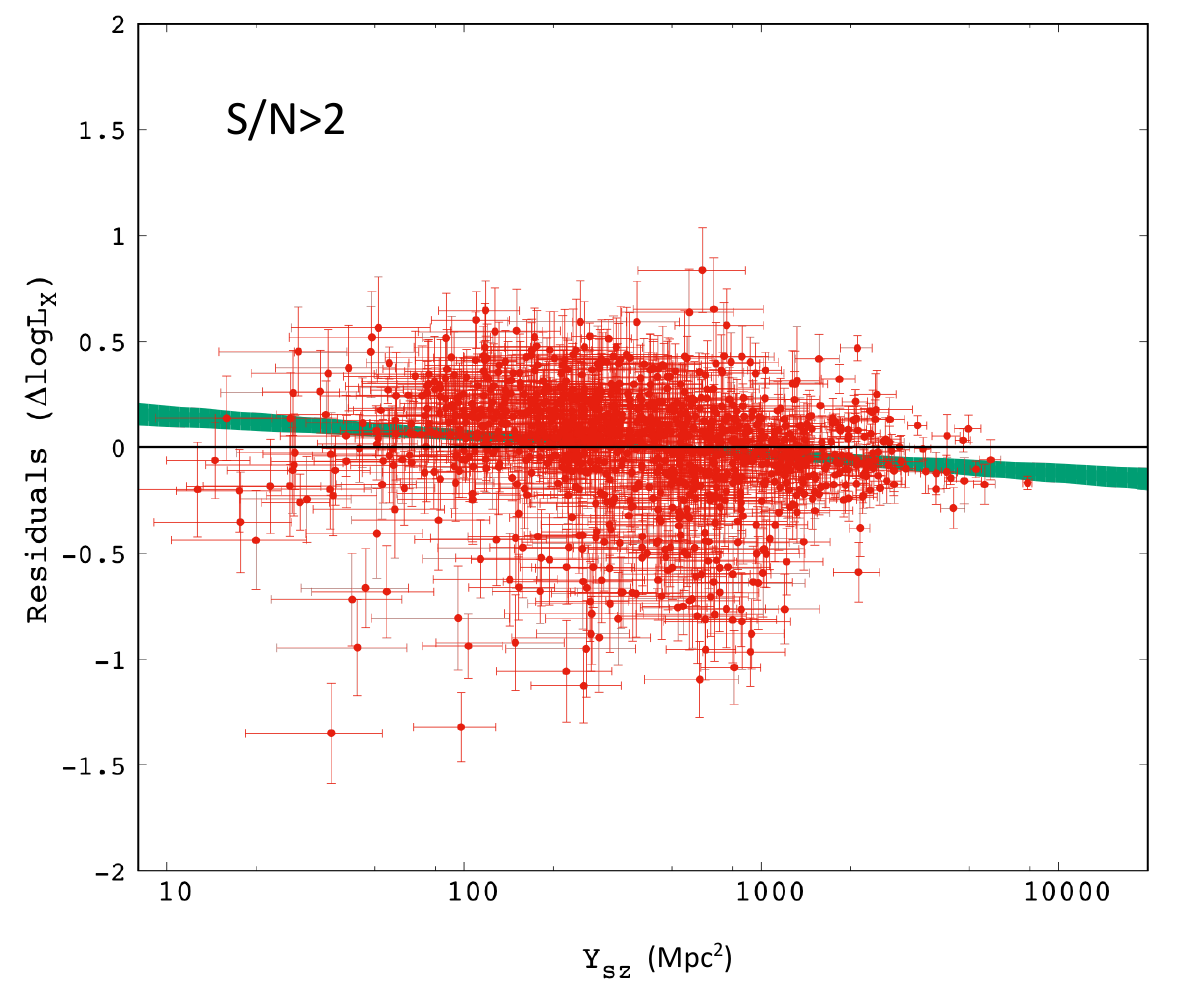}
                \includegraphics[width=0.33\textwidth, height=5cm]{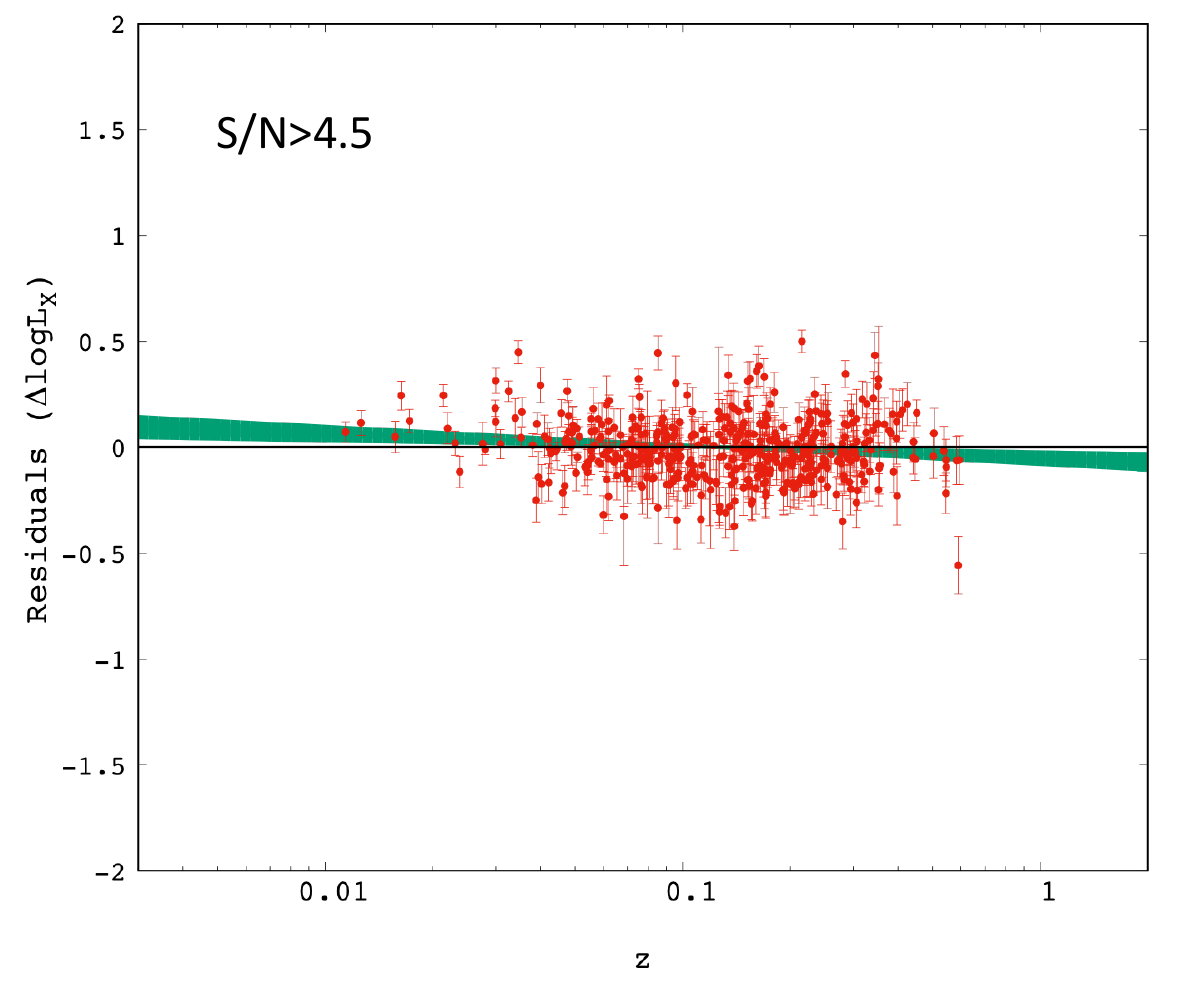}
                \includegraphics[width=0.33\textwidth, height=5cm]{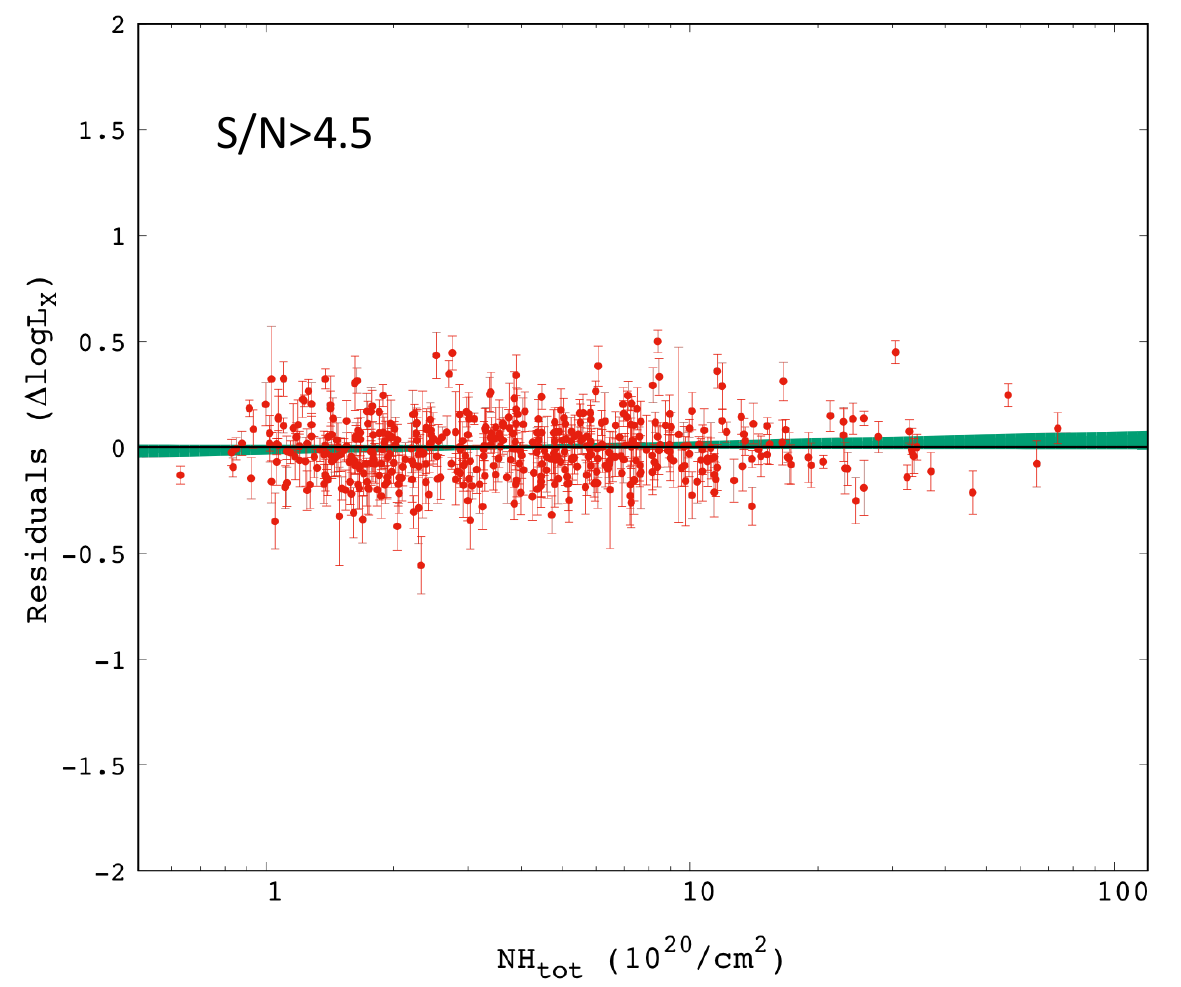}
                \includegraphics[width=0.33\textwidth, height=5cm]{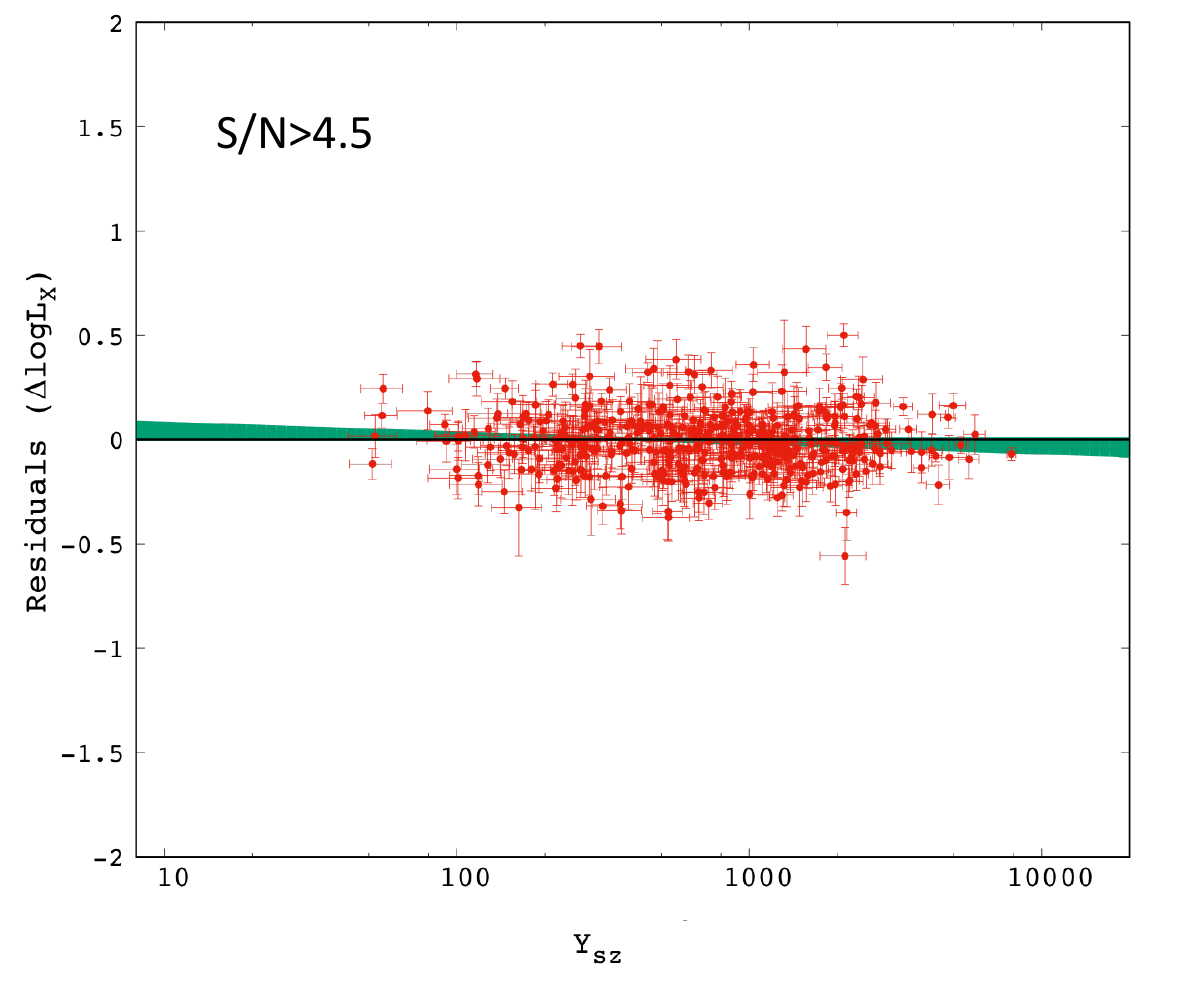}
               \caption{\Lx\ residuals for the \LY\ relation for S/N$>2$ (top) and S/N$>4.5$ (bottom), as a function of $z$ (left), \nhtot\ (middle), and \Ysz\ (right). The green strips display the best-fit line with their $1\sigma$ uncertainties. It is evident that for S/N$>4.5$, the slope is always consistent with zero, indicating no dependence of the residuals on the cluster physical properties.}
        \label{LY-resid}
\end{figure*}

For the \LY\ relation, a lower \Ysz\ threshold of S/N$\geq 4.5$ was applied in the default analysis. This resulted in 460 clusters. If we consider two lower thresholds instead, namely S/N$\geq 2$ and S/N$\geq 3$, we have 1095 and 747 clusters respectively. The $3\sigma$ \LY\ contours for the three different cases are compared to Fig. \ref{LY-ab-SN}. There is a $>5\sigma$ shift in the best-fit \LY\ relation going from S/N$>4.5$ to S/N$>2$, while the scatter increases by $\sim 100\%$. The S/N$>2$ fit however, is dominated by several systematics. To begin with, the \Lx\ residuals are strongly correlated with $z$, \nhtot, and \Ysz, as can be seen in Fig. \ref{LY-resid}. These probably point toward biases in the X-ray selection process, and the need of a broken power law to describe \LY. In the same figure, one can see that these systematic behaviors are not present for the S/N$>4.5$ case. 

\begin{figure}[hbtp]
                \includegraphics[width=0.49\textwidth, height=4.5cm]{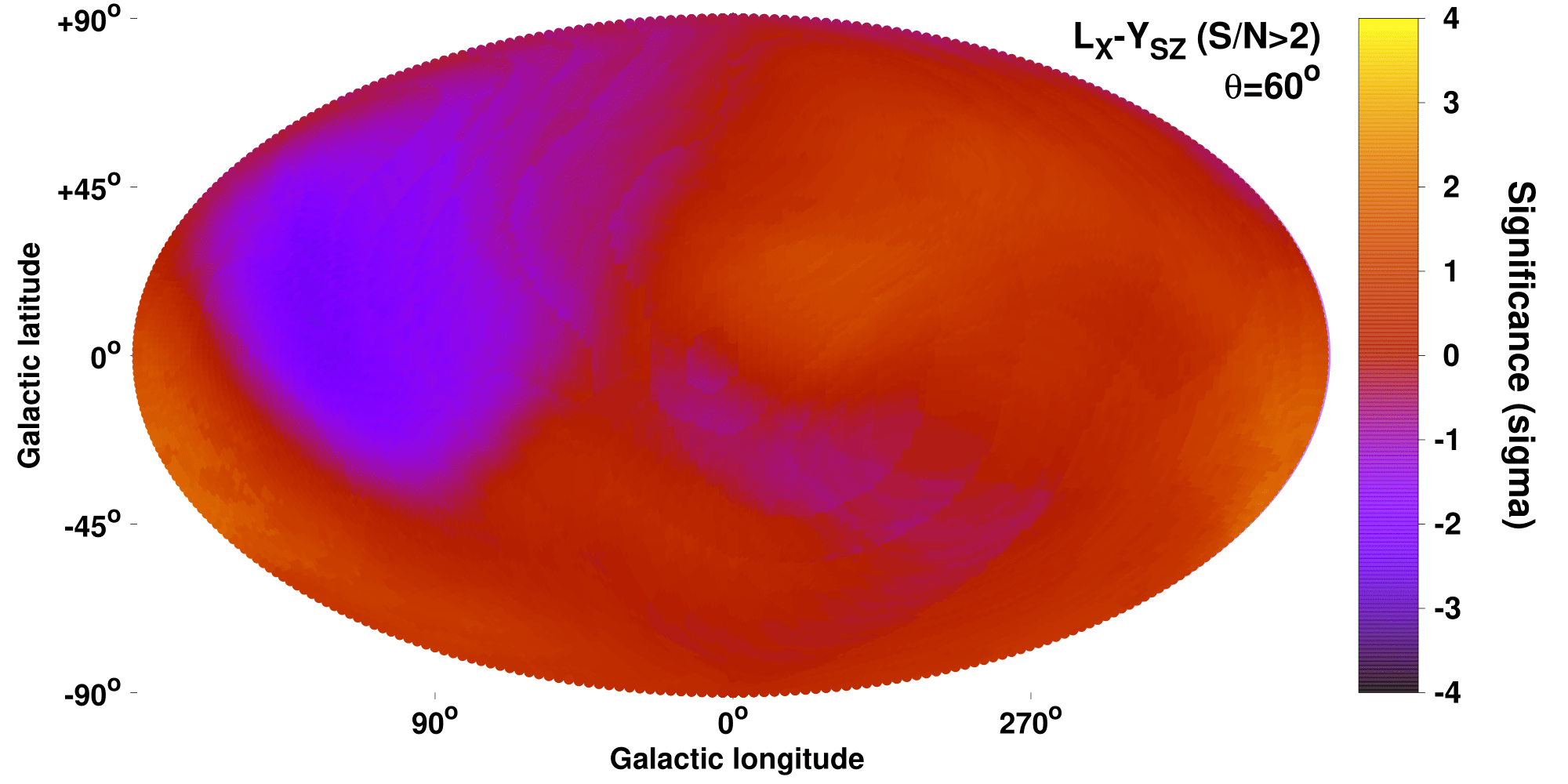}
               \caption{Statistical significance map of the \LY\ anisotropies for S/N$>2$.}
        \label{LY-SN-2-anis}
\end{figure}

Despite these issues, when we scan the sky to detect \LY\ anisotropies, we get a similar map as for the default case. The anisotropy significance map (Fig. \ref{LY-SN-2-anis}) illustrates that there are no excess X-ray absorption issues toward $(l,b)\sim (280^{\circ}, -15^{\circ})$ where the cosmological anisotropies are found. At the same time, the only anisotropic region again appears to be toward $(l,b)\sim (128^{\circ}, +21^{\circ})$, being $\sim 14\%$ fainter than the rest of the sky at a $2.5\sigma$ level, similar to the default case.

\subsection{The \YT\ relation}\label{YT-cuts}

\begin{figure}[hbtp]
                \includegraphics[width=0.49\textwidth, height=4.5cm]{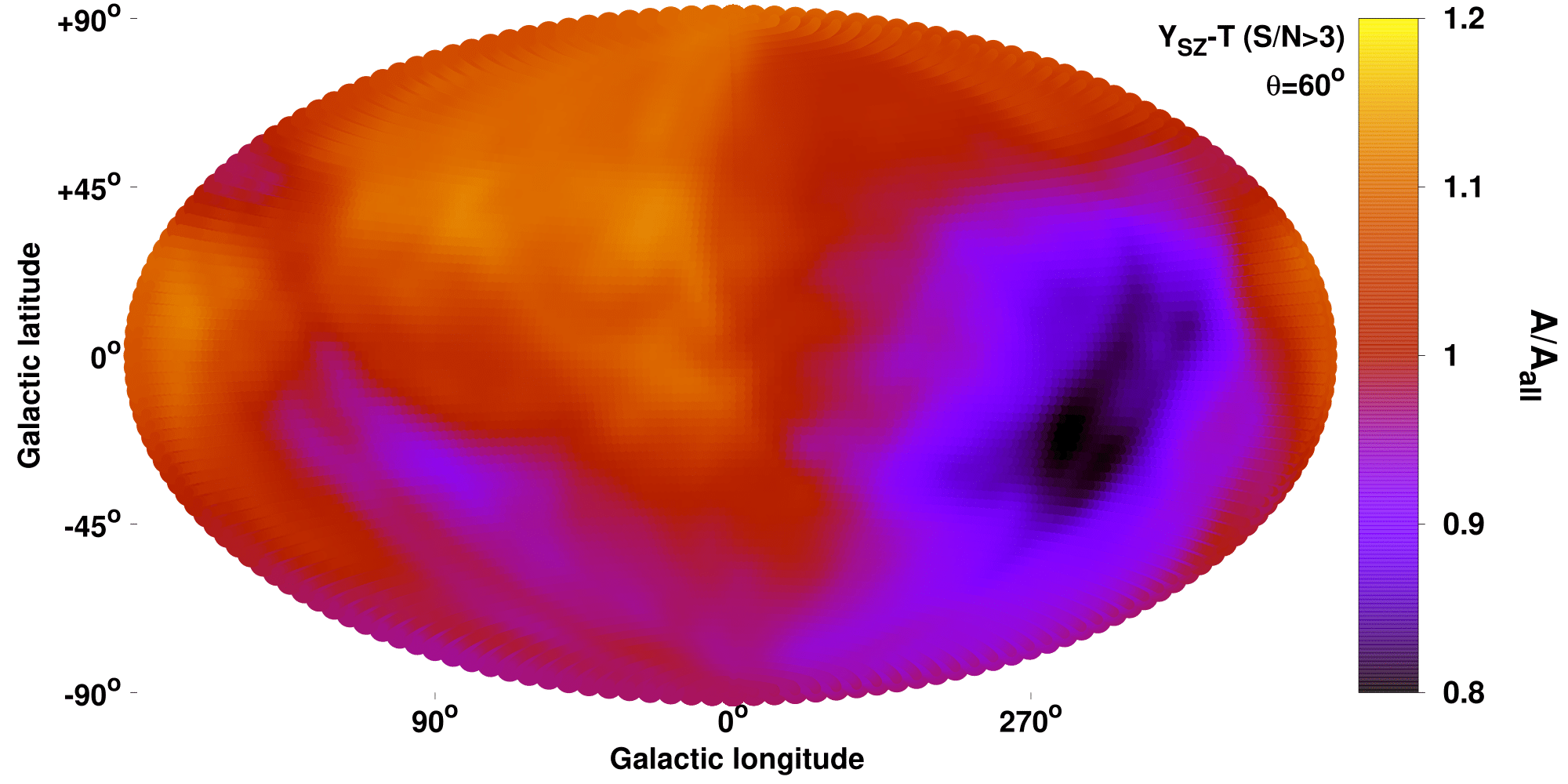}
               \caption{Normalization anisotropy map of the \YT\ relation for S/N$>3$.}
        \label{YT-SN-3-anis}
\end{figure}

For the \YT\ relation, we used a lower \Ysz\ threshold of S/N$\geq 2$ in the default analysis, resulting in 263 clusters. If we increase this lower threshold to S/N$\geq 3$ or S/N$\geq 4.5$, we are left with 242 and 190 clusters respectively. The best-fit \YT\ does not change as a function of S/N, as can be seen in the middle panel of Fig. \ref{LY-ab-SN}. The \Ysz\ residuals remain independent of the cluster physical properties as well. When we repeat the anisotropy analysis for S/N$\geq 3$ and for S/N$\geq 4.5$, we obtain almost identical anisotropy results as for the default case. The $A_{YT}$ variance map is displayed in Fig. \ref{YT-SN-3-anis}. Thus, the observed anisotropies are independent of \Ysz\ selection effects.

\subsubsection{The \YT\ relation using the \textit{Planck} values instead}\label{psz2-test}

\begin{figure}[hbtp]
                \includegraphics[width=0.49\textwidth, height=6cm]{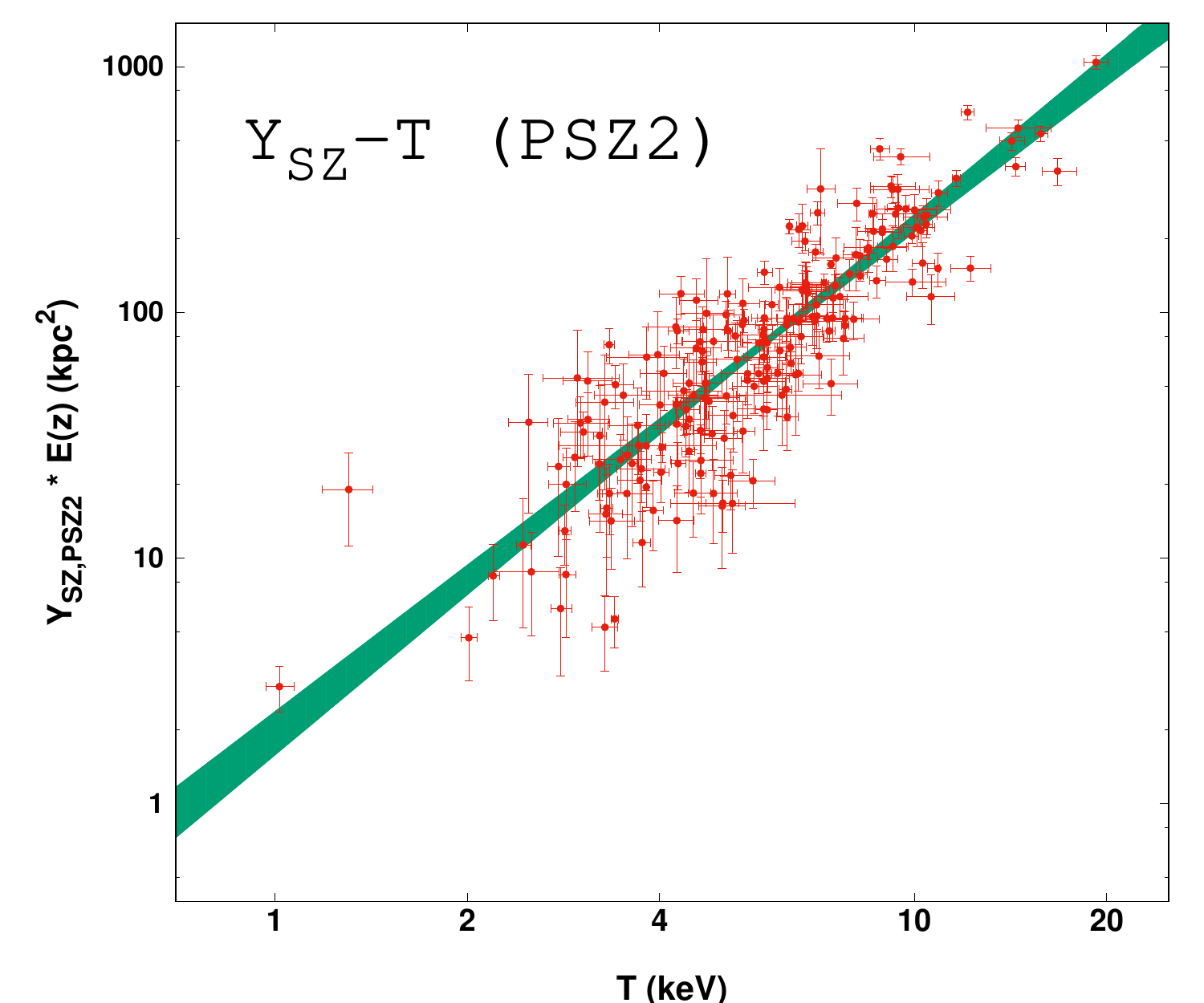}
                \includegraphics[width=0.49\textwidth, height=4.5cm]{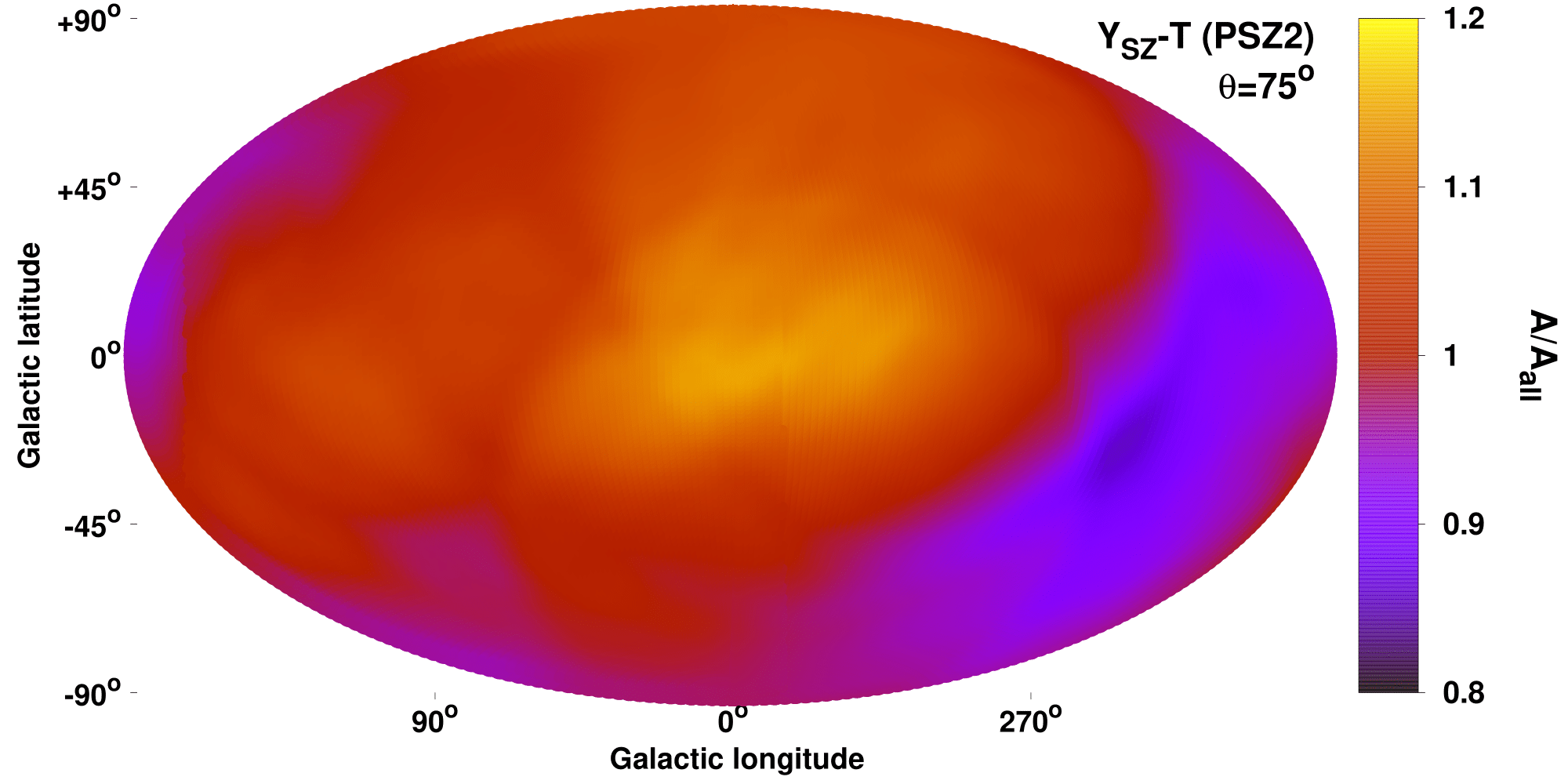}
               \caption{\textit{Top}: The \YT\ relation when using the PSZ2 \Ysz\ values, together with its $1\sigma$ best-fit function (green). \textit{Bottom}: Normalization anisotropy map of the \YT\ relation when the PSZ2 values for \Ysz\ are used.}
        \label{Y-T-planck}
\end{figure}

To ensure that the observed \YT\ anisotropies do not emerge due to some unknown directional bias in our own \Ysz\ measurements, we repeat the analysis using the \Ysz\ values from the PSZ2 catalog. Due to the higher \Ysz\ S/N$>4.5$ threshold that PSZ2 applied, 206 clusters were matched with our M20 sample. This S/N cut also leads to higher median $T$ and \Ysz\ compared to our default analysis. Thus, we adopt $C_Y=85$ kpc$^2$ and $C_X=6$ keV. For the best-fit parameters, we find $A_{YT}=0.940\pm 0.041$, $B_{YT}=2.074\pm 0.095$, $\sigma_{\text{int}}=0.184\pm 0.019$, and $\sigma_{\text{int}}=0.232\pm 0.024$. The scatter is $\sim 35\%$ larger than in our case (when \Ysz\ S/N$>4.5$),  due to the different adopted $R_{500}$ between the two independent analyses. This highlights the need to use our own measurement, where we increase the number of clusters, while decreasing the scatter. The slope differs by $2.2\sigma$. The \YT\ scaling relation is plotted in the top panel of Fig. \ref{Y-T-planck}.

Performing the \YT\ sky scanning, one sees that we obtain roughly the same anisotropies with the PSZ2 \Ysz\ values, as we did with our measurements. The $A_{YT}$ map is shown in the bottom panel of Fig. \ref{Y-T-planck}. The maximum anisotropy is found toward $(l,b)=({254^{\circ}}^{+32^{\circ}}_{-44^{\circ}} ,{-22^{\circ}}^{+51^{\circ}}_{-29^{\circ}})$, with $H_0=64.6\pm 2.5$ km/s, at a $2.9\sigma$ tension with the rest of the sky. This region is only $15^{\circ}$ from the most anisotropic region as found in our default analysis, which demonstrates that the \YT\ anisotropies are independent of the adopted \Ysz\ catalog.

\subsection{The \LbcgT\ relation}

\begin{figure}[hbtp]
                \includegraphics[width=0.49\textwidth, height=6cm]{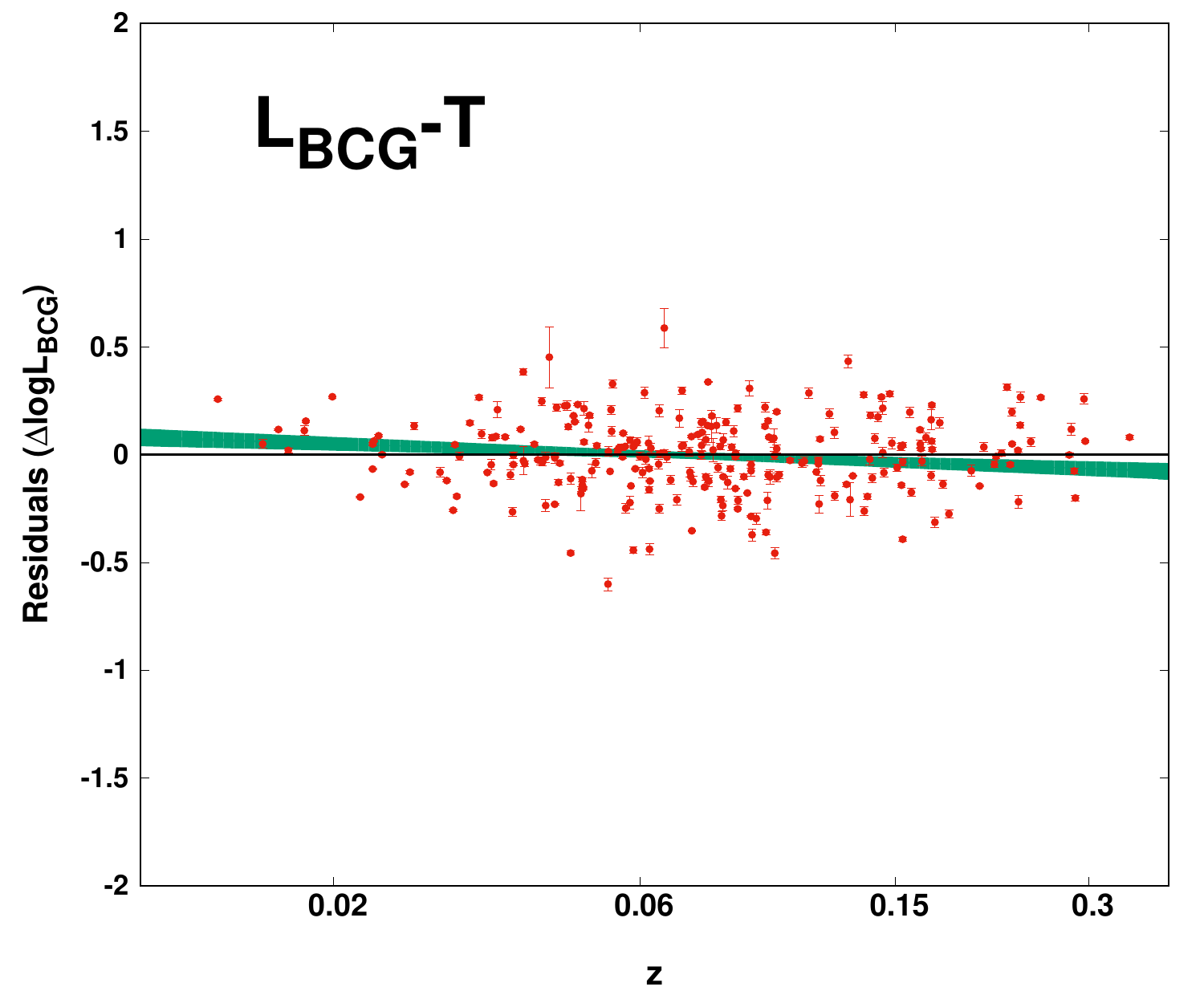}
                \includegraphics[width=0.49\textwidth, height=6cm]{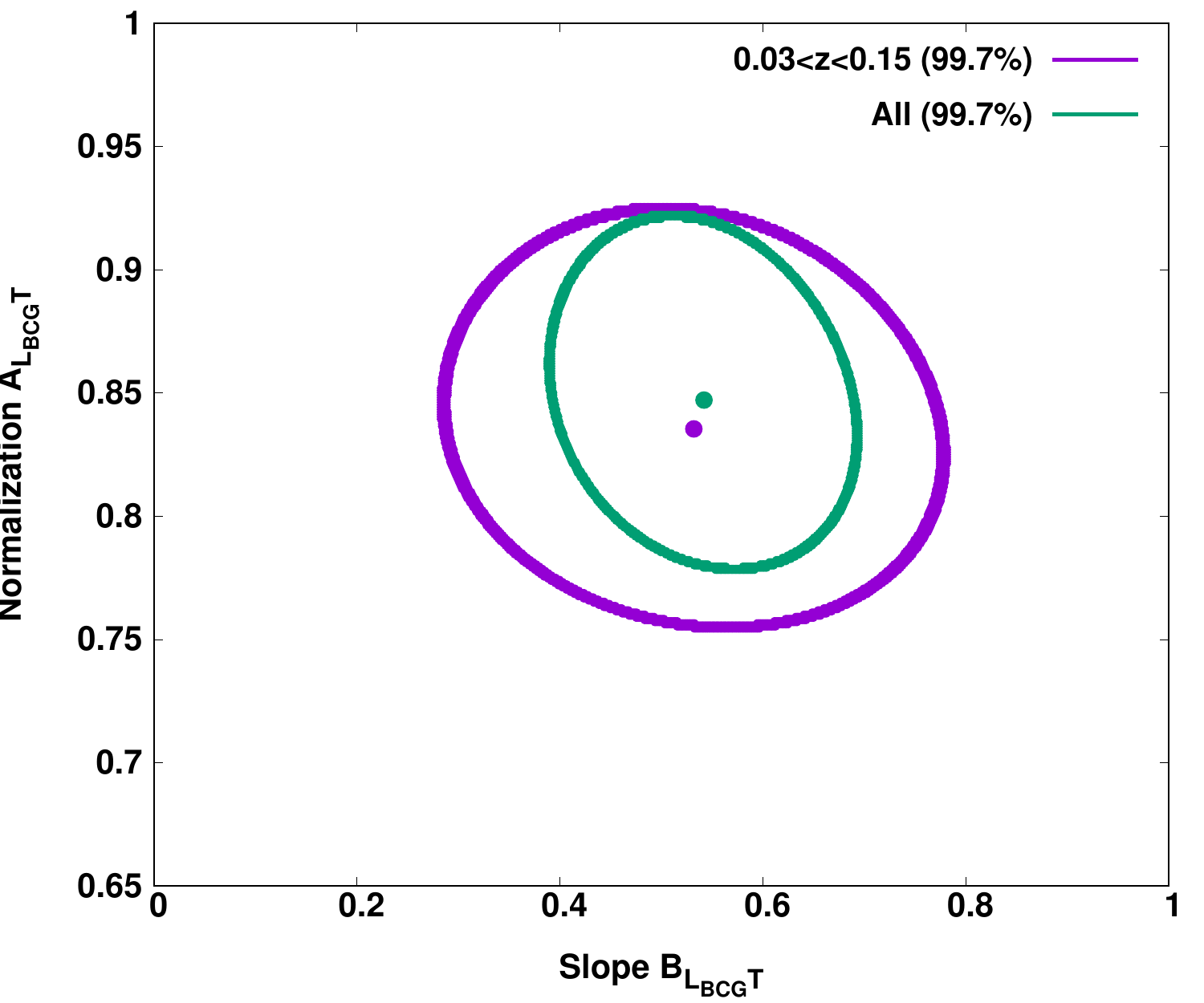}
                 \includegraphics[width=0.49\textwidth, height=4.5cm]{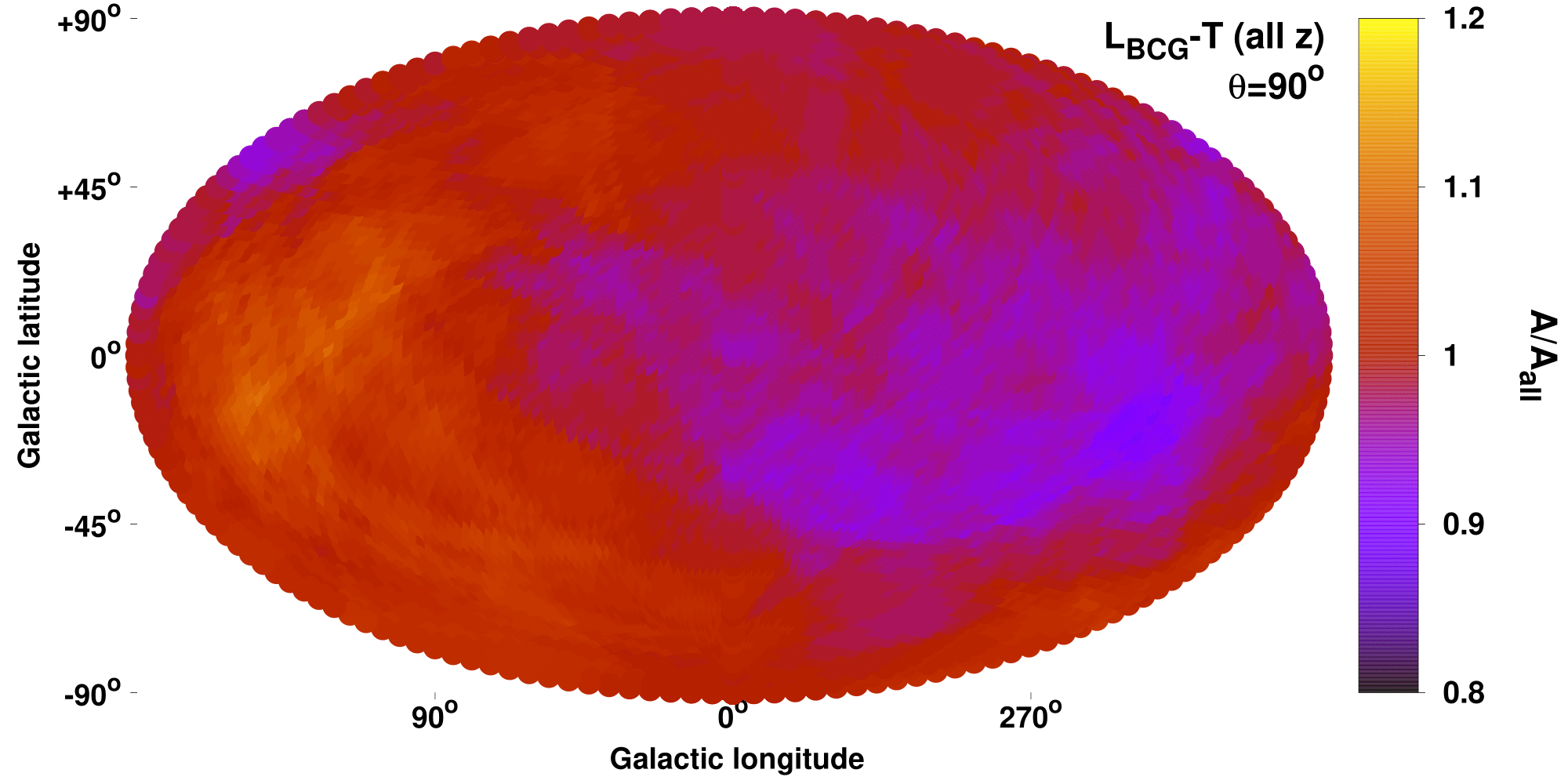}
               \caption{\textit{Top}: \Lbcg\ residuals of the \LbcgT\ fit, as a function of the BCG redshift. The green stripe corresponds to the best-fit function within $1\sigma$. \textit{Middle}: $3\sigma$ (99.7\%) parameter space of the normalization and slope of the \LbcgT\ relation, for all clusters (green), and for $0.03<z<0.15$ clusters (purple). \textit{Bottom}: Normalization anisotropy map of the \LbcgT\ relation when all clusters are considered independent of their redshift.}
        \label{LbcgT-tests}
\end{figure}

For thel \Lbcg\ scaling relations, we removed clusters at $z<0.03$ since they appear systematically brighter than expected. We have also excluded all the $z>0.15$ clusters, since the redshift evolution of \Lbcg\ remains unknown. We applied these cuts to all \Lbcg\ scaling relations in order to be consistent. However, these redshift issues are not as important in the \LbcgT\ relation. One can see that in Fig. \ref{LbcgT-tests}, where the \Lbcg\ residuals are plotted as a function of $z$. Only the seven clusters with $z<0.02$ are systematically upscattered, while $z>0.15$ show the same behavior as the less distant clusters. Thus, we repeat our analysis using all the 259 clusters, regardless of their redshift. The best-fit \LbcgT\ relation remains similar to the default analysis, with the parameter uncertainties being naturally smaller, due to the larger number of clusters. The anisotropic behavior of the relation remains the same as before. Interestingly, the statistical significance of the \LbcgT\ anisotropy signal rises from $1.9\sigma$ to $2.1\sigma$. Both of these results are displayed in Fig. \ref{LbcgT-tests}.

\section{More details about performed tests}\label{more_tests}

\subsection{ZoA of Avoidance bias} \label{zoa_bias}

As discussed in Sect. \ref{zoa-sect}, the ZoA gap does not introduce any significant bias to our results. Here we provide some plots related to that discussion. For simplicity, we only provide the relative images for the \LT\ relation, since results from the other relations are similar. This is due to the fact that the used samples are simulated and isotropic, and that the cluster positions are the same across scaling relations. Thus, any effects come only from the applied cluster weighting during the sky scanning. We stress again that the observed anisotropies remain unaffected when this weighting is omitted, and the ZoA gap is invisible to the used algorithm, highlighting the lack of bias coming from ZoA in the real data. 

\begin{figure}[hbtp]
                \includegraphics[width=0.49\textwidth, height=6cm]{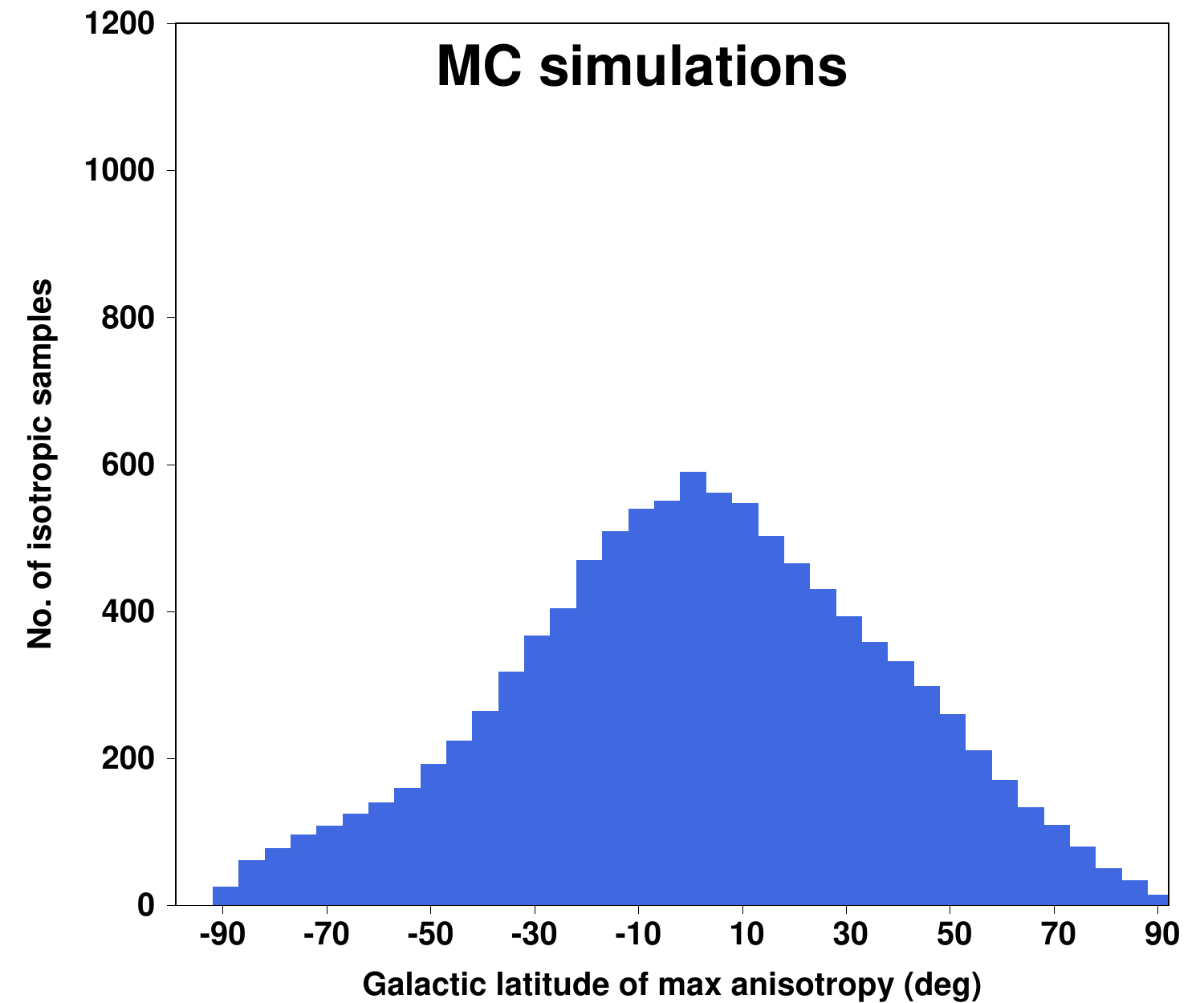}
                \includegraphics[width=0.49\textwidth, height=6cm]{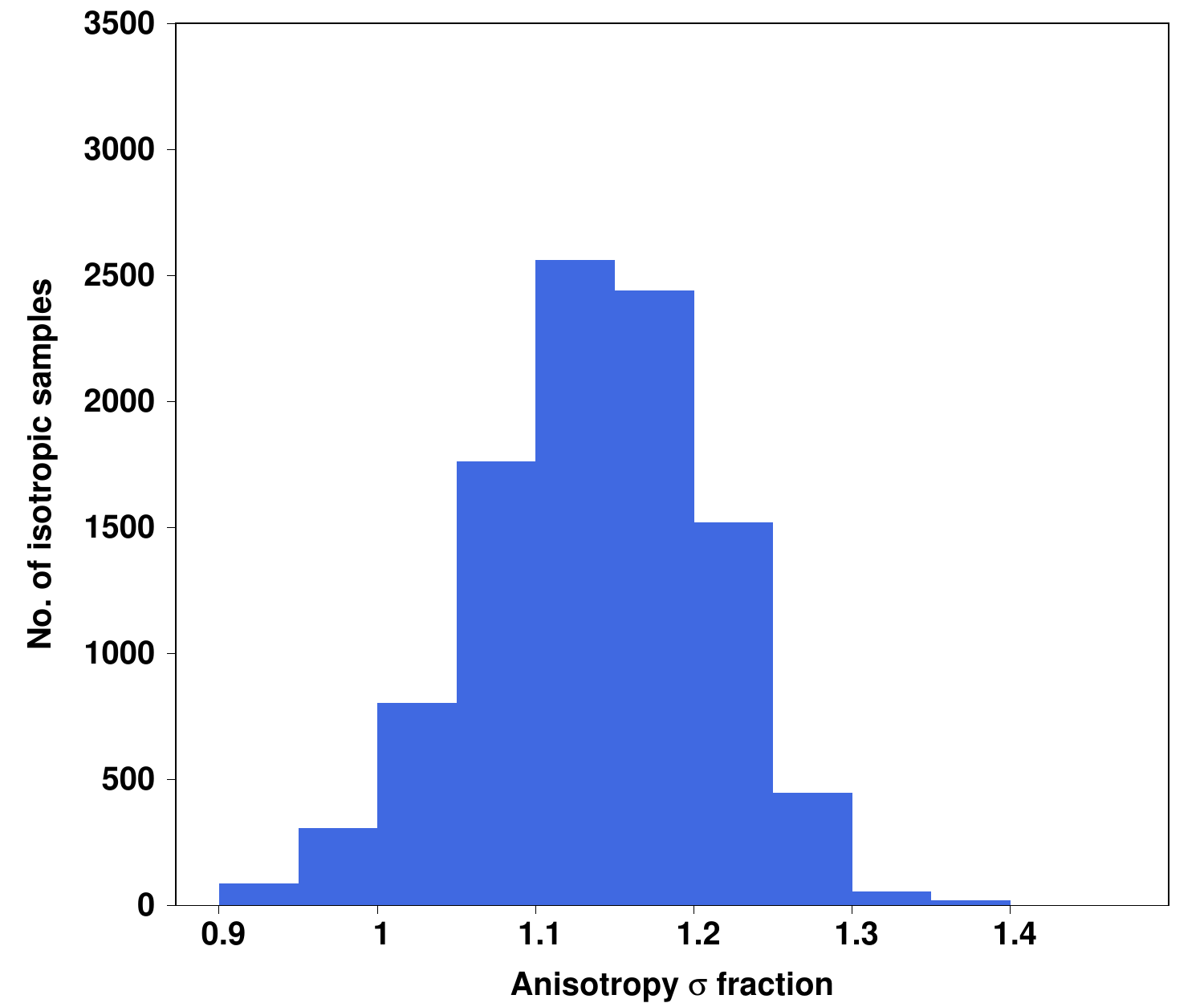}
               \caption{\textit{Top}: Distribution of the Galactic latitude of the most anisotropic regions as detected in the 10000 isotropic simulated samples for the \LT\ relation. The distance weighting during the \LT\ fitting was used here. Note that bins close to the ZoA cover a larger portion of the sky, and naturally more anisotropies are expected to be detected there even if no bias existed. }
        \label{zoa-tests}
\end{figure}

In the top panel of Fig. \ref{zoa-tests}, the Galactic latitude of the most anisotropic region for every simulated sample is shown. As explained before, $\sim 43\%$ of these regions lie within ZoA, instead of the $\sim33\%$ which would be the expectation if no bias was present. Here a nonflat distribution of the galactic latitudes is expected even in the completely bias-free case, since the sky area covered by each bin is not the same. Hence, the probability of the most anisotropic region to be located within each bin changes proportionally. 

In the bottom panel of Fig. \ref{zoa-tests}, we compare the relative difference of the observed anisotropy (in terms of $\sigma$) for two cases. The first is when the results from the analysis above are considered, without any data in the ZoA. The second is when we fill ZoA with simulated samples, as described in Sect. \ref{zoa-sect}. As discussed there, the average maximum anisotropy signal is increased by $14\pm 8\%$ when the ZoA clusters are excluded. This is mostly due to the different number of available data in the two cases. 

\begin{figure}[hbtp]
                \includegraphics[width=0.49\textwidth, height=4.5cm]{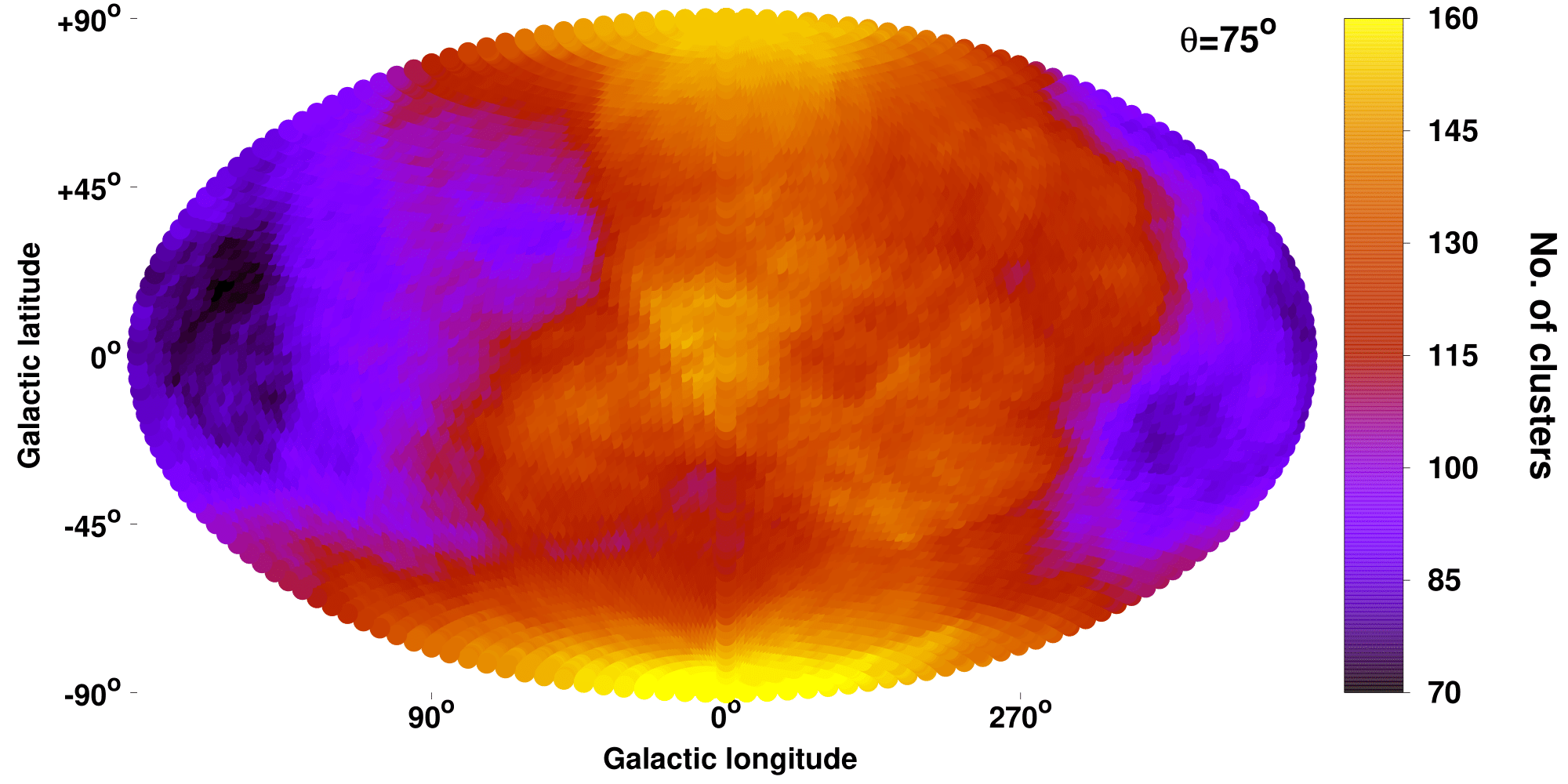}
               \caption{Number of clusters per $\theta=75^{\circ}$ cone for the 313 clusters used in the \LT\ relation. }
        \label{no-clusters}
\end{figure}

Finally, in Fig. \ref{no-clusters} the number of clusters within each cone with $\theta=75^{\circ}$ is shown. As expected, close to the Galactic center fewer clusters are included in the cones, while the most clusters are found for the Galactic pole cones. The main anisotropic region of our analysis at $(l,b)\sim (280^{\circ}, -15^{\circ})$ shows an average number of clusters.

\subsection{MCMC fitting}\label{mcmc_details}
In Sect. \ref{mc_sim} we discussed the results obtained by the MCMC fitting. Here we provide some more details and plots about that test. Due to the high number of free parameters, we performed $2\times 10^7$ iterations of the chain, with a burn in period of $10^4$. A variable step size was used for every parameter, randomly drawn from the same Gaussian distribution with a zero mean, and a standard deviation of $\sigma=0.05$. Every $10^4$ iterations, $\sigma=0.5$ was used a single time, to fully explore the possibility of multiple likelihood maxima within the complicated parameter space. An acceptance rate of $18\%$ was reached. The chain was run multiple times from varying initial positions and step size source distributions, to ensure that the same results were reached each time. 

\begin{figure}[hbtp]
                \includegraphics[width=0.49\textwidth, height=6cm]{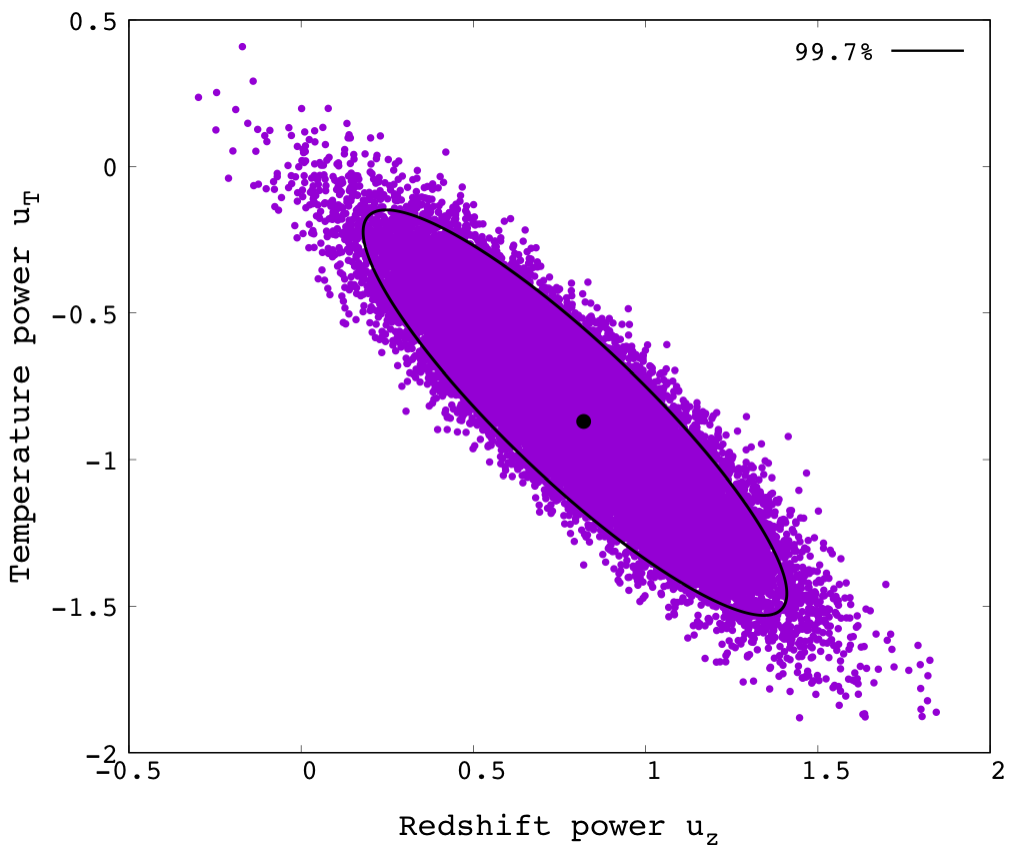}
                \includegraphics[width=0.49\textwidth, height=6cm]{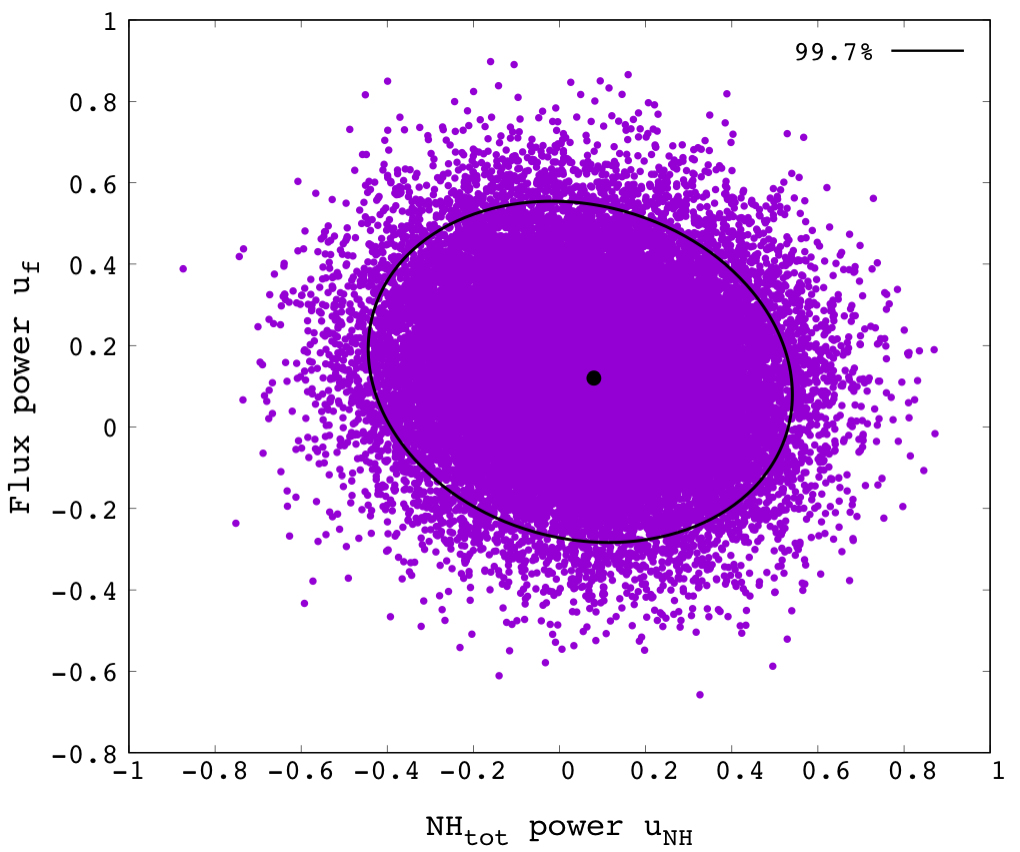}
               \caption{\textit{Top}: Distribution of the galactic latitude of the most anisotropic regions as detected in the 10000 isotropic simulated samples for the \LT\ relation. The distance weighting during the \LT\ fitting was used here. Note that bins close to the ZoA cover a larger portion of the sky, and naturally more anisotropies are expected to be detected there even if no bias existed. }
        \label{MC-tests}
\end{figure}

In Fig. \ref{MC-tests}, we plot the $3\sigma$ parameter space for $u_z$ and $u_T$ (redshift and temperature power indexes), which are the only two parameters with a significant impact on the best-fit normalization. As discussed before, their anticorrelation is strong, which generally cancels out any strong effects in the normalization (due to the flux-limited sample and the Eddington bias). The $3\sigma$ parameter space for $u_{NH}$ and $u_f$ (\nhtot\ and flux) is also plotted as a representative example for the rest of the free parameters. There is no significant effect from these cluster properties ($u_{NH}\sim u_f\sim 0$), and no correlation between them. This demonstrates that the rest of the cluster parameters do not induce any biases in the observed normalization of a cluster subsample.

\subsection{Sky fraction of Chandra and XMM-Newton clusters in our sample}

\begin{figure}[hbtp]
                \includegraphics[width=0.49\textwidth, height=4.5cm]{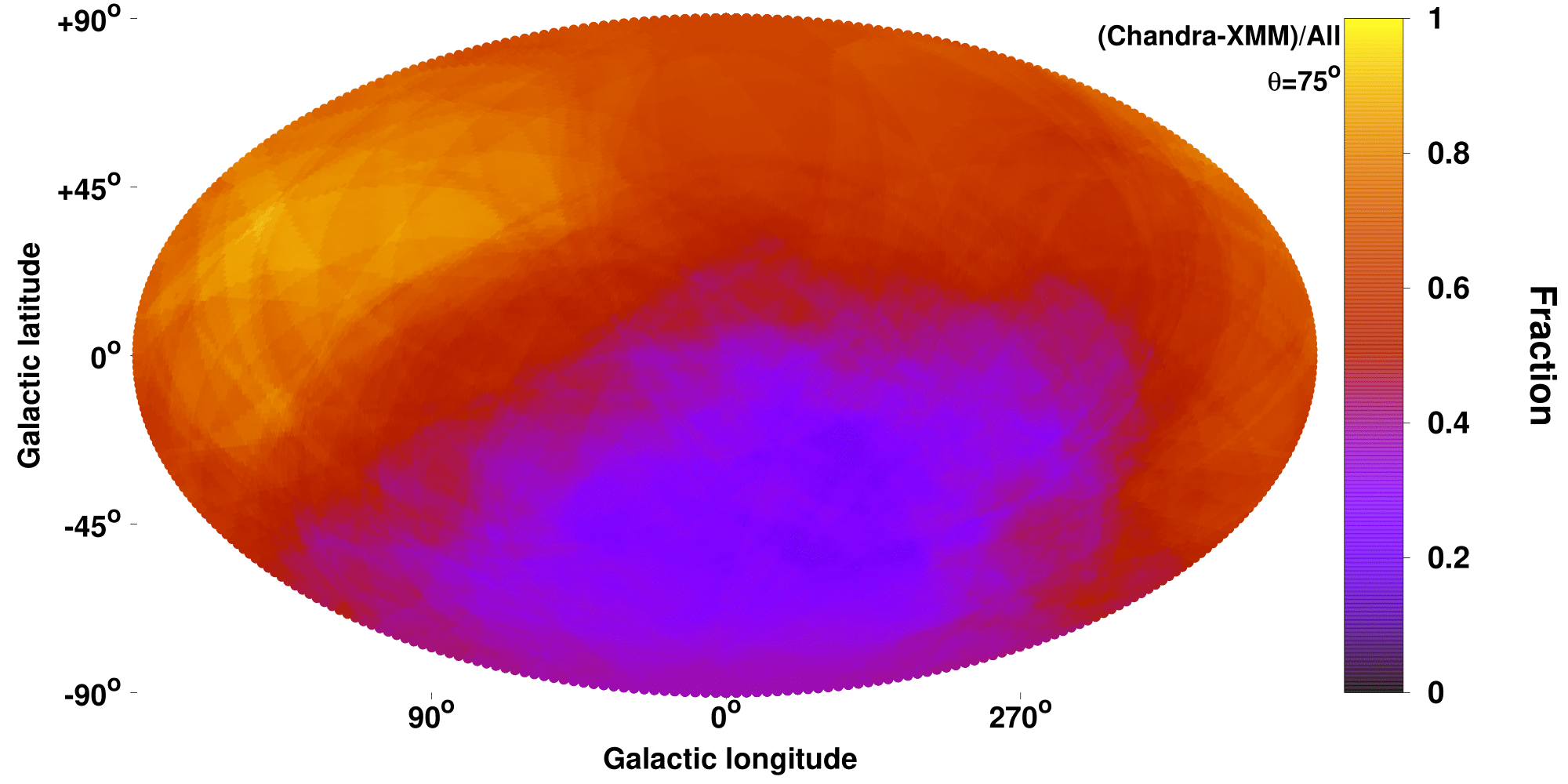}
               \caption{Spatial variation of fraction of Chandra clusters minus XMM-Newton clusters, over the sum.}
        \label{xmm-chandra}
\end{figure}

In Fig. \ref{xmm-chandra}, we display the fraction of the clusters for which Chandra or XMM-Newton were used to determine $T$. Specifically, the fraction is defined as the Chandra clusters minus the XMM-Newton clusters over the sum. The results are discussed in Sect. \ref{temp-calib}.

\subsection{Effects of bulk flows to adopted apparent $R_{500}$}
If a BF is present, then the distance of a cluster is miscalculated. As a result, its \Lx, and its mass $M_{500}$, its physical $R_{500}$ (Mpc), and its apparent ${\theta}_{500}$ (arcmin) are also misinterpreted (see \citealt{mcxc} and Sect. 2.4 of M20). Thus, the region within which we measure $T$, \Ysz, and \Lx, will also change, possibly yielding different results. This however is a negligible effect and not taken into account. In M20 (Appendix B), we showed that for $H_0$ anisotropies, ${\theta}_{500}$ remains practically unchanged. That is because the opposite changes of $R_{500}$ and $D_A$ cancel each other out. Here, we test if a BF can cause significant changes in the adopted ${\theta}_{500}$ and how this would affect our used parameter values. We combine the \Lx$\sim E(z)^{7/3} M_{500}^{1.64}$ relation of \citet{arnaud2}, with the fact that $M_{500}\sim R_{500}^3 \ H_0^2E(z)^2$, and that $D_L=(1+z)^2D_A$. Then, ${\theta}_{500}$ reads as
\begin{equation}
\begin{aligned}
{\theta}_{500}&=\dfrac{R_{500}}{D_A}\sim \frac{1}{D_A}\left(\dfrac{M_{500}}{E(z)^2}\right)^{1/3}\sim \frac{1}{D_A} \dfrac{L_{\text{X}}^{0.203}}{E(z)^{1.14}}\\
&\sim \dfrac{1}{D_A}\dfrac{D_L^{0.406}}{E(z)^{1.14}}\sim \dfrac{(1+z)^{0.812}}{D_A^{0.592}E(z)^{1.14}}.
\label{r500}
\end{aligned}
\end{equation}
We typically find BFs of $\sim 1000$ km/s. For clusters at $z=0.05$, this would cause a $\lesssim 6\%$ change in ${\theta}_{500}$. For $z=0.1$, this change would be $\lesssim 2\%$. For $T$, the percent changes are roughly the same as in ${\theta}_{500}$ (see Appendix B in M20). Therefore, for the BF case we considered, and for low-$z$ clusters, we expect a $\lesssim 6\%$ bias. As an example, let us estimate how this would affect the anisotropies of the \LT\ relation. If clusters appear less luminous (or closer) than expected due to a BF (as for the main anisotropic region of our results), this would mean that the distance is underestimated, hence ${\theta}_{500}$ is slightly overestimated. Thus, the measured $T$ is actually slightly underestimated, since it was measured in an annulus further from the center than planned, where clusters are generally cooler. If we indeed measured the correct, higher $T$, then these clusters would appear even fainter than before compared to the expectations. Therefore, the observed anisotropies would be amplified. Consequently, any small bias that is introduced to ${\theta}_{500}$ due to an existing BF, would eventually suppress the BF signal, which is the opposite of what we would need to explain the anisotropies. 

\begin{figure}[hbtp]
                \includegraphics[width=0.48\textwidth, height=6.4cm]{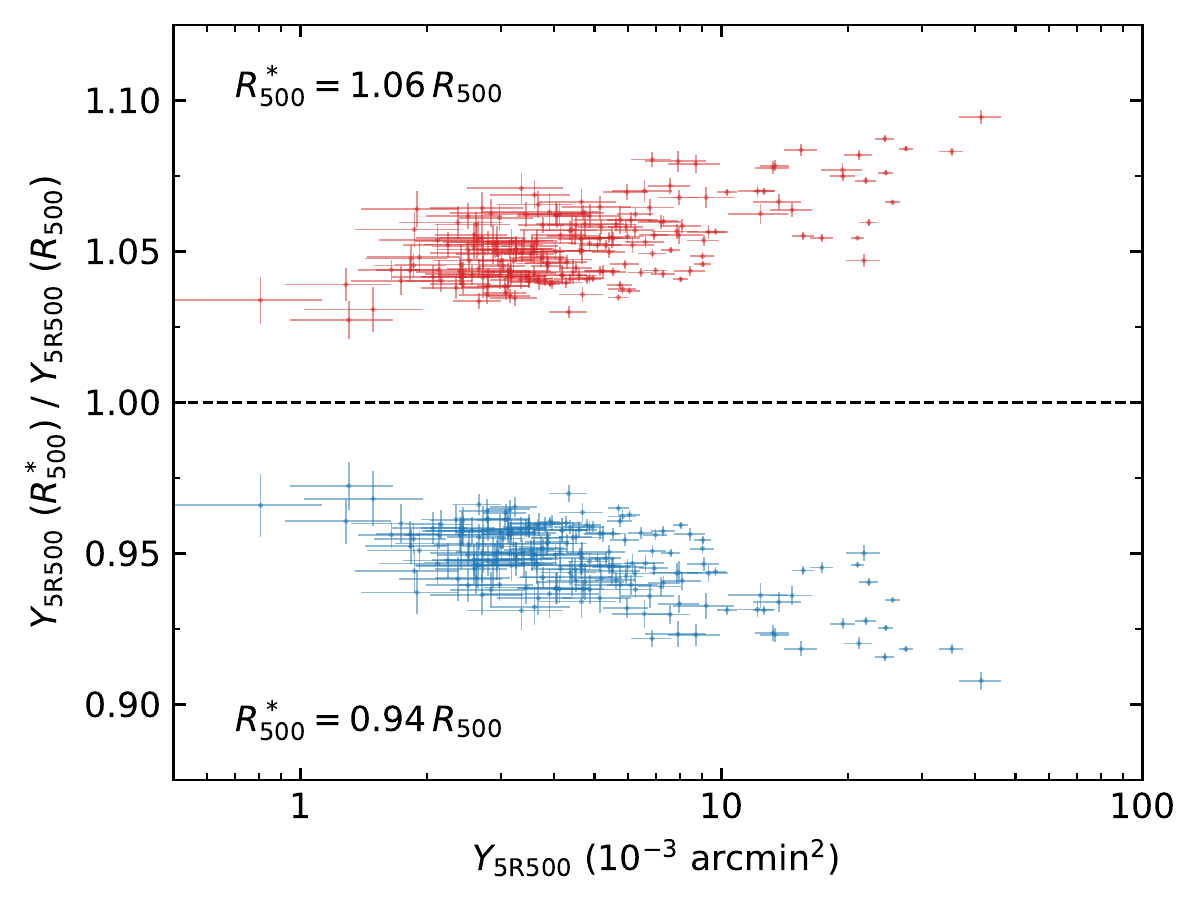}
               \caption{Relative change of the measured $Y_{5R500}$ when the input $R_{500}$ (and ${\theta}_{500}$) is increased (red) or decreased (blue) by $6\%$.}
        \label{Ysz-change}
\end{figure}

The \Ysz\ measurements also suffer the same percent changes as ${\theta}_{500}$ (Fig. \ref{Ysz-change}). One sees that for the same BF case as above, the employed MMF technique leads to a slight overestimation of \Ysz, which again suppress the BF signal. However, this effect is insignificant for two reasons. Firstly, due to the slope of the \YT\ relation ($B_{YT}=2.546$), the changes of \Ysz\ are less important than changes in $T$ (which we already saw that can mildly smooth out the anisotropies). Secondly, the \YT\ anisotropies are much larger than the \Ysz\ changes due to a falsely assumed ${\theta}_{500}$.

Moreover, the measured $T$ value goes into the measurement of \Ysz\ when relativistic effects are considered. However, this dependance is very weak, since a 1 keV change in $T$ would only lead to a $\sim 1\%$ change in \Ysz, and thus a BF would practically not affect \Ysz\ in that regard.

For the \Lx\ measurements we cannot make quantitive predictions of the change they will suffer for a different ${\theta}_{500}$ input, since we did not conduct the measurements. However, we expect that \Lx\ will remain almost unchanged, since only the outskirt area of the assumed cluster size will slightly change in case of a BF. The vast majority of X-ray cluster emission though comes from well within this area. Even if the change was not negligible, the same effect as for $T$ and \Ysz\ is expected, where the miscalculation of \Lx\ actually makes us underestimate the BF signal rather than creating it.

It is evident that these changes are only minimal compared to the observed anisotropy amplitudes, not trivial to be accounted for (iterative \Ysz\ and $T$ measurements and BF estimations would be needed), opposite than explaining the anisotropies, and thus we ignore them.

\end{document}